\documentclass[11pt,english]{article}

\usepackage[margin=1in]{geometry}
\usepackage{bbm}
\usepackage{graphicx,color}
\usepackage{amsmath,amssymb,amsthm}
\usepackage{enumerate}
\usepackage{enumitem}
\usepackage{fullpage}
\usepackage{algorithm}
\usepackage{mathrsfs}
\usepackage[noend]{algcompatible}
\usepackage[noblocks]{authblk}
\usepackage{varwidth}
\usepackage{mathrsfs}
\usepackage{babel}
\usepackage{accents}
\usepackage{mathtools}
\usepackage{soul}
\usepackage{tipa}
\usepackage[most]{tcolorbox}
\usepackage{wrapfig}
\usepackage{blindtext}
\usepackage{thm-restate}
\usepackage{longtable}
\usepackage[backref=page, pagebackref=true]{hyperref}
\usepackage[capitalize]{cleveref} 
\Crefname{enumtry}{try1}{}   
\newcommand{\rdrm}{\textsf{Word RAM~}}
\newcommand{\srt}{\mathsf{SORT}}
\newcommand{\ssa}{\mathsf{SSA}}
\newcommand{\sso}{\mathsf{SSO}}
\newcommand{\gsso}{\mathsf{GSSO}}
\newcommand{\take}{\mathsf{take}}
\newcommand{\redunt}{\mathsf{redunt}}
\newcommand{\reduce}{\mathsf{reduce}}

\DeclarePairedDelimiter{\nor2}{\lVert}{\rVert}

\newtheorem{question}{Question}
\newtheorem{goal}{Goal}

\def\denseformat{
\setlength{\textheight}{9in}
\setlength{\textwidth}{6.9in}
\setlength{\evensidemargin}{-0.2in}
\setlength{\oddsidemargin}{-0.2in}
\setlength{\headsep}{10pt}
\setlength{\topmargin}{-0.3in}
\setlength{\columnsep}{0.375in}
\setlength{\itemsep}{0pt}
}

\newtheorem{theorem}{Theorem}[section]

\newtheorem{definition}[theorem]{Definition}
\newtheorem{claim}[theorem]{Claim}
\newtheorem{lemma}[theorem]{Lemma}

\newtheorem{corollary}[theorem]{Corollary}

\newtheorem{remark}[theorem]{Remark}

\newtheorem{observation}[theorem]{Observation}

\def\boldhead#1:{\par\vskip 7pt\noindent{\bf #1:}\hskip 10pt}
\def\ithead#1:{\par\vskip 7pt\noindent{\it #1:}\hskip 10pt}

\def\inline#1:{\par\vskip 7pt\noindent{\bf #1:}\hskip 10pt}
\def\midinline#1:{\par\noindent{\bf #1:}\hskip 10pt}
\def\dnsinline#1:{\par\vskip -7pt\noindent{\bf #1:}\hskip 10pt}
\def\ddnsinline#1:{\newline{\bf #1:}\hskip 10pt}
\def\largeinline#1:{\par\vskip 7pt\noindent{\large\bf #1:}\hskip 10pt}
\long\def\commhide #1\commhideend{}
\long\def\commfull #1\commend{#1}
\long\def\commabs #1\commenda{}
\long\def\commtim #1\commendt{#1}
\long\def\commb #1\commbend{}
\long\def\commedit #1\commeditend{} 

\long\def\commB #1\commBend{}       

\long\def\commex #1\commexend{}     

\long\def\commsiena #1\commsienaend{}  

\long\def\commBI #1\commBIend{}  

\long\def\CProof #1\CQED{}
\def\qed{\mbox{}\hfill $\Box$\\}
\def\blackslug{\hbox{\hskip 1pt \vrule width 4pt height 8pt
    depth 1.5pt \hskip 1pt}}
\def\QED{\quad\blackslug\lower 8.5pt\null\par}

\long\def\PPP#1{\noindent{\bf Proof:}{ #1}{\quad\blackslug\lower 8.5pt\null}}

\long\def\denspar #1\densend
{#1}

\newif\ifnotesw\noteswtrue
   {\ifnotesw\marginpar[\hfill\(\top\)]{\(\top\)}\fi}%
   {\ifnotesw\marginpar[\hfill\(\bot\)]{\(\bot\)}\fi}

\newcommand{\mnote}[1]%
    {\ifnotesw\marginpar%
        [{\scriptsize\it\begin{minipage}[t]{\marginparwidth}
        \raggedleft#1%
                        \end{minipage}}]%
        {\scriptsize\it\begin{minipage}[t]{\marginparwidth}
        \raggedright#1%
                        \end{minipage}}%
    \fi}

\def\MathF{\hbox{\rm I\kern-2pt F}}
\def\MathP{\hbox{\rm I\kern-2pt P}}
\def\MathR{\hbox{\rm I\kern-2pt R}}
\def\MathZ{\hbox{\sf Z\kern-4pt Z}}
\def\MathN{\hbox{\rm I\kern-2pt I\kern-3.1pt N}}
\def\MathC{\hbox{\rm \kern0.7pt\raise0.8pt\hbox{\footnotesize I}
\kern-4.2pt C}}
\def\MathQ{\hbox{\rm I\kern-6pt Q}}

\newsavebox{\ttop}\newsavebox{\bbot}

\def\eps{\epsilon}
\def\epsi{\varepsilon}

\newcommand{\mst}{\mathrm{MST}}

\newcommand{\oracle}{\mathsf{Oracle}}

\newcommand{\be}{\delta}

\newcommand{\tw}{\mathsf{tw}}
\newcommand{\pathg}{\mathsf{path~greedy}}
\newcommand{\greedy}{\mathsf{greedy}}
\newcommand{\treeClustering}{\textsc{TreeClustering}}
\newcommand{\msttilde}{\widetilde{\mst}}
\newcommand{\Ftilde}{\widetilde{F}}
\newcommand{\Ttilde}{\widetilde{T}}
\newcommand{\Ptilde}{\widetilde{P}}
\newcommand{\Qtilde}{\widetilde{Q}}

\newcommand{\Fbar}{\overline{F}}
\newcommand{\Tbar}{\overline{T}}
\newcommand{\Pbar}{\overline{P}}
\newcommand{\Qbar}{\overline{Q}}
\newcommand{\Ibar}{\overline{I}}

\newcommand{\bmx}{\overline{\mathcal{X}}}
\newcommand{\uncontract}{\mathsf{uctrt}}

\newcommand{\tm}{\mathsf{Time}}
\newcommand{\source}{\mathsf{source}}
\newcommand{\lt}{\mathsf{Light}}

\newcommand{\high}{\mathsf{high}}
\newcommand{\lowp}{\mathsf{low}^+}
\newcommand{\lowm}{\mathsf{low}^-}
\newcommand{\highp}{\mathsf{high}^+}

\newcommand{\cone}{\mathsf{Cone}}
\newcommand{\prune}{\mathsf{pruned}}
\newcommand{\geom}{\mathsf{Geom}}
\newcommand{\gen}{\mathsf{Gen}}
\newcommand{\minor}{\mathsf{Minor}}
\newcommand{\internal}{\mathsf{intrnl}}
\newcommand{\prefix}{\mathsf{pref}}

\def\eps{\epsilon}

\DeclareMathAlphabet{\mathpzc}{OT1}{pzc}{m}{it}
\newcommand{\light}{\mathsf{light}}

\newcommand{\dm}{\mathsf{Dm}}
\newcommand{\adm}{\mathsf{Adm}}

\newcommand {\ignore} [1] {}

\denseformat
\newcommand{\doverline}[1]{\widehat{#1}}

\newcommand{\ma}{\mathcal{A}}
\newcommand{\mb}{\mathcal{B}}
\newcommand{\mc}{\mathcal{C}}
\newcommand{\md}{\mathcal{D}}
\newcommand{\mh}{\mathcal{H}}

\newcommand{\mg}{\mathcal{G}}
\newcommand{\mv}{\mathcal{V}}
\newcommand{\me}{\mathcal{E}}
\newcommand{\ms}{\mathcal{S}}
\renewcommand{\mp}{\mathcal{P}}

\newcommand{\mk}{\mathcal{K}}
\newcommand{\mt}{\mathcal{T}}
\newcommand{\mx}{\mathcal{X}}
\newcommand{\my}{\mathcal{Y}}
\newcommand{\mz}{\mathcal{Z}}

\newcommand{\mi}{\mathcal{I}}

\newcommand{\wsp}{\mathtt{Ws}}
\newcommand{\ssp}{\mathtt{Ss}}

\newcommand{\mbe}{\mathbf{e}}

\newcommand{\real}{\mathbb{R}}
\DeclareMathOperator{\MST}{\mathrm{MST}}

\DeclareMathOperator{\defi}{\overset{\mathrm{def.}}{=} }
\DeclareMathOperator{\pr}{\mathtt{Pr}}

\newcommand{\bnu}{\bar\nu}
\newcommand{\bmu}{\bar\mu}

\newcommand{\dblind}{the authors~}

\newcommand{\arc}[1]{{%
		\setbox9=\hbox{#1}%
		\ooalign{\resizebox{\wd9}{\height}{\texttoptiebar{\phantom{A}}}\cr#1}}}

\date{}

\title{A Unified Framework for Light Spanners}
\author{Hung Le}
\affil{University of Massachusetts Amherst}
\author{Shay Solomon}
\affil{Tel Aviv University}

\begin{document}
\pagenumbering{gobble}
\maketitle
\begin{abstract}
Seminal works on {\em light} spanners over the years provide spanners with optimal {\em lightness} in various graph classes,\footnote{The {\em lightness} is a normalized notion of weight: a graph's lightness is the ratio of its weight to the MST weight.} 
such as in general graphs~\cite{CW16}, Euclidean spanners \cite{das1994fast} and minor-free graphs~\cite{BLW17}.
Three shortcomings of previous works on light spanners are: (i) The runtimes of these constructions are almost always sub-optimal, and usually far from optimal.
(ii) These constructions are optimal in the standard and crude sense, but not in a refined sense that takes into account a wider range of involved parameters.
(iii) {\em The techniques are {\bf ad hoc} per graph class, and thus can't be applied broadly}.

This work aims at addressing these shortcomings by presenting  a {\em unified framework} of light spanners in a variety of graph classes. Informally, the framework boils down to a {\em transformation} from sparse spanners to light spanners; since the state-of-the-art for sparse spanners is much more advanced than that for light spanners, such a transformation is powerful.  First, we apply our framework to design {\em fast} constructions with {\em optimal lightness} for several graph classes.
Among various applications, we highlight the following (for simplicity assume $\eps> 0$ is fixed):
\begin{itemize}
\item
In \emph{low-dimensional Euclidean spaces}, we present an $O(n\log n)$-time construction  of $(1+\eps)$-spanners with lightness and degree both bounded by constants in the {\em algebraic computation tree (ACT)} (or {\em real-RAM}) model, which is the basic model used in Computational Geometry. 
The previous state-of-the-art runtime in this model for constant lightness (even for unbounded degree) was $O(n \log^2 n / \log\log n)$,
whereas $O(n \log n)$-time spanner constructions with constant degree (and  $O(n)$ edges) are known for years.
Our construction is optimal with respect to all the involved quality measures --- runtime, lightness and degree --- and it resolves a major problem in the area of geometric spanners, which was open for three decades (cf.\ \cite{CDNS92,ADMSS95,GLN02,NS07}).  
\end{itemize}

Second, we apply our framework to achieve {\em more refined} optimality  bounds for several graph classes, i.e., the bounds remain optimal when taking into account a \emph{wider range of involved parameters},
	most notably $\eps$. Our new constructions are significantly better than the state-of-the-art {\em for every examined graph class}. 
Among various applications, we highlight the following (now $\eps > 0$ is any parameter):
\begin{itemize}
\item For \emph{ $K_r$-minor-free graphs}, we provide a  $(1+\epsilon)$-spanner with lightness $\tilde{O}_{r,\epsilon}( \frac{r}{\epsilon} + \frac{1}{\epsilon^2})$,
where $\tilde{O}_{r,\epsilon}$ suppresses $\mathsf{polylog}$ factors of $1/\epsilon$ and $r$,
improving the lightness bound $\tilde{O}_{r,\epsilon}( \frac{r}{\epsilon^3})$ of Borradaile, Le and Wulff-Nilsen~\cite{BLW17}.
We complement our upper bound with a highly nontrivial lower bound construction, for which any $(1+\epsilon)$-spanner must have lightness $\Omega(\frac{r}{\epsilon} + \frac{1}{\epsilon^2})$.
Interestingly, our lower bound is realized by a geometric graph in $\mathbb{R}^2$.
Also, the quadratic dependency on $1/\eps$ that we prove is surprising,
as   prior work suggested that the dependency on $\eps$ should be around $1/\eps$. 
\end{itemize}

\end{abstract}

\pagebreak

\tableofcontents
\pagebreak
\pagenumbering{arabic}

\clearpage

\section{Introduction}\label{sec:intro}

For a weighted graph $G = (V,E,w)$ and a {\em stretch parameter} $t \ge 1$, a subgraph $H = (V,E')$ of $G$
is called a \emph{$t$-spanner} if $d_H(u,v) \le t \cdot d_G(u,v)$, for every $e = (u,v) \in E$,
where $d_G(u,v)$ and $d_H(u,v)$ are the distances between $u$ and $v$ in $G$ and $H$, respectively.
Graph spanners were introduced in two celebrated papers from 1989 \cite{PS89,PU89} for unweighted graphs,
where it is shown that for any $n$-vertex graph $G = (V,E)$ and integer $k \ge 1$, there is an $O(k)$-spanner with $O(n^{1+ 1/k})$ edges.
We shall sometimes use a normalized notion of size, {\em sparsity}, which is the ratio of the size of the spanner to the size of a spanning tree, namely $n-1$.
Since then, graph spanners have been extensively studied, both for general weighted graphs and for restricted graph families,
such as Euclidean spaces and minor-free graphs.  
In fact, spanners for Euclidean spaces---{\em Euclidean spanners}---were studied implicitly already in the pioneering SoCG'86 paper of Chew~\cite{Chew86}, who showed that any 2-dimensional Euclidean space admits a spanner of $O(n)$ edges and stretch $\sqrt{10}$, and later improved the stretch to 2~\cite{Chew89}.

As with the sparsity parameter, its weighted variant---lightness---has been extremely well-studied; the \emph{lightness} is the ratio of the weight of the spanner to $w(MST(G))$. 
Seminal works on {\em light} spanners over the years provide spanners with optimal {\em lightness} in various graph classes, such as in general graphs~\cite{CW16}, Euclidean spanners \cite{das1994fast} and minor-free graphs~\cite{BLW17}.
{\bf Despite the large body of work on light spanners,  the stretch-lightness tradeoff is not nearly as well-understood as the stretch-sparsity tradeoff}, and the intuitive reason behind that is clear: Lightness seems inherently more challenging to optimize than sparsity, since different edges may contribute disproportionately to the overall lightness due to differences in their weights.  The three shortcomings of light spanners that emerge, when considering the large body of work in this area, are: 
(i) The runtimes of these constructions are usually far from optimal.
(ii) These constructions are optimal in the standard and crude sense, but not in a refined sense that takes into account a wider range of involved parameters,
	most notably $\eps$, but also other parameters, such as the dimension (in Euclidean spaces) or the minor size (in minor-free graphs).
(iii) {\bf The techniques are ad hoc per graph class, and thus can't be applied broadly} 
(e.g., some require large stretch and are thus suitable to general graphs, while others are naturally suitable to stretch $1 + \eps$). 

In this work, we are set out to address these shortcomings by presenting  a {\em unified framework} of light spanners in a variety of graph classes. Informally, the framework boils down to a {\em transformation} from sparse spanners to light spanners; since the state-of-the-art for sparse spanners is much more advanced than that for light spanners, such a transformation is powerful.

Our ultimate goal is to bridge the gap in the understanding between light spanners and sparse spanners. 
This gap is prominent when considering (i) the construction time of light versus sparse spanners, and (ii) a {\em fine-grained} optimality of the lightness. In terms of (ii), the state-of-the-art spanner constructions for general graphs, 
	as well as for most restricted graph families, incur a (multiplicative) $(1+\eps)$-factor slack on the stretch with a {\em suboptimal} dependence on $\eps$ as well as other parameters in the lightness bound.   In this work, we present new spanner constructions, all of which are derived as applications and implications from a unified framework developed in this paper. 
	
\begin{itemize}
\item In terms of (i), i.e., runtime, our constructions are significantly faster than the state-of-the-art {for every examined graph class}; moreover, our runtimes are near-linear or linear and usually optimal. Our main result in this context is an $O(n\log n)$ time algorithm in the ACT model for constructing a Euclidean spanner with constant lightness and degree.
\item In terms of (ii), i.e., fine-grained optimality,  our constructions are significantly better than the state-of-the-art for every examined graph class; our main result in this context is for minor-free graphs, where we achieve tight dependencies on both $\eps$ and the minor size -- the upper bound follows as an application of the unified framework, and the lower bound is obtained by different means.
\end{itemize}

We now highlight three completely different yet well-studied settings to which our framework applies.

\paragraph{Fast construction of Euclidean spanners in the algebraic computation tree (ACT) model.~} 
Spanners have had special success in geometric settings, especially in low-dimensional Euclidean spaces. 
The reason Euclidean spanners have been extensively studied over the years --- in both theory and practice --- is that one can achieve stretch arbitrarily close to 1. The {\em algebraic computation tree (ACT)} introduced by Ben-Or~\cite{BenOr83}  model is used extensively in computational geometry, and in the area of Euclidean spanners in particular; it is intimately related to the real RAM model. (The reader can refer to \cite{BenOr83} and Chapter 3 in the book \cite{NS07} for a detailed description of ACT model.) Computing $(1+\epsilon)$-spanners for point sets in $\mathbb{R}^d$, $d = O(1)$, in ACT model requires $\Omega(n\log n)$ time~\cite{CDS01,FP16}. Despite a large body of work on light Euclidean spanners~\cite{LL89,CDNS92,DHN93,das1994fast,DNS95,das1996constructing,ADMSS95,RS98,GLN02,NS07,elkin2015optimal,LS19} since the late 80s, the following problem has been open for nearly three decades:

\begin{question} [Question 22 in \cite{NS07}] \label{ques:algebraic2}
	Can one   construct a Euclidean $(1+\epsilon)$-spanner with constant lightness and degree (and thus constant sparsity)
	in optimal time $O(n \log n)$ in the ACT model, for any fixed $\eps < 1$? 
\end{question}
While \Cref{ques:algebraic2} asks for both constant lightness and degree, it is even not known how to achieve constant lightness only in $O(n\log n)$ time in the ACT model. The best-known algorithm has running time  $O(n\frac{\log^2n}{\log \log n})$~\cite{GLN02}. If one assumes \emph{indirect addressing}, then there is an algorithm with running time $O(n\log n)$~\cite{GLN02}. But indirect addressing is a very strong operation: the lower bound of $\Omega(n \log n)$ in ACT model for Euclidean spanners mentioned earlier no longer applies.  Some applications of light spanners~\cite{RS98,czumaj2000fast} require that they can be computed in $O(n\log n)$ time. Euclidean spanners of bounded degree have applications in designing routing schemes (see, e.g., ~\cite{CGMZ16,DBLP:conf/soda/GottliebR08,DBLP:conf/cocoon/BrankovicGR20}), and more generally, the degree of the spanner determines the local memory constraints when using spanners also for other purposes, such as constructing network synchronizers and efficient broadcast protocols. 

\paragraph{Fast construction of general weighted graphs.~} Alth\"{o}fer et al~\cite{ADDJS93} shown that for every $n$-vertex \emph{weighted} graph $G = (V,E,w)$ and integer $k \ge 1$, there is a {\em greedy} algorithm for constructing a $(2k-1)$-spanner with $O(n^{1+1/k})$ edges, which is optimal under Erd\H{o}s' girth conjecture. Moreover, there is an $O(m)$-time algorithm for constructing $(2k-1)$-spanners in unweighted graphs with sparsity $O(n^{\frac{1}{k}})$~\cite{HZ96}.
Therefore, not only is the stretch-sparsity tradeoff in general graphs optimal (up to Erd\H{o}s' girth conjecture), one can achieve it in optimal time. For weighted graphs, one can construct $(2k-1)$-spanners with sparsity $O(k n^{\frac{1}{k}})$ within time $O(k m)$~\cite{baswana2007simple,roditty2005deterministic}. On the other hand, the best running time for achieving lightness bound $O(n^{1/k})$ for stretch $(2k-1)(1+\eps)$ for a \emph{fixed $\eps$} is super-quadratic in $n$: $O(n^{2 + 1/k + \epsilon'})$~\cite{ADFSW19} for any fixed constant $\epsilon' < 1$. Other faster constructions have a worst dependency on $n$ and $k$~\cite{ENS14}.

\begin{question}\label{ques:gen}
	Can one construct a $(2k-1)\cdot(1+\eps)$-spanner in general weighted graphs with
	lightness $O(n^{1/k})$, within (nearly) linear time for any fixed $\eps < 1$?
\end{question}

\paragraph{Fine-grained lightness bound for minor-free graphs.~} The gap between sparsity and lightness is prominent in minor-free graphs, for stretch $1+\eps$.
Indeed, minor-free graphs are sparse to begin with, and the sparsity is trivially $\tilde \Theta(r)$.  On the other hand, for lightness, bounds are much more interesting.
Borradaile, Le, and Wulff-Nilsen~\cite{BLW17} showed that the greedy $(1+\eps)$-spanners of $K_r$-minor-free graphs have lightness $\tilde{O}_{r,\epsilon}(\frac{r}{\epsilon^3})$, where the notation $\tilde{O}_{r,\epsilon}(.)$ hides polylog factors of $r$ and $\frac{1}{\epsilon}$. Moreover, this is the state-of-the-art lightness bound also in some sub-classes of minor-free graphs, particularly bounded treewidth graphs. Past works provided strong evidence that the dependence of lightness on $1/\epsilon$ of $(1+\epsilon)$-spanners
should be \emph{linear}: $O(\frac{1}{\epsilon})$ in planar graphs by Alth\"{o}fer et al.~\cite{ADDJS93}, 
$O(\frac{g}{\epsilon})$ in bounded genus graphs by Grigni~\cite{Grigni00}, and $\tilde{O}_{r}(\frac{r \log n}{\epsilon})$ in $K_r$-minor-free graphs by Grigni and Sissokho~\cite{GS02}. 

\begin{question} \label{ques:minor-light}
	Is there a $(1+\eps)$-spanner of lightness $O(1/\eps)$ for $K_r$-minor-free graphs for a fixed $r$?
\end{question}

\paragraph{Other open problems.~} There are several settings where the \emph{fast constructions} of light spanners for \emph{fixed $\eps$} remains open.

\begin{itemize}
	\item \textbf{Unit disk graph.~}  There is a significant gap between the fastest constructions of sparse versus light spanners in UDGs.   F{\"u}rer and Kasiviswanathan~\cite{FK06} showed that {\em sparse} $(1+\epsilon)$-spanners for UDGs can be built in nearly linear time when $d = 2$, and in subquadratic time when $d$ is a constant of value at least $3$. However , no $o(n^2)$-time $(1+\eps)$-spanner construction for UDGs with a nontrivial lightness bound is known, even for $d = 2$. Can we construct a light $(1+\eps)$-spanner for UDGs in $O(n\log n)$ time for $d = 2$, and in truly subquadratic time for general $d$?
	\item  \textbf{Minor-free graphs.~} The fastest algorithm for constructing light spanners in $K_r$-minor-free graphs is greedy~\cite{ADDJS93} with quadratic running time $\tilde{O}_r(n^2 r^2)$. Can we construct a light $(1+\eps)$-spanner  for $K_r$-minor-free graphs in nearly linear time?
\end{itemize}

For \emph{fine-grained lightness bounds}, there are two additional settings where the lightness bounds are not well-understood. 

\begin{itemize}
	\item \textbf{General graphs.~}  While the  stretch-sparsity tradeoff for spanners of general graphs is resolved up to the girth conjecture, the stretch-lightness tradeoff, on the other hand, is still far from being resolved. A long line of research~\cite{ADDJS93,CDNS92,ENS14,CW16,FS16} over the past three decades leads to a $(2k-1)\cdot(1+\eps)$-spanner with lightness $O(n^{1/k} (1/\eps)^{3+2/k})$~\cite{CW16,FS16}. While the dependence on $n$  and $k$ are optimal assuming Erd\H{o}s' girth conjecture, the  dependence on $1/\epsilon$ is  super-cubic. Can we reduce the dependency of the lightness on $\eps$ to linear? 
	\item \textbf{Euclidean Steiner spanners in low dimensional spaces.~} Le and Solomon~\cite{LS19} studied {\em Steiner spanners}, namely, spanners that are allowed to use 
	{\em Steiner points}, which are additional points that are not part of the input point set.
	It was shown there that Steiner points can be used to improve the sparsity quadratically, i.e., to $O(\epsilon^{\frac{-d+1}{2}})$,
	which was shown to be tight for dimension $d = 2$ in \cite{LS19}, and for any $d = O(1)$ by Bhore and T\'{o}th~\cite{BT21B}. An important question left open in~\cite{LS19} is whether one could use Steiner points to improve the lightness bound \emph{quadratically} to  $O(\epsilon^{-d/2})$ for any dimension $d$. Previous results either have a dependency on the \emph{spread} of the metric~\cite{LS20} which could be huge, or only work for $d=2$~\cite{BT20}.
\end{itemize}

\subsection{Research Agenda: From Sparse to Light Spanners}

Thus far we exemplified the statement that the stretch-lightness tradeoff is not as well-understood as the stretch-sparsity tradeoff. 
As we showed, this lack of understanding is prominent when considering (i) the \emph{construction time}, and (ii) fine-grained dependencies. This statement is not to underestimate in any way the exciting line of work on light spanners,
	but rather to call for attention to the important research agenda of narrowing this gap and ideally closing it.

All questions regarding fast constructions of light spanners ask the same thing: Can one achieve {\em fast constructions} of {\em light} spanners that {match} the analogous results for {\em sparse} spanners? 

\begin{goal} \label{g1}
Achieve {\em fast constructions} of light spanners that {\em match} the corresponding constructions of sparse spanners. 
In particular, achieve (nearly) linear-time constructions of spanners with optimal lightness for basic graph families, such as the ones covered in the aforementioned questions. 
\end{goal}
 A fine-grained optimization of the stretch-lightness tradeoff, which takes into account the exact dependencies on $\eps$ and the other involved parameters, is a highly challenging goal.
 
\begin{goal} \label{g2}
Achieve fine-grained optimality for light spanners in basic graph families. 
\end{goal}

Some of the papers on light spanners employ inherently different techniques than others, e.g., the technique of \cite{CW16} requires large stretch while others are naturally suitable to stretch $1+\eps$.
Since the techniques in this area are ad hoc per graph class, they can't be applied broadly.
A unified framework for light spanners would be of both theoretical and practical merit.
\begin{goal} \label{g3}
Achieve a unified framework of light spanners.
\end{goal}

\begin{comment}
Establishing a thorough understanding of light spanners by meeting (some of) the above goals is not only of theoretical interest, but is also of practical importance, due to the wide applicability of spanners.  Perhaps the most prominent applications of light spanners are to efficient broadcast protocols in the message-passing model of distributed computing \cite{ABP90,ABP91},
to network synchronization and computing global functions \cite{Awerbuch85,PU89,ABP90,ABP91,Peleg00}, and to the TSP \cite{Klein05,Klein06,RS98,GLN02,BLW17,Gottlieb15}.
There are many more applications, such as to data gathering and dissemination tasks in overlay networks \cite{BKRCV02,VWFME03,KV01}, 
to VLSI circuit design \cite{CKRSW91,CKRSW292,CKRSW92,SCRS01},
to wireless and sensor networks \cite{RW04,BDS04,SS10}, 
to routing \cite{WCT02,PU89,PU89b,TZ01}, 
to compute almost shortest paths \cite{Cohen98,RZ11,Elkin05,EZ06,FKMSZ05},
and to computing distance oracles and labels \cite{Peleg00Prox,TZ01b,RTZ05}. 
\end{comment}

\subsection{Our Contribution}\label{subsec:contribution}

Our work aims at meeting the above goals (\Cref{g1}---\Cref{g3}) by presenting a unified framework for optimal and fast constructions of light spanners in a variety of graph classes.
Basically, we strive to translate results --- in a unified manner --- from sparse spanners to light spanners, without significant loss in any of the parameters.  Our paper achieves \Cref{g1} and \Cref{g3} or achieves \Cref{g2} and \Cref{g3};
achieving all three goals simultaneously is left open by our work.

We also answer almost all the aforementioned open problems, either positively or negatively. In particular, we answer \Cref{ques:algebraic2} and \Cref{ques:gen} positively, and \Cref{ques:minor-light} negatively.  For other open problems, we completely resolve them in the affirmative.   

Two of our results are particularly surprising. First, we show that the optimal lightness bound of $(1+\eps)$ for $K_r$-minor-free graphs is $\tilde{\Theta}_{r,\eps}(\frac{r}{\eps} + \frac{1}{\eps^2})$ (\Cref{thm:minor-free-opt-lightness}). That is, the lightness dependency on $1/\eps$ is quadratic, despite ample evidence~\cite{ADDJS93,Grigni00,GS02} of a linear dependency on $1/\eps$ in subclasses of minor-free graphs. 
In particular, our result negatively settles \Cref{ques:minor-light}.  
Second, we construct light spanners in general graphs with near-optimal lightness in $O(m\alpha(m,n))$ time (\Cref{thm:general-fast}); our algorithm is significantly faster than the best algorithms for sparse spanners with the same sparsity bound.

\paragraph{Fast constructions.~} We present a spanner construction that achieves constant lightness and degree, within  optimal time of $O(n \log n)$ in   ACT model; this
 proves the following theorem, which affirmatively resolves~\Cref{ques:algebraic2} which was open for   three decades.

\begin{theorem}\label{thm:ACT} For any set $P$ of $n$ points in $R^d$, any $d = O(1)$ and any  fixed $\eps >0$, one can construct in the ACT model   a $(1+\epsilon)$-spanner for $P$ with constant degree and lightness  within optimal time $O(n \log n)$.
\end{theorem}

For general graphs we provide a nearly linear-time spanner construction with optimal lightness, assuming Erd\H{o}s' girth conjecture (and up to the $\eps$-dependency), thus answering~\Cref{ques:gen}.

\begin{theorem}\label{thm:general-fast}
	For any edge-weighted graph $G(V,E)$, a stretch parameter $k \geq 2$ and an arbitrary small fixed $\epsilon < 1$, there is a deterministic algorithm that constructs a $(2k-1)(1+\epsilon)$-spanner of $G$ with lightness $O(n^{1/k})$ in $O(m\alpha(m,n))$ time, where $\alpha(\cdot,\cdot)$ is the inverse-Ackermann function. 
\end{theorem}

We remark that $\alpha(m,n) = O(1)$ when $m = \Omega(n\log^{*}n)$. Thus, the running time in \Cref{thm:general-fast} is linear in $m$ in almost the entire regime of graph densities, i.e., except for very sparse graphs. The previous state-of-the-art runtime for the same lightness bound is super-quadratic~\cite{ADFSW19}. Surprisingly, the result of~\Cref{thm:general-fast} outperforms the analog result for sparse spanners in weighted graphs: for stretch $2k-1$, the only spanner construction with sparsity $O(n^{1/k})$ is the greedy spanner, whose runtime is $O(mn^{1+\frac{1}{k}})$. Other results~\cite{ADFSW19,EN18} with stretch $(2k-1)(1+\eps)$ have (nearly) linear running time, but the sparsity is $O(n^{1/k}\log(k))$, which is worse than our lightness bound by a factor of $\log(k)$. 

\paragraph{Subsequent work.~}  In a subsequent and consequent follow-up to this work, \dblind~\cite{LS22} used our framework here to present a fast construction of spanners with {near-optimal} {\em sparsity} and \emph{lightness} for general graphs \cite{LS22}. We also adapted and simplified our construction here to construct a \emph{sparse} spanner (with unbounded lightness) in $O(m\alpha(m,n) + \srt(m))$ time in the pointer-machine model, where $\srt(m)$ is the time to sort $m$ integers. Even in a stronger \rdrm model, the best known algorithm for sorting $m$ integers takes $O(m\sqrt{\log \log m})$~\cite{HT02} expected time. Thus, the running time of the sparse spanner algorithm is still inferior to our running time in \Cref{thm:general-fast}. In the  \rdrm model, a linear time algorithm for constructing a sparse spanner was presented; we do not consider this model in our work here.

Our framework also resolves two open problems regarding fast constructions of light spanners in two different settings. In particular, we get an $O(n\log n)$ time algorithm for UDGs; the running time is optimal in the ACT model. For minor-free graphs, we get the first linear time algorithm, which significantly improves  over the best known algorithm for this problem. 

\begin{theorem}\label{thm:UDG-fast} For any set $P$ of $n$  points in $R^d$,  any $d = O(1)$ and any  fixed $\eps >0$, one can construct a $(1+\epsilon)$-spanner of the UDG for $P$ with constant sparsity and lightness. 
	For $d = 2$, the construction runtime  is  $O(n \log n)$ in the ACT model; for $d = 3$, the runtime is $\tilde{O}(n^{4/3})$; and for $d\geq 4$, the runtime is $O(n^{2-\frac{2}{(\lceil d/2 \rceil+1)} + \delta})$ for any constant $\delta > 0$. 
\end{theorem}
\begin{theorem}\label{thm:minor-free-fast}
	For any $K_r$-minor-free graph $G$  and any fixed $\eps >0$, one can construct a $(1+\epsilon)$-spanner of $G$ with lightness  $O(r\sqrt{\log r})$ in $O(nr\sqrt{\log r})$ time.  
\end{theorem}

\paragraph{Fine-grained lightness bounds.~} The most important implication of our framework   in terms of fine-grained lightness bounds is to minor-free graphs,
where we obtain a tight dependence on $\eps$ in the lightness.

\begin{restatable}{theorem}{MinorFree}
	\label{thm:minor-free-opt-lightness}
	Any $K_r$-minor-free graph admits a $(1+\epsilon)$-spanner with lightness $\tilde{O}_{r,\epsilon}(\frac{r}{\epsilon} + \frac{1}{\epsilon^2})$ for any $\epsilon< 1$ and $r\geq 3$. \\
	Furthermore, for any fixed $r\geq 6$, any  $\epsilon < 1$  and $n \geq r + (\frac{1}{\epsilon})^{\Theta(1/\epsilon)}$, there is an $n$-vertex graph $G$ excluding $K_r$ as a minor for which any $(1+\eps)$-spanner must have lightness $\Omega(\frac{r}{\epsilon} + \frac{1}{\epsilon^2})$.
\end{restatable}

The $\tilde{O}_{\eps,r}(.)$ notation in \Cref{thm:minor-free-opt-lightness} hides a poly-logarithmic factor of $1/\eps$ and $r$.   \Cref{thm:minor-free-opt-lightness} resolves \Cref{ques:minor-light} negatively. We remark that, in \Cref{thm:minor-free-opt-lightness}, the exponential dependence on $1/\epsilon$ in the lower bound on $n$ is unavoidable since, if $n = \mathrm{poly}(1/\epsilon)$, the result of ~\cite{GS02} yields a lightness of
$\tilde{O}_r(\frac{r}{\epsilon}\log(n)) = \tilde{O}_{r,\epsilon}(\frac{r}{\epsilon})$.

Interestingly, our lower bound  applies to a geometric graph, where the vertices correspond to points in $\mathbb R^2$ and the edge weights are the Euclidean distances between the points. The construction is recursive. We start with a basic gadget and then recursively ``stick'' many copies of the same basic gadgets in a fractal-like structure. We use geometric considerations to show that any $(1+\eps)$-spanner must take every edge of this graph, whose total edge weight is $\Omega(1/\eps^2)w(\mst)$. 

A prominent application of light spanners for $K_r$-minor-free graphs is to the Traveling Salesperson Problem (TSP). \Cref{thm:minor-free-opt-lightness} implies a PTAS (polynomial time approximation scheme) with approximation $1+\eps$ and running time $2^{1/\eps^3}n^{O(1)}$, improving upon the algorithm by Borradaile, Le, and Wulff-Nilsen~\cite{BLW17} with running time $2^{O(1/\eps^4)}n^{O(1)}$.  Our lower bound of \Cref{thm:minor-free-opt-lightness} implies that to further improve the runtime for TSP {\em one has to significantly deviate from the standard technique~\cite{DHK11} that relies on light spanners}.

Using our framework, we obtain near-optimal lightness bounds in two different settings: general graphs (\Cref{thm:light-general-spanner}) and Steiner Euclidean spanners (\Cref{thm:light-Steiner}). Both results  resolve two open problems mentioned above.

\begin{theorem}\label{thm:light-general-spanner}
	Given an edge-weighted graph $G(V,E)$ and two parameters $k \geq 1, \epsilon < 1$, there is a $(2k-1)(1+\epsilon)$-spanner of $G$ with lightness $O(g(n,k)/{\epsilon})$ where $g(n,k)$ is the minimum sparsity of $n$-vertex graphs with girth $2k+1$. As $g(n,k) = O(n^{1/k})$, the lightness is  $O(n^{1/k}/{\epsilon})$. 
\end{theorem}
The Erd\H{o}s' girth conjecture implies that $g(n,k) = \Omega(n^{1/k})$. While the conjecture is very commonly used in the computer science community as evidence for spanners' optimality, the combinatorics community is quite skeptical about it~\cite{Bodwin22,DS91,CFSZ21,HL04}; in particular, a bipartite version of the conjecture was refuted~\cite{DS91,CFSZ21}. Consequently, the fact that our \Cref{thm:light-general-spanner} gives a near-optimal lightness bound that does not rely on Erd\H{o}s' girth conjecture is a significant advantage. We are not aware of any prior work showing the existence of a near optimal spanner without the Erd\H{o}s' girth conjecture.

\begin{theorem}\label{thm:light-Steiner}
	For any $n$-point set $P \in \mathbb{R}^d$ and any $d \ge 3$, $d = O(1)$, there is a Steiner $(1+\epsilon)$-spanner for $P$ with lightness 
	$\tilde{O}(\epsilon^{-(d+1)/2})$ that is constructable in polynomial time.
\end{theorem}

We also obtain improved lightness bounds for light spanners in high dimensional Euclidean spaces. The literature on spanners in high-dimensional Euclidean spaces is surprisingly sparse. Har-Peled, Indyk and Sidiropoulos~\cite{HIS13} showed that for any set of $n$-point Euclidean space (in any dimension) 
and any parameter $t \geq 2$, there is an $O(t)$-spanner with sparsity $O(n^{1/t^2} \cdot (\log n \log t))$.  Filtser and Neiman~\cite{FN18} gave an analogous but weaker result for lightness, achieving a lightness bound of $O(t^3 n^{\frac{1}{t^2}}\log n)$. They also generalized their results to any $\ell_p$ metric, for $p \in (1,2]$, achieving a lightness bound of $O(\frac{t^{1+p}}{\log^2 t}n^{\frac{\log^2 t}{t^p}}\log n)$.  Our results improve all of these results.

\begin{theorem}\label{thm:Euclidean-high} \begin{itemize}
		\item For any $n$-point set $P$ in a Euclidean space and any given $t \ge 2$, there is an $O(t)$-spanner for $P$ with lightness 
		$O(tn^{\frac{1}{t^2}}\log n)$ that is constructible in polynomial time.
		\item  For any $n$-point $\ell_p$ normed space $(X,d_X)$ with $p \in (1,2]$ and any $t \ge 2$, there is an $O(t)$-spanner for $(X,d_X)$ with lightness $O(t n^{\frac{\log^2 t}{t^p}}\log n)$.
	\end{itemize}
\end{theorem}

\subsection{Our Unified Framework: Technical and Conceptual Highlights}\label{subsec:framework-intro}
In this section, we give a high-level overview of our framework for constructing light spanners with stretch $t(1+\eps)$, for some parameter $t$ that depends on the examined graph class; e.g., for Euclidean spaces $t  = 1+\eps$, while for general graphs $t = 2k-1$. We shall construct spanners with stretch $t(1+O(\eps))$ and   assume w.l.o.g.\ that $\eps$ is sufficiently smaller than $1$; a stretch of $t(1+\eps)$, for any $0 \le \eps \le 1$,   can be achieved by scaling.

Let $L$ be a positive parameter, and let $H_{< L}$ be a $t(1+\gamma\eps)$-spanner for all edges in $G = (V,E,w)$ 
of weight $< L$, for some constant $\gamma \geq 1$.  That is, $V(H_{< L}) = V$ and for any edge $(u,v)\in E$ with $w(u,v)<  L$:
\begin{equation}\label{eq:Stretch-HL}
	d_{H_{< L}}(u,v) \leq t(1+\gamma \eps)w(u,v) .
\end{equation}

Note that by the triangle inequality, $H_{< L}$ is also a $t(1+\gamma\eps)$-spanner for every pair of vertices of distance $< L$. Our framework relies on the notion of a {\em cluster graph}, defined as follows.

\begin{definition}[$(L,\eps,\beta,\Upsilon)$-Cluster Graph]\label{def:ClusterGraph-Param} An edge-weighted graph $\mg= (\mv,\me,\omega)$ is called an \emph{$(L,\eps,\beta)$-cluster graph} with respect to spanner $H_{< L}$, for positive parameters $L, \eps, \beta, \Upsilon > 1$, if it satisfies the following conditions:
	\begin{enumerate}
		\item Each node $\varphi_C \in \mv$ corresponds to a subset of vertices $C \in V$, called a \emph{cluster},
		 in the original graph $G$. For any pair $\varphi_{C_1}, \varphi_{C_2}$ of distinct nodes 
		in $\mv$, we have $C_1\cap C_2 = \emptyset$.
		\item Each edge $(\varphi_{C_1},\varphi_{C_2})\in \me$ corresponds to an edge $(u,v)\in E$, such that $u \in C_1$ and $v\in C_2$. Furthermore, $\omega(\varphi_{C_1},\varphi_{C_2}) = w(u,v)$.
		\item $L \leq \omega(\varphi_{C_1},\varphi_{C_2}) < \Upsilon L$, for every edge $(\varphi_{C_1},\varphi_{C_2})\in \me$.
		\item $\dm(H_{< L}[C]) \leq \beta \eps L$, for any cluster $C$ corresponding to a node $\varphi_C \in \mv$.  
	\end{enumerate} 
Here $\dm(X)$ denotes the {\em diameter} of a graph $X$, i.e., the maximum pairwise distance in $X$. 
\end{definition} 

Condition (1) asserts that clusters corresponding to nodes of $\mg$ are vertex-disjoint. Condition (3) asserts that edges in $\me$ have the same weight up to a factor of $\Upsilon$, which is always at most $2$ in our construction..  Furthermore, Condition (4) asserts that they induce subgraphs of low diameter in $H_{< L}$. In particular, if $\beta$ is constant, then the diameter of clusters is roughly $\eps$ times the weight of edges in the cluster graph. That is, the diameter of the clusters is much smaller than the weight of the edges when $\eps$ is sufficiently small. 

In our framework, we use the cluster graph to compute a subset of edges in $G$ of weights in $[L,\Upsilon L)$ to add to the spanner $H_{< L}$, so as to obtain a spanner, denoted by $H_{<\Upsilon L}$, for all edges in $G$ of weight less than  $\Upsilon L$. As a result, we extend the set of edges whose endpoints' distances are preserved (to within the required stretch bound) by the spanner.  By repeating the same construction for edges of higher and higher weights, we eventually obtain a spanner that preserves all pairwise distances in $G$. 

There are two values that $\Upsilon$ can take, depending on whether we wish to optimize the running time or the fine-grained dependence on $\eps$ and other parameters such as the size of the excluded minor.  In the former case we set $\Upsilon = 1+\eps$, whereas in the latter we set $\Upsilon = 2$.

Note that a single edge of $\mg$ may correspond to multiple edges of $G$;
to facilitate the transformation of edges of $\mg$ to edges of $G$,  we assume access to a function $\source(\cdot)$ that supports the following operations in $O(1)$ time: (a) given a node $\varphi_C$, $\source(\varphi_C)$ returns a vertex $r(C)$ in cluster $C$, called the {\em representative} of $C$, (b) given an edge $(\varphi_{C_1}, \varphi_{C_2})$ in $\me$,  $\source(\varphi_{C_1}, \varphi_{C_2})$ returns the corresponding edge $(u,v)$ of  $(\varphi_{C_1}, \varphi_{C_2})$, which we refer to as the {\em source edge} of $(u,v)$, where $u \in C_1$ and $v\in C_2$; we note that $u$ (resp., $v$) need not be $r(C_1)$ (resp., $r(C_2)$). 
Constructing the function $\source(\cdot)$ efficiently is straightforward; the details are in \Cref{sec:FastProof}.

For {\em optimizing the construction time}, our framework assumes the existence of the following algorithm, hereafter the \emph{sparse spanner algorithm ($\ssa$)}, which computes a subset of edges   in $\mg$, whose source edges are added to $H_{< L}$. 
Recall that the parameter $\Upsilon$ is set as $\Upsilon = 1+\eps$ in this case.
\begin{tcolorbox}
	\hypertarget{SPSSA}{}
	$\ssa$: Given an $(L,\eps,\beta, \Upsilon = 1+\eps)$-cluster graph $\mg(\mv,\me,\omega)$ and function $\source(\cdot)$ as defined above, 
		the $\ssa$ outputs a subset of edges  $\me^{\prune}\subseteq \me$ such that: 
	\begin{enumerate}[noitemsep]
	 	\item \textbf{(Sparsity)~} \hypertarget{SparsitySSA}{}  $|\me^{\prune}| \leq \chi |\mv|$ for some $\chi > 0$. 
		\item \textbf{(Stretch)~} \hypertarget{StretchSSA}{} For each edge $(\varphi_{C_u},\varphi_{C_v})\in \me$, $d_{H_{<(1+\eps)L}}(u,v)\leq t(1+s_{\ssa}(\beta)\eps)w(u,v)$  where $(u,v) = 
		\source(\varphi_{C_u}, \varphi_{C_v})$ and $s_{\ssa}(\beta)$ is some constant that depends on $\beta$  only, and $H_{< (1+\eps)L}$  is the graph obtained by adding the source edges of $\me^{\prune}$ to $H_{< L}$. 
	\end{enumerate}
	Let  $\tm_{\ssa} = O((m'+n')\tau(m',n'))$ be the running time of the $\ssa$, where $\tau$ is a monotone non-decreasing function,   $n' = |\mv|$ and $m' = |\me|$.
\end{tcolorbox}

Intuitively, the $\ssa$ can be viewed as an algorithm that constructs a \emph{sparse spanner for an unweighted graph}, as edges of $\mg$ have the same weights up to a factor of $(1+\eps)$ and the only requirement from the edge set $\me^{\prune}$ returned by the $\ssa$, besides achieving small stretch, is that it would be of small size.   While the interface to the $\ssa$ remains the same across all graphs,  its exact implementation may change from one graph class to another;  
informally, for each graph class, the $\ssa$ is akin to the state-of-the-art unweighted spanner construction for that class, and this part of the framework is pretty simple. 

For {\em optimizing the fine-grained dependence on $\eps$ and other parameters (such as minor size) in the lightness bound}, our framework assumes the existence of the following algorithm, called \emph{sparse spanner oracle ($\sso$)}, which computes a subset of edges in $G$ to add to $H_{< L}$. Recall that the parameter $\Upsilon$ is set as $\Upsilon = 2$ in this case.

\begin{tcolorbox}
	\hypertarget{SPSSO}{}
	\textbf{$\sso$:} Given an $(L,\eps,\beta, \Upsilon = 2)$-cluster graph $\mg(\mv,\me,\omega)$, 
	the $\sso$ outputs a  subset of edges  $F\subseteq E$  in polynomial time such that: 
	\begin{enumerate}[noitemsep]
		\item \textbf{(Sparsity)~} \hypertarget{SparsitySSO}{} $w(F) \leq \chi|\mv| L$  for some $\chi> 0$. 
		\item \textbf{(Stretch)~} \hypertarget{StretchSSO}{} For each edge $(\varphi_{C_u},\varphi_{C_v})\in \me$, $d_{H_{<2L}}(u,v)\leq t(1+s_{\sso}(\beta)\eps)w(u,v)$ 		where $(u,v)$ is the corresponding edge of $(\varphi_{C_u}, \varphi_{C_v})$ and $s_{\sso}(\beta)$ is some constant that depends on $\beta$  only, and $H_{< 2L}$  is the graph obtained by adding $F$ to $H_{< L}$. 
	\end{enumerate}	
	
\end{tcolorbox}

We can interpret the $\sso$ as a construction of a {\em sparse spanner} in the following way: If $F$ contains only edges of $G$ corresponding to a subset of $\me$, say $\me^{\prune} \subseteq \me$, then, $w(e)\geq L$ for every $e \in F$; in this case $|F| \leq \chi|\mv|$.   The edges in the set $F$  produced by the $\sso$ may not correspond to edges in $\me$ of $\mg$. This allows for more flexibility in choosing the set of edges to add to $H_{< L}$, and is the key to obtaining a fine-grained optimal dependencies on $\eps$ and the other parameters, such as the Euclidean dimension or the minor size. Importantly, for all classes of graphs considered in this paper, the implementation of $\sso$ is very simple, as we show in \Cref{sec:applications-light}.

The highly nontrivial part of the framework is given by the following theorem, which provides  a {\em black-box transformation} (i)  from an $\ssa$ to an efficient (in terms of running time) {\em meta-algorithm} for constructing light spanners and (ii) from an $\sso$ to an efficient (in terms of fine-grained dependencies) {\em meta-algorithm} for constructing light spanners.  We note that this transformation remains the same across all graphs.

\begin{restatable}{theorem}{Framework}
	\label{lm:framework} Let $L,\eps, t, \gamma, \beta$ be non-negative parameters where $\gamma, \beta \geq 1$ only take on constant values, and $\eps \ll 1$. 
	Let $\mathcal{F}$ be an arbitrary graph class.
	If, for any graph $G$ in $\mathcal{F}$:
	
	\begin{enumerate}
		\item[(1)] the $\ssa$ can take any $(L,\eps,\beta, \Upsilon=1+\eps)$-cluster graph $\mg(\mv,\me,\omega)$ corresponding to $G$  as input and return as output a subset of edges $\me^{\prune}\subseteq \me$ satisfying the aforementioned two properties of (\hyperlink{SparsitySSA}{Sparsity}) and (\hyperlink{StretchSSA}{Stretch}),
		then for any graph in $\mathcal{F}$ we can construct a spanner with stretch $t(1+(s_{\ssa}(O(1))+O(1))\eps)$, lightness $O((\chi \eps^{-3} + \eps^{-4})\log(1/\eps))$, and in time $O(m\eps^{-1}(\alpha(m,n) + \tau(m,n) + \eps^{-1})\log(1/\eps))$.  
		
		\item[(2)] the $\sso$ can take any $(L,\eps,\beta, \Upsilon=2)$-cluster graph $\mg(\mv,\me,\omega)$ corresponding to $G$  as input and return as output a subset of edges $F$ of $G$ satisfying the aforementioned two properties of (\hyperlink{SparsitySSO}{Sparsity}) and (\hyperlink{StretchSSO}{Stretch}),
		then for any graph in $\mathcal{F}$ we can construct a spanner with stretch $t(1+(2s_{\sso}(O(1))+O(1))\eps)$, lightness $\tilde{O}_{\eps}((\chi\eps^{-1} + \eps^{-2}))$ when $t = 1+\eps$, and lightness  $\tilde{O}_{\eps}((\chi\eps^{-1}))$ when $t\geq 2$. 
	\end{enumerate}
\end{restatable}

See \Cref{fig:unifedFull} for an illustration of how \Cref{lm:framework} is used to derive various results in our paper.  We remark the following regarding \Cref{lm:framework}.

\begin{remark}\label{remark:ACTIntro}	If the \hyperlink{SPSSA}{$\ssa$} can be implemented in the ACT model with the stated running time, then the construction of light spanners provided by \Cref{lm:framework} can also be implemented in the ACT model in the stated running time.
\end{remark}

\begin{figure}[!htb]
	\centering
	\input{figs/Unified.tex}
	\caption{Applications of our framework in obtaining fast constructions of light spanners (on the left) and spanners with truly optimal lightness (on the right). The notion of \emph{general sparse spanner oracle} ($\gsso$) is another abstraction that we will formally introduce in \Cref{subsec:oracle-intro}.}
	\label{fig:unifedFull}
\end{figure}

In the implementations of \hyperlink{SPSSA}{$\ssa$} for Euclidean spaces and UDGs, we rely on the condition that $H_{< L}$ preserves distances smaller than $L$ within a factor of $t(1+\gamma \eps)$. However, we do not need this condition to hold for general graphs and minor-free graphs; for them all we need is Condition 4 in \Cref{def:ClusterGraph-Param}.

For fast constructions, the transformation provided by \Cref{lm:framework} --- from sparsity in almost unweighted graphs (as captured by the $\ssa$)  to lightness --- has a constant loss on lightness (for constant $\eps$) and a small running time overhead.  In \Cref{sec:applications-fast}, we provide simple implementations of the $\ssa$ for several classes of graphs in time $O(m+n)$, for a constant $\eps$; \Cref{lm:framework} thus directly yields a running time of $O((m+n)\alpha(m,n))$. For minor-free graphs, with an additional effort, we remove the factor $\alpha(m,n)$ from the runtime. For Euclidean spaces and UDGs, we apply the transformation not on the input space but rather on a sparse spanner, with $O(n)$ edges, hence the runtime $O((m+n)\alpha(m,n))$ of the transformation is not the bottleneck, as it is dominated by the time $\Theta(n \log n)$ needed for building Euclidean spanners.

For obtaining fine-grained lightness bounds,  the transformation from sparsity to lightness in \Cref{lm:framework} only looses a factor of $1/\eps$ for stretch $t\geq 2$, and, in addition, another additive term of $+\frac{1}{\eps^2}$ is lost for stretch $t =  1+\eps$. 
Later, we complement this upper bound by  a lower bound (\Cref{sec:lowerbounds}) showing that for $t = 1+\eps$, the additive  term  of $+\frac{1}{\eps^2}$ is unavoidable in the following sense: There is a graph class --- the class of bounded treewidth graphs --- where we can implement an $\sso$ with $\chi = O(1)$ for stretch $(1+\eps)$, and hence the lightness of the transformed spanner is $O(1/\eps^2)$ due to the additive term of $+\frac{1}{\eps^2}$, but any light $(1+\eps)$-spanner for this class of graphs must have lightness $\Omega(1/\eps^2)$. (We modify this construction to obtain the lower bound for minor-free graphs in \Cref{thm:minor-free-opt-lightness}.)

Despite the clean conceptual message behind \Cref{lm:framework} --- in providing a transformation from sparse to light spanners --- its proof is technical and highly intricate. This should not be surprising as our goal is to have a single framework that can be applied to basically any graph class. The applicability of our framework goes far beyond the specific graph classes considered in the current paper, which merely aim at capturing several very different conceptual and technical hurdles, e.g.,  complete vs.\ non-complete graphs, geometric vs.\ non-geometric graphs, stretch $1+\eps$ vs.\ large stretch, etc. The heart of our framework is captured by \Cref{lm:framework}, whose proof appears in \Cref{partII}. The starting point of our proof of \Cref{lm:framework}  is a basic hierarchical partition, which dates back to the early 90s \cite{ALGP89,CDNS92}, and was used by most if not all of the works on light spanners (see, e.g.,~\cite{ES16,ENS14,CW16,BLW17,BLW19,LS19}).
The current paper takes this hierarchical partition approach to the next level, by proposing  a unified framework that reduces the problem of \emph{efficiently} constructing a \emph{light} spanner to the conjunction of two problems: (1) efficiently constructing a hierarchy of clusters with several \emph{carefully chosen properties}, and (2) efficiently constructing a {\em sparse spanner};
these two problems are intimately related, in the sense 
that the ``carefully chosen properties'' of the clusters are set so that we can efficiently apply the sparse spanner construction.

To minimize the dependency on $\eps$ in the transformation in \Cref{lm:framework}, we construct clusters in such a way that (1) a cluster at a higher level should contain as many clusters as possible, called subclusters, at lower levels, and (2) the augmented diameter of the cluster must be within a restricted bound. Condition (1) implies that each cluster has a large potential change, which is used to ``pay'' for spanner edges that the algorithm adds to the spanner, while condition (2) implies that the constructed spanner has the desired stretch. The two conditions are in conflict with each other, since the more subclusters we have in a single cluster, the larger the diameter of the cluster gets. Achieving the right balance between these two conflicting conditions is the main technical contribution of this paper.

Another significant technical contribution of our paper in this context is in introducing the notion of \emph{augmented diameter} of a cluster. The definition of augmented diameter appears in Section~\ref{sec:prelim}, but at a high level, the idea is to consider weights on both nodes and edges in a cluster, where the node weights are determined by the {\em potential values} of clusters computed (via simple recursion) in previous levels of the hierarchy.
The main advantage of augmented diameter over the standard notion of diameter is that it can be computed efficiently, while the computation of diameter is much more costly. Informally, the augmented diameter can be computed efficiently since (i) we can upper bound the hop-diameter of clusters, and (ii) the clusters at each level are computed on top of some underlying tree;
roughly speaking, that means that all the distance computations are carried out on top of subtrees of bounded hop-diameter (or depth), hence the source of efficiency.

{\bf We next argue that our approach is inherently different than previous ones}. First, the very fact that our approach is unified makes it inherently different than previous approaches, which, as mentioned, are {ad hoc} per graph class. Second, our approach is not just a unified framework for reproving known results --- we employ it to break through the state-of-the-art. To this end, we highlight one concrete result --- on Euclidean spanners in the ACT model --- which breaks a longstanding barrier in the area of geometric spanners, by using an inherently {\em non-geometric} approach. All the previous algorithms for light Euclidean spanners were achieved via the greedy and approximate-greedy spanner constructions. The greedy algorithm is non-geometric but slow, whereas the approximate-greedy algorithm is geometric and can be implemented much more efficiently.
The analysis of the lightness in both algorithms is done via the so-called {\em leapfrog property} \cite{DHN93,das1994fast,DNS95,das1996constructing,GLN02,NS07}, which is a geometric property. The fast spanner construction of GLN \cite{GLN02} implements the approximate-greedy algorithm by constructing a hierarchy of clusters with $O(\frac{\log n }{\log \log n})$ levels and, for each level, Dijkstra's algorithm is used for the construction of clusters for the next level. The GLN construction incurs an additional $O(n\log n)$ factor for each level to run  Dijkstra's algorithm in the ACT model, which ultimately leads to a runtime of $O( n \frac{\log^2 n}{\log \log n})$. Our approach is inherently different, and in particular we do not need to run	Dijkstra's algorithm or any other single-source shortest (or approximately shortest) path algorithm. The key to our efficiency is in a careful usage of the new notion of augmented diameter, as well as its interplay with the potential function argument and the hierarchical partition that we use. We stress again that our approach is non-geometric, and the only potential usage of geometry is in the sparse spanner construction that we apply. (Indeed, the sparse spanner construction that we chose to apply is geometric, but this is not a must.)

\section{Preliminaries}\label{sec:prelim}

Let $G$ be an arbitrary weighted graph. We denote by $V(G)$ and $E(G)$ the vertex set and edge set of $G$, respectively. We denote by $w: E(G)\rightarrow \mathbb{R}^+$ the weight function on the edge set.  Sometimes we write $G = (V,E)$ to clearly explicate the vertex set and edge set of $G$, and  $G =(V,E,w)$ to further indicate the weight function $w$ associated with $G$. 
We use $\mst(G)$ to denote a minimum spanning tree of $G$; when the graph is clear from context, we simply use $\mst$  as a shorthand for $\mst(G)$.

For a subgraph $H$ of $G$, we use $w(H) \defi \sum_{e\in E(H)}w(e)$ to denote the total edge weight of $H$.   The {\em distance} between two vertices $p,q$ in $G$, denoted by $d_G(p,q)$, is the minimum weight of a path between them in $G$. The diameter of $G$,
denoted by $\dm(G)$, is the maximum pairwise distance in $G$. A \emph{diameter path} of $G$ is a shortest (i.e., of minimum weight) path in $G$ realizing the diameter of $G$, that is, 
it is a shortest path between some pair $u,v$ of vertices in $G$ such that $\dm(G) = d_G(u,v)$.

Sometimes we shall consider graphs with weights on both \emph{edges and vertices}. We define the \emph{augmented weight} of a path to be the total weight of all edges and vertices along the path. The \emph{augmented distance} between two vertices in $G$ is defined as the minimum augmented weight of a path between them in $G$. 
Likewise, the \emph{augmented diameter} of $G$, denoted by $\adm(G)$,
 is the maximum pairwise augmented distance in $G$; 
since we will focus on non-negative weights, the augmented distance and augmented diameter
are no smaller than the (ordinary notions of) distance and diameter. 
An \emph{augmented diameter path} of $G$ is a path of minimum augmented weight realizing the augmented diameter of $G$.

Given a subset of vertices $X\subseteq V(G)$, we denote by $G[X]$ the subgraph of $G$ \emph{induced by $X$}: $G[X]$ has $V(G[X]) = X$ and $E(G[X]) = \{(u,v)\in E(G) ~\vert~   u,v \in X\}$. Let $F\subseteq E(G)$ be a subset of edges of $G$; we denote by $G[F]$ the subgraph of $G$ with $V(G[F]) = V(G)$ and $E(G[F]) = F$.

Let $S$ be a \emph{spanning} subgraph of $G$; weights of edges in $S$ are inherited from $G$. The \emph{stretch} of $S$  is  given by $\max_{x,y \in V(G)} \frac{d_S(x,y)}{d_G(x,y)}$, and it is realized by some edge $e$ of $G$. We say that $S$ is a \emph{$t$-spanner} of $G$ if the stretch of $S$ is at most $t$. There is a simple greedy algorithm,  called $\pathg$ (or shortly $\greedy$), to find a $t$-spanner of a graph $G$: Examine the edges $e = (x,y)$ in  $G$ in nondecreasing order of weights, and add to the spanner edge $(x,y)$ iff the distance between $x$ and $y$ in the {\em current} spanner is larger than $t\cdot w(x,y)$.
 
  We say that a subgraph $H$ of $G$ is a $t$-spanner for a \emph{subset of edges $X\subseteq E$} if $\max_{(u,v) \in X} \frac{d_H(u,v)}{d_G(u,v)} \leq t$.

In the context of minor-free graphs, we denote by $G/e$ the graph obtained from $G$ by contracting $e$, where $e$ is an edge in $G$. If $G$ has weights on edges, then every edge in $G/e$ inherits its weight from $G$.

In addition to general and minor-free graphs, this paper studies {\em geometric graphs}.  Let $P$ be a set of $n$ points in $\mathbb{R}^d$. We denote by $\nor2{p,q}$ the Euclidean distance between two points $p,q\in \mathbb{R}^d$.   A  \emph{geometric graph} $G$ for $P$ is a graph where the vertex set corresponds to the point set, i.e., $V(G) = P$, and the edge weights are the Euclidean distances, i.e.,  $w(u,v) = \nor2{u,v}$ for every edge $(u,v)$ in $G$. Note that $G$ need not be a complete graph. If $G$ is a complete graph, i.e., $G = (P, {P \choose 2},\nor2{\cdot})$, then $G$ is equivalent to the {\em Euclidean space} induced by the point set $P$.
For geometric graphs, we use the term \emph{vertex} and \emph{point} interchangeably.

We use $[n]$ and $[0,n]$ to denote the sets $\{1,2,\ldots,n\}$ and $\{0,1,\ldots,n\}$, respectively.

\section{Lightness Lower Bounds}\label{sec:lowerbounds}

In this section, we provide lower bounds on light $(1+\epsilon)$ spanners to prove the lower bound in  \Cref{thm:minor-free-opt-lightness}.  Interestingly, our lower bound construction draws a connection between geometry and graph spanners: we construct a fractal-like geometric graph of weight $\Omega(\frac{\mst}{\epsilon^2})$ such that it has treewidth at most $4$ and any $(1+\epsilon)$-spanner of the graph must take all the edges.

\begin{theorem}\label{thm:treewdith}
	For any $n = \Omega(\epsilon^{\Theta(1/\epsilon)})$ and $\epsilon < 1$, there is an $n$-vertex graph $G$ of treewidth at most $4$ such that any light $(1+\epsilon)$-spanner of $G$ must have lightness $\Omega(\frac{1}{\epsilon^2})$.
\end{theorem}

\noindent Before proving \Cref{thm:treewdith}, we show its implications  to  the lower bound in \Cref{thm:minor-free-opt-lightness}.

\begin{proof}[Proof of the lower bound in \Cref{thm:minor-free-opt-lightness}]
	First, construct a complete graph $H_1$ on $r-1$ vertices for which any $(1+\eps)$-spanner has lightness $\Omega(\frac{r}{\epsilon})$ as follows: Let $X_1\subseteq V(H_1)$ be a subset of $r/2$ vertices and $X_2 = V(H_1)\setminus X_1$. We assign weight $2\epsilon$ to every edge with both endpoints in $X_1$ or $X_2$, and weight $1$ to every edge between $X_1$ and $X_2$. Clearly $\mst(H_1)  = 1 + (r-2)2\epsilon$. We claim that any $(1+\epsilon)$-spanner $S_1$ of $H_1$ must take every edge between $X_1$ and $X_2$; otherwise, if $e = (u,v)$ is not taken where $u\in X_1,v\in X_2$, then $d_{S_1}(u,v) \geq d_{H_1\setminus e}(u,v)=1+2\epsilon > (1+\epsilon)d_G(u,v)$. Thus, $w(S_1)\geq |X_1||X_2| = \Omega(r^2)$. This implies $w(S_1) = \Omega(\frac{r}{\epsilon})w(\mst(H_1))$.

	Let $H_2$ be an $(n-r+1)$ vertex graph of treewidth 4 guaranteed by \Cref{thm:treewdith}; $H_2$ excludes $K_r$ as a minor for any $r\geq 6$. We scale edge weights of $H_1$ appropriately so that $w(\mst(H_2)) = w(\mst(H_1))$.  Connect $H_1$ and $H_2$ by a single edge of weight $2w(\mst(H_1))$ to form a graph $G$. Then $G$ excludes $K_r$ as minor (for $r\geq 5$) since $H_1$ and $H_2$ both exclude $K_r$ as a minor. Furthermore, any $(1+\epsilon)$-spanner of $G$ must have lightness at least $\Omega(\frac{r}{\epsilon} + \frac{1}{\epsilon^2})$ as $w(\mst(G)) = 4 w(\mst(H_1))$.\qed
\end{proof}

\begin{figure}
	\centering
	\vspace{-20pt}
	\includegraphics[scale = 0.8]{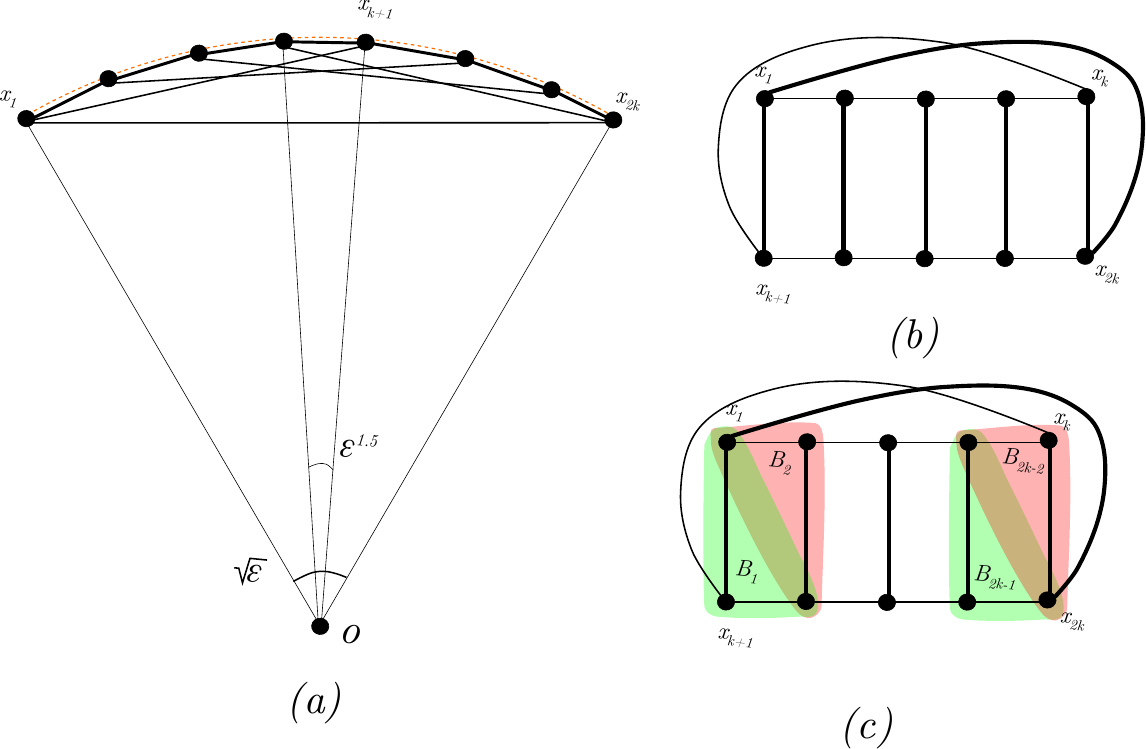}
	\caption{\footnotesize{(a) The core gadget. (b) A different view of the core gadget. (c) A tree decomposition of the core gadget. }}
	\label{fig:core}
\end{figure}

We now focus on proving \Cref{thm:treewdith}. The core gadget in our construction is depicted in Figure~\ref{fig:core}.  Let $C_r$ be a circle on the plane centered at a point $o$ of radius $r$. We use $\arc{ab}$ to denote an arc of $C_r$ with two endpoints $a$ and $b$. We say \emph{$\arc{ab}$ has angle $\theta$} if $\angle aob = \theta$.We use $|\arc{ab}|$ to denote the (arc) length of $\arc{ab}$, and $||a,b||$ to denote the Euclidean length between $a$ and $b$.

By elementary geometry and Taylor's expansion, one can verify that if $\arc{ab}$ has angle $\theta$, then:

\begin{equation}\label{eq:arc-chord-length}
\begin{split}
|\arc{ab}| &= \theta r \\
||a,b|| &= 2r \sin(\theta/2) = r\theta(1-\theta^2/24 + o(\theta^3)) \\
||a,b|| &= \frac{2\sin(\theta/2)}{\theta} |\arc{ab}| = (1-\theta^2/24 + o(\theta^3))|\arc{ab}|
\end{split}
\end{equation}

\paragraph{Core Gadget.~}  The construction starts with an arc $ab$ of angle $\sqrt{\epsilon}$ of a circle $C_r$.~W.l.o.g., we assume that $\frac{1}{\epsilon}$ is an odd integer.  Let $k = \frac{1}{2}(\frac{1}{\epsilon}+1)$.  Let $\{a \equiv x_1, x_2, \ldots, x_{2k} \equiv b\}$ be the set of points, called \emph{break points}, on the arc $ab$ such that $\angle x_iox_{i+1} = \epsilon^{3/2}$ for any $1\leq i \leq 2k-1$.

Let $H_r$ be a graph with vertex set $V(H_r) = \{x_1,\ldots, x_{2k}\}$. We call $x_1$ and $x_{2k}$ two \emph{terminals} of $H_r$. For each $i \in [2k-1]$, we add an edge $x_ix_{i+1}$ of weight $w(x_ix_{i+1}) = ||x_i,x_{i+1}||$ to $E(H_r)$. We refer to edges between $x_ix_{i+1}$ for $i \in [2k-1]$ as \emph{short edges}.  For each $i \in [k]$, we add an edge $x_ix_{i+k}$ of weight $||x_{i},x_{i+k}||$.  We refer to these edges as \emph{long edges}. Finally, we add edge $||x_1,x_k||$ of $E(H_r)$, that we refer to as the \emph{terminal edge} of $H_r$.   We call $H_r$ a \emph{core gadget} of scale $r$.  See Figure~\ref{fig:core}(a) for a geometric visualization of $H_r$ and Figure~\ref{fig:core}(b) for an alternative view of $H_r$.

\noindent We observe that:

\begin{observation}\label{obs:edge-length} $H_r$ has the following properties:
	\begin{enumerate}
		\item  For any edge $e \in E(H_r)$, we have:
		\begin{equation}
		w(e) = \begin{cases} 2r\sin(\epsilon^{3/2}/2) &\text{if $e$ is a short edge}\\
		2r\sin(k\epsilon^{3/2}/2) &\text{if $e$ is a long edge}\\ 2r\sin(\sqrt{\epsilon}/2) &\text{if $e$ is the terminal edge}
		\end{cases}
		\end{equation}
		\item $w(\mst(H_r)) \leq r\sqrt{\epsilon}$.
		\item $w(H_r) \geq \frac{r}{6\sqrt{\epsilon}}$ when $\epsilon \ll 1$.
	\end{enumerate}
	\begin{proof}
		We only verify (3); other properties can be seen by direct calculation. By Taylor's expansion, each long edge of $H_r$ has weight $w(e) = 2\sin(\frac{1}{4}(\sqrt{\epsilon} +\epsilon^{3/2})) =  \frac{r}{2}(\sqrt{\epsilon} + o(\epsilon)) \geq r\sqrt{\epsilon}/3$ when $\epsilon \ll 1$. Since $H_r$ has $k$ long edges, $w(H_r) \geq k r \sqrt{\epsilon}/3 \geq  \frac{r}{6\sqrt{\epsilon}}$.\qed
	\end{proof}
\end{observation}

\noindent Next, we claim that $H_r$ has small treewidth.

\begin{claim} \label{clm:treewidth} $H_r$ has treewidth at most $4$.
\end{claim}
\begin{proof}
	We construct a tree decomposition of width $4$ of $H_r$. In fact, we can construct a path decomposition of width $4$ for $H_r$. Let $B_1,\ldots, B_{2k-2}$ be set of vertices where $B_{2i-1}= \{x_{2i-1},x_{2i+k-1}, x_{2i+k}\}$  and $B_{2i}= \{x_{2i-1}, x_{2i+k}, x_{2i}\}$  for each $i \in [k-1]$ (see Figure~\ref{fig:core}(c)). We then add $x_1$ and $x_{k}$ to every $B_i$. Then, $\mathcal{P} = \{B_1,\ldots, B_{2k-2}\}$ is a path decomposition of $H_r$ of width $4$. \qed
\end{proof}
	\begin{wrapfigure}{r}{0.4\textwidth}
	\vspace{-30pt}
	\begin{center}
		\includegraphics[width=0.4\textwidth]{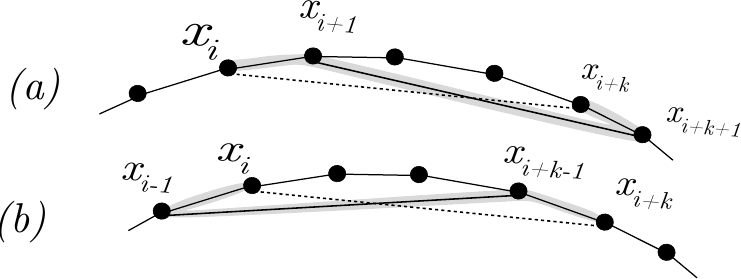}
	\end{center}
	\caption{\footnotesize{Paths $P_e$ between $x_i$ and $x_{i+k}$ are highlighted.}}
	\vspace{-5pt}
	\label{fig:long-edge}
\end{wrapfigure}

\noindent \textbf{Remark:} It can be seen that $H_r$ has $K_4$ as a minor, thus has treewidth at least $3$. Showing that $H_r$ has treewidth at least $4$ needs more work.

\begin{lemma}\label{lm:spanner}
	There is a constant $c$ such that any $(1+\epsilon/c)$-spanner of $H_r$ must have weight at least $$ \frac{w(\mst(H_r))}{6\epsilon}.$$
\end{lemma}
\begin{proof}
	Let $e$ be a long edge of $H_r$ and $G_e = H_r\setminus \{e\}$. We claim that the shortest path between $e$'s endpoints in $G_e$ must have length at least $(1+\epsilon/c)w(e)$ for some constant $c$. That implies any $(1+\epsilon/c)$-spanner of $H_r$ must include all long edges. The lemma then follows from Observation~\ref{obs:edge-length} since $H_r$ has at least $1/2\epsilon$ long edges, and each has length at least $w(\mst(H_r))/3$ for $\epsilon \ll 1$.
	
	Suppose that $e = x_{i}x_{i+k}$.  Let $P_e$ is a shortest path between $x_i$ and $x_{i+k}$ in $G_e$. Suppose that $w(P_e) \leq (1+\epsilon/c)w(e)$.  Since the terminal edge has length at least $3/2 w(e)$, $P_e$ cannot  contain the terminal edge. For the same reason, $P_e$ cannot contain two long edges. It remains to consider two cases:
	
	\begin{enumerate}
		\item $P_e$ contains exactly one long edge. Then, it must be that $P_e = \{x_{i},x_{i+1},x_{i+k+1},x_{i+k}\}$\footnote{indices are mod $2k$.} (Figure~\ref{fig:long-edge}(a)) or $P_e = \{x_{i},x_{i-1},x_{i+k-1},x_{i+k}\}$ (Figure~\ref{fig:long-edge}(b)). In both case, $w(P_i) = w(e) + 4r\sin(\epsilon^{3/2}/2)  \geq w(e)(1  + 2\frac{\sin(\epsilon^{3/2}/2)}{\sin(k\epsilon^{3/2}/2)}) \geq (1+2\epsilon)w(e)$.
		\item $P_e$ contains no long edge. Then, $P_e = \{x_i,x_{i+1}, \ldots,x_{i+k}\}$. Thus we have:
		\begin{equation*}
		\begin{split}
		\frac{w(P_e)}{w(e)} ~=~ \frac{2kr\sin(\epsilon^{3/2}/2)}{2r\sin(k \epsilon^{3/2}/2)} ~=~ 1 + \epsilon/96 + o(\epsilon)  ~\geq~ 1+ \epsilon/100
		\end{split}
		\end{equation*}
	\end{enumerate}
	Thus,  by choosing $c = 100$, we derive a contradiction.\qed
\end{proof}

\begin{figure}[h]
	\centering
	\vspace{0pt}
	\includegraphics[scale = 1.0]{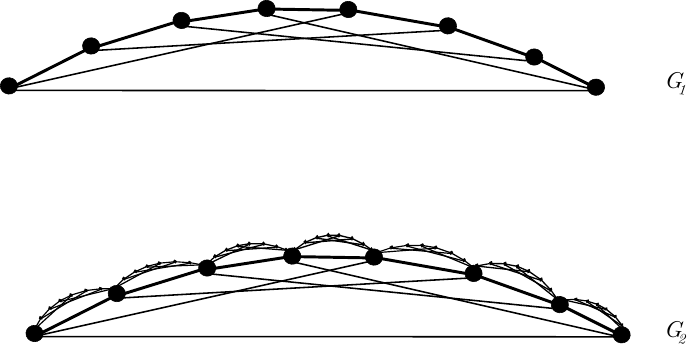}
	\caption{\footnotesize{An illustration of the recursive construction of $G_L$ with two levels.}}
	\label{fig:fractal}
\end{figure}

\paragraph{Proof of \Cref{thm:treewdith}.~} The construction is recursive. Let $H_1$ the core gadget of scale $1$.  Let $s_1$ ($\ell_1$) be the length of short  edges (long edges) of $H_1$. Let $x^1_1,\ldots, x^1_k$ be break points of $H_1$. Let $\be$ be the ratio of  the length of a short edge to the length of the terminal edge. That is:
\begin{equation}
\be = \frac{||x^1_1,x^1_2||}{||x^1_1,x^1_{2k}||} = \frac{\sin(\epsilon^{3/2}/2)}{\sin(\sqrt{\epsilon}/2)} = \epsilon + o(\epsilon)
\end{equation}
 Let $L = \frac{1}{\epsilon}$. We construct a set of graphs $G_1,\ldots, G_L$ recursively; the output graph is $G_L$. We refer to $G_i$ is the level-$i$ graph.

\noindent \textbf{Level-$1$ graph} $G_1 = H_1$. We refer to breakpoints of $H_1$ as breakpoints of $G$.

\noindent \textbf{Level-$2$ graph} $G_2$ obtained from $G_1$ by: (1) making $2k-1$ copies of the core gadget $H_{\be}$ at scale $\be$ (each $H_\delta$ is obtained by scaling every edge the core gadget by $\delta$), (2) for each $i \in [2k-1]$, attach each copy of $H_{\be}$ to $G_1$ by identifying the terminal edge of $H_{\be}$ and the edge between two consecutive breakpoints $x^1_ix^1_{i+1}$ of $G_1$. We then refer to breakpoints of all $H_{\be}$ as breakpoints of $G_2$. (See Figure~\ref{fig:fractal}.) Note that by definition of $\be$, the length of the terminal edge of $H_{\be}$ is equal to $||x^1_i,x^1_{i+1}||$. We say two adjacent breakpoints of $G_2$ \emph{consecutive} if they belong to the same copy of $H_{\be}$ in $G_2$ and are connected by one short edge of $H_{\be}$.

\noindent \textbf{Level-$j$ graph} $G_j$ obtained from $G_{j-1}$ by: (1) making $(2k-1)^j$ copies of the core gadget $H_{\be^{j-1}}$ at scale $\be^{j-1}$, (2) for every two consecutive breakpoints of $G_{j-1}$, attach each copy of $H_{\be^{j-1}}$ to $G_{j-1}$ by identifying the terminal edge of $H_{\be^{j-1}}$ and the edge between the two consecutive breakpoints. This completes the construction.

\noindent We now show some properties of $G_L$. We first claim that:

\begin{claim}\label{clm:tw-GL} $G_L$ has treewidth at most $4$.
\end{claim}
\begin{proof}
	Let $T_1$ be the tree decomposition of $G_1$ of width $5$, as guaranteed by Claim~\ref{clm:treewidth}. Note that for every pair of consecutive breakpoints $x^1_i,x^1_{i+1}$ of $G_1$, there is a bag, say $X_i$, of $T_1$ contains both $x^1_i$ and $x^1_{i+1}$. Also, there is a bag of $T_1$ containing both terminals of $T_1$.
	
	We extend the tree decomposition $T_1$ to a tree decomposition $T_2$  of $G_2$ as follows. For each gadget $H_{\be}$ attached to $G_1$ via consecutive breakpoints $x_1^i,x^1_{i+1}$, we add a bag $B = \{ x_1^i,x^1_{i+1}\}$, connect $B$ to $X_i$ of $T_1$ and to the bag containing terminals of the tree decomposition of $H_{\be}$. Observe that the resulting tree decomposition $T_2$ has treewidth at most $4$. The same construction can be applied recursively to construct a tree decomposition of $G_L$ of width at most $4$.\qed
\end{proof}

\begin{claim}\label{clm:mst-GL} $w(\mst(G_L)) = O(1) w(\mst(H_1))$.
\end{claim}
\begin{proof}
	Let $r(\epsilon)$ be the ratio between $\mst(H_1)$ and the length of the terminal edge of $H_1$.  Note that $\mst(H_1)$ is a path of short edges between $x_1^1$ and $x^1_{2k}$. By Observation~\ref{obs:edge-length}, we have:
	\begin{equation}
	r(\epsilon)  \leq \frac{r\sqrt{\epsilon}}{2r\sin(\sqrt{\epsilon}/2)}  = 1+\epsilon/24 + o(\epsilon) \leq  1+\epsilon
	\end{equation}
	when $\epsilon \ll 1$. When we attach copies of $H_{\be}$ to edges between two consecutive breakpoints of $G_1$, by re-routing each edge of $\mst(H_1)$ through the path $\mst(H_{\be})$ between $H_{\be}$'s terminals, we obtain a spanning tree of $G_2$ of weight at most $r(\epsilon)w(\mst(H_1)) \leq (1+\epsilon)w(\mst(H_1))$. By induction, we have:
	\begin{equation*}
	w(\mst(G_j)) \leq (1+\epsilon)w(\mst(G_{j-1})) \leq (1+\epsilon)^{j-1} w(\mst(H_1))
	\end{equation*}
This implies that $w(\mst(G_L)) \leq (1+\epsilon)^{L-1}w(\mst(H_1)) = O(1)w(\mst(H_1))$. \qed
\end{proof}

Let $S$ be an $(1+\epsilon/100)$-spanner of $G_L$ ($c = 100$ in Lemma~\ref{lm:spanner}). By Lemma~\ref{lm:spanner}, $S$ includes every long edge of all copies of $H_r$ at every scale $r$ in the construction. Recall that $||x^1_1,x^{1}_{2k}||$ is the terminal edge of $G_1$. Let $L_j$ be the set of long edges of all copies of $H_{\be^{j-1}}$ added at level $j$. Since $\frac{\mst(G_1)}{||x^1_1,x^{1}_{2k}||} = r(\epsilon)$,  we have:
\begin{equation}
\begin{split}
w(\mst(G_1) &= \frac{r(\epsilon)}{r(\epsilon)-1}\left(w(\mst(G_1)) - ||x^1_1,x^1_{2k}||\right)\geq \frac{24}{\epsilon}\left(w(\mst(G_1)) - ||x^1_1,x^1_{2k}||\right)
\end{split}
\end{equation}
By Lemma~\ref{lm:spanner}, we have:
\begin{equation}
\begin{split}
w(L_1) &\geq \frac{1}{6\epsilon}w(\mst(G_1)) \geq \frac{4}{\epsilon^2}(w(\mst(G_1)) - ||x^1_1,x^1_{2k}||)\\
w(L_2) &\geq   \frac{4}{\epsilon^2}(w(\mst(G_2)) - \mst(G_1))\\
&\ldots \\
w(L_j) &\geq \frac{4}{\epsilon^2}(w(\mst(G_{j})) - w(\mst(G_{j-1})))
\end{split}
\end{equation}
Thus, we have:
\begin{equation*}
w(S) \geq \sum_{j=1}^L w(L_j) \geq \frac{1}{4\epsilon^2}(w(\mst(G_L)) - ||x^1_1,x^1_{2k}||) = \Omega(\frac{1}{\epsilon^2})w(\mst(G_L)
\end{equation*}

By setting $\epsilon \leftarrow \epsilon/100$, we complete the proof of \Cref{thm:treewdith}. The condition on $n$ follows from the fact that  $G_L$ has $|V(G_L)| = O( (2k-1)^{L}) = O((\frac{1}{\epsilon})^{\frac{1}{\epsilon}})$ vertices. \qed

\newpage
\part{Our Unified Framework: Applications (\Cref{sec:applications-fast} and \Cref{sec:applications-light})}

In this part, we show applications of our unified framework described in \Cref{lm:framework} in obtaining results in \Cref{sec:intro}. 

\section{Applications of the Unified Framework: Fast Constructions}\label{sec:applications-fast}
In this section we implement the \hyperlink{SPSSA}{$\ssa$} for each of the graph classes.  
By plugging the $\ssa$ on top of the general transformation, as provided by \Cref{lm:framework},
we shall prove all theorems stated in \Cref{sec:intro}. We assume that $\eps \ll 1$, and this is without loss of generality since we can remove this assumption by scaling $\eps \leftarrow \eps'/c $ for \emph{any} $\eps' \in (0,1)$ and $c$ is sufficiently large constant. The scaling will incur a constant loss on lightness and runtime, as the dependency on $1/\eps$ is polynomial in all constructions below. 
 
\subsection{Euclidean Spanners and UDG Spanners}\label{subsec:EuclideanUDG}

In this section we prove the following theorem.

\begin{restatable}{theorem}{ACTUDGClustering}
	\label{thm:ACTUDGSpanner}
	Let $G = (V,E,w)$ be a $(1+\eps)$-spanner  either for a set of $n$ points $P$  or for  the unit ball graph $U$ of $P$ in $\mathbb{R}^d$. There is an algorithm that can  compute a $(1 + O(\eps))$-spanner $H$ of $G$ in the ACT model with lightness  $O((\eps^{-(d+2)} + \eps^{-4})\log(1/\eps))$  in time  $O(m\eps^{-1}(\alpha(m,n) + \eps^{1-d})\log(1/\eps))$.
\end{restatable}

We now show that \Cref{thm:ACTUDGSpanner} implies \Cref{thm:ACT} and \Cref{thm:UDG-fast}. Our construction for UDGs relies on the following result by F{\"u}rer and Kasiviswanathan~\cite{FK06}.

\begin{lemma}[Corollary 1 in~\cite{FurerK12}]\label{lm:UDG-sparse-sp} Given a set of $n$  points $P$ in $R^d$, there is an algorithm that constructs  a $(1+\epsilon)$-spanner of the unit ball graph for $P$ with $O(n\epsilon^{1-d})$ edges. For $d = 2$, the running time  is  $O(n(\epsilon^{-2}\log n))$; for $d = 3$, the running time is $\tilde{O}(n^{4/3}\eps^{-3})$; and for $d\geq 4$, the running time is $O(n^{2-\frac{2}{(\lceil d/2 \rceil+1)} + \delta}\epsilon^{-d+1} + n\epsilon^{-d})$ for any constant $\delta > 0$.
\end{lemma}

\begin{proof}[Proofs of \Cref{thm:ACT} and \Cref{thm:UDG-fast}]
	
It is known that a Euclidean $(1+\eps)$-spanner for a set of $n$ points $P$ in $\mathbb{R}^d$ with degree $O(\epsilon^{1-d})$ can be constructed in $O(n\log n)$ time in the ACT model (cf.\ Theorems 10.1.3 and 10.1.10 in \cite{NS07}). 
Furthermore, when $m = O(n\eps^{1-d})$, we have that: $$\alpha(m,n) ~=~ \alpha(nO(\eps^{-d}), n) ~=~ O(\alpha(n) + \log(\eps^{-d})) ~=~ O(\alpha(n) + d\log(1/\eps)).$$
Thus,  \Cref{thm:ACT} follows from \Cref{thm:ACTUDGSpanner}.

By \Cref{lm:UDG-sparse-sp}, we can construct sparse $(1+\epsilon)$-spanners for unit ball graphs with $m = O(n\eps^{1-d})$ edges in $O(n(\epsilon^{-2}\log n)$ time when $d = 2$, $\tilde{O}(n^{4/3}\eps^{-3})$ time when $d = 3$, and  $O(n^{2-\frac{2}{(\lceil d/2 \rceil+1)} + \delta}\epsilon^{-d+1} + n\epsilon^{-d})$ time for any constant $\delta > 0$ when $d\geq 4$.  Thus, \Cref{thm:UDG-fast} follows from \Cref{thm:ACTUDGSpanner}.\qed

\end{proof}

By \Cref{lm:framework}, in order to prove \Cref{thm:ACTUDGSpanner},  it suffices to implement the \hyperlink{SPSSA}{$\ssa$} for Euclidean and UDG spanners.
Next, we give a detailed geometric implementation of the \hyperlink{SPSSA}{$\ssa$}, hereafter $\ssa_{\geom}$; note that the stretch parameter $t$ in the geometric setting is $1+\eps$.  The idea is to use a Yao-graph like construction: For each node $\varphi_C \in \mv$, we construct a collection of cones of angle $\eps$ around the representative $r(C) = \source(\varphi_{C})$ of the cluster $C$ corresponding to $\varphi_C$. Recall that we have access to a $\source$ function that returns the representative of each cluster in $O(1)$ time.  Then for each cone, we look at all the representatives of the neighbors (in $\mg$) of $C$ that fall into that cone, and pick  to $\me^{\prune}$ the edge that connects $r(C)$ to the representative that is closest to it.

\begin{tcolorbox}
	\hypertarget{SPHEuclidean}{}
	\textbf{$\ssa_{\geom}$ (Euclidean and UDG):} The input is a $(L,\eps,\beta)$-cluster graph $\mg(\mv,\me,\omega)$ that corresponds to a Euclidean or UDG spanner. The output is $\me^{\prune}$; initially, $\me^{\prune} = \emptyset$.
	\begin{quote}
		 For each node $\varphi_{C_u} \in \mv$, do the following:
		\begin{itemize}
		\item Let $\mathcal{N}(\varphi_{C_u})$ be the set of neighbors of $\varphi_{C_u}$ in $\mg$.  We construct a collection of $\tau = O(\epsilon^{1-d})$  cones $\cone(C_u) = \{Q_1, Q_2,\ldots, Q_{\tau}\}$ that partition $\mathbb{R}^d$, each of angle $\eps$ and with apex at $r(C_u)$, the representative of  $C_u$. It is   known (see, e.g. Lemma 5.2.8 in~\cite{NS07}) that we can construct $\cone(C_u)$ in time $O(\epsilon^{1-d})$ in the ACT model. 
		
		\item For each $j\in [\tau]$:
		\begin{itemize}
		\item Let $R_j = \{r(C'): \varphi_{C'} \in \mathcal{N}(\varphi_{C_u}) \wedge (r(C') \in Q_j)\}$ 
		be the set of representatives that belong to the cone $Q_j \in \cone(C_u)$.  
		 Let $r_j^{*} = \arg\min_{r\in R_j} \nor2{r(C_u),r}$ be the representative in $R_j$ that is closest to $r(C_u)$. 
		\item Let $\varphi_{C_v}$ be the node of $\mg$ whose cluster $C_v$  has $r^*_j$ as the representative. By the definition of $R_j$, $(\varphi_{C_u},\varphi_{C_v})$ is an edge in $\me$. Add $(\varphi_{C_u},\varphi_{C_v})$ to $\me^{\prune}$.  
					\end{itemize}
			/* We add at most one edge to $\me^{\prune}$ incident on $\varphi_{C_u}$ for each of the $\tau$ cones.	*/	
		\end{itemize}
	\end{quote}
\end{tcolorbox}

We next analyze the running time of $\ssa_{\geom}$, and also show that it satisfies the two properties of (\hyperlink{SparsitySSA}{Sparsity}) and (\hyperlink{StretchSSA}{Stretch}) 
required by the abstract \hyperlink{SPSSA}{$\ssa$}; these properties are described in \Cref{subsec:framework-intro}. Recall that $H_{<(1+\eps)L}$ is the graph obtained by adding the source edges of $\me^{\prune}$ to $H_{< L}$, which is the spanner for all edges in $G$ of weight $< L$. Note that the stretch of $H_{< L}$ is $t(1+\gamma \eps)$ for $t = 1+\eps$, where $\gamma$ is a constant. Furthermore, as mentioned, we assume w.l.o.g.\ that $\eps$ is sufficiently smaller than $1$.

\begin{figure}[!h]
	\begin{center}
		\includegraphics[width=0.3\textwidth]{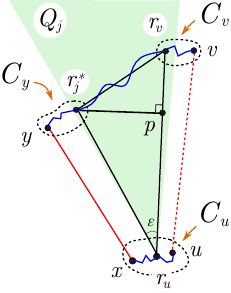}
	\end{center}
	\caption{Illustration for the stretch bound proof of \Cref{lm:ACT-UDG}. Black dashed curves represent three clusters $C_u,C_v, C_y$. 
	The solid red edge $(x,y)$ corresponds to an edge added to $\me^{\prune}$, while the dashed red edge  $(u,v)$ is not added. The green shaded region represents cone $Q_j$ of angle $\epsilon$ with apex at $r_u$.}
	\label{fig:ACT-stretch}
\end{figure}

\begin{lemma}\label{lm:ACT-UDG} \hyperlink{SPHEuclidean}{$\ssa_{\geom}$} can be implemented in $O((|\mv| + |\me|)\eps^{1-d})$ time in the ACT model. Furthermore, 1. (Sparsity) $|\me^{\prune}| = O(\eps^{1-d})|\mv|$, and 2. (Stretch) 
For each edge $(\varphi_{C_u},\varphi_{C_v})\in \me$, $d_{H_{<(1+\eps)L}}(u,v)\leq t(1+s_{\ssa_{\geom}}(\beta)\eps)w(u,v)$, where $(u,v) = 
		 \source(\varphi_{C_u}, \varphi_{C_v})$, $s_{\ssa_{\geom}}(\beta) = 2(19\beta + 14)$ and $\eps \leq \min\{\frac{1}{\gamma},\frac{1}{8\beta + 6}\}$.
\end{lemma}
\begin{proof} We first analyze the running time. We observe that, since we can construct $\cone(C_u)$ for a single node $\varphi_{C_u}$ in $O(\epsilon^{1-d})$ time in the ACT model,  the running time to construct all sets of cones $\{\cone(C_u)\}_{\varphi_{C_u} \in \mv}$ is $O(|\mv|\epsilon^{1-d})$. Now consider a specific node $\varphi_{C_u}$. For each neighbor $\varphi_{C'} \in \mathcal{N}(\varphi_{C_u})$ of $\varphi_{C_u}$, finding the cone $Q_j \in \cone(C_u)$ such that $r(C') \in Q_j$ takes $O(\tau) = O(\epsilon^{1-d})$ time. 
	Thus, $\{R_j\}_{j=1}^{\tau}$ can be constructed in $O(|\mathcal{N}(\varphi_{C_u})|\eps^{1-d})$ time. Finding the set of representatives $\{r^*_j\}_{j=1}^{\tau}$ takes $O(|\mathcal{N}(\varphi_{C_u})|)$ time by calling function $\source(\cdot)$. Thus, the total running time to implement \hyperlink{SPHEuclidean}{$\ssa_{\geom}$} is:
	\begin{equation*}
		O(|\mv|\epsilon^{1-d}) +  \sum_{\varphi_{C_u} \in \mv} O(|\mathcal{N}(\varphi_{C_u})|\eps^{1-d}) = O((|\mv| + |\me|)\eps^{1-d})~,
	\end{equation*}
	as claimed.	 
	
	 By the construction of the algorithm, for each node $\varphi_{C}\in \mv$, we add at most $\tau = O(\eps^{1-d})$ incident edges in $\me$ to $\me^{\prune}$; this implies Item 1. 
	
It remains to prove Item 2: For each edge $(\varphi_{C_u},\varphi_{C_v})\in \me$, the stretch  in $H_{<(1+\eps)L}$ of the corresponding edge $(u,v) = \source(\varphi_{C_u},\varphi_{C_v})$  is at most $(1+s_{\ssa_{\geom}}(\beta)\eps)$ with $s_{\ssa_{\geom}}(\beta)= 2(19\beta + 14)$. Let $r_u \defi r(C_u)$ and $r_v \defi r(C_v)$ be the representatives of $C_u$ and $C_v$, respectively. Let $Q_j$ be the cone in $\cone(C_u)$ such that $r_v \in Q_j$ for some $j \in [\tau]$ (we are using the notation in \hyperlink{SPHEuclidean}{$\ssa_{\geom}$}).  If $r_v = r^*_j$, then $(u,v) \in H_{<(1+\eps)L}$ by the construction in \hyperlink{SPHEuclidean}{$\ssa_{\geom}$}, and so the stretch is $1$. Otherwise, let $C_y$ be the level-$i$ cluster that contains the representative $r^*_j$. By the construction in \hyperlink{SPHEuclidean}{$\ssa_{\geom}$}, there is an edge $(x,y)\in H_{<(1+\eps)L}$ where $x\in C_u$ and $y\in C_y$.  (See Figure~\ref{fig:ACT-stretch}.) By property 4 of $\mg$ in \Cref{def:ClusterGraph-Param},  $\max\{\dm(H_{<(1+\eps)L}[C_u]), \dm(H_{<(1+\eps)L}[C_v]), \dm(H_{<(1+\eps)L}[C_y])\} \leq \beta \eps L$. Note that edges in $\me$ have weights in $[L,(1+\eps)L)$ by property 3 in \Cref{def:ClusterGraph-Param}. By the triangle inequality:
	\begin{equation}\label{eq:ACT-triangle}
		\begin{split}
			\nor2{r_u,r_v} &\leq \nor2{u,v} + 2\beta\eps L \leq (1+(1+2\beta)\epsilon)L\\
			\nor2{r_u,r^*_j}  &\leq \nor2{x,y} + 2\beta L \leq (1+(1+2\beta)\epsilon)L\\
			\nor2{u,v} \leq \nor2{r_u,r_v} &+ 2\beta\epsilon L  \qquad \mbox{ and} \nor2{x,y}  \leq \nor2{r_u,r^*_j} + 2\beta\epsilon L
		\end{split}
	\end{equation}
	
	Furthermore, since $L \leq  \nor2{u,v},  \nor2{x,y}\leq  (1+\eps)L$, it follows that:
	\begin{equation}\label{eq:ACT-uvxy}
		\begin{split}
			\nor2{u,v} \leq (1+\eps) \nor2{x,y} \quad \mbox{and}\quad
			\nor2{x,y} \leq (1+\eps) \nor2{u,v} 
		\end{split}
	\end{equation}

	\begin{claim}\label{clm:ACT-rvrj} $\nor2{r_v,r^*_j} \leq (8\beta+6)\eps L $. 
	\end{claim}
	\begin{proof} Recall that $\nor2{r_u,r^*_j} \leq \nor2{r_u,r_v}$. Let $p$ be the projection of $r_j^*$ onto the segment $r_ur_v$ (see \Cref{fig:ACT-stretch}). Since $\angle r_vr_ur^*_j \leq \epsilon$, $\nor2{r_j^*,p} \leq \sin(\eps) \nor2{r_u,r^*_j} \leq \sin(\eps) \nor2{r_u,r_v} ~\leq~ \eps(1+(1+2\beta)\eps)L$. We have:
		\begin{equation}\label{eq:ACT-rvrj}
			\begin{split}
				\nor2{r_v,r^*_j} &\leq  \nor2{p,r^*_j} + \nor2{r_v,p}   \leq \nor2{p,r^*_j} +   \nor2{r_u,r_v} - (\nor2{r_u,r_j^*} - \nor2{r_j^*,p})\\
				&\leq (\nor2{r_u,r_v} - \nor2{r_u,r_j^*}) + 2\eps(1+(1+2\beta)\eps)L
			\end{split}
		\end{equation}

		We now bound $(\nor2{r_u,r_v} - \nor2{r_u,r_j^*})$. By \Cref{eq:ACT-triangle} and \Cref{eq:ACT-uvxy}, it holds that:
		\begin{equation} \label{addedeq}
			\begin{split}
				\nor2{r_u,r_v} - \nor2{r_u,r_j^*}  &\leq \nor2{u,v} + 2\beta\epsilon L  - (\nor2{x,y}- 2\beta\epsilon L) \leq (4\beta+1+\eps)\eps L
			\end{split}
		\end{equation}
		
		Plugging \Cref{addedeq} into \Cref{eq:ACT-rvrj}, we get:
		\begin{equation*}
			\begin{split}
			\nor2{r_v,r_j^*} &\leq (4\beta+1+\eps)\eps L +  2\eps(1+(1+2\beta)\eps)L \leq (8\beta +6)\eps L~\qquad \mbox{(since }\eps \leq 1),
			\end{split}
		\end{equation*}
	as claimed.	This completes the proof of \Cref{clm:ACT-rvrj}. \qed
	\end{proof}
	
Next we continue with the proof of \Cref{lm:ACT-UDG}.	By \Cref{clm:ACT-rvrj}, $\nor2{r_v,r^*_j}  < L$ when $\eps < 1/(8\beta+ 6)$. If the input graph is a UDG, then $\me  \not=\emptyset$ only if $L \leq 1$. Thus,  $\nor2{r_v,r^*_j}\leq 1$ and hence, there is an edge $(r_v,r^*_j)$ of length $\nor2{r_v,r^*_j}$ in the input UDG. (This is the only place, other than starting our construction with a $(1+\eps)$-spanner for the input UDG, where we exploit the fact that the input graph is a UDG.)
	
	Since $\nor2{r_v,r^*_j} < L$, the distance between $r_v$ and $r^*_j$ is preserved up to a factor of $(1+\gamma \eps)$ in $H_{< L}$.	That is, $d_{H_{<(1+\eps)L}}(r_v,r^*_j) \leq (1+\gamma \eps)\nor2{r_v,r^*_j}$.

	Note that $r_u,r_v,r_j^*$ are all in the input point set $P$ by the definition of representatives.  By the triangle inequality, it follows that:
	\begin{equation}\label{eq:ACT-e1}
		\begin{split}
			d_{H_{<(1+\eps)L}}(u,v) &\leq d_{H_{<(1+\eps)L}}(u, x) +  \nor2{x,y} +  d_{H_{<(1+\eps)L}}(y,r_j^*)+ d_{H_{<(1+\eps)L}}(r_j^*, r_v) +  d_{H_{<(1+\eps)L}}(r_v, v)\\
			&\leq \beta\epsilon L +  \nor2{x,y} + \beta\eps L + (1+\gamma \eps)\nor2{r_v,r^*_j}  + \beta\eps L  \\
			&\leq  \nor2{x,y} +3\beta\eps L + \underbrace{(1+\gamma \eps)}_{\leq~ 2 \text{ since } \eps ~\leq~ 1/\gamma}(8\beta+6)\eps L \qquad \mbox{(by \Cref{clm:ACT-rvrj})}\\
			&\leq   \nor2{x,y} + (19\beta + 12)\eps L 
		\end{split}
	\end{equation}
	By \Cref{eq:ACT-uvxy},   $\nor2{x,y}~\leq~ (1+\eps)\nor2{u,v}\leq \nor2{u,v} + (1+\eps)\eps L \leq \nor2{u,v} + 2\eps L $. Thus, by Equation~\eqref{eq:ACT-e1}:
	$$	d_{H_{<(1+\eps)L}}(u,v) \leq \nor2{u,v} + (19\beta + 14)\eps L \stackrel{\nor2{u,v}\geq L/2}{\leq} (1+2(19\beta + 14)\eps)\nor2{u,v}.$$
	That is, the stretch of  $(u,v)$  in $H_{<(1+\eps)L}$ is at most $1+s_{\ssa_{\geom}}(\beta)\eps$ with $s_{\ssa_{\geom}}(\beta)= 2(19\beta + 14)$.
	\qed
\end{proof}

\begin{remark}\label{remark:ACT-Faster} \hyperlink{SPHEuclidean}{$\ssa_{\geom}$} can be implemented slightly faster, within time $O(|\mv|\eps^{1-d} + |\me|\log(1/\eps))$, by using a data structure that allows us to search for the cone that a representative belongs to in $O(\log(1/\eps))$ time. Such a data structure is described in Theorem 5.3.2 in the book by Narasimhan and Smid~\cite{NS07}.
\end{remark}

\begin{proof}[Proof of \Cref{thm:ACTUDGSpanner}] We use $\ssa_{\geom}$ in place of the abstract \hyperlink{SPSSA}{$\ssa$} in \Cref{lm:framework} to construct the light spanner. By \Cref{lm:ACT-UDG}, we have $s_{\ssa}(\beta) = 2(19\beta + 14)$, $\chi = O(\eps^{1-d})$ and $\tau(m',n') = O(\eps^{1-d})$. Thus, by plugging in the values of $\chi$ and $\tau$, we obtain the lightness and the running time as required by \Cref{thm:ACTUDGSpanner}. The stretch of the spanner is $(1+\eps)(1 + (s_{\ssa}(O(1)) + O(1))\eps)  = (1 + O(\eps))$ when $\eps \leq 1$.
 \qed
\end{proof}

\subsection{General Graphs}\label{subsec:general-fast}

In this section, we prove \Cref{thm:general-fast} by giving a detailed implementation of \hyperlink{SPSSA}{$\ssa$} for general graphs, hereafter $\ssa_{\gen}$. Here we have $t = 2k-1$ for an integer parameter $k \ge 2$. We will use as a black-box the linear-time construction of sparse spanners in general unweighted graphs by Halperin and Zwick~\cite{HZ96}. 

\begin{theorem}[Halperin-Zwick~\cite{HZ96}]\label{thm:unweighted-2k-1} Given an unweighted $n$-vertex graph $G$ with $m$ edges, a $(2k-1)$-spanner of $G$ with $O(n^{1+\frac{1}{k}})$ edges can be constructed deterministically in $O(m + n)$ time, for any $k \ge 2$.
\end{theorem}

\begin{tcolorbox}
	\hypertarget{SPHGeneral}{}
	\textbf{$\ssa_{\gen}$ (General Graphs):} The input is a $(L,\eps,\beta)$-cluster graph $\mg(\mv,\me,\omega)$. The output is $\me^{\prune}$; initially, $\me^{\prune} = \emptyset$.
	\begin{quote}
		We construct a new \emph{unweighted} graph $J = (V_J,E_J)$ as follows. For each node in $\varphi \in \mv$, we add a vertex $v_{\varphi}$ to $V_J$. For each edge $(\varphi_1,\varphi_2) \in \mv$, we add an edge $(v_{\varphi_1}, v_{\varphi_2})$ to $E_J$. 
		
		Next, we run  Halperin-Zwick's algorithm (\Cref{thm:unweighted-2k-1}) on $J$ to construct a $(2k-1)$-spanner $S_J$ for $J$. Then for each edge $(v_{\varphi_1},v_{\varphi_2})$ in $E(S_J)$, we add the corresponding edge $(\varphi_1,\varphi_2)$ to $\me^{\prune}$. 
	\end{quote}
\end{tcolorbox}

We next analyze the running time of $\ssa_{\gen}$, and also show that it satisfies the two properties of (\hyperlink{SparsitySSA}{Sparsity}) and (\hyperlink{StretchSSA}{Stretch}) 
required by the abstract \hyperlink{SPSSA}{$\ssa$}; these properties are described in \Cref{subsec:framework-intro}.

\begin{lemma}\label{lm:App-Gen-SSA} \hyperlink{SPHGeneral}{$\ssa_{\gen}$} can be implemented in $O(|\mv| + |\me|)$ time. Furthermore, 1. (Sparsity) $\me^{\prune} = O(n^{1/k})|\mv|$, and 2. (Stretch) For each edge $(\varphi_{C_u},\varphi_{C_v})\in \me$, $d_{H_{<(1+\eps)L}}(u,v)\leq (2k-1)(1+s_{\ssa_{\gen}}(\beta)\eps)w(u,v)$, where $(u,v) = 
		 \source(\varphi_{C_u}, \varphi_{C_v})$, $s_{\ssa_{\gen}}(\beta) = (2\beta+1)$ and $\eps \leq 1$. 
\end{lemma}
\begin{proof} The running time of $\ssa_{\gen}$ follows directly from \Cref{thm:unweighted-2k-1}. Also, by \Cref{thm:unweighted-2k-1}, $|\me^{\prune}|= O(|\mv|^{1+1/k}) = O(n^{1/k}|\mv|)$; this implies Item 1. 
		
	It remains to prove Item 2: 
	For each edge $(\varphi_{C_u},\varphi_{C_v})\in \me$, the stretch in $H_{<(1+\eps)L}$ (constructed as described in \hyperlink{SPSSA}{$\ssa$})
	of the corresponding edge $(u,v) = 		 \source(\varphi_{C_u}, \varphi_{C_v})$ is
at most  $(2k-1)(1+ (2\beta+1)\eps) w(u,v)$.	Recall that $H_{<(1+\eps)L}$ is the graph obtained by adding the source edges of $\me^{\prune}$ to $H_{< L}$.
	
	Let $(u_{1},v_1)$ be the edge in $E_J$ that corresponds to the edge $(\varphi_{C_u},\varphi_{C_v})$. By \Cref{thm:unweighted-2k-1}, there is a path $P$ between $u_1$ and $v_1$ in $J$ such that $P$ contains at most $2k-1$ edges. We write $P = (u_1=x_0, (x_0,x_1), x_1, (x_1,x_2), \ldots, x_{p} = v_1)$ as an alternating sequence of vertices and edges. Let $\mp = ( \varphi_0, (\varphi_0,\varphi_1), \varphi_1, (\varphi_1,\varphi_2), \ldots, \varphi_{p})$ be a path of $\mg$, written as an alternating sequence of vertices and edges, that is obtained from $P$ where $\varphi_j$ corresponds to $x_j$, $1\leq j\leq p$. Note that $\varphi_1 = \varphi_{C_u}$ and $\varphi_{p} = \varphi_{C_v}$.
	
	\begin{figure}[!h]
		\begin{center}
			\includegraphics[width=0.9\textwidth]{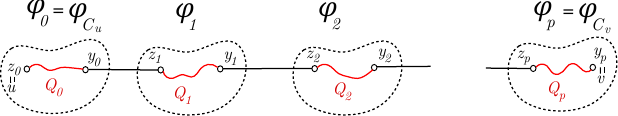}
		\end{center}
		\caption{A path from $u$ to $v$.}
		\label{fig:path}
	\end{figure}
	Let $\{y_i\}_{i=0}^p$ and $\{z_i\}_{i=0}^p$ be two sequences of vertices of $G$ such that (a) $z_0 = u$ and $y_p = v$, and (b) $(y_{i-1}, z_i)$ is the edge in $G$ corresponding to edge $(\varphi_{i-1},\varphi_i)$ in $\mathcal{P}$, for $1\leq i\leq p$. Let $Q_i$, $0\leq i \leq p$, be a shortest path in $H_{< L}[C_i]$ between $z_i$ and $y_i$, where $C_i$ is the cluster corresponding to $\varphi_i$. See \Cref{fig:path} for an illustration. Observe that $w(Q_i) \leq \beta \eps L$ by property 4 in \Cref{def:ClusterGraph-Param}. Let $P' = Q_0\circ (y_0,z_1)\circ \ldots\circ Q_p$ be a (possibly non-simple) path from $u$ to $v$ in $H_{<(1+\eps)L}$; here $\circ$ is the path concatenation operator. 
		\begin{equation}
			\begin{split}
				w(P') &\leq (2k-1)(1+\eps)L + (2k)\beta\eps L \leq (2k-1)(1+\eps +  2\beta\eps)L\\
				 &\leq (2k-1)(1+(2\beta + 1)\eps)w(u,v) \qquad \mbox{(since $w(u,v)\geq L$)}\\
			\end{split}
		\end{equation}
	 
Thus, the stretch of edge $(u,v)$ is at most $(2k-1)(1+(2\beta + 1)\eps)$, as required.\qed 
\end{proof}
 \begin{proof}[Proof of \Cref{thm:general-fast}]   We use algorithm $\ssa_{\gen}$ in place of the abstract \hyperlink{SPSSA}{$\ssa$} in \Cref{lm:framework} to construct the light spanner. By \Cref{lm:ACT-UDG}, we have $s_{\ssa}(\beta) = (2\beta + 14)$, $\chi = O(n^{1/k})$ and $\tau(m',n') = O(1)$. Thus, by plugging in the values of $\chi$ and $\tau$, we obtain the lightness and the running time as required by \Cref{thm:general-fast}. The stretch of the spanner is $(2k-1)(1 + (s_{\ssa}(O(1)) + O(1))\eps)  = (2k-1)(1 + O(\eps))$. By scaling, we get the required stretch of $(2k-1)(1+\eps)$. \qed
 \end{proof}

 \subsection{Minor-free Graphs}\label{subsec:minor-free}

 In this section, we prove a weaker version of \Cref{thm:minor-free-fast}, where the running time is $O(nr\sqrt{r}\alpha(nr\sqrt{r},n))$. In \Cref{sec:minor-linear} we show how to achieve a linear running time, via an adaptation of our framework (described in detail in \Cref{sec:framework}) to minor-free graphs.

The implementation of the abstract algorithm \hyperlink{SPSSA}{$\ssa$} for minor-free graphs, hereafter $\ssa_{\minor}$, simply outputs the edge set $\me$. Note that the stretch in this case is $t = 1+\eps$.

 \begin{tcolorbox}
 	\hypertarget{SPHMinor}{}
 	\textbf{$\ssa_{\minor}$ (Minor-free Graphs):} The input is a  $(L,\eps,\beta)$-cluster graph $\mg(\mv,\me,\omega)$. The output is $\me^{\prune}$.
 	\begin{quote}
 		The algorithm returns $\me^{\prune} = \me$.
 	\end{quote}
 \end{tcolorbox}
 
We next analyze the running time of $\ssa_{\minor}$, and also show that it satisfies the two properties of (\hyperlink{SparsitySSA}{Sparsity}) and (\hyperlink{StretchSSA}{Stretch})
required by the abstract \hyperlink{SPSSA}{$\ssa$}. 
 
 \begin{lemma}\label{lm:App-Minor-Fast}\hyperlink{SPHMinor}{$\ssa_{\minor}$} can be implemented in $O((|\mv| + |\me|))$ time. 
Furthermore, 1. (Sparsity) $\me^{\prune} = O(r\sqrt{\log r})|\mv|$, and 2. (Stretch) For each edge $(\varphi_{C_u},\varphi_{C_v})\in \me$, $d_{H_{<(1+\eps)L}}(u,v)\leq (1+\eps)(1+s_{\ssa_{\minor}}(\beta)\eps)w(u,v)$, where $(u,v) = 
\source(\varphi_{C_u}, \varphi_{C_v})$, $s_{\ssa_{\minor}}(\beta) = 0$ and $\eps \leq 1$. 
 \end{lemma}
 \begin{proof} The running time of $\ssa_{\minor}$ follows trivially from the construction. Noting that $\mg$ is a minor of the input graph $G$, $\mg$ is $K_r$-minor-free. Thus, $|\me| = O(r\sqrt{\log r})|\mv|$ by the sparsity of minor-free graphs~\cite{Kostochka82,Thomason84}; this implies Item 1.  Since we take every edge of $\me$ to $\me^{\prune}$, the stretch is $1$ and hence $s_{\ssa_{\minor}}(\beta) = 0$, yielding Item 2. \qed 
 \end{proof}
 
  We are now ready to prove a weaker version of \Cref{thm:minor-free-fast} for minor-free graphs, where the running time is $O(nr\sqrt{r}\alpha(nr\sqrt{r},n))$.
 \begin{proof}[Proof of \Cref{thm:minor-free-fast}]   We use algorithm $\ssa_{\minor}$ in place of the abstract \hyperlink{SPSSA}{$\ssa$} in \Cref{lm:framework} to construct the light spanner. By \Cref{lm:ACT-UDG}, we have $s_{\ssa}(\beta) = 0$, $\chi =  O(r\sqrt{\log r})$ and $\tau(m',n') = O(1)$. Thus, by plugging in the values of $\chi$ and $\tau$, we obtain the lightness claimed in \Cref{thm:minor-free-fast} and a running time of $O(nr\sqrt{r}\alpha(nr\sqrt{r},n))$, for a constant $\eps$. The stretch of the spanner is:
 	\begin{equation*}
 		(1+\eps)(1 + (s_{\ssa}(O(1))+ O(1))\eps)  = (1 + O(\eps))
 	\end{equation*}
 By scaling, we get a stretch of $(1+\eps)$. \qed
 \end{proof}

\section{Applications of the Unified Framework: Fine-Grained Optimality}\label{sec:applications-light}

In this section, we use the framework outlined in  \Cref{subsec:framework-intro} to obtain all results regarding fine-grained lightness bounds stated in \Cref{subsec:contribution}: \Cref{thm:minor-free-opt-lightness}, \Cref{thm:light-general-spanner}, \Cref{thm:light-Steiner}, and \Cref{thm:Euclidean-high}. We do so by introduce another layer of abstraction via an object that we call \emph{general sparse spanner oracle} ($\gsso$) in \Cref{subsec:oracle-intro}: we show that the existence of $\gsso$ implies the existence of light spanners.   In \Cref{subsec:Oracle}, we construct $\gsso$es for different class of graphs: general graphs, high dimensional Euclidean spanners, and Steiner Euclidean spanners. Finally, in \Cref{subsec:minor-light}, we construct a light spanner for minor-free graphs by directly implementing \hyperlink{SPSSO}{$\sso$}. See \Cref{fig:unifedFull} for relationships between theorems/lemmas.

\subsection{General Sparse Spanner Oracles}\label{subsec:oracle-intro}

We introduce the notion of a {\em general sparse spanner} oracle ($\gsso$). Our $\gsso$ for stretch $t = 1+\eps$ coincides with a notion called {\em spanner oracle}, introduced by Le~\cite{LS20}; nonetheless, our goal is much more ambitious: First we wish to optimize the fine-grained dependencies  and second we wish to do so while considering a much wider regime of the stretch parameter $t$, which may even depend on $n$. 

\begin{definition}[General Sparse Spanner Oracle]\label{def:oracle} Let $G$ be  an edge-weighted graph and let $t > 1$ be a stretch parameter. A general sparse spanner oracle ($\gsso$) of $G$ for a given stretch $t$ is an algorithm that, given a subset of vertices  $T\subseteq V(G)$ and a distance parameter $L > 0$, outputs in \emph{polynomial time} a subgraph $S$ of $G$ such that for every pair of vertices $x,y \in T, x\not= y$ with $L \leq d_G(x,y) < 2L$:
	\begin{equation}
		d_{S}(x,y)\leq t\cdot d_G(x,y).
	\end{equation}	
	We denote a $\gsso$ of $G$ with stretch $t$ by  $\mathcal{O}_{G,t}$, and its output subgraph is denoted by $\mathcal{O}_{G,t}(T,L)$, given two parameters $T\subseteq V(G)$ and $L >0$.
\end{definition}

\begin{definition}[Sparsity]\label{def:sparsity} Given a $\gsso$ $\mathcal{O}_{G,t}$ of a graph $G$, we define weak sparsity and strong sparsity of $\mathcal{O}_{G,t}$, denoted by $\wsp_{\mathcal{O}_{G,t}}$ and $\ssp_{\mathcal{O}_{G,t}}$ respectively, as follows:
	\begin{equation}\label{eq:wsp-ssp}
		\begin{split}
			\wsp_{\mathcal{O}_{G,t}} &= \sup_{T\subseteq V, L \in \real^+}\frac{w\left(\mathcal{O}_{G,t}(T,L)\right)}{|T|L}\\
			\ssp_{\mathcal{O}_{G,t}} &=  \sup_{T\subseteq V, L \in \real^+} \frac{|E\left( \mathcal{O}_{G,t}(T,L)\right)|}{|T|}
		\end{split}
	\end{equation}	
\end{definition}
\noindent We observe that:
\begin{equation}\label{wsp-vs-ssp}
	\wsp_{\mathcal{O}_{G,t}} \leq t\cdot \ssp_{\mathcal{O}_{G,t}},
\end{equation}
since every edge $E\left( \mathcal{O}_{G,t}(T,L)\right)$ must have weight at most $t\cdot L$; indeed, otherwise we can remove it from $ \mathcal{O}_{G,t}(T,L)$ without affecting the stretch.  Thus, when $t$ is constant, strong sparsity implies weak sparsity; note, however, that this is not necessarily the case when $t$ is super-constant. 

Our main result in this section is to show that for stretch $t\geq 2$, we can construct a light spanner with lightness bound roughly $O(\frac{1}{\eps})$ times the sparsity of the spanner oracle (\Cref{thm:general-stretch-2}).   For stretch $t = 1+\eps$, we can construct a light spanner with lightness bound roughly $O(\frac{1}{\eps})$ times the sparsity of the spanner oracle plus \emph{an additive factor $1/\eps^2$}.

\begin{restatable}{theorem}{GeneralStretchT}
	\label{thm:general-stretch-2} Let $G$ be an arbitrary edge-weighted graph that admits a $\gsso$ $\mathcal{O}_{G,t}$ of weak sparsity $\wsp_{\mathcal{O}_{G,t}}$ for $t\geq 2$. Then for any $\eps > 0$, we can construct in polynomial time a $t(1+\epsilon)$-spanner for $G$ with lightness $\tilde{O}_{\epsilon}\left(\frac{\wsp_{\mathcal{O}_{G,t}}}{\epsilon}\right)$
\end{restatable}

\begin{restatable}{theorem}{GeneralStretchE}
	\label{thm:general-stretch-1eps} Let $G$ be an arbitrary edge-weighted graph  that admits a $\gsso$ $\mathcal{O}_{G,1+\eps}$ of weak sparsity $\wsp_{\mathcal{O}_{G,1+\eps}}$  for any $\eps > 0$. Then there exists an $(1+O(\epsilon))$-spanner for $G$ with lightness $\tilde{O}_{\epsilon}\left(\frac{\wsp_{\mathcal{O}_{G,t}}}{\epsilon} + \frac{1}{\epsilon^2}\right)$.
\end{restatable}

In both \Cref{thm:general-stretch-2} and~\Cref{thm:general-stretch-1eps}, $\tilde{O}_{\epsilon}(.)$ hides a factor of $\log \frac{1}{\epsilon}$. The proofs of these theorem are presented in \Cref{subsec:B-Oracle}.

The bound in \Cref{thm:general-stretch-1eps} improves over the lightness bound due to Le~\cite{Le20} by a factor of $\frac{1}{\epsilon^2}$. The stretch of $S$ in \Cref{thm:general-stretch-1eps} is $1+O(\epsilon)$, but we can scale it down to $(1+\eps)$ while increasing the lightness by a constant factor. 
Moreover, this bound is optimal, as we shall assert next.
First, the additive factor $\frac{\wsp_{\mathcal{O}_{G,t}}}{\epsilon}$ is unavoidable: \dblind showed in~\cite{LS19} that there exists a set of $n$ points in $\mathbb R^d$ such that any $(1+\epsilon)$-spanner for it must have lightness $\Omega(\epsilon^{-d})$, while the result of Le~\cite{Le20} implies that  point sets in $\mathbb R^d$ have $\gsso$es  with weak sparsity $O(\epsilon^{1-d})$. 
Second, the additive factor $\frac{1}{\epsilon^2}$ is tight by the following theorem.

\begin{theorem}\label{thm:lb-oracle-1eps}  
	For any $\epsilon < 1$  and $n \geq (\frac{1}{\epsilon})^{\Theta(\frac{1}{\epsilon})}$, there is an $n$-vertex graph $G$  admitting a  $\gsso$ of stretch $(1+\eps)$ with weak sparsity $O(1)$ such that any $(1+\epsilon)$-spanner of $G$ must have lightness $\Omega(\frac{1}{\epsilon^2})$.  
\end{theorem}
\begin{proof} Le (Theorem 1.3 in~\cite{Le20}), building upon the work of Krauthgamer, Nguy$\tilde{\hat{\mbox{e}}}$n and Zondiner~\cite{KNZ14}, showed that graphs with treewidth $\tw$ have a $1$-spanner oracle with weak sparsity $O(\tw^4)$. Since the treewidth of $G$ in \Cref{thm:treewdith} is $4$, it has a $1$-spanner oracle with weak sparsity $O(1)$; this implies \Cref{thm:lb-oracle-1eps}. \qed
\end{proof}
\paragraph{Light spanners from GSSO.~}\label{subsec:B-Oracle} We now turn to proving \Cref{thm:general-stretch-2} and \Cref{thm:general-stretch-1eps}. We do so by providing an implementation of \hyperlink{SPSSO}{$\sso$} using a $\gsso$. We assume that we are given a $\gsso$ $\mathcal{O}_{G,t}$ with weak sparsity $\wsp_{\mathcal{O}_{G,t}}$. We denote the algorithm by $\sso_{\oracle}$. We assume that every edge in $G$ is a shortest path between its endpoints; otherwise, we can safely remove them from the graph.

	\begin{tcolorbox}
		\hypertarget{SPHOracle}{}
		\textbf{$\sso_{\oracle}$:} The input is an $(L,\eps,\beta)$-cluster graph $\mg=(\mv,\me,\omega)$. The output is a set of edges $F$.
		\begin{quote}
			For each node $\varphi_C \in \mv(\mathcal{G})$ corresponding to a cluster $C$, we choose a $v \in C$. Let $S$ be the set of chosen vertices. Let 
			\begin{equation} \label{eq:F-oracle}
				F = E(\mathcal{O}_{G,t}(S,L/2))\cup E(\mathcal{O}_{G,t}(S,L))\cup  E(\mathcal{O}_{G,t}(S,2L))
			\end{equation}
			be the edge set of the spanner returned by the oracle.   We then return $F$.
		\end{quote}
	\end{tcolorbox}

	We now show that $\sso_{\oracle}$ has all the properties as described in the abstract  \hyperlink{SPSSO}{$\sso$}.
	
	\begin{lemma}\label{lm:App-Oracle} Let $F$ be the output of \hyperlink{SPHOracle}{$\sso_{\oracle}$}. Then $w(F) = O(\wsp_{\mathcal{O}_{G,t}})L\cdot |\mv|$. Furthermore,  $d_{H_{<2L}}(u,v) \leq t(1+ s_{\sso_{\oracle}}(\beta)\eps)w(u,v)$ for every edge $(u,v)$ corresponding to an edge in $\me$, where  $s_{\sso_{\oracle}}(\beta) = 4\beta$ and $\eps$ is sufficiently smaller than $1$, in particular $\eps \leq 1/(4\beta)$. 
	\end{lemma}
	\begin{proof} Since we only choose exactly one vertex in $S$ per node in $\mg$, $|S| = |\mv|$. By the definition of the sparsity of an oracle (\Cref{def:sparsity}), $w(F) \leq \wsp_{\mathcal{O}_{G,t}} (L/2)\cdot |S| + \wsp_{\mathcal{O}_{G,t}} L\cdot |S| + \wsp_{\mathcal{O}_{G,t}} 2L\cdot |S| = O(\wsp_{\mathcal{O}_{G,t}})L\cdot |\mv|$; this implies the first claim. 
		
		Let $(u,v)$ be an edge in $G$ corresponding to an edge $(\varphi_{C_u}, \varphi_{C_v}) \in \me$. We have that $L \leq w(u,v)< 2L$ by property 3 in \Cref{def:ClusterGraph-Param}. By the construction of $S$ in \hyperlink{SPHOracle}{$\sso_{\oracle}$}, there are two vertices $u_1 \in C_u$ and $v_1 \in C_v$ that are in $S$. Let $P_{u_1,u}$ ($P_{v_1,v}$) be the shortest path in $H_{<L}[C_u]$ ($H_{<L}[C_v]$) between $u$ and $u_1$ ($v$ and $v_1$). By property 4 in \Cref{def:ClusterGraph-Param}, we have that $\max\{w(P_{u_1,u}), w(P_{v_1,v})\} \leq \beta \eps L$. By the triangle inequality, we have:
		\begin{equation}\label{eq:oracleStretch-uv-up}
			d_G(u_1,v_1)\leq w(u,v) + 2\beta\eps L < (2+2\beta\eps) L \leq 4L,
		\end{equation}
		since $\eps \leq 1/\beta$. Also by the triangle equality, it follows that:
		\begin{equation}\label{eq:oracleStretch-uv-low}
			d_G(u_1,v_1) \geq w(u,v) - 2\beta\eps L\geq (1 - 2\beta\eps L) \geq L/2,
		\end{equation}
		since $\eps \leq \frac{1}{4\beta}$. Thus, $d_G(u_1,v_1) \in [L/2, 2L)$. It follows by the definition of $\gsso$ (\Cref{def:oracle}) that there is a path, say $P_{u_1,v_1}$, of weight at most $t\cdot d_G(u_1,v_1)$  between $u_1$ and $v_1$ in the graph induced by $F$. Let $P_{u,v} = P_{u_1,u}\circ P_{u_1,v_1}\circ P_{v,v_1}$ be the path between $u$ and $v$  obtained by concatenating $P_{u_1,u}, P_{u_1,v_1}, P_{v,v_1}$. By the triangle inequality, it follows that:
		\begin{equation}
			\begin{split}
				w(P_{u,v}) &\leq w(P_{u_1,v_1}) + w(P_{u_1,u}) + w(P_{v_1,v}) \leq t\cdot d_G(u_1,v_1) + 2\eps \beta L \\
				&\stackrel{\mbox{\footnotesize{\cref{eq:oracleStretch-uv-up}}}}{=} t\cdot ( w(u,v) + 2\eps\beta  L) + 2\eps\beta L \\
				&\leq t\cdot (w(u,v) + 4\eps\beta L) \leq t\cdot (1 + 4\eps  \beta)w(u,v) \qquad \mbox{(since $w(u,v)\geq L$ and $t\geq 1$)},
			\end{split}
		\end{equation}
		as desired. \qed 
	\end{proof}
	\begin{proof}[Proof of \Cref{thm:general-stretch-2}]
		By \Cref{lm:framework} and \Cref{lm:App-Oracle}, we can construct in polynomial time a spanner $H$ with stretch $t(1 + (2s_{\sso_{\oracle}}(O(1)) +  O(1))\eps)$ where $s_{\sso_{\oracle}}(\beta) = 8\beta$. Thus, the stretch of $H$ is $t(1 + O(\eps))$; we then can recover stretch $t(1+\eps)$ by scaling.  The lightness of $H$ is  $\tilde{O}_{\eps}((\chi \eps^{-1}))$ with  $\chi = O(\wsp_{\mathcal{O}_{G,t}})$. That implies a lightness of  $\tilde{O}_{\eps}((\wsp_{\mathcal{O}_{G,t}} \eps^{-1}))$  as claimed. \qed
	\end{proof}
	
	\begin{proof}[Proof of \Cref{thm:general-stretch-1eps}] The proof follows the same line of the proof of \Cref{thm:general-stretch-2}. The difference is that we  apply \Cref{lm:App-Oracle} and \Cref{lm:framework} with $t =  1+\eps$ to construct $H$. Thus, the stretch of $H$ is $t(1 + O(\eps)) = 1 + O(\eps)$. Since   $\chi  = \wsp_{\mathcal{O}_{G,1+\eps}}$, the lightness is $\tilde{O}_{\epsilon}\left(\frac{\wsp_{\mathcal{O}_{G,t}}}{\epsilon} + \frac{1}{\epsilon^2}\right)$ as claimed.	\qed
	\end{proof}

\subsection{Constructing General Sparse Spanner Oracles}\label{subsec:Oracle}

 We construct $\gsso$es for different class of graphs: general graphs, high dimensional metric spanners, and Steiner Euclidean spanners. This together with \Cref{thm:general-stretch-2} and \Cref{thm:general-stretch-1eps}
 give  \Cref{thm:light-general-spanner}, \Cref{thm:light-Steiner}, and \Cref{thm:Euclidean-high}.

 \subsubsection{General graphs and high dimensional metric spaces: Proof of \Cref{thm:light-general-spanner} and \Cref{thm:Euclidean-high}}\label{subsec:general}

 \begin{theorem}\label{thm:graph-oracles}The following $\gsso$es exist.
 	\begin{enumerate}[noitemsep]
 		\item For any weighted graph $G$ and any $k\geq 2$, $\wsp_{\mathcal{O}_{G,2k-1}} = O(g(n,k))$.
 		\item For the complete weighted graph $G$ corresponding to any Euclidean space (in any dimension) and for any $t\geq 1$, $\wsp_{\mathcal{O}_{G,O(t)}} = O(tn^{\frac{1}{t^2}}\log n)$.
 		\item For the complete weighted graph $G$ corresponding to any finite $\ell_p$ normed space for $p \in (1,2]$  and for any $t \ge 1$,  $\wsp_{\mathcal{O}_{G,O(t)}} = O(tn^{\frac{\log t}{t^p}}\log n)$.
 	\end{enumerate}
 \end{theorem}
 
 \Cref{thm:light-general-spanner} follows directly from \Cref{thm:general-stretch-2} and Item (1) of \Cref{thm:graph-oracles}; \Cref{thm:Euclidean-high}  follows directly from \Cref{thm:general-stretch-2} and Item (2) and Item (3) of \Cref{thm:graph-oracles} with  $\epsilon = 1/2$; any constant $\epsilon < 1$ works. See \Cref{fig:unifedFull} for a graphical illustration of the relationships between these theorems. We now focus on proving \Cref{thm:graph-oracles}.
 
 \paragraph{General graphs.~}  For a given graph $G(V,E)$ and $T\subseteq V$, we construct another weighted graph $G_T(T, E_T,w_T)$ with vertex set $T$ such that for every two vertices $u,v$ that form a critical pair, we add an edge $(u,v)$ with weight $w_T(u,v) = d_G(u,v)$.
 
 We apply the greedy algorithm~\cite{ADDJS93} to $G_T$ with $t = 2k-1$ and return the output of the greedy spanner, say $S_T$, (after replacing each artificial edge by the shortest path between its endpoints) as the output of the oracle $\mathcal{O}_{G,2k-1}$.  We now bound the weak sparsity of $\mathcal{O}_{G,2k-1}$.
 
 It was shown (Lemma 2 in~\cite{ADDJS93}) that $S_T$ has girth $2k+1$ and hence has at most $g(|T|,k)|T| \leq g(n,k)|T|$ edges.  It follows that $w(S_T) ~\leq~ |g(n,k)|T|2L ~=~ O(g(n,k))|T|L$. That implies:
 
 \begin{equation*}
 	\wsp_{\mathcal{O}_{G,2k-1}} = \sup_{T\subseteq V, L \in \mathcal{R}^+} \frac{O(g(n,k))|T|L}{|T|L} = O(n^{1/k}).
 \end{equation*}
 This implies Item (1) of \Cref{thm:graph-oracles}.

\paragraph{High dimensional metric spaces.~} Let $(X,d_X)$ be a metric space and $\mathcal{P}$ be a partition  of $(X,d_X)$ into clusters. We say  that $\mathcal{P}$ is  \emph{$\Delta$-bounded} if $\dm(P) \leq \Delta$ for every $P \in \mathcal{P}$.  For each $x \in X$, we denote the cluster containing $x$ in $\mathcal{P}$ by $\mathcal{P}(x)$. The following notion of $(t,\Delta,\delta$)-decomposition was introduced by Filtser and Neiman~\cite{FN18}.

\begin{definition}[($t,\Delta,\eta$)-decomposition] Given parameters $t \geq 1, \Delta > 0, \eta \in [0,1]$, a distribution $\mathcal{D}$ over partitions of $(X,d_X)$ is a $(t,\Delta,\eta)$-decomposition if:
	\begin{itemize}
		\item[(a)] Every partition $\mathcal{P}$ drawn from $\mathcal{D}$ is $t\cdot\Delta$-bounded.
		\item[(b)] For every $x\not= y \in X$ such that $d_X(x,y) \leq \Delta$, $\pr\limits_{\mathcal{P}\sim \mathcal{D}}[\mathcal{P}(x) = \mathcal{P}(y)] \geq \eta$
	\end{itemize}
\end{definition}

$(X,d)$ is $(t,\eta)$-decomposable if it has a ($t,\Delta,\eta$)-decomposition for any $\Delta > 0$.

\begin{claim}\label{clm:strong-sparse-decomposable} If $(X,d_X)$ is $(t,\eta)$-decomposable, it has a  $\gsso$ $\mathcal{O}_{X,O(t)}$ with sparsity $\wsp_{\mathcal{O}_{X,O(t)}} = O(\frac{t \log |X|}{\eta})$. Furthermore, there is a polynomial time Monte Carlo algorithm constructing  $\mathcal{O}_{X,O(t)}$ with constant success probability.
\end{claim}
\begin{proof}
	Let $T$ be a set of terminals given to the oracle $\mathcal{O}_{X,O(t)}$. Let $\mathcal{D}$ be a $(t, 2L,\eta)$-decomposition of $(X,d_X)$.
	
	Initially the spanner $S$ has $V(S) = T$ and $E(S) = \emptyset$. We sample $\rho = \frac{2\ln |T|}{\eta}$ partitions from $\mathcal{D}$, denoted by $\mathcal{P}_1, \ldots, \mathcal{P}_\rho$. For each $i \in [\rho]$ and each cluster $C \in \mathcal{P}_i$, if $|T\cap C| \geq 2$, we pick a terminal $t\in C$ and add to $S$ edges from $t$ to all other terminals in $C$. We then return $S$ as the output of the oracle.
	
	For each partition $\mathcal{P}_i$, the set of edges added to $S$ forms a forest. That implies we add to $S$ at most $|T|-1$ edges per partition. Thus, $|E(S)| \leq (|T|-1) \rho = O(\frac{|T| \log |T|}{\eta})$. Observe that $w(S) \leq |E(S)| \cdot t 2L = (\frac{2|T| t L \log |T| }{\eta})$ since each edge has weight at most $t\cdot (2L)$. Thus, $\wsp_{\mathcal{O}} = O(\frac{t \log |T|}{\eta})  =  O(\frac{t\log |X|}{\eta})$.
	
	It remains to show that with constant probability, $d_{S}(x,y)  \leq O(t)d_X(x,y)$ for every $x\not= y \in T$  such that $L \leq d_X(x,y) < 2L$. Observe by construction that if $x$ and $y$ fall into the same cluster in any partition, there is a $2$-hop path of length at most $4tL = O(t)d_X(x,y)$. Thus, we only need to bound the probability that $x$ and $y$ are clustered together  in some partition. Observe that the probability that there is no cluster containing both  $x$ and $y$ in $\rho$ partitions is at most:
	\begin{equation*}
		(1-\eta)^\rho  = (1-\eta)^{ \frac{2\ln |T|}{\eta}} \leq \frac{1}{|T|^2}
	\end{equation*}
	Since there are at most $\frac{|T|^2}{2}$ distinct pairs, by union bound, the desired probability is at least $\frac{1}{2}$.\qed
\end{proof}

Filtser and Neiman~\cite{FN18} showed that any $n$-point Euclidean metric is $(t,n^{-O(\frac{1}{t^2})})$-decomposable for any given $t > 1$; this implies Item (2) in \Cref{thm:graph-oracles}.  If $(X,d_X)$ is an $\ell_p$ metric with $p \in (1,2)$,  Filtser and Neiman~\cite{FN18} showed that it is $(t,n^{-O(\frac{\log t}{t^2})})$-decoposable for any given $t > 1$; this implies Item (3) in \Cref{thm:graph-oracles}.

\subsubsection{Steiner Euclidean Spanners}\label{subsec:Euclidean}

To prove \Cref{thm:light-Steiner}, we allow the oracle to include Steiner points, i.e., points in $\mathbb{R}^d\setminus P$ in the construction of $\gsso$ (\Cref{thm:general-stretch-1eps} remains true for $\gsso$ with Steiner points). Formally, a $\gsso$ with Steiner points, given a subset of points $T\subseteq P$ and a distance parameter $L > 0$,  outputs a Euclidean graph $S(V_S,E_S)$ with $T\subseteq V_S$ such that $d_S(x,y) \le (1+\epsilon) ||x,y||$ for any $x\not=y$ in $T$,\footnote{$||x,y||$ is the Euclidean distance between two points $x,y\in \mathbb{R}^d$.}  where $||x,y||\in [L,2L]$. We denote the oracle by $\mathcal{O}_{P,1+\epsilon}$. Our construction of the $\gsso$ with Steiner points uses the sparse Steiner $(1+\epsilon)$-spanner from our previous work~\cite{LS19} (in the full version) as a black-box.

\begin{theorem}[Theorem 1.3~\cite{LS19}]\label{thm:sparse-Steiner}  Given an $n$-point set $P \in \mathbb{R}^d$, there is a Steiner $(1+\epsilon)$-spanner for $P$ with  $\tilde{O}_{\epsilon}(\epsilon^{-(d-1)/2} |P|)$ edges.
\end{theorem}

\begin{theorem}\label{thm:Euclidean-oracle} Any point set $P$ in $\mathbb{R}^d$ admits a $\gsso$ with Steiner points that has  weak sparsity $\wsp_{\mathcal{O}_{P,t+\epsilon}} = \tilde{O}_{\epsilon}(\epsilon^{-(d-1)/2})$.
\end{theorem}

We note that  \Cref{thm:light-Steiner} follows directly from \Cref{thm:Euclidean-oracle} and \Cref{thm:general-stretch-1eps}.

\begin{proof}
	Let $T\subseteq P$ be a subset of points given to the oracle and $L$ be the distance parameter. By \Cref{thm:sparse-Steiner}, we can construct a Steiner $(1+\epsilon)$-spanner $S$ for $T$ with $|E(S)| = \tilde{O}_{\epsilon}(\epsilon^{-(d-1)/2} |T|)$. We observe that:
	
	\begin{observation}\label{obs:remove-heavy-edge} Let $x\not= y$ be two points in $T$ such that $||x,y||\leq 2L$, and $Q$ be a shortest path between $x$ and $y$ in $S$. Then, for any edge $e$ such that $w(e)\geq 4L$, $e\not\in P$ when $\epsilon < 1$.
	\end{observation}
	\begin{proof}
		Since $S$ is a $(1+\epsilon)$-spanner, $w(P)\leq (1+\epsilon)||x,y|| \leq (1+\epsilon)2 L < 4L$.\qed
	\end{proof}
	Let $\mathcal{O}_{P,(1+\epsilon)}(T,L)$ be the graph obtained from $S$ by removing every edge $e\in E(S)$ such that $w(e)\geq 4L$. By Observation~\ref{obs:remove-heavy-edge}, $\mathcal{O}_{P,(1+\epsilon)}(T,L)$ is a $(1+\epsilon)$-spanner for $T$. Observe that $$w(\mathcal{O}_{P,(1+\epsilon)}(T,L))~\leq~ 4L |E(\mathcal{O}_{P,(1+\epsilon)}(T,L))| \leq 4L |E(S)| ~=~ \tilde{O}_{\epsilon}(\epsilon^{-(d-1)/2} |T| L).$$ It follows that $\wsp_{\mathcal{O}_{P,1+\epsilon}} = \tilde{O}_{\epsilon}(\epsilon^{-(d-1)/2})$. This completes the proof of \Cref{thm:Euclidean-oracle}.
\end{proof}

\subsection{Light Spanners for Minor-Free Graphs}\label{subsec:minor-light}

In this section, we provide an implementation of \hyperlink{SPSSO}{$\sso$} for minor-free graphs, which we denote by $\sso_{\minor}$. The algorithm simply outputs the edge set $\me$. Note that in this case, we set $t = 1+\eps$.

\begin{tcolorbox}
	\hypertarget{SPHMinorSSO}{}
	\textbf{$\sso_{\minor}$:} The input is an $(L,\eps,\beta)$-cluster graph $\mg=(\mv,\me,\omega)$. The output is a set of edges $F$. 
	\begin{quote}
		 Let $F$ be the subset of edges of $G$ that correspond to edges in $\me$. We then return $F$.
	\end{quote}
\end{tcolorbox}

We now show that $\sso_{\minor}$ has all the properties as described in the abstract  \hyperlink{SPSSO}{$\sso$}, which implies \Cref{thm:minor-free-opt-lightness}. 

\MinorFree*
\begin{proof}  
	Since we add every edge corresponds to an edge in $\me$ in \hyperlink{SPHMinor}{$\sso_{\minor}$}, $s_{\sso_{\minor}}(\beta) = 0$.
		By \Cref{lm:framework} and \Cref{lm:App-Oracle}, we can construct in polynomial time a spanner $H$ with stretch $t(1 + (2s_{\sso_{\minor}}(O(1)) +  O(1))\eps) = (1 + O(\eps))$; note that $t = (1+\eps)$ in this case. We then can recover stretch $(1+\eps)$ by scaling. 
	
	We observe that $\mg$ is a minor of $G$ and hence is $K_r$-minor-free. Thus, by the sparsity of minor-free graphs, $|\me| = O(r\sqrt{\log r})|\mv|$. It follows that $w(F) = O(r\sqrt{\log r})L\cdot |\mv|$ since every edge in $\mg$ has weight at most $2L$. This gives $\chi = O(r\sqrt{\log r})$. By \Cref{lm:framework} for the case $t = 1+\eps$,
	The lightness of $H$ is  $\tilde{O}_{\eps}((\chi \eps^{-1}) + \eps^{-2}) = \tilde{O}_{\eps,r}(r\eps^{-1} + \eps^{-2})$  as claimed. \qed
	\end{proof}

\newpage
\part{Our Unified Framework: The Proof (\Cref{sec:framework} --- \Cref{sec:stretch1E})}\label{partII}
In this part, we present the proof of \Cref{lm:framework} in detail.  We start by setting in a technical framework on which the proof rests.

\section{Unified Framework: Technical Setup}\label{sec:framework}
 
In \Cref{subsec:framework-details}, we outline a technical framework that we use to prove \Cref{lm:framework}. The proof of \Cref{lm:framework} boils down to constructions of clusters and associated subgraphs. In \Cref{sec:FastProof}, we show how to design a fast algorithm to find the clusters and the subgraphs.   In \Cref{sec:LightProof}, we construct the clusters and the subgraphs that have a small dependency on $1/\eps$.

\subsection{The Framework}\label{subsec:framework-details}
Our starting point is a basic hierarchical partition, which dates back to the early 90s \cite{ALGP89,CDNS92}, and was used by most if not all of the works on light spanners (see, e.g.,~\cite{ES16,ENS14,CW16,BLW17,BLW19,LS19}).
The current paper takes this hierarchical partition approach to the next level by proposing a unified framework.  

Let $\MST$ be a minimum spanning tree of the input $n$-vertex $m$-edge graph $G = (V,E,w)$. Let $T_{\MST}$ be the running time needed to construct $\MST$. By scaling, we shall assume w.l.o.g.\ that the minimum edge weight is $1$. Let $\bar{w} = \frac{w(\mst)}{m}$. We remove from $G$ all edges of weight larger than $w(\MST)$; such edges do not belong to any shortest path, hence 
removing them does not affect the distances between vertices in $G$.  We define two sets of edges, $E_{light}$ and $E_{heavy}$, as follows:
\begin{equation}\label{Eprimedef}
	E_{light} = \{e\in E: w(e) \leq \frac{\bar{w}}{\eps}\} \qquad \& \qquad E_{heavy} = E\setminus E_{light}
\end{equation}

It could be that $\frac{\bar{w}}{\eps} < 1$; in this case, $E_{light} = \emptyset$. 
The next observation follows from the definition of $\bar{w}$.

\begin{observation}\label{obs:Eprime-weight} $w(E_{light}) \leq  \frac{w(\mst)}{\epsilon}$.
\end{observation}

Recall that the parameter $\eps$ is in the stretch $t(1+\eps)$ in \Cref{lm:framework}. It controls the stretch blow-up in \Cref{lm:framework}, and ultimately, the stretch of the final spanner. There is an inherent trade-off between the stretch blow-up (a factor of $1+\eps$) and the blow-up of the other parameters, including runtime and lightness, by at least a factor of $1/\eps$. 

By \Cref{obs:Eprime-weight}, we can safely add $E_{light}$ to our final spanner, while paying only an additive $+\frac{1}{\epsilon}$ factor to the lightness bound. Hence, as the stretch of a spanner is realized by some edge of the graph, in the spanner construction that follows, it suffices to focus on the stretch for edges in $E_{heavy}$.
 Next, we partition the edge set $E_{heavy}$ into subsets of edges, such that for any two edges $e, e'$ in the same subset, their weights are either \emph{almost the same} (up to a factor of $1+\psi$)
or they are \emph{far apart} (by at least a factor of $\frac{1}{\eps(1+\psi)}$), where $\psi$ is a parameter to be optimized later. In fast constructions (\Cref{sec:FastProof}), we choose $\psi = \eps$ and in optimal lightness constructions (\Cref{sec:LightProof}), we choose $\psi = 1/250$. 

\begin{definition}[Partitioning $E_{heavy}$]\label{def:refineEdprime} Let $\psi$ be any parameter
in the range  $(0,1]$. Let $\mu_{\psi} = \lceil \log_{1+\psi}\frac{1}{\eps}\rceil$. We partition  $E_{heavy}$ into subsets $\{E^{\sigma}\}_{\sigma \in [\mu_\psi] }$ such that $ E^{\sigma} = \cup_{i\in \mathbb{N}^+} E^{\sigma}_i$ where:
	\begin{equation}\label{eq:Esigmaixdef}
		E^{\sigma}_i = \left\{e : \frac{L_i}{1+\psi} \leq w(e) < L_i \right\} \mbox{ with } L_i = L_{0}/\eps^i, L_0 = (1+\psi)^{\sigma}\bar{w}~. 
	\end{equation}
\end{definition}

By definition, we have $L_i = L_{i-1}/\eps$ for each $i \ge 1$. Readers may notice that if $\log_{1+\psi}\frac{1}{\eps}$ is not an integer, by the definition of $E^{\sigma}$, it could be that $E^{\mu_{\psi}}\cap E^{1} \not= \emptyset$, in which case $\{E^{\sigma}\}_{\sigma \in [\mu_\psi] }$ is not really a partition of $E_{heavy}$. This can be fixed by taking to $E^{\mu_\psi}$ edges that are not in $\cup_{1\leq \sigma \leq \mu_{\psi}-1} E^{\sigma}$. We henceforth assume that  $\{E^{\sigma}\}_{\sigma \in [\mu_\psi] }$  is a partition of $E_{heavy}$.  The following lemma shows that it suffices to focus on the stretch of edges in $E^{\sigma}$, for an arbitrary $\sigma \in [\mu_{\psi}]$.

\begin{lemma}\label{lm:ReductionToEsigma} If for every $\sigma \in [\mu_{\psi}]$, we can construct a $k$-spanner $H^{\sigma}\subseteq G$ for $E^{\sigma}$ with lightness at most $\lt_{H^{\sigma}}$ in time $\tm_{H^{\sigma}}(m,n)$ (where $\lt_{H^{\sigma}}$ and $\tm_{H^{\sigma}}(m,n)$ do not depend on $\sigma$), then we can construct a $k$-spanner for $G$ with lightness $O\left(\frac{\lt_{H^{\sigma}}\log(1/\eps)}{\psi} + \frac{1}{\eps}\right)$ in time $O\left(\frac{\tm_{H^{\sigma}}(m,n) \log(1/\eps)}{\psi} + T_{\mst}\right)$.
\end{lemma}
\begin{proof}
Let $H$ be a graph with $V(H) = V(G)$ and $E(H) = E_{light} \cup \left( \cup_{\sigma \in [\mu_{\psi}]}H^{\sigma}\right)$. The fact that $H$ is a $k$-spanner of $G$ follows directly from the fact that the stretch of a spanner is realized by some edge of the graph. The lightness bound follows from the fact that $\mu_{\psi} = O(\frac{\log(1/\eps)}{\log(1+\psi)}) = O(\log(1/\eps)/\psi)$ and \Cref{obs:Eprime-weight}.

To bound the running time, note that the time needed to construct $E_{light}$ is $T_{\mst} + O(m) = O(T_{\mst})$. Since we remove edges of weight at least $\mst$ from $G$ and every edge in $E_{heavy}$ has a weight at least $\frac{\bar{w}}{\eps} = \frac{w(\mst)}{\eps m}$, the number of sets that each $E^{\sigma}$ is partitioned to is $O(\log_{1/((1+\psi)\eps)}(\eps m)) = O(\log(m))$ for any $\eps \leq 1/2$. Thus, the partition of $E_{heavy}$ can be trivially constructed in $O(m)$ time. The running time bound now follows. \qed
\end{proof}

 We shall henceforth focus on constructing a spanner for $E^{\sigma}$, for an arbitrarily fixed $\sigma \in [\mu_{\psi}]$. In what follows we present a clustering framework for constructing a spanner $H^{\sigma}$ for $E^{\sigma}$  with \emph{stretch $t(1+\epsilon)$}. We will assume that $\epsilon$ is sufficiently smaller than $1$.

\paragraph{Subdividing $\mst$.~} We subdivide each edge $e \in \mst$ of weight more than $\bar{w}$ into $\lceil \frac{w(e)}{\bar{w}} \rceil$ edges of weight (of at most $\bar{w}$ and at least $\bar{w}/2$ each) that sums to $w(e)$. (New edges do not have to have equal weights.)  Let $\widetilde{\mst}$ be the resulting subdivided $\mst$.
We refer to vertices that are subdividing the $\mst$  edges as \emph{virtual vertices}. Let $\tilde V$ be the set of vertices in $V$ and virtual vertices; we call $\tilde{V}$  the {\em extended set} of vertices. Let $\tilde G = (\tilde V,\tilde E)$ be the graph that consists of the edges in $\widetilde{\mst}$ and $E^{\sigma}$. 
\begin{observation}\label{obs:Edge-Gtilde}
	$|\tilde{E}|  = O(m)$.
\end{observation}
\begin{proof}
It suffices to show that
$|E(\widetilde{\mst})| = O(m)$. Indeed, since $w(\widetilde{\mst}) = w(\mst)$ and each edge of $\widetilde{\mst}$ has weight at least $\bar{w}/2$, we have $|E(\widetilde{\mst})| \le 2m$.	\qed
\end{proof}

The $t(1+\eps)$-spanner that we construct for $E^{\sigma}$ is a subgraph of $\tilde{G}$ containing all edges of $\widetilde{\mst}$; we can enforce this assumption by adding the edges of $\widetilde{\mst}$ to the spanner. By replacing the edges of $\widetilde{\mst}$ by those of $\mst$, we can transform any subgraph of $\tilde{G}$ that contains the entire tree $\widetilde{\mst}$ to a subgraph of $G$ that contains the entire tree $\mst$. We denote  by $\tilde H^{\sigma}$ the $t(1+\eps)$-spanner of $E^{\sigma}$ in $\tilde{G}$;
by abusing the notation, we will write $H^{\sigma}$ rather than $\tilde H^{\sigma}$ in the sequel,
under the understanding that in the end we transform $H^{\sigma}$ to a subgraph of $G$.

Recall that $E^{\sigma} = \cup_{i\in \mathbb{N}^+} E^{\sigma}_i$ where $E^{\sigma}_i$ is the set of edges defined in \Cref{eq:Esigmaixdef}. We refer to   edges in $E^{\sigma}_i$ as   \emph{level-$i$ edges}. We say that a level $i$ is empty if the set $E^{\sigma}_i$  of level-$i$ edges is empty;
in the sequel, we shall only consider the nonempty levels. 

\begin{claim}\label{clm:numLevels} The number of (nonempty) levels is $O(\log m)$.
\end{claim}
\begin{proof} The claim follows from the fact that every edge of $E^{\sigma}$ has weight at least $\frac{\bar{w}}{\epsilon}$ and at most $w(\mst) = m\bar{w}$, and the weight of edges in $E^\sigma_{i+1}$ is at least $\frac{1}{(1+\psi)\epsilon} $ times the weight of edges	$E^\sigma_{i}$.\qed 
\end{proof}

Our construction crucially relies on a \emph{hierarchy of clusters}. A {\em cluster} in a graph is simply a subset of vertices in the graph.  Nonetheless, as will become clear soon, we care also about {\em edges} connecting vertices in the cluster, and of the properties that these edges possess. Our hierarchy of clusters, denoted by $\mathcal{H} = \{\mathcal{C}_1,\mathcal{C}_2, \ldots \}$ satisfies the following properties:  
	\begin{itemize}  [noitemsep] 

		\item \textbf{(P1)~} 	\hypertarget{P1}{} For any $i\geq 1$, each $\mathcal{C}_i$ is a partition of $\tilde{V}$. When $i$ is large enough, $\mathcal{C}_i$ contains a single set $\tilde{V}$ and $\mathcal{C}_{i+1} = \emptyset$.
		\item \textbf{(P2)~} \hypertarget{P2}{} $\mathcal{C}_i$ is an \emph{$\Omega(\frac{1}{\eps})$-refinement} of $\mathcal{C}_{i+1}$, i.e., every cluster $C\in \mathcal{C}_{i+1}$ is obtained as the union of $\Omega(\frac{1}{\epsilon})$ clusters in $\mathcal{C}_i$ for $i\geq 1$.
		\item \textbf{(P3)~} \hypertarget{P3}{} For each cluster $C\in \mathcal{C}_i$, we have $\dm(H^{\sigma}[C]) \leq g L_{i-1}$, for a sufficiently large constant $g$ to be determined later. (Recall that $L_i$ is defined in \Cref{eq:Esigmaixdef}.) 
	\end{itemize}

\begin{remark}\label{remark:diameter} (1) We construct $H^{\sigma}$ along with the cluster hierarchy. Suppose that at some step $s$ of the algorithm, we construct a level-$i$ cluster $C$. Let $H^{\sigma}_s$ be $H^{\sigma}$ at step $s$. We shall maintain (\hyperlink{P3}{P3}) by maintaining the invariant that $\dm(H^{\sigma}_s[C]) \leq g L_{i-1}$; indeed,  adding more edges in later steps of the algorithm does not increase the diameter of the subgraph induced by $C$. 

(2) It is time-expensive to compute the diameter of a cluster {\em exactly}.  Thus, we explicitly associate with each cluster $C\in \mathcal{C}_i$ a proxy parameter of the diameter  during the course of the construction. This proxy parameter has two properties: (a) it is at least the diameter of the cluster, and (b) it is lower-bounded by $\Omega(L_{i-1})$. Property (a) is crucial in arguing for the stretch of the spanner. Property (b) is crucial to  have an upper bound on the number of level-$i$ clusters contained in a level-$(i+1)$ cluster, which speeds up its (the level-$(i+1)$ cluster's)  construction.
\end{remark}

 When $\eps$ is sufficiently small, specifically smaller than the constant hiding in the $\Omega$-notation in property (\hyperlink{P2}{P2}) by at least a factor of 2, it holds that $|\mc_{i+1}| \leq |\mc_i|/2$, yielding a geometric decay in the number of clusters at each level of the hierarchy. This geometric decay is crucial to our fast constructions. 

Our construction of the cluster hierarchy $\mathcal{H}$ will be carried out level by level, starting from level $1$. After we construct the set  of level-$(i+1)$ clusters, we compute a subgraph $H^{\sigma}_{i}\subseteq G$ as stated in \Cref{lm:framework}. The final spanner $H^{\sigma}$ is obtained as the union of all subgraphs $\{H^{\sigma}_i\}_{i \in \mathbb{N}^+}$.  To bound the weight of $H^{\sigma}$, we rely on a potential function $\Phi$ that is formally defined as follows:

\begin{definition}[Potential Function $\Phi$]\label{def:Potential}  We use a potential function $\Phi: 2^{\tilde{V}}\rightarrow \mathbb{R}^+$ that maps each cluster $C$ in the hierarchy $\mathcal{H}$ to a potential value $\Phi(C)$, such that the total potential of clusters at level $1$ satisfies:
	\begin{equation}\label{eq:Phi1}
		 \sum_{C\in\mathcal{C}_1}\Phi(C) ~\leq~ w(\MST)~.
	\end{equation}
 Level-$i$ potential is defined as $\Phi_i = \sum_{C\in \mathcal{C}_i} \Phi(C)$ for any $i\geq 1$. The \emph{potential change} at level $i$, denoted by $\Delta_i$ for every $i \geq 1$, is defined as:
\begin{equation}\label{eq:PotentialReduction}
	 \Delta_i ~=~ \Phi_{i-1} - \Phi_{i}~. 
\end{equation}
\end{definition} 

The key to our framework is the following lemma.

\begin{restatable}{lemma}{FrameworkTechnical}
	\label{lm:framework-technical} Let $\psi \in (0,1], t \geq 1, \eps \in (0,1)$ be  parameters, and $E^{\sigma}= \cup_{i\in \mathbb{N}^+} E^{\sigma}_i$ be the set of edges defined in Equation~\eqref{eq:Esigmaixdef}. Let $\{a_i\}_{i \in \mathbb{N}^+}$ be a sequence of positive real numbers such that $\sum_{i \in \mathbb{N}^+} a_i \leq A\cdot w(\mst)$ for some $A\in \mathbb{R}^+$. Let $H_0 = \mst$. For any level $i\geq 1$, if we can compute all subgraphs  $H_1,\ldots,H_i\subseteq G$  as well as the cluster sets $\{\mathcal{C}_{1},\ldots,\mathcal{C}_{i},\mathcal{C}_{i+1}\}$ in total runtime  $O(\sum_{j=1}^i(|\mathcal{C}_j| + |E^{\sigma}_j|)f(n,m) + m)$ for some function $f(\cdot,\cdot)$ such that:
	\begin{enumerate}[noitemsep]
		\item[(1)] $w(H_i) \leq  \lambda \Delta_{i+1} + a_i$ for some $\lambda \geq 0$,
		\item[(2)] for every $(u,v)\in E^{\sigma}_i$, $d_{H_{< L_i}}(u,v)\leq t(1+ \rho\cdot \epsilon)w(u,v)$ when $\eps \in (0,\eps_0)$ for some constants $\rho$ and $\eps_0$, where $H_{< L_i}$ is the spanner constructed for edges of $G$ of weight less than $L_i$. 
	\end{enumerate}
	Then  we  can construct a $t(1+ \rho \eps)$-spanner for $G(V,E)$ with lightness  $O(\frac{\lambda + A + 1}{\psi}\log \frac{1}{\epsilon} + \frac{1}{\eps})$ in time $ O(\frac{mf(n,m)}{\psi}\log \frac{1}{\epsilon} + T_{\mst})$ when $\eps \in (1,\eps_0)$.
\end{restatable}

\begin{proof} Let $H^{\sigma} = \cup_{i\in \mathbb{N}} H_i$.  The stretch bound $t(1+\rho\eps)$ follows directly from the fact that $E^{\sigma}= \cup_{i\in \mathbb{N}^+}E^{\sigma}_{i}$, Item (2), and \Cref{lm:ReductionToEsigma}. By condition (1) of Lemma~\ref{lm:framework} and \Cref{eq:Phi1}, we have: 
	\begin{equation*}
		\begin{split}
			w(H^{\sigma}) \leq \lambda \sum_{i \in \mathbb{N}^+}\Delta_i + \sum_{i \in \mathbb{N}^+} a_i + w(\mst) ~\leq~ 
			\lambda\cdot \Phi_1 + A\cdot w(\mst) + w(\mst) \leq (\lambda + A+1)w(\mst)~. 
		\end{split}
	\end{equation*}
	
	This and \Cref{lm:ReductionToEsigma} implies the lightness upper bound; here $\lt_{H^{\sigma}} = (O(\lambda) + A+1)$. To bound the running time, we note that $\sum_{i \in \mathbb{N}^+}|E^{\sigma}_i| \leq m$ and by property (\hyperlink{P2}{P2}), we have $\sum_{i \in \mathbb{N}^+} |\mathcal{C}_i| = |\mathcal{C}_1|\sum_{i \in \mathbb{N}^+}\frac{O(1)}{\epsilon^{i+1}} = O(|\mathcal{C}_1|) = O(m)$. Thus, by the assumption of  Lemma~\ref{lm:framework}, the total running time to construct $H^{\sigma}$ is $\tm_{H^{\sigma}}(m,n) = O\left((\sum_{i \in \mathbb{N}^+}(|\mathcal{C}_i|) + |E_i|)f(m,n) + m\right) = O\left( mf(m,n)\right)$. 
	Plugging this runtime bound on top of Lemma~\ref{lm:ReductionToEsigma} yields the required runtime bound in Lemma~\ref{lm:framework}. \qed
\end{proof}

\begin{remark}\label{remark:stretchLevel-i}In \Cref{lm:framework-technical}, we construct spanners for edges of $G$ level by level, starting from level $1$. By Item (2), when constructing spanners for edges in $E^{\sigma}_i$, we could assume by induction that all edges of weight less than $L_i/(1+\psi)$ already have stretch $t(1+\rho\eps)$ in the spanner constructed so far, denoted by $H_{< L_i/(1+\psi)}$. By defining $H_{<L_i} = H_{< L_i/(1+\psi)}\cup H_i$, we get a spanner for edges of length less than $L_i$. 
\end{remark}

In summary, two important components in our spanner construction is a hierarchy of clusters and a potential function as defined in \Cref{def:Potential}.  In \Cref{subsec:DesignPotential}, we present  a construction of level-$1$ clusters and a general principle for assigning potential values to clusters. The construction of clusters at any level $i+1$ for $i\geq 1$, which basically gives the proof of \Cref{lm:framework}, is presented in \Cref{sec:FastProof} and \Cref{sec:LightProof}.

 \subsection{Designing A Potential Function}\label{subsec:DesignPotential}
 
In this section, we present in detail the underlying principle used to design the potential function $\Phi$ in \Cref{def:Potential}. We start by constructing and assigning potential values for level-$1$ clusters.

 \begin{lemma}\label{lm:level1Const} In time $O(m)$, we can construct a set of level-$1$ clusters $\mathcal{C}_1$ such that, for each cluster $C\in \mathcal{C}_1$, the subtree $\msttilde[C]$
 	of $\msttilde$ induced by $C$ satisfies $L_0 \leq \dm(\msttilde[C]) \leq 14L_0$. 
 \end{lemma}
 \begin{proof} We apply a simple greedy construction to break $\msttilde$ into a set $\mathcal{S}$ of subtrees of diameter at least $L_0$ and at most $5L_0$ as follows. (1) Repeatedly pick a vertex $v$ in a component $T$ of diameter at least $4L_0$, break a minimal subtree of radius at least $L_0$ with center $v$ from $T$, and add the minimal subtree to $\mathcal{S}$. (2) For each remaining component $T'$ after step (1), there must be an $\msttilde$ edge $e$ connecting $T'$ and  a subtree $T\in \mathcal{S}$ formed in step (1); we add $T'$ and $e$ to $T$. Finally, we form $\mathcal{C}_1$ by taking the vertex set of each subtree in $\mathcal{S}$ to be a level-$1$ cluster.   The running time bound follows directly from the construction.
 	
 	We now bound the diameter of each subtree in $\mathcal{S}$. In step (1), the diameter is at most 
 	$2(L_0+\bar{w})$. In step (2), each subtree $T$ is augmented by subtrees of diameter at most $4L_0$ via $\widetilde{\mst}$ edges in a star-like way. Thus, the diameter of the resulting subtrees is at most $2(L_0+\bar{w}) + 2(4L_0 + \bar{w}) \le 14L_0$, as required.  \qed
 \end{proof}
 
 By choosing $g\geq 14$, clusters in $\mathcal{C}_1$ satisfy properties (\hyperlink{P1}{P1}) and (\hyperlink{P3}{P3}). Note that (\hyperlink{P2}{P2}) is not applicable to level-$1$ clusters by definition. As for (\hyperlink{P3}{P3}), $\dm(H^{\sigma}[C]) \leq 14 L_{0}$, for each $C \in \mathcal{C}_1$.
 
 Next, we assign a potential value for each level-$1$ cluster as follows:
 
 \begin{equation}\label{eq:Level1Poten}
 	\Phi(C) = \dm(\msttilde[C]) \qquad \forall C \in \mathcal{C}_1
 \end{equation}
 We now claim that the total potential of all clusters at level $1$ is at most $w(\mst)$ as stated in \Cref{def:Potential}.
 
 \begin{lemma}\label{lm:Level1Poten} $\Phi_1 \leq w(\mst)$. 
 \end{lemma}
 \begin{proof}
 	By definition of $\Phi_1$, we have: $$\Phi_1 ~=~ \sum_{C\in \mathcal{C}_1} \Phi(C) ~=~ \sum_{C\in \mathcal{C}_1}\dm(\msttilde[C]) ~\leq~ \sum_{C\in \mathcal{C}_1} w(\msttilde[C]) ~\leq~ w(\msttilde) ~=~ w(\mst)~.$$
 	The penultimate inequality holds since level-$1$ clusters induce vertex-disjoint subtrees of $\msttilde$. \qed
 \end{proof}
 
 While the potential of a level-1 cluster is the diameter of the subtree induced by the cluster, the potential assigned to a cluster at level at least $2$ need not be the diameter of the cluster. Instead, it is an {\em overestimate} of the cluster's diameter, as imposed by the following {\em potential-diameter (PD) invariant}.
 
\hypertarget{PD}{}
 \begin{quote}
 	\textbf{PD Invariant:} For every cluster $C \in \mathcal{C}_{i}$ and any $i\geq 1$, $\dm(H_{< L_{i-1}}[C]) \leq \Phi(C)$. (Recall that $H_{< L_{i-1}}$ is the spanner constructed for edges of $G$ of weight less than $L_{i-1}$, as defined in \Cref{lm:framework-technical}.)
 \end{quote}
 
 \begin{remark}\label{remark:potential-diameter} As discussed in \Cref{remark:diameter}, it is time-expensive to compute the diameter of each cluster. By the \hyperlink{PD}{PD Invariant}, we can use the potential $\Phi(C)$ of a cluster $C \in \mathcal{C}_i$ as an upper bound on the diameter of $H_{< L_{i-1}}[C]$. As we will demonstrate in \Cref{sec:FastProof}, $\Phi(C)$ can be computed efficiently.
 \end{remark}
 
 To define potential values for clusters at levels at least $2$, we introduce a {\em cluster graph}, in which the nodes correspond to clusters.
 We shall derive the potential values of clusters via their \emph{structure} in the cluster graph, as described next.

 \begin{definition}[Cluster Graph]\label{def:ClusterGraphNew} A cluster graph at level $i \geq 1$, denoted by $\mg_i = (\mv_i, \me'_i, \omega)$, is a \emph{simple graph} where each node corresponds to a cluster in $\mc_i$ and each inter-cluster edge corresponds to an edge between vertices that belong to the corresponding clusters. We assign  weights to both \emph{nodes and edges} as follows:  for each node $\varphi_C \in \mv_i$ corresponding to a cluster $C \in \mathcal{C}_i$, $\omega(\varphi_C) = \Phi(C)$, and for each edge $\mbe = (\varphi_{C_u},\varphi_{C_v}) \in \me'_i$ corresponding to an edge $(u,v)$ of $\tilde{G}$, $\omega(\mbe) = w(u,v)$.   
 \end{definition} 
 
 \begin{remark}The  notion of cluster graphs in \Cref{def:ClusterGraphNew} is slightly different from  that of $(L,\eps,\beta)$-cluster graphs defined in \Cref{def:ClusterGraph-Param}. In particular, cluster graphs in \Cref{def:ClusterGraphNew} have weights on both edges and nodes, while $(L,\eps,\beta)$-cluster graphs in \Cref{def:ClusterGraph-Param} have weights on edges only. 
 \end{remark}
 
 In our framework, we want the cluster graph $\mg_i$ to have the following basic properties. 
 
 \begin{definition}[Properties of $\mg_i$]\label{def:GiProp} \begin{enumerate}
 		\item[(1)] The edge set $\me'_i$ of $\mg_i$ is the union $\msttilde_{i}\cup \me_i$, where $\msttilde_{i}$ is the set of edges corresponding to edges in $\msttilde$ and $\me_i$ is the set of edges corresponding to \emph{a subset of} edges in $E^{\sigma}_i$.
 		\item[(2)] $\msttilde_{i}$ induces a spanning tree of $\mg_i$. We abuse notation by using $\msttilde_{i}$ to denote the induced spanning tree.
  	\end{enumerate}
 \end{definition}

 At the outset of the construction of \emph{level-$(i+1)$ clusters},
 	we construct a cluster graph $\mg_i$. We assume that the spanning tree $\msttilde_{i}$ of $\mg_i$ is given, as we construct the tree  by the end of the construction of level-$i$ clusters. After we complete the construction of level-$(i+1)$ clusters, we construct $\msttilde_{i+1}$ for the next level.

 \begin{observation}\label{obs:Level1MST} At level $1$, both $\mv_1$ and $\msttilde_1$ can be constructed in $O(m)$ time.
 \end{observation} 
 \begin{proof} Edges of $\msttilde_1$ correspond to the edges of $\msttilde$ that do not belong to
 	any level-1 cluster, i.e., to any $\msttilde[C]$, where $C \in \mathcal{C}_1$. Thus, the observation follows from \Cref{obs:Edge-Gtilde} and \Cref{lm:level1Const}. \qed
 \end{proof}

 \paragraph{The structure of level-$(i+1)$ clusters.~} 
 Next, we describe how to construct the level-$(i+1)$ clusters via the cluster graph $\mg_i$. 
 We shall construct a collection of subgraphs $\mathbb{X}$ of $\mg_i$, and then map each subgraph $\mx \in \mathbb{X}$ to a cluster $C_{\mx} \in \mathcal{C}_{i+1}$ as follows:
 \begin{equation}\label{eq:XtoCluster}
 	C_{\mathcal{X}}  = \cup_{\varphi_C\in \mv(\mx)} C~.
 \end{equation}

That is, $C_{\mathcal{X}}$ is the union of all level-$i$ clusters that correspond to nodes in $\mathcal{X}$. 
 
 For any subgraph $\mx$ in a cluster graph, we denote by $\mv(\mx)$ and $\me(\mx)$ the vertex and edge sets of $\mx$, respectively. To guarantee properties (\hyperlink{P1}{P1})-(\hyperlink{P3}{P3}) defined before \Cref{remark:diameter} for clusters in $\mathcal{C}_{i+1}$, we will make sure that subgraphs in $\mathbb{X}$  satisfy the following properties:
 
 \begin{itemize}[noitemsep]
 	\item \textbf{(P1').~} \hypertarget{P1'}{}  $\{\mv(\mx)\}_{\mx \in \mathbb{X}}$ is a partition of $\mv_i$.
 	\item \textbf{(P2').~} \hypertarget{P2'}{} $|\mv(\mx)| = \Omega(\frac{1}{\eps})$.
 	\item \textbf{(P3').~} \hypertarget{P3'}{} $L_i \leq \adm(\mx) \leq gL_{i}$.
 \end{itemize}
 
 Recall that $\adm(\mx)$ is the augmented diameter of $\mx$, a variant of diameter defined for graphs with weights on both nodes and edges, see \Cref{sec:prelim}. Recall that the augmented diameter of $\mx$ is at least the diameter of the corresponding cluster $C_{\mx}$. 
 
 We then set the potential of cluster $C_{\mx}$ corresponding to subgraph $\mx$ as:
 	\begin{equation}\label{eq:SetPotential-i}
 		\Phi(C_{\mx}) = \adm(\mx).
 	\end{equation}
 
  Thus, the augmented diameter of any such subgraph $\mx$ will be the weight of the corresponding node in the level-$(i+1)$ cluster graph $\mg_{i+1}$.   Our  goal is to construct $H_i$ along with $\mathcal{C}_{i+1}$ as guaranteed by \Cref{lm:framework}.  $H_i$ consists of a subset of the edges in $E^{\sigma}_i$ (and in the case of optimal lightness constructions, some edges of $G$ as well). We can assume that the vertex set of $H_i$ is just the entire set $V$. Up to this point, we have not explained yet how $H_i$ is constructed, since the exact construction of $H_i$ depends on specific incarnations of our framework, which may change from one graph class to another. 
 
While properties (\hyperlink{P1'}{P1'}) and (\hyperlink{P2'}{P2'}) directly imply properties (\hyperlink{P1}{P1}) and (\hyperlink{P2}{P2}) of $C_{\mx}$, property (\hyperlink{P3'}{P3'}) does not directly imply property (\hyperlink{P3}{P3}); although the diameter of any weighted subgraph (with edge and vertex weights) is upper bounded by its augmented diameter, we need to guarantee that the (corresponding) edges of $\mx$ belong to $H_{< L_i}$. Indeed, without this condition, the diameter of $H_{< L_i}$ could be much larger than the augmented diameter of $\mx$.
 
 \begin{lemma}\label{lm:PropEquiv} Let $\mx \in \mathbb{X}$ be a subgraph of $\mg_i$ satisfying properties (\hyperlink{P1'}{P1'})-(\hyperlink{P3'}{P3'}). Suppose that for every edge $(\varphi_{C_u},\varphi_{C_v})\in \me(\mx)$, $(u,v) \in H_{< L_i}$.  By setting the potential value of $C_{\mx}$ to be $\Phi(C_{\mx}) = \adm(\mx)$ for every $\mx \in \mathbb{X}$, the \hyperlink{PD}{PD Invariant} is satisfied, and that  $C_{\mx}$ satisfies all properties (\hyperlink{P1}{P1})-(\hyperlink{P3}{P3}).
 \end{lemma}
 \begin{proof} It can be seen directly that properties (\hyperlink{P1'}{P1'}) and (\hyperlink{P2'}{P2'}) of $\mx$ directly imply properties (\hyperlink{P1}{P1}) and (\hyperlink{P2}{P2}) of $C_{\mx}$, respectively. We prove, by induction on $i$,
that property (\hyperlink{P3}{P3}) holds and that the  \hyperlink{PD}{PD Invariant} is satisfied. The basis $i=1$ is trivial.  For the induction step, we assume inductively that for each cluster $C\in \mathcal{C}_{i}$, $\dm(H_{< L_{i-1}})[C] \leq gL_{i-1}$ and that the \hyperlink{PD}{PD Invariant} is satisfied: $\Phi(C)\geq \dm(H_{< L_{i-1}})[C]$. Consider any level-$(i+1)$ cluster $C_{\mx}$ corresponding to a subgraph $\mx \in \mathbb{X}$. Let $H_{C_{\mx}}$ be the graph obtained by first taking the union $\cup_{\varphi_{C} \in \mv(\mx)}H_{< L_{i-1}}[C]$  and then adding in the edge set  $ \{(u,v)\}_{(\varphi_{C_u},\varphi_{C_v}) \in \me(\mx)}$.  Observe that $H_{C_{\mx}}$ is a subgraph of $H_{< L_i}$ by the assumption that  $(u,v) \in H_{< L_i}$ for every edge $(\varphi_{C_u},\varphi_{C_v})\in \me(\mx)$. We now show that $\dm(H_{C_{\mx}}) \leq \adm(\mx)$, which is at most $gL_i$ by property (\hyperlink{P3'}{P3'}). This would imply both property (\hyperlink{P3}{P3}) and the \hyperlink{PD}{PD Invariant} for $C_{\mx}$ since $\Phi(C_{\mx}) = \adm(\mx)$, which would complete the proof of the induction step.
 	
 	Let $u,v$ be any two vertices in $H_{C_{\mx}}$ whose shortest distance in $H_{C_{\mx}}$ realizes $\dm(H_{C_{\mx}})$. Let $\varphi_{C_u}, \varphi_{C_v}$ be the two nodes in $\mx$ that correspond to two clusters $C_u,C_v$ containing $u$ and $v$, respectively. Let $\mathcal{P}_{u,v}$ a path in $\mg_i$ of minimum augmented weight between $\varphi_{C_u}$ and $\varphi_{C_v}$. Observe that $\omega(\mathcal{P}_{u,v}) \leq \adm(\mx)$. We now construct a path $P_{u,v}$ between $u$ and $v$ in $H_{C_{\mx}}$ as follows. We write $\mathcal{P}_{u,v} \equiv (\varphi_{C_u} =\varphi_{C_1}, \mbe_1,\varphi_{C_2},\mbe_2, \ldots, \varphi_{C_\ell} = \varphi_{C_v})$ as an alternating sequence of nodes and edges. For every $1\leq p \leq \ell-1$, let $(u_p,v_p)$ be the edge in $E^{\sigma}_i$ that corresponds to $\mbe_p$. We then define
	$v_0 = u, u_\ell = v$ and
 		\begin{equation*}
 			P_{u,v} =Q_{H_{< L_{i-1}}[C_1]}(v_0,u_1) \circ (u_1,v_1)\circ Q_{H_{< L_{i-1}}[C_2]}(v_1,u_2) \circ (u_2,v_2)\circ \ldots \circ  Q_{H_{< L_{i-1}}[C_{\ell}]}(v_{\ell-1},u_\ell)~,
 		\end{equation*}
 	where $Q_{H_{< L_{i-1}}[C_p]}(v_{p-1},u_p)$ for $1\leq p \leq \ell$ 
	denotes  the shortest path in the corresponding subgraph (between the endpoints of the respective edge, as specified in all the subscripts), and $\circ$ is the path concatenation operator. By the induction hypothesis for the \hyperlink{PD}{PD Invariant} and $i$, 
	$w(Q_{H_{< L_{i-1}}[C_p]}(v_{p-1},u_p)) \leq \omega(\varphi_{C_p})$ for each $1\leq p \leq \ell$. Thus, $w(P_{u,v}) \leq \omega(\mathcal{P}_{u,v}) \leq \adm(\mx)$. It follows that $\dm(H_{C_{\mx}}) ~\leq~ w(P_{u,v}) \leq  \adm(\mx)$ as desired. 	\qed
 \end{proof}
 
 \paragraph{Local potential change.~}  For each subgraph $\mx \in \mathbb{X}$, we define the \emph{local potential change} of $\mx$, denoted by $\Delta_{i+1}(\mx)$ as follows:
 \begin{equation}\label{eq:LocalPotential}
 	\Delta_{i+1}(\mx) \stackrel{\mbox{\tiny{def.}}}{=}   \left(\sum_{\varphi_C\in \mv(\mx)} \Phi(C) \right) -  \Phi(C_{\mx}) = \left(\sum_{\varphi_C\in \mv(\mx)} \omega(\varphi_C) \right) - \adm(\mx). 
 \end{equation} 
 
 \begin{claim}\label{clm:localPotenDecomps}$\Delta_{i+1} = \sum_{\mx \in \mathbb{X}}\Delta_{i+1}(\mx)$.
 \end{claim}
 \begin{proof} By property (\hyperlink{P1}{P1}), subgraphs in $\mathbb{X}$ are vertex-disjoint and cover the vertex set $\mv_i$,  hence $\sum_{\mx \in \mathbb{X}}(\sum_{\varphi_C\in \mv(\mx)} \Phi(C)) = \sum_{C\in \mathcal{C}_i} \Phi(C) = \Phi_i$. Additionally, by the construction of level-$(i+1)$ clusters,  $\sum_{\mx \in \mathbb{X}}  \Phi(C_{\mx}) = \sum_{C'\in \mathcal{C}_{i+1}} \Phi(C') = \Phi_{i+1}$. Thus,  we have:
 	\begin{equation*}
 			\sum_{\mx \in \mathbb{X}}\Delta_{i+1}(\mx) = \sum_{\mx \in \mathbb{X}}\left(\left(\sum_{\varphi_C\in \mv(\mx)} \Phi(C) \right) -  \Phi(C_{\mx}) \right) = \Phi_i - \Phi_{i+1} = \Delta_{i+1}, 
 	\end{equation*}
	as claimed.\qed
 \end{proof}

 The decomposition of the (global) potential change into local potential changes makes the task of analyzing the spanner weight (Item (1) in \Cref{lm:framework}) easier as we can do so locally. Specifically, we often construct $H_i$ by considering each node in $\mv_i$ and taking a subset of (the corresponding edges of) the edges incident to the node to $H_i$. We then calculate the number of edges taken to $H_i$ incident to all nodes in $\mx$, and bound their total weight by the local potential change of $\mx$. By summing up over all $\mx$, we obtain a bound on $w(H_i)$ in terms of the (global) potential change $\Delta_{i+1}$. 
 
 \subsection{Summary}
 
 We have introduced the technical framework (\Cref{lm:framework-technical}) for constructing light spanners that we will use to both design fast construction of light spanners (\Cref{sec:FastProof}) and spanners with optimal lightness (\Cref{sec:LightProof}). The construction boils down to constructing two objects: (a) clusters for level $i$ satisfying all properties (\hyperlink{P1}{P1})-(\hyperlink{P3}{P3}) and (b) a spanner $H_{i}$ for $E)i^{\sigma}$ whose weight is bounded by potential change at level $i$ (Item (1) in \Cref{lm:framework-technical}). The cluster construction is based on a cluster graph \Cref{def:ClusterGraphNew}: each level $i+1$ cluster $\mx$ corresponds to a subgraph of the cluster graph $\mg_i$ satisfying properties (\hyperlink{P1'}{P1'})-(\hyperlink{P3'}{P3'}). The detailed construction of level $i+1$ clusters for fast algorithms is different from the construction for optimal lightness, and is deferred to the corresponding sections (\Cref{sec:FastProof} and \Cref{sec:LightProof}). \Cref{table:notation} below summarizes the notation introduced in this section. 
 
 \renewcommand{\arraystretch}{1.3}
 \begin{longtable}{| l | l|} 
 	\hline
 	\textbf{Notation} & \textbf{Meaning} \\ \hline
 	$E^{light}$ &$ \{e \in E(G) : w(e)\le w/\varepsilon\}$\\ \hline 
 	$E^{heavy}$ & $E \setminus E^{light}$ \\\hline
 	$E^{\sigma} $ & $\bigcup_{i \in \mathbb{N}^{+}}E_{i}^{\sigma}$\\\hline
 	$E_{i}^{\sigma} $ & $\{e \in E(G) : \frac{L_i}{1+\psi} \leq w(e) < L_i\}$\\\hline
 	$g$ & constant in \hyperlink{P3}{property (P3)}. \\\hline
 	$\mathcal{G}_i = (V_i, \msttilde_{i} \cup \mathcal{E}_i, \omega)$ & cluster graph; see \Cref{def:ClusterGraphNew}. \\\hline
 	$\me_i$ & corresponds to a subset of edges of $E^{\sigma}_i$\\\hline
 	$\mathbb{X}$ & a collection of subgraphs of $\mathcal{G}_i$\\\hline
 	$\mx, \mv(\mx), \me(\mx)$ & a subgraph in $\mathbb{X}$, its vertex set, and its edge set\\\hline
 	$\Phi_i$ & $\sum_{c \in C_i}\Phi(c)$ \\\hline
 	$\Delta_{i+1} $&$ \Phi_i - \Phi_{i+1}$\\\hline
 	$\Delta_{i+1}(\mx)$ & $(\sum_{\phi_C\in \mx }\Phi(C) ) - \Phi(C_{\mx})$\\\hline
 	$C_\mx$ & $\bigcup_{\phi_C \in \mx}C$ \\\hline
 	\caption{Notation introduced in \Cref{sec:framework}.}
 	\label{table:notation}
 \end{longtable}
 \renewcommand{\arraystretch}{1}

\section{Fast Construction: Proof of \Cref{lm:framework}(1)}\label{sec:FastProof}

In this section, we give the detailed construction of level $i+1$ clusters and graph $H_i$, thereby proving Item (1) in \Cref{lm:framework}. We set $\psi  = \eps$ where $\psi$ is the parameter in \Cref{eq:Esigmaixdef}.

We guarantee that the cluster graph $\mg_i$ introduced in \Cref{subsec:DesignPotential} satisfies an additional property, which we will exploit for  efficient construction. 

 \begin{definition}[Additional Properties of $\mg_i$]\label{def:GiPropAdd} $\mg_i$ satisfies properties (1) and (2) in \Cref{def:GiProp}, and the following property:
 	 \begin{enumerate}
		\item[(3)] $\mg_i$ has no \emph{removable edge}: an edge $(\varphi_{C_u},\varphi_{C_v}) \in \me_i$ is removable if (3a) the path $\msttilde_i[\varphi_{C_u},\varphi_{C_v}]$ between $\varphi_{C_u}$ and $\varphi_{C_v}$ only contains nodes in $\msttilde_{i}$ of degree at most $2$ and (3b) $\omega(\msttilde_i[\varphi_{C_u},\varphi_{C_v}]) \leq t(1 + 6g\eps)\omega(\varphi_{C_u},\varphi_{C_v})$.
	\end{enumerate}
\end{definition}
As we will show in the sequel, if an edge $(\varphi_{C_u},\varphi_{C_v})$ satisfies property (3b), there is a path of stretch at most $t(1+6g\eps)$ in $H_{< L_{i-1}}$ between $u$ and $v$ and hence, we do not need to consider edge $(u,v)$ in the construction of $H_i$.  To meet the required lightness bound, it turns out that it suffices to remove edges satisfying both properties (3a) and (3b), rather than removing all edges satisfying property (3b). 

\subsection{Constructing Level-$(i+1)$ Clusters}\label{subsec:LeveIplus1Construction}

To obtain a fast spanner construction, we will maintain for each cluster $C \in \mathcal{C}_i$ a {\em representative} vertex $r(C) \in C$. If $C$ contains at least one original vertex, then $r(C)$ is one original vertex in $C$; otherwise, $r(C)$ is a virtual vertex. (Recall that virtual vertices are those subdividing $\mst$ edges.)  For each vertex $v \in C$, we designate $r(C)$ as the {\em representative} of $v$, i.e., we set $r(v) = r(C)$ for each $v \in C$. We use the \textsc{Union-Find} data structure to maintain these representatives. Specifically, the representative of $v$ will be given as \textsc{Find}($v$). Whenever a level-$(i+1)$ cluster is formed from level-$i$ clusters, we call \textsc{Union} (sequentially on the level-$i$ clusters) to construct a new representative for the new cluster. 

\paragraph{A careful usage of the Union-Find data structure.~} We will use the \textsc{Union-Find} data structure~\cite{Tarjan75}  for grouping subsets of clusters to larger clusters (via the \textsc{Union} operation) and checking whether two given vertices belong to the same cluster (via the \textsc{Find} operation). To reduce the amortized time to $O(\alpha(m,n))$, we only store original vertices in the \textsc{Union-Find} data structure. To this end, for each virtual vertex, say $x$, which subdivides an edge $(u,v) \in \mst$, we store a pointer, denoted by $p(x)$, which points to one of the endpoints, say $u$, in the same cluster with $x$, \emph{if there is at least one endpoint in the same cluster with $x$}. In particular, any virtual vertex has at most two optional clusters that it can belong to at each level of the hierarchy.  Hence, we can apply every \textsc{Union-Find} operation to $p(x)$ instead of $x$. For example,  to check whether two virtual vertices  $x$ and $y$ are in the same cluster,  we compare $r(p(x)) \stackrel{?}{=} r(p(y))$ via two \textsc{Find} operations. The total number of \textsc{Union} and \textsc{Find} operations in our construction remains $O(m)$ while the number of vertices that we store in the data structure is reduced to $n$. Thus, the amortized time of each operation reduces to $O(\alpha(m,n))$ and the total runtime due to all these operations is $O(m \alpha(m,n))$.

Following the approach in \Cref{subsec:DesignPotential},  we construct a graph $\mathcal{G}_i$ satisfying all properties in \Cref{def:GiProp} and \Cref{def:GiPropAdd}. Then we construct a set $\mathbb{X}$ of subgraphs of $\mg_i$ satisfying the three properties (\hyperlink{P1'}{P1'})-(\hyperlink{P3'}{P3'}) and a subgraph $H_i$ of $G$ (and of $\tilde{G}$ as well). Each subgraph $\mx\in \mathbb{X}$ is then converted to a level-$(i+1)$ cluster by \Cref{eq:XtoCluster}. 

\paragraph{Constructing $\mathcal{G}_i$.~}  We shall assume inductively on $i, i \ge 1$ that:
\begin{itemize}[noitemsep]
	\item The set of edges $\widetilde{\mst}_i$ is given by the construction of the previous level $i$ in the hierarchy; for the base case $i = 1$ (see \Cref{subsec:DesignPotential}), $\widetilde{\mst}_1$ is simply a set of edges of $\widetilde{\mst}$ that are not in any level-$1$ cluster. 
	\item The weight $\omega(\varphi_C )$ on each node $\varphi_C \in \mv_i$ is the potential value of cluster $C \in \mathcal{C}_i$; for the base case $i = 1$, the  potential values of level-$1$ clusters were computed in  $O(m)$ time by \Cref{subsec:DesignPotential}.
\end{itemize}

By the end of this section, we will have constructed  
the edge set $\widetilde{\mst}_{i+1}$ and the weight function on nodes  of $\mathcal{G}_{i+1}$, in  time $O(|\mathcal{V}_i|\alpha(m,n))$. Computing the weight function on nodes of $\mathcal{G}_{i+1}$ is equivalent to computing the augmented diameter of $\mx$, which in turn, is related to the potential function. The fact that we can compute all the weights efficiently in almost linear time is the crux of our framework.

Note that we make no inductive assumption regarding the set of edges ${E^{\sigma}_i}$, which can be computed once in $O(m)$  overall time at the outset for all levels $i \ge 1$, since the edge sets $E^{\sigma}_{1}, E^{\sigma}_{2}, \ldots$ are pairwise disjoint and the number of levels is $O(m)$ by \Cref{clm:numLevels}.

\begin{lemma}\label{lm:G_i-construction}$\mathcal{G}_i = (\mathcal{V}_i,\mathcal{E}_i\cup \widetilde{\mst}_i,\omega)$ 
	can be constructed in $O\left(\alpha(m,n)(|\mv_i|  + |E^{\sigma}_i|)\right)$ time, where $\alpha(\cdot,\cdot)$ is the inverse-Ackermann function.
\end{lemma}
\begin{proof} Note that  $\widetilde{\mst}_i$  and $E^{\sigma}_i$ are given at the  outset of the construction of $\mg_i$.  To construct the edge set $\mathcal{E}_i$, we do the following. 	For each edge $e = (u,v) \in E^{\sigma}_i$, we compute the representatives $r(u), r(v)$; this can be done in $O(\alpha(m,n))$ amortized time over all the levels up to $i$ using the \textsc{Union-Find} data structure. 
	Equipped with the representatives, it takes $O(1)$ time to check whether $e$'s endpoints lie in the same level-$i$ cluster  and check in $O(1)$ time whether edges $e = (u,v)$ and $e' = (u',v')$ are parallel in the cluster graph. Next, we remove all removable edges from $\mg_i$ as specified by property (3b) in \Cref{def:GiPropAdd}. First we find in $O(|\mv_i|)$ time a collection $\mathbb{P}$ of \emph{maximal paths} in $\msttilde_{i}$ that only contain degree-$2$ vertices. We then find for each path $\mathcal{P}\in \mathbb{P}$ a subset of edges $\me_{\mp} \subseteq \me_i$ whose both endpoints belong to $\mathcal{P}$.  Finally, for each path $\mathcal{P} \in \mathbb{P}$ and each edge $(\varphi_{C_u},\varphi_{C_v}) \in \me_{\mp}$, we can compute $\omega(\mathcal{P}[\varphi_{C_u},\varphi_{C_v}])$ in $O(1)$ time, after an $O(|\mv(\mathcal{P})|)$ preprocessing tim by fixing an endpoint $\varphi_C \in \mp$ and for every node $\varphi_{C'} \in \mp$, we compute $\omega(\mathcal{P}[\varphi_{C},\varphi_{C'}])$ in total $O(|\mv(\mp)|)$ time. Given $\omega(\mathcal{P}[\varphi_{C_u},\varphi_{C_v}])$, we can check in $O(1)$ time whether $(\varphi_{C_u},\varphi_{C_v})$ is removable and if so, we remove it from $\me_i$.	The total running time is $O(|\mv_{i}| + |E^{\sigma}_i|)$. \qed
\end{proof}

The following key lemma states all the properties of clusters constructed in our framework; the details of the construction are deferred to \Cref{sec:ClusteringDetails}.

\begin{restatable}{lemma}{Clustering}
	\label{lm:ClusteringFast}  Given $\mg_i$, we can construct in time $O((|\mv_i| + |\me_i|)\eps^{-1})$ (i) a partition of $\mv_i$ into three sets $\{\mv_i^{\high}, \mv_i^{\lowp},\mv_i^{\lowm}\}$ and (ii) a collection $\mathbb{X}$ of subgraphs of $\mg_i$ and their augmented diameters, such that:
	\begin{enumerate}
		\item[(1)]   For every node $\varphi_C \in \mv_i$: If $\varphi_C \in \mv_i^{\high}$, then $\varphi_C$ is incident to $\Omega(1/\eps)$ edges in $\me_i$; otherwise ($\varphi_C \in \mv_i^{\lowp} \cup \mv_i^{\lowm}$), the number of edges in $\me_i$ incident to $\varphi_C$ is $O(1/\eps)$.
		
		\item[(2)] If a subgraph $\mx$ contains at least one node in $\mv^{\lowm}_{i}$, then every node of $\mx$ is in $\mv^{\lowm}_i$. Let $\mathbb{X}^{\lowm} \subseteq \mathbb{X}$ be a set of sugraphs whose nodes are in $\mv_i^{\lowm}$ only.
		\item[(3)] Let $\Delta_{i+1}^+(\mx) = \Delta(\mx) + \sum_{\mbe \in \msttilde_i\cap \me(\mx)}w(\mbe)$.  Then, $\Delta_{i+1}^+(\mx) \geq 0$ for every $\mx \in \mathbb{X}$, and
		\begin{equation}\label{eq:averagePotential}
			\sum_{\mx \in \mathbb{X}\setminus \mathbb{X}^{\lowm}} \Delta_{i+1}^+(\mx) = \sum_{\mx \in \mathbb{X}\setminus \mathbb{X}^{\lowm}} \Omega(|\mv(\mx)|\eps^2 L_i). 
		\end{equation}
		\item[(4)] There is no edge in $\me_i$ between a node in $\mv^{\high}_i$ and a node in $\mv^{\lowm}_i$. Furthermore, if there exists an edge   $(\varphi_{C_u},\varphi_{C_v}) \in \me_i$ such that both $\varphi_{C_u}$ and $\varphi_{C_v}$ are in 
		$\mv_i^{\lowm}$, then  $\mv_i^{\lowm} = \mv_i$ and $|\me_i| = O(\frac{1}{\epsilon^2})$; we call this case the \emph{degenerate case}.
		\item[(5)] For every subgraph $\mx \in \mathbb{X}$, $\mx$ satisfies the three properties (\hyperlink{P1'}{P1'})-(\hyperlink{P3'}{P3'}) with constant $g=31$ and $\eps \leq \frac{1}{8(g+1)}$,  and $|\me(\mx)\cap \me_i| = O(|\mv(\mx)|)$.
	\end{enumerate}	
	Furthermore, the construction of $\mathbb{X}$ can be constructed in the \emph{pointer-machine model} with the same running time.
\end{restatable}

\begin{comment}
We remark the following points regarding subgraphs in $\mathbb{X}$ constructed by \Cref{lm:ClusteringFast}.

\begin{remark}\label{remark:Clustering} 
	\begin{enumerate}
		\item It is possible for a subgraph   $\mx \in \mathbb{X}$ to contain nodes in $\mv_i^{\high}$ and in $\mv_i^{\lowp}$. 
		\item $\Delta_{i+1}(\mx)$ could be negative, but $\Delta_{i+1}^+(\mx)$ is always non-negative by Item (2) in \Cref{lm:ClusteringFast}. We could view $\sum_{\mbe \in \msttilde_i\cap \me(\mx)}w(\mbe)$ as a \emph{corrective term} to $\Delta_{i+1}(\mx)$ (to make it non-negative). We call  $\Delta_{i+1}^+(\mx)$  \emph{corrected potential change}.
		\item  \Cref{eq:averagePotential} implies that the \emph{average} amount of corrected potential change \emph{per subgraph} $\mx \in \mathbb{X}\setminus \mathbb{X}^{\lowm}$ is $\Omega(|\mv(\mx)|\eps^2 L_i)$.  On the other hand, there is no guarantee, other than non-negativity, on the corrected potential change of $\mx$ if $\mx \in \mathbb{X}^{\lowm}$.
	\end{enumerate}
\end{remark}
\end{comment}

We observe the  following observations about subgraphs of $\mathbb{X}$ in \Cref{lm:ClusteringFast}.

\begin{observation}\label{obs:XhighXlowp} If a subgraph $\mx \in \mathbb{X}$ has $\mv(\mx)\cap (\mv^{\high}_{i}\cup \mv^{\lowp}_i) \not= \emptyset$, then $\mv(\mx)\subseteq (\mv^{\high}_{i}\cup \mv^{\lowp}_i)$.
\end{observation}
\begin{proof}
	Follows from Item (2)  in \Cref{lm:ClusteringFast} and the fact that $\{\mv_i^{\high}, \mv_i^{\lowp},\mv_i^{\lowm}\}$  is a partition of $\mv_i$.\qed
\end{proof}

\begin{observation}\label{obs:LowmStructure} Unless the degenerate case happens, for every edge $(\varphi_{C_u},\varphi_{C_v})$ with one endpoint in $\mv_i^{\lowm}$, w.l.o.g.\ $\varphi_{C_v}$, the other endpoint $\varphi_{C_u}$ must be in $\mv_i^{\lowp}$. 
	As a result, $\me(\mx)\cap \me_i = \emptyset$ if $\mx \in \mathbb{X}^{\lowm}$.  
\end{observation}
\begin{proof} If the degenerate case does not happen, by Item (4) in \Cref{lm:ClusteringFast}, any edge incident to a node in $\mv_i^{\lowm}$ must be incident to a node in $\mv_i^{\lowp}$. By Item (2), if $\mx \in \mathbb{X}^{\lowm}$, then $\mv(\mx) \subseteq \mv_i^{\lowm}$ and hence, there is no edge between two nodes in $\mx$. Thus, $\me(\mx)\cap \me_i = \emptyset$. \qed
\end{proof}

Next, we show how to construct $\msttilde_{i+1}$ for the construction of the next level.

\begin{lemma}\label{lm:MSTiPlus1} Given the collection of subgraphs $\mathbb{X}$ of $\mg_i$ and their augmented diameters, we can construct the set of nodes $\mv_{i+1}$, and their weights, and the cluster tree $\msttilde_{i+1}$ of $\mg_{i+1}$ in $O(|\mv_{i}|\alpha(m,n))$ time.
\end{lemma}
\begin{proof} For each subgraph $\mx \in \mathbb{X}$, we call \textsc{Union} operations sequentially on the set of clusters corresponding to the nodes of $\mx$ to create a level-$(i+1)$ cluster $C_{\mx} \in \mathcal{C}_{i+1}$. Then we create a set of nodes $\mv_{i+1}$ for $\mg_{i+1}$: each node $\varphi_{C_{\mx}}$ corresponds to a cluster $C_{\mx} \in \mc_{i+1}$ (and also subgraph $\mx \in \mathbb{X}$). Next, we set the weight $\omega(\varphi_{C_{\mx}}) = \adm(\mx)$. The total running time of this step is $O(|\mv_i|\alpha(m,n))$.
	
	We now construct $\msttilde_{i+1}$. Let $\msttilde^{out}_{i} = \msttilde_{i}\setminus (\cup_{\mx \in \mathbb{X}}(\me(\mx)\cap \msttilde_{i}))$ be the set of $\msttilde_{i}$ edges that are not contained in any subgraph $\mx \in \mathbb{X}$. Let $\msttilde_{i+1}'$ be the graph with vertex set $\mv_{i+1}$ and there is an edge between two nodes $(\mx,\my)$ in $\mv_{i+1}$ of there is at least one edge in $\msttilde^{out}_{i}$ between two nodes in the two corresponding subgraphs $\mx$ and $\my$. Since $\msttilde_{i}$ is  a spanning tree of $\mg_i$, $\msttilde_{i+1}'$ must be connected. $\msttilde_{i+1}$ is then a spanning tree of $\msttilde_{i+1}'$. \qed
\end{proof}

\subsection{Constructing $H_i$: Proof of \Cref{lm:framework}(1)} \label{subsec:ConstructHi}

Recall that to obtain a fast algorithm for constructing a light spanner, \Cref{lm:framework-technical} requires a fast construction of clusters at every level and a fast construction of $H_i$, the spanner for level-$i$ edges $E^{\sigma}_i$.  In \Cref{subsec:LeveIplus1Construction}, we have designed an efficient construction of level-$i$ clusters (\Cref{lm:MSTiPlus1}). In this section, we show to construct $H_i$ efficiently with stretch $t(1+\max\{s_{\ssa}(2g) + 4g, 10g\}\eps)$; that is parameter $\rho$ in \Cref{lm:framework-technical} is $\rho = \max\{s_{\ssa}(2g) + 4g, 10g\}$. By induction, we assume that the stretch of every edge of weight less than $L_i/(1+\psi)$ in $H_{<L_i/(1+\psi)}$ is $t(1+\max\{s_{\ssa}(2g) + 4g, 10g\}\eps)$.  Note that $H_{< L_i} = H_{<L_i/(1+\psi)} \cup H_i$; see \Cref{remark:stretchLevel-i}. 

Our construction of $H_i$ assumes the existence of \hyperlink{SPSSA}{$\ssa$}. Since edges of the input graph to \hyperlink{SPSSA}{$\ssa$} must have weights in $[L,(1+\eps)L)$ for some parameter $L$, we set parameter $\psi$ in \Cref{lm:framework-technical} to be $\eps$. Thus, level-$i$ edges $E^\sigma_{i}$ (and hence edges in $\me_i$ of $\mg_i$) have weights in $[L_i/(1+\eps),L_i)$.

We  now go into the details of the construction of $H_i$. We assume that we are given the collection $\mathbb{X}$ of subgraphs as described in \Cref{lm:ClusteringFast}. Define:
\begin{equation}\label{eq:ACT-XLowHighdef}
	\begin{split}
		\mathbb{X}^{\high} &= \{\mx \in \mathbb{X}: \mv(\mx)\cap \mv^{\high}_{i} \not=\emptyset\}\\
		\mathbb{X}^{\lowp} &= \{\mx \in \mathbb{X}: \mv(\mx)\cap \mv^{\lowp}_{i} \not=\emptyset\}\\
	\end{split}
\end{equation}
It could be that $\mathbb{X}^{\high}\cap \mathbb{X}^{\lowp}\not= \emptyset$. By \Cref{obs:XhighXlowp}, $\{\mathbb{X}^{\high}\cup \mathbb{X}^{\lowp},\mathbb{X}^{\lowm}\}$ is a partition of $\mathbb{X}$.

Recall that each edge $(\varphi_{C_u},\varphi_{C_v}) \in \me_i$ has a corresponding edge $(u,v)\in E^{\sigma}_i$ where $u$ and $v$ are in two level-$i$ clusters $C_u$ and $C_v$, respectively.  Our goal in this section is to prove the following lemma.

\begin{restatable}{lemma}{HiConstructionFast}
	\label{lm:ConstructH_i} Given \hyperlink{SPSSA}{$\ssa$}, we can construct  $H_i$ in total time $O((|\mv_i| + |\me_i|)\tau(m,n))$	satisfying \Cref{lm:framework-technical}  with  $\lambda = O(\chi \eps^{-2} + \epsilon^{-3})$, and $A = O(\chi\eps^{-2} + \eps^{-3})$, when $\eps \leq 1/(2g)$. Furthermore, the stretch of every edge in $E_i^{\sigma}$ in $H_{<L_i}$ is $t(1+\max\{s_{\ssa}(2g) + 4g, 10g\}\eps)$. 
\end{restatable}

We apply $\ssa$ to $\mv^{\high}_i$ that has size at most $n$ since every level-$i$ cluster corresponding to a node in  $\mv^{\high}_i$ contains at least one original vertex in $G$. Furthermore, $|\me^{\high}_i|$ is bounded by $m$ and hence, $\tau(|\me^{\high}_i|, |\mv^{\high}_i|)\leq \tau(m,n)$.

\begin{remark}\label{remark:ACT} If $\ssa$ can be implemented in the ACT model in time $O((|\mv^{\high}_i| + |\me_i^{\high}|)\tau(m,n))$, then the construction of $H_i$ can be implemented in the ACT model  in time $O((|\mv_i| + |\me_i|)\tau(m,n))$.
\end{remark}

\paragraph{Constructing $H_i$.~} We construct $H_i$ in three steps, as briefly described in the construction overview above. Initially $H_i$ contains no edges.

\begin{itemize}
	\item \textbf{(Step 1).~} For every sugraph $\mx \in\mathbb{X}$ and every edge $\mbe = (\varphi_{C_u},\varphi_{C_v}) \in \me(\mx)$ such that $\mbe \in \me_i$, we add the corresponding edge $(u,v)$ to $H_i$. (Note that if $\mbe \not\in \me_i$, it is in $\msttilde_i$ and hence $(u,v)$ belongs to $H_0$).
	
	\item \textbf{(Step 2).~} For each node $\varphi_{C_u} \in \mv_i^{\lowp} \cup \mv_{i}^{\lowm}$, and for each edge $(\varphi_{C_u},\varphi_{C_v})$ in $\me_i$ incident to $\varphi_{C_u}$, we add the corresponding edge $(u,v)$ to $H_i$,
	
	\item \textbf{(Step 3).~} Let $\me_i^{\high}\subseteq \me_i$ be the set of edges whose both endpoints are in $\mv_i^{\high}$, and $\mathcal{K}_i = (\mv^{\high}_i, \me_i^{\high},\omega)$ be a subgraph of $\mg_i$. We run \hyperlink{SPSSA}{$\ssa$} on $\mathcal{K}_i$ to obtain $\me^{\prune}_i$. For every edge $(\varphi_{C_u},\varphi_{C_v})\in \me_i^{\prune}$, we add the corresponding edge $(u,v)$ to $H_i$.
\end{itemize}

\paragraph{Analysis.~} In \Cref{clm:Hi-Time}, \Cref{clm:Hi-Stretch}, and \Cref{clm:Hi-Weight} below,  we bound the running time to construct $H_i$, the stretch of edges in $E^{\sigma}_i$, and the weight of $H_i$, respectively. The following claims follows directly from the construction. 

\begin{claim}\label{clm:Hi-Time} $H_i$ can be constructed in time $O((|\mv_i| + |\me_i|)\tau(m,n))$.
\end{claim}

We bound the stretch of edges in $E^\sigma_{i}$. We first show that the input to \hyperlink{SPSSA}{$\ssa$} satisfies its requirement.

\begin{claim}\label{clm:Ki-clustergraph}$\mathcal{K}_i = (\mv^{\high}_i, \me_i^{\high},\omega)$ is a $(L, \eps, \beta, \Upsilon = 1+\eps)$-cluster graph with $L = L_i/(1+\eps)$, $\beta = 2g$,  and  $H_{< L} = H_{< L_i/(1+\eps)}$, where $H_{< L_i/(1+\eps)}$ is the spanner constructed for edges of weight less than $L_i/(1+\eps)$ (see \Cref{remark:stretchLevel-i} with $\psi = \eps$). Furthermore, the stretch of $H_{< L}$ for edges of weight less than $L$ is $t(1+ \max\{s_{\ssa}(2g)+4g,10g\}\eps)$.  
\end{claim} 
\begin{proof}
	We verify all properties in \Cref{def:ClusterGraph-Param}. Properties (1) and (2) follow directly from the definition of $\mathcal{K}_i$. Since we set $\psi = \eps$, every edge $(u,v) \in E^{\sigma}_i$  has $L_i/(1+\eps) \leq w(u,v)\leq L_i$. As $L = L_i/(1+\eps)$, property (3) follows. By property \hyperlink{P3}{(P3)}, $\dm(H_{<L_i/(1+\eps)}[C]) \leq gL_{i-1} = g(1+\eps) \eps L \leq 2g \eps L = \beta \eps L$ when $\eps < 1$. Thus, $\mk_i$ is a   $(L, \eps, \beta)$-cluster graph. By induction, the stretch of $H_{< L}$ is $t(1+ \max\{s_{\ssa}(2g)+4g,10g\}\eps)$.   
	\qed 
\end{proof}

\begin{claim}\label{clm:Hi-Stretch}$\forall (u,v) \in E^\sigma_{i}$, $d_{H_{< L_i}}(u,v) \leq t(1+\max\{s_{\ssa}(2g) + 4g, 10g\}\eps)w(u,v)$ when $\eps \leq 1/(2g)$.
\end{claim}
\begin{proof}
	Let $F^{\sigma}_i = \{(u,v) \in E^{\sigma}_i: \exists (\varphi_{C_u},\varphi_{C_v}) \in \me_i\}$ be the set of edges in $E^{\sigma}_i$ that correspond to the edges in $\me_i$. We first show that:
	\begin{equation}\label{eq:stertchF}
		d_{H_{< L_i}}(u,v) \leq t(1+ s_{\ssa}(2g)\eps)w(u,v) \qquad \forall (u,v) \in F^{\sigma}_i.
	\end{equation}
	
	To that end, let $(\varphi_{C_u},\varphi_{C_v}) \in \me_i$ be the edge corresponding to $(u,v)$ where $(u,v) \in F^{\sigma}_i$. If at least one of the endpoints of $(\varphi_{C_u},\varphi_{C_v})$ is in $\mv^{\lowp}_i \cup \mv^{\lowm}_i$, then $(u,v) \in H_i$ by the construction in Step 2, hence \Cref{eq:stertchF} holds. Otherwise, $\{\varphi_{C_u},\varphi_{C_v}\}\subseteq \mv_i^{\high}$, which implies that $(\varphi_{C_u},\varphi_{C_v}) \in \me^{\high}_i$. Since we add all edges of $\me_i^{\prune}$ to $H_i$, by property (2) of \hyperlink{SPSSA}{$\ssa$} and \Cref{clm:Ki-clustergraph}, the stretch of $(u,v)$ is $t(1+s_{\ssa}(2g)\eps)$. 
	
	It remains to bound the stretch of any edge $(u',v') \in E^{\sigma}_i\setminus F^{\sigma}_i$. Recall that  $(u',v')$ is not added to $\me_i$ because (a) both $u'$ and $v'$ are in the same level-$i$ cluster in the construction of the cluster graph in \Cref{lm:G_i-construction} , or (b) $(u',v')$ is parallel with another edge $(u,v)$ also in \Cref{lm:G_i-construction}, or (c) the edge $(\varphi_{C_{u'}},\varphi_{C_{v'}})$ corresponding to $(u',v')$ is a removable edge (see \Cref{def:GiProp}).

	In case (a), since the level-$i$ cluster containing both $u'$  and $v'$ has diameter at most $gL_{i-1}$ by property (\hyperlink{P3}{P3}), we have a path from $u'$ to $v'$ in $H_{< L_{i-1}}$ of diameter at most $gL_{i-1} ~=~ g\eps L_i ~\leq \frac{L_i}{1+\psi}~\leq~w(u',v')$ when $\eps \leq  1/(2g)$. Thus, the stretch of edge $(u',v')$ is $1$. For case (c), the stretch of $(u',v')$ in $H_{< L_{i-1}}$ is $t(1+6g\eps)$ since $\eps \leq 1$.  Thus, in both cases, we have:
	\begin{equation}\label{eq:stertch-uvprime}
		d_{H_{< L_i}}(u',v') \leq t(1+ 6g\eps)w(u',v') 
	\end{equation}

	We now consider case (b). 	Let $C_u$ and $C_v$ be two level-$i$ clusters containing $u$ and $v$, respectively. W.l.o.g, we assume that $u' \in C_u$  and $v' \in C_v$. Since we only keep the edge of minimum weight among all parallel edges, $w(u,v) \leq w(u',v')$. Since the level-$i$ clusters that contain $u$  and $v$ have diameters at most $gL_{i-1} = g\eps L_i$ by property (\hyperlink{P3}{P3}), it follows that $\dm(H_{< L_i}[C_u]),\dm(H_{< L_i}[C_v]) \le g\epsi L_i$. 	We have:
	\begin{equation*}
		\begin{split}
			d_{H_{< L_i}}(u',v') &\leq 	d_{H_{< L_i}}(u,v) + \dm(H_{< L_i}[C_u]) + \dm(H_{< L_i}[C_v])\\ &\leq t(1+ \max\{s_{\ssa}(2g),6g\}\eps)w(u,v) +   2g\epsi L_i \qquad \mbox{(by \Cref{eq:stertchF} and \Cref{eq:stertch-uvprime})}\\
			&\leq  t(1+ \max\{s_{\ssa}(2g),6g\}\eps)w(u',v')  +  2g\epsi L_i \leq  t(1+ \max\{s_{\ssa}(2g)+4g,10g\}\eps)w(u',v') 
		\end{split}
	\end{equation*}
Since $w(u',v') \geq L_i/(1+\eps) \geq L_i/2$ and $t\geq 1$. The lemma now follows. \qed
\end{proof}

\begin{claim}\label{clm:Hi-Weight} Let $\msttilde^{in}_i = \cup_{\mx \in \mathbb{X}}(\me(\mx)\cap \msttilde_i)$ be the set of $\msttilde_i$ edges that are contained in subgraphs in $\mathbb{X}$. Then, $w(H_i) \leq \lambda \Delta_{i+1} + a_i$ for $\lambda = O(\chi \eps^{-2} + \eps^{-3})$  and $a_i = (\chi\eps^{-2}) \cdot w(\msttilde^{in}_i) + O(L_i/\eps^2)$. 
\end{claim}
\begin{proof} Let $\msttilde^{in}_i(\mx) = \me(\mx)\cap \msttilde_i$ for each subgraph $\mx \in \mathbb{X}$.  By the definition of $ \mathbb{X}^{\lowp}$ and $ \mathbb{X}^{\high}$ (see~\Cref{eq:ACT-XLowHighdef}), it holds that:
	\begin{equation} \label{eq:mvilowhigh}
		\begin{split}
			|\mv_{i}^{\high}| \leq \sum_{\mx \in \mathbb{X}^{\high}}|\mv(\mx)|  \quad &\mbox{and} \quad	|\mv_{i}^{\lowp}| \leq \sum_{\mx \in \mathbb{X}^{\lowp}}|\mv(\mx)| 
		\end{split}
	\end{equation}
	
	First, we consider the non-degenerate case where $\mv_i^{\lowm} \not = \mv_i$. By \Cref{obs:LowmStructure}, any edge in $\me_i$ incident to a node in $\mv_i^{\lowm}$ is also incident to a node in $\mv_i^{\lowp}$. We bound the total weight of the edges added to $H_i$ by considering each step in the construction of $H_i$ separately. Let $F^{(a)}_i\subseteq E^{\sigma}_i$ be the set of edges added to $H_i$ in the construction in Step $a$, $a\in \{1,2,3\}$.
	
	By \Cref{obs:LowmStructure}, $\me(\mx)\cap \me_i = \emptyset$ if $\mx \in \mathbb{X}^{\lowm}$. Recall that $\mathbb{X}^{\high}\cup \mathbb{X}^{\lowp} = \mathbb{X}\setminus \mathbb{X}^{\lowm}$. By Item (5) in \Cref{lm:ClusteringFast}, the total weight of the edges added to $H_i$ in Step 1 is:
	\begin{equation}\label{eq:Fi1}
		\begin{split}
			w(F^{(1)}_i)  &=  \sum_{\mx \in \mathbb{X}^{\high} \cup \mathbb{X}^{\lowp}} O(|\mv(\mx)|) L_i \stackrel{{\tiny {\mbox{Eq.~}}(\ref{eq:averagePotential})}}{=}  O(\frac{1}{\eps^2})\sum_{\mx \in \mathbb{X}^{\high} \cup \mathbb{X}^{\lowp}} \Delta^+_{i+1}(\mx) = O(\frac{1}{\eps^2})\sum_{\mx \in\mathbb{X}} \Delta^+_{i+1}(\mx) \\
			&=  O(\frac{1}{\eps^2})\sum_{\mx \in \mathbb{X}} \left(\Delta_{i+1}(\mx) + w(\msttilde^{in}_i(\mx))\right) = O(\frac{1}{\eps^2})(\Delta_{i+1} + w(\msttilde^{in}_i)) \qquad \mbox{(by \Cref{clm:localPotenDecomps})}~.
		\end{split}
	\end{equation}   
	
	Next, we bound $w(F^{(2)}_i)$. Let $(u,v)$ be an edge added to $H_i$ in Step 2  and let $(\varphi_{C_u},\varphi_{C_v})$ be the corresponding edge of $(u,v)$. Since $\mv^{\lowm}_i \not= \mv_i$, at least one of the endpoints of $(\varphi_{C_u},\varphi_{C_v})$, w.l.o.g.\ $\varphi_{C_u}$,  	is in $\mv^{\lowp}_{i}$ by \Cref{obs:LowmStructure}. Recall by Item (1) of \Cref{lm:ClusteringFast} that all nodes in $\mv_i^{\lowp}$ have low degree, i.e., incident to  $O(1/\eps)$ edges in $\me_i$.	Thus, $|F^{(2)}_i| = O(\frac{1}{\eps}) |\mv_i^{\lowp}|$. We have: \begin{equation}\label{eq:Fi2}
		\begin{split}
			w(F^{(2)}_i)  &=  O(\frac{1}{\eps}) |\mv_i^{\lowp}|  L_i \stackrel{{\tiny \mbox{Eq.~}(\ref{eq:mvilowhigh})}}{=} O(\frac{1}{\eps})\sum_{\mx \in \mathbb{X}^{\lowp}}|\mv(\mx)| L_i ~=~ O(\frac{1}{\eps})\sum_{\mx \in \mathbb{X}^{\high}\cup \mathbb{X}^{\lowp}}|\mv(\mx)| L_i\\
			&\stackrel{{\tiny {\mbox{Eq.~}}(\ref{eq:averagePotential})}}{=}  O(\frac{1}{\eps^3})\sum_{\mx \in \mathbb{X}^{\high}\cup \mathbb{X}^{\lowp}} \Delta^+_{i+1}(\mx) = O(\frac{1}{\eps^3})(\Delta_{i+1} + w(\msttilde^{in}_i)) ~.
		\end{split}
	\end{equation}  
	
	By property (1) of \hyperlink{SPSSA}{$\ssa$}, the number of edges added to $H_i$ in Step 3 is at most $\chi|\mv^{\high}_i|$. Thus: 
	\begin{equation}\label{eq:Fi3}
		\begin{split}
			w(F^{(3)}_i)  &~\leq~  \chi |\mv_i^{\high}|  L_i \stackrel{{\tiny \mbox{Eq.~}(\ref{eq:mvilowhigh})}}{\leq}  \chi\sum_{\mx \in \mathbb{X}^{\high}}|\mv(\mx)| L_i ~\leq~ \chi\sum_{\mx \in \mathbb{X}^{\high}\cup \mathbb{X}^{\lowp}}|\mv(\mx)| L_i\\
			&\stackrel{{\tiny {\mbox{Eq.~}}(\ref{eq:averagePotential})}}{=}  O(\chi \eps^{-2})\sum_{\mx \in \mathbb{X}^{\high}\cup \mathbb{X}^{\lowp}} \Delta^+_{i+1}(\mx)   = O(\chi \eps^{-2})(\Delta_{i+1} + w(\msttilde^{in}_i))~.
		\end{split}
	\end{equation} 
	
	By \Cref{eq:Fi1,eq:Fi2,eq:Fi3}, we conclude that:
	\begin{equation}\label{eq:Hi-nondegen}
		\begin{split}
			w(H_i) &=  O( \chi \eps^{-2}  + \eps^{-3}) (\Delta_{i+1} + w(\msttilde^{in}_i)) \leq \lambda(\Delta_{i+1} + w(\msttilde^{in}_i))
		\end{split}
	\end{equation} 
	for some $\lambda =  O(\chi \eps^{-2} + \eps^{-3})$.

	It remains to consider the degenerate case where $\mv_i^{\lowm} = \mv_i$. Even if we add every single edge that corresponds to an edge in $\me_i$ to $H_i$, Item (3) in \Cref{lm:ClusteringFast} implies that the number of such edges is at most $O(\frac{1}{\eps^2})$. Thus, we have:
	\begin{equation}\label{eq:Hi-degen}
		w(H_i) = O(\frac{L_i}{\eps^2}) \leq  \lambda\cdot (\Delta_{i+1} + w(\msttilde^{in}_i)) + O(\frac{L_i}{\eps^2}) 
	\end{equation} 
	where in the last equation, we use the fact that:
	\begin{equation*}
		\Delta_{i+1} + w(\msttilde^{in}_i) \stackrel{\text{\footnotesize{\Cref{clm:localPotenDecomps}}}}{=} \sum_{\mx \in \mathbb{X}} (\Delta_{i+1}(\mx) + \msttilde^{in}_i(\mx))  = \sum_{\mx \in \mathbb{X}} \Delta^+_{i+1}(\mx)\geq 0~
	\end{equation*}
	by Item (3) of \Cref{lm:ClusteringFast}. Thus, the claim follows from \Cref{eq:Hi-degen,eq:Hi-nondegen}. \qed
\end{proof}

\begin{proof}[Proof of \Cref{lm:ConstructH_i}] The running time follows from \Cref{clm:Hi-Time}.  By \Cref{clm:Hi-Stretch}, the stretch is $t(1+ \max\{s_{\ssa}(2g) + 4g, 10g\}\eps)$. By \Cref{clm:Hi-Weight}, we have
	$\sum_{i \in \mathbb{N}^+}a_i = \sum_{i\in \mathbb{N}^+}(\lambda  \msttilde^{in}_i + O(L_i/\eps^2))$. Observe by the definition that the sets of corresponding edges of $\msttilde^{in}_i$ and $\msttilde^{in}_j$ are disjoint for any $i\not=j \geq 1$. Thus, $\sum_{i\in \mathbb{N}^+} \msttilde^{in}_i\leq w(\mst)$. Observe that:
	
	\begin{equation*}
		\sum_{i\in \mathbb{N}^+}O(\frac{L_i}{\eps^2}) ~=~  O(\frac{1}{\epsilon^2}) \sum_{i=1}^{i_{\max}} \frac{L_{i_{\max}}}{\epsilon^{i_{\max}-i}} ~=~ O(\frac{L_{i_{\max}}}{\epsilon^2(1-\epsilon)}) ~=~ O(\frac{1}{\epsilon^2}) w(\mst)~;
	\end{equation*}
	here $i_{\max}$ is the maximum level. The last equation is due to that $\eps \leq 1/2$  and every edge has weight at most $w(\mst)$ (by the removal step in the construction of $\tilde{G}$). Thus, $A = \lambda + O(\eps^{-2}) = O(\chi \eps^{-2} + \eps^{-3}) +  O(\eps^{-2}) =  O(\chi \eps^{-2} + \eps^{-3})$ as claimed.  \qed
\end{proof}

We are now ready to prove Item (1) of \Cref{lm:framework}.

\paragraph{Proof of Item (1) of \Cref{lm:framework}.~} By \Cref{lm:ClusteringFast} and \Cref{lm:MSTiPlus1}, level-$(i+1)$ clusters can be constructed in time $O((|\mv_i| + |\me_i|)\eps^{-1} + |\mv_i|\alpha(m,n)) = O((|\mathcal{C}_i| + |E^{\sigma}_i|)(\alpha(m,n) + \eps^{-1})$ when $\eps \ll 1$. By \Cref{lm:ConstructH_i}, $H_i$ can be constructed in time $O((|\mv_i| + |\me_i|)\tau(m,n))= O((|\mathcal{C}_i| + |E^{\sigma}_i|)\tau(m,n)$.

We can construct a minimum spanning tree in time  $T_{\mst} = O((n+m)\alpha(m,n))$ by using Chazelle's algorithm~\cite{Chazelle00}. Thus, by \Cref{lm:framework-technical}, the construction time of the light spanner is $$O(m\eps^{-1}(\tau(m,n) + \alpha(m,n) + \eps^{-1})\log(1/\eps) +  T_{\mst}) = O(m\eps^{-1}(\tau(m,n) + \alpha(m,n) + \eps^{-1})\log(1/\eps)~.$$

By \Cref{lm:ConstructH_i} and \Cref{lm:framework-technical}, the lightness of the spanner is $$O(\frac{\lambda + A + 1}{\eps}\log \frac{1}{\epsilon} + \frac{1}{\eps}) = O((\chi \eps^{-3} + \eps^{-4})\log(1/\eps)).$$
Note that we set $\psi = \eps$ in this case. Since $g = 31$, by \Cref{lm:ConstructH_i} and \Cref{lm:framework-technical},  the stretch of the spanner is $$t(1+\max\{s_{\ssa}(2g) + 4g, 10g\}\eps) \leq t(1+(s_{\ssa}O(1)) + O(1))\eps)~.$$ This completes the proof of the theorem.\qed

\section{Clustering: Proof of \Cref{lm:ClusteringFast}}\label{sec:ClusteringDetails}

In this section, we construct the set of subgraph $\mathbb{X}$ of the cluster graph $\mg_i= (\mv_i, \msttilde_{i}\cup \me_i,\omega)$ as claimed in \Cref{lm:ClusteringFast}. See \Cref{table:notation} for a summary of notation we introduced in \Cref{sec:framework}.   Our construction builds upon the construction of Borradaile, Le and Wulff-Nilsen (BLW)~\cite{BLW17}. However, unlike their construction, which is inefficient, our main focus here is on having a \emph{linear-time} construction. Using the augmented diameter, we could bound the size of subgraphs (specifically in the construction of Step 4) arising during the course of our algorithm, and compute  the augmented diameters of clusters efficiently. We note that in Borradaile, Le and Wulff-Nilsen~\cite{BLW17}, the efficiency of the construction is not relevant since they use the cluster hierarchy to \emph{analyze the greedy algorithm}, not in the construction of the spanner. 

Our construction  has five main steps (Steps 1-5). In Step 1, we group all vertices of $\mv^{\high}_{i}$ and their neighbors into subgraphs of $\mathbb{X}$; see \Cref{lm:Clustering-Step1}. In Step 2, we deal with  \emph{branching nodes} of $\msttilde_{i}$; see \Cref{lm:Clustering-Step2}. In Step 3, we augment existing subgraphs  formed in Steps 1 and 2, to guarantee a special structure of the ungrouped nodes.  In Step 4, we group subpaths of  $\msttilde_{i}$ connected by an edge $\mbe$ in $\msttilde_{i}$  into clusters; see \Cref{lm:Clustering-Step4}. Finally, in Step 5, we deal with the remaining nodes of $\mv_i$.

Recall that $g$ is a constant defined in \hyperlink{P3}{property (P3)} (by \Cref{lm:ClusteringFast}, $g=31$), and that $\msttilde_{i}$ is a spanning tree of $\mg_i$. We refer readers to \Cref{table:notation} for a summary of the notation.

\begin{lemma}[Step 1]\label{lm:Clustering-Step1} Let $\mv^{\high}_i = \{\varphi_{C} \in \mv: \varphi_{C} \mbox{ is incident to at least $\frac{2g}{\eps}$ edges in } \me_i\}$. Let $\mv_i^{\high+}$ be obtained from $\mv^{\high}_i$  by adding all neighbors that are connected to nodes in $\mv^{\high}_i$ via edges in $\me_i$. We can construct in $O(|\mv_i| + |\me_i|)$ time a collection of node-disjoint subgraphs $\mathbb{X}_1$ of $\mg_i$ such that:
		\begin{enumerate}[noitemsep]
			\item[(1)] Each subgraph $\mx \in \mathbb{X}_1$ is a tree.
			\item[(2)] $\cup_{\mx \in \mathbb{X}_1}\mv(\mx) = \mv^{\high+}_i$.
			\item[(3)] $L_i \leq \adm(\mx) \leq 13L_i$, assuming that $\eps \leq 1/g$. 
			\item[(4)] $|\mv(\mx)|\geq \frac{2g}{\eps}$.
		\end{enumerate}
\end{lemma}
\begin{proof} Let $\mathcal{J} = (\mv_i,\me_i)$ be the subgraph of $\mg_i$ with the same vertex set and with edge set $\me_i$. Let $\mathcal{N}_{\mathcal{J}}(\varphi)$ be the set of neighbors of a node $\varphi$ in $\mathcal{J}$, and  $\mathcal{N}_{\mathcal{J}}[\varphi] = \mathcal{N}_{\mathcal{J}}(\varphi)\cup \{\varphi\}$.   We construct $\mathbb{X}_1$ in three steps; initially, $\mathbb{X}_1 = \emptyset$.
	\begin{enumerate}
		\item[(1)] Let $\mathcal{I}$ be  a \emph{maximal} set of nodes in $\mathcal{V}^{\high}$ such that for any two nodes $\varphi_{1},\varphi_{2} \in \mathcal{I}$,  $\mathcal{N}_{\mathcal{J}}[\varphi_{1}] \cap \mathcal{N}_{\mathcal{J}}[\varphi_{2}] = \emptyset$.   For each node $\varphi \in \mathcal{I}$, we form a subgraph $\mx$ that consists of $\varphi$, its neighbors $\mathcal{N}_{\mathcal{J}}[\varphi]$, and all incident edges in $\me_i$ of $\varphi$. We then add $\mx$ to $\mathbb{X}_1$.
		
		\item[(2)]  We iterate over all nodes of $\mv^{\high}_i\setminus \mathcal{I}$ that are not grouped yet to any subgraph. For each such node  $\varphi \in \mv^{\high}_i \setminus \mathcal{I}$, there must be a neighbor $\varphi'$ that is already grouped to a subgraph, say $\mx \in \mathbb{X}_1$; if there are multiple such neighbors, we pick one of them arbitrarily. We add $\varphi$ and the edge $(\varphi,\varphi')$ to $\mathcal{X}$. Observe that every node in $\mv^{\high}_i$ is grouped to some subgraph at the end of this step.
		\item[(3)]  For each node $\varphi$ in $\mv_i^{\high+}$ that has not grouped to a subgraph in steps (1) and (2), there must be at least one neighbor, say $\varphi'$, of $\varphi$ that is grouped in step (1) or step (2) to a subgraph $\mx \in \mathbb{X}_1$; if there are multiple such nodes, we pick one of them arbitrarily. We then add $\varphi$ and the edge $(\varphi,\varphi')$ to $\mx$. 
	\end{enumerate}
This completes the construction of $\mathbb{X}_1$. We now show that subgraphs in $\mathbb{X}_1$ have all desired properties.

Observe that Items (1) and (2) follow directly from the construction. For Item (4), we observe that every subgraph $\mx \in \mathbb{X}_1$ is created in step (1) and hence, contains a node $\varphi \in \mv^{\high}_i$ and all of its neighbors (in $\mathcal{J}$) by the definition of $\mi$. Thus, $|\mv(\mx)|\geq 2g/\eps$ since $\varphi$ has at least $2g/\eps$ neighbors.   For Item (3), we observe that each subgraph $\mx\in \mathbb{X}_1$ after step (3) has hop-diameter at least $2$ and at most $6$. Thus, $\adm(\mx) \leq 7 g\epsilon L_i + 6L_i ~\leq~ 13L_i$. Furthermore, since every edge $\mbe\in \me_{i}$ has a weight of at least $L_i/(1+\psi)\geq L_i/2$ and $\mx$ has at least two edges in $\me_i$,  $\adm(\mx)\geq 2(L_i/2) = L_i$. The construction time follows straightforwardly from the algorithm. 
\qed
\end{proof}

Given a tree $T$, we say that a node $x\in T$ is \emph{$T$-branching} if it has degree at least 3 in $T$.  For brevity, we shall omit the prefix $T$ in ``$T$-branching'' whenever this does not lead to  confusion.  Given a forest $F$, we say that $x$ is \emph{$F$-branching} if it is $T$-branching for some tree $T\subseteq F$. The construction of Step 2 is described in the following lemma.

\begin{lemma}[Step 2]\label{lm:Clustering-Step2} Let $\Ftilde^{(2)}_i$ be the forest obtained from $\msttilde_{i}$ by removing every node in $\mv^{\high+}_i$ (defined in \Cref{lm:Clustering-Step1}). We can construct in $O(|\mv_i|)$ time a collection $\mathbb{X}_2$ of subtrees of $\Ftilde^{(2)}_i$ such that for every $\mx\in \mathbb{X}_2$:
	\begin{enumerate}[noitemsep]
		\item[(1)] $\mx$ is a tree and has an $\mx$-branching node.
		\item[(2)] $L_i \leq \adm(\mx)\leq 2L_i$.
		\item[(3)] $|\mv(\mx)| = \Omega(\frac{1}{\epsilon})$  when $\epsilon \leq 1/g$. 
		\item[(4)] Let $\Ftilde^{(3)}_i$ be obtained from $\Ftilde^{(2)}_i$ by removing every node contained in subgraphs of $\mathbb{X}_2$. Then, for every tree $\Ttilde \subseteq \Ftilde^{(3)}_i$, either (4a) $\adm(\Ttilde)\leq 6L_i$ or (4b) $\Ttilde$ is a path.
	\end{enumerate}
\end{lemma}
\begin{proof}
	 We say that a tree $\Ttilde \in \Ftilde^{(2)}_i$ is  \emph{long} if $\adm(\Ttilde)\geq 6L_i$ and \emph{short} otherwise. We construct $\mathbb{X}_2$, initially empty, as follows:
	 \begin{itemize}
	 	\item Pick a long tree  $\Ttilde$ of $\Ftilde^{(2)}_i$ that has at least one $\Ttilde$-branching node, say $\varphi$. We traverse $\Ttilde$ starting from $\varphi$ and {\em truncate} the traversal at nodes whose augmented distance from $\varphi$ is at least $L_i$, which will be the leaves of the subtree. (The exact implementation details are delayed until the end of this proof.) 
	 	As a result, the augmented radius (with respect to the center $\varphi$) of the subtree induced by the visited (non-truncated) nodes is at least $L_i$ and at most $L_i + \bar{w} + g\epsilon L_i$.  We then form a subgraph, say $\mx$, from the subtree induced by the visited nodes, add $\mx$ to $\mathbb{X}_2$, remove every node of $\mx$ from $\Ttilde$, and repeat this step until it no longer applies. 
	 \end{itemize}
 	We observe that Item (1) follows directly from the construction. Since the algorithm only stops when every long tree has no branching node, meaning that it is a path, Item (4) is satisfied.  By construction, $\mx$ is a tree of augmented radius at least $L_i$ and at most $L_i + g\epsilon L_i +\bar{w}$, hence $L_i ~\le~ \adm(\mx) ~\leq~ 2(L_i + g\epsilon L_i +\bar{w}) ~\leq~ 6L_i$ since $\bar{w} < L_i$ and $\epsilon \leq 1/g$; this implies Item (2). Let $\md$ be an augmented diameter path of $\mx$; $\adm(\md)\geq L_i$ by construction. Note that every edge has a weight of at most $\bar{w} \leq L_{i-1}$ and every node has a weight of in $[L_{i-1},gL_{i-1}]$ by \hyperlink{P3'}{property (P3')}. Thus,  $\md$ has at least $\frac{\adm(\md)}{2gL_{i-1}} ~\geq~ \frac{L_i}{2g\eps L_i} = \Omega(\frac{1}{\epsilon})$ nodes; this implies Item (3). The construction of $\mathbb{X}_2$ can be implemented efficiently in  $O(|\mv_i|)$ by simply maintaining  a list $\mathcal{B}$ of branching nodes of $\Ftilde^{(2)}_i$.	\qed
\end{proof}

 The goal of constructing a subgraph from a branching node $\varphi$ is to guarantee that there must be at least one neighbor, say $\varphi'$, of $\varphi$ that does not belong to the augmented diameter path of $\mx$. Thus, we could show that the amount of corrected potential change $\Delta^{+}_{i+1}(\mx)$ is at least $ \omega(\varphi')\geq L_{i-1} = \eps L_i$. This will ultimately help us show that the corrected potential change $\Delta^{+}_{i+1}(\mx)$  is $\Omega(\eps^{2} |\mv(\mx)|L_i)$.

\paragraph{Step 3: Augmenting $\mathbb{X}_1\cup \mathbb{X}_2$.~}   Let $\Ftilde^{(3)}_i$ be the forest obtained in Item (4b) in \Cref{lm:Clustering-Step2}. Let $\mathcal{A}$ be the set of all nodes $\varphi$ in $\Ftilde^{(3)}_i$ such that $\varphi$ is in a tree $\Ttilde \in \Ftilde^{(3)}_3$ of augmented diameter at least $6L_i$ and $\varphi$ is a branching node in $\msttilde_i$.  For each node $\varphi \in \mathcal{A}$ such that $\varphi$ is connected to a node, say $\varphi'$,  in a subgraph $\mx \in \mathbb{X}_1\cup \mathbb{X}_2$ via an $\msttilde_{i}$ edge $\mathbf{e}$, we add $\varphi$ and $\mathbf{e}$ to $\mx$.  We note that $\varphi'$ exists since $\varphi$ has degree at least $3$ in  $\msttilde_i$. (If there are many such nodes $\varphi'$, we choose an arbitrary one.)

The following lemma follows directly from the construction.

\begin{lemma}\label{lm:Clustering-Step3} The augmentation in Step 3 can be implemented in $O(|\mv_i|)$ time, and increases the augmented diameter of each subgraph in  $\mathbb{X}_1\cup \mathbb{X}_2$ by at most $4L_i$ when $\eps \leq 1/g$. \\
Furthermore, let $\Ftilde^{(4)}_i$ be the forest obtained from $\Ftilde^{(3)}_i$ by removing every node in $\mathcal{A}$. Then, for every tree $\Ttilde \subseteq \Ftilde^{(4)}_i$, either:
	\begin{enumerate}[noitemsep]
		\item[(1)]$\adm(\Ttilde)\leq 6L_i$ or
		\item[(2)] $\Ttilde$ is a path such that (2a)  every node in $\Ttilde$ has \emph{degree at most $2$} in $\msttilde_{i}$ and (2b) at least one endpoint $\varphi$ of  $\Ttilde$ is connected via an $\msttilde_{i}$ edge to a node $\varphi'$ in a subgraph of $\mathbb{X}_1\cup \mathbb{X}_2$, unless $\mathbb{X}_1\cup \mathbb{X}_2 = \emptyset$. 
	\end{enumerate}
\end{lemma}

The main intuition behind Step 3 is to guarantee properties (2a) and (2b) for every long path $\Ttilde \in \Ftilde^{(4)}_i$. Recall that in Item (3) of \Cref{def:GiProp}, we guarantee that $\mg_i$ has no removable edge. Thus, any edge between two nodes in $\Ttilde$ is not removable. Later, we use this property to argue that the corrected potential change $\Delta^+_{i+1}(\mx)$ is non-trivial for every subgraph $\mx$ formed in the construction of Step 4 below.

\paragraph{Required definitions/preparations for Step 4.~} Let $\Ftilde^{(4)}_i$ be the forest obtained from $\Ftilde^{(3)}_i$ as described in \Cref{lm:Clustering-Step3}. By Item (2b) in  \Cref{lm:Clustering-Step3}, every tree of augmented diameter at least $6L_i$ of $\Ftilde^{(4)}_i$ is a simple path, which we call a \emph{long path}.

\begin{quote}
	\textbf{Red/Blue Coloring.~}\hypertarget{RBColoring}{}  Given a path $\Ptilde\subseteq \Ftilde^{(4)}_i$, we color their nodes red or blue. If a node has augmented distance at most $L_i$ from at least one of the path's endpoints, we color it red; otherwise, we color it blue. Observe that each red node belongs to the suffix or prefix of $\mathcal{P}$; the other nodes are colored blue. 
\end{quote}

\begin{lemma}[Step 4]\label{lm:Clustering-Step4} Let $\Ftilde^{(4)}_i$ be the forest obtained from $\Ftilde^{(3)}_i$ as described in \Cref{lm:Clustering-Step3}. We can construct in $O((|\mv_i| + |\me_i|)\epsilon^{-1})$ time a collection $\mathbb{X}_4$ of subgraphs of $\mg_i$ such that every $\mx\in \mathbb{X}_4$:
	\begin{enumerate}[noitemsep]
		\item[(1)] $\mx$ contains a single edge in $\me_i$.
		\item[(2)] $L_i \leq \adm(\mx)\leq 5L_i$.
		\item[(3)]  $|\mv(\mx)| = \Theta(\frac{1}{\epsilon})$ when $\epsilon \leq 1/(8(g+1))$. 
		\item[(4)] $\Delta_{i+1}^{+}(\mx) = \Omega(\eps^2 |\mv(\mx)| L_i)$.
		\item[(5)] Let $\Ftilde^{(5)}_i$ be obtained from $\Ftilde^{(4)}_i$ by removing every node contained in subgraphs of $\mathbb{X}_4$. If we apply \hyperlink{RBColoring}{Red/Blue Coloring} to each path of augmented diameter at least $6L_i$ in $\Ftilde^{(5)}_i$, then there is no edge in $\me_i$ that connects two blue nodes in $\Ftilde^{(5)}_i$.
	\end{enumerate}
\end{lemma}
\begin{proof} We only apply the construction to paths  of augmented diameter at least $6L_i$ in  $\Ftilde^{(4)}_i$, called \emph{long paths}.   Let $\Ptilde$ be a long path. For each blue node $\varphi \in \Ptilde$, we assign a subpath $\mathcal{I}(\varphi)$ of $\Ptilde$,
	called the \emph{interval of $\varphi$}, which contains every node within an augmented distance (in $\Ptilde$) at most $L_i$ from $\varphi$. By definition, we have:
	
	\begin{claim}\label{clm:Interval-node}
		For any blue node $\nu$, it holds that
		\begin{itemize}[noitemsep]
			\item[(a)] $ (2 - (3g+2)\epsilon)L_i \leq  \adm(\mathcal{I}(\nu))\leq 2L_i $.
			\item[(b)]   	Denote by  $\mi_1$ and $\mi_2$  the two subpaths obtained by removing $\nu$ from the path $\mi(\nu)$. 
			Each of these subpaths has $\Theta(\frac{1}{\epsilon})$ nodes and augmented diameter at least $(1-2(g+1)\epsilon)L_i$.
		\end{itemize}
	\end{claim}
	
	We keep track of a list $\mathcal{B}$ of edges in $\me_i$ with both \emph{blue endpoints}. We then construct $\mathbb{X}_4$, initially empty, as follows:
	
	\begin{itemize}
		\item  Pick an edge $(\nu,\mu)$ with both blue endpoints,  form a subgraph $\mx = \{(\nu,\mu)\cup \mathcal{I}(\nu) \cup \mathcal{I}(\mu)\}$, and add $\mx$ to $\mathbb{X}_4$. We then remove  all nodes in  $\mathcal{I}_{\nu} \cup \mathcal{I}_{\mu} $ from the path or two paths containing $\nu$ and $\mu$, update the color of nodes in the new paths to satisfy \hyperlink{RBColoring}{Red/Blue Coloring} and the edge set $\mathcal{B}$, and repeat this step until it no longer applies.
	\end{itemize}
	
	We observe that Items (1) and (5) follow directly from the construction. For Item (2), we observe by Claim~\ref{clm:Interval-node}  that $\mathcal{I}(v)$ has augmented diameter at most $2L_i$ and at least $L_i$ when $\epsilon \leq \frac{1}{8(g+1)}$, and  the weight of the edge $(\mu,\nu)$ is at most $L_i$. Thus, $L_i \leq \adm(\mx)\leq L_i + 2\cdot 2L_i = 5L_i$, as claimed. Item (3) follows directly from  Claim~\ref{clm:Interval-node} since $|\mathcal{I}(v)| = \Theta(\frac{1}{\epsilon})$ and $|\mathcal{I}(\mu)| = \Theta(\frac{1}{\epsilon})$. 
	
	Next, we show that the construction of $\mathbb{X}_4$ can be implemented efficiently.  Since the interval $\mi(\nu)$ assigned to each blue node $\nu$ consists of $O(\frac{1}{\epsilon})$ nodes by Claim~\ref{clm:Interval-node}(b), it takes $O(|\mathcal{E}_i|\epsilon^{-1})$ time to construct  $\mathcal{B}$. For each edge $(\nu,\mu) \in \mathcal{B}$ picked in the construction of $\mathbb{X}_4$, forming $\mx = \{(\nu,\mu)\cup \mathcal{I}(\nu) \cup \mathcal{I}(\mu)\}$ takes $O(1)$ time. When removing any such interval $\mathcal{I}(\nu)$ from a path $\Ptilde$, we may create two new sub-paths $\Ptilde_1,\Ptilde_2$, and then need to recolor  the nodes following  \hyperlink{RBColoring}{Red/Blue Coloring}. Specifically, some blue nodes in the prefix and/or suffix of $\Ptilde_1,\Ptilde_2$ are colored red; importantly, a node's color may only change from blue to red, but it may not change in the other direction. Since the total number of nodes to be recolored as a result of removing such an interval $\mathcal{I}(\nu)$ is $ O(\frac{1}{\epsilon})$,  the total recoloring running time is $O(|\mv(\Ftilde^{(4)}_i)|\epsilon^{-1}) = O(|\mv_i|\eps^{-1})$.
	To bound the time required for updating the edge set $\mathcal{B}$ throughout this process, we
	note that edges are never added to $\mathcal{B}$ after its initiation. Specifically, when a blue node $\nu$ is 	recolored as red, we remove all incident edges of $\nu$ from  $\mathcal{B}$, and none of these edges will be considered again; this can be done in $O(\frac{1}{\epsilon})$ time per node $\nu$, since $\nu$ is incident to at most $\frac{2g}{\eps} = O(\frac{1}{\eps})$ edges in $\me_i$ due to the construction of Step 1 (\Cref{lm:Clustering-Step1}). Once a node is added to $\mx$, it will never be considered again. It follows that the total running time required for implementing Step 3 is  $O(|\mv_i|\eps^{-1})$, as claimed.
	
	We now prove Item (4). We consider two cases.
	
	\paragraph{Case 1: $\mi(\nu) \cap \mi(\mu) =\emptyset$.~} Let $\mx = (\nu,\mu)\cup \mi(\nu) \cup \mi(\mu)$ where $\mbe = (\nu,\mu)$ is the only edge in $\me_i$ contained in $\mx$.  For any subgraph $\mz$ of $\mx$, we define: 
	\begin{equation}\label{eq:Potential-Y-D}
		\Phi^+(\mz) = \sum_{\alpha \in \mz}\omega(\alpha) + \sum_{\mbe' \in \widetilde{\mst}_i\cap \me(\mz)} \omega(\mbe)
	\end{equation}
	to be the total weight of nodes and $\widetilde{\mst}_i$ edges in $\mz$.  Let $\md$ be an augmented diameter path of $\mx$, and $\my = \mx \setminus \mv(\md)$ be the subgraph obtained from $\mx$ by removing nodes on $\md$. Let $\mi(\nu)$ and $\mi(\mu)$ be two intervals  in the construction on Step 4 that are connected by an edge $\mbe = (\nu,\mu)$.

	\begin{claim}\label{clm:PotentialY-bound} $\Phi^+(\my) =  \frac{5 L_i}{4} +\Omega(|\mv(\my)|\epsilon L_i)$.
	\end{claim}
	\begin{proof}
		Let $\ma = \my \setminus (\mi(\nu)\cup \mi(\mu))$ be the subgraph of $\my$ obtained by removing every node in $\mi(\nu)\cup \mi(\mu)$ from $\my$, and $\mb = \my \cap (\mi(\nu)\cup \mi(\mu))$ be the subgraph of $\my$ induced by nodes of $\my$ in $(\mi(\nu)\cup \mi(\mu))$. Observe that $\Phi^+(\ma) \geq |\mv(\ma)| L_{i-1} = |\mv(\ma)| \epsilon L_i. $. If $\md$ does not contain the edge $(\nu,\mu)$ (see Figure~\ref{fig:Step3-diam}(a)), then $\mathcal{I}(\nu)\cap\md =\emptyset$, say, which implies  $\Phi^+(\mb)\geq \adm(\mathcal{I}(\nu)) \geq (2-(3g+2)\epsilon) L_i$ by Claim~\ref{clm:Interval-node}. If $D$ contains the edge $(\nu,\mu)$ (see Figure~\ref{fig:Step3-diam}(b)), then at least two sub-intervals, say $\mi_1,\mi_2$, are disjoint from $\md$.  
			By Claim~\ref{clm:Interval-node}, $\Phi^+(\mb)\geq \adm(\mi_1) + \adm(\mi_2) \geq (2- 4(g+1)\epsilon) L_i$.	In both cases, $\Phi^+(\mb) \geq (2- 4(g+1)\epsilon) L_i \geq \frac{3 L_i}{2} $ when $\epsilon  \leq \frac{1}{8(g+1)}$.  Thus:
		\begin{equation*}
			\begin{split}
				\Phi^+(\my) &\geq \Phi^+(\ma) + \frac{3 L_i}{2}~=~  \frac{5 L_i}{4}+ \Omega((|\mv(\ma)| + |\mv(\mb)|)\epsilon L_i) =\frac{5 L_i}{4}+ \Omega(|\mv(\my)|\epsilon L_i),
			\end{split}
		\end{equation*}
		which concludes the proof of Claim~\ref{clm:PotentialY-bound}.\qed		 	
	\end{proof}
	
	\begin{wrapfigure}{r}{0.5\textwidth}
		\vspace{-15pt}
		\begin{center}
			\includegraphics[width=0.35\textwidth]{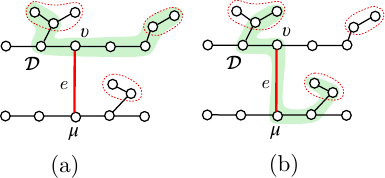}
		\end{center}
		\caption{\footnotesize{$\mathcal{D}$ is the diameter path and enclosed trees are augmented to a Step-4 subgraph in Step 5A. The green shaded regions contain nodes in $\mathcal{D}$.  (a) $\mathcal{D}$ does not contain $\mathbf{e}$. (b) $\mathcal{D}$ contains $\mathbf{e}$.}}
		\vspace{-35pt}
		\label{fig:Step3-diam}
	\end{wrapfigure}
	
	Note that $\mv(\md) \leq \frac{gL_i}{L_{i-1}}  = O(\frac{1}{\epsilon})$ since every node has weight at least $L_{i-1}$ by \hyperlink{P3'}{property (P3')}. Thus, we have:
	\begin{equation*}
		\begin{split}
			\Delta_{i}^+(\mx)  &= \Phi^+(\md) + \Phi^+(\my) - \adm(\mx) = \Phi(\my) - \omega(\mbe)\\ & \geq L_i/4 +  \Omega(|\mv(\my)|\epsilon L_i) \qquad \mbox{(by Claim~\ref{clm:PotentialY-bound})}\\
			&= \Omega( |\mv(\md)| \epsilon L_i) + \Omega(|\mv(\my)|\epsilon L_i)= \Omega(|\mv(\mx)|\epsilon L_i)~.
		\end{split}
	\end{equation*}
	
	Thus, Item (4) of \Cref{lm:Clustering-Step4} follows.
	
	\paragraph{Case 2: $\mi(\nu) \cap \mi(\mu) \not=\emptyset$.~}
	Let $\md$ be a diameter path of $\mx$, and $\my = \mx \setminus \mv(\md)$. Recall that $\mx$ contains only one edge $\mbe = (\nu,\mu) \in \me_i$. Let $\mathcal{P}_{\mbe} = (\nu,\mbe,\mu)$ be the path that consists of only edge $\mbe$ and its endpoints. Let  $\mathcal{P}[\nu,\mu]$ be the subpath of $\msttilde_{i}$ between $\nu$ and $\mu$.
	
	\begin{wrapfigure}{r}{0.4\textwidth}
		\vspace{-20pt}
		\begin{center}
			\includegraphics[width=0.4\textwidth]{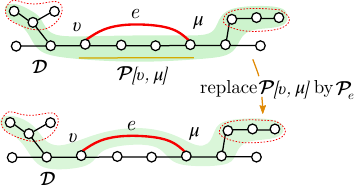}
		\end{center}
		\caption{Nodes enclosed in dashed red curves are augmented to $\mx$ in Step 4.}
		\vspace{-15pt}
		\label{fig:replace-e-vsPe}
	\end{wrapfigure}
	
	We observe that $\mbe$ is not removable by Item (3) of \Cref{def:GiProp}. Then it follows that:

	\begin{equation}\label{eq:p-vs-pe}
		\begin{split}
			\omega(\mathcal{P}[\nu,\mu]) - 	\omega(\mathcal{P}_e) ) &> 6g\epsilon  \cdot \omega(\mbe)    - w(\nu) - w(\mu) \\
			&> 6g\epsilon L_i/2 - 2g\epsilon L_i= g\epsilon L_i
		\end{split}
	\end{equation}
	In particular, this means that $	\omega(\mathcal{P}(\nu,\mu)) \geq 	\omega(\mbe)$.

	Thus, if $\md$ contains both $\nu$ and $\mu$, then it must contain $\mbe$, since otherwise, $\md$ must contain $\mathcal{P}[\nu,\mu]$ and by replacing $\mathcal{P}[\nu,\mu]$ by $\mathcal{P}_{\mbe}$ we obtain a shorter path by Equation~\eqref{eq:p-vs-pe} (see Figure~\ref{fig:replace-e-vsPe}).  Observe that

	\begin{observation}\label{clm:D-P-size}
		$|\mv(\mp[\nu,\mu])|\leq \frac{4}{\epsilon}$ and $|\mv(\md)|\leq \frac{g}{\epsilon}$.
	\end{observation}
	
	\noindent We consider two cases:
	\begin{itemize}
		\item \emph{Case 1} If $\md$ does not contain edge $\mbe$, then (a) $\md\subseteq \widetilde{\mst}_i$  and (b) $|\{\nu,\mu\} \cap D| \leq 1$. From (a) and \Cref{clm:D-P-size}, we have:
		\begin{equation}
			\begin{split}
				\Delta_{i+1}^+(\mx) &\geq \adm(\md) + \Phi^+(\my) - \adm(\mx) =  \Phi^+(\my) \\
				&\geq \adm(\mp[\mu,\nu]) + \Phi^+(\my\setminus \mp[\mu,\nu])\\
				&\geq w(\mbe) 		+  |\mv(\my\setminus \mp[\mu,\nu])|L_{i-1} \geq L_i/2 + |\mv(\my\setminus \mp[\mu,\nu])|\epsilon L_{i} \\
				&= \Omega(\eps(|\mv(\mp[\mu,\nu])| + |\mv(\md)|)L_i) +   |\mv(\my\setminus \mp[\mu,\nu])|\epsilon L_{i} = \Omega(|\mv(\mx)|\epsilon L_i)
			\end{split}
		\end{equation}
		
		\item \emph{Case 2} If $\md$ contains $\mbe$, then $\md \cap \mp(\nu,\mu) = \emptyset$; here $\mp(\nu,\mu)$ is the path obtained from $\mp[\nu,\mu]$ by removing its endpoints. It follows that
		\begin{equation}
			\begin{split}
				\Delta_{i+1}^+(\mx) &\geq \adm(\md) + \Phi^+(\my) - \adm(\mx) =  \Phi(\my) - w(\mbe) \\
				&\geq \adm(\mp[\mu,\nu]) + \Phi^+(\my\setminus \mp[\mu,\nu]) - w(\mbe)\\
				&\geq g\epsilon L_i 		+  |\mv(\my\setminus \mp[\mu,\nu])|L_{i-1} \qquad\mbox{(by Equation~\eqref{eq:p-vs-pe})} \\
				&= \Omega((|\mv(\mp[\mu,\nu])| + |\mv(\md)|)\epsilon^2 L_i) +   |\mv(\my\setminus \mp[\mu,\nu])|\epsilon L_{i}  = \Omega(|\mv(\mx)|\epsilon^2 L_i)
			\end{split}
		\end{equation}
	\end{itemize}
	In both cases, we have $\Delta^+_{i+1}(\mx) = \Omega(|\mv(\mx)|\epsilon^2 L_i)$ as claimed in Item (4) of \Cref{lm:Clustering-Step4}.  \qed		
\end{proof}

\begin{observation}\label{obs:Clustering-F5} For every tree $\Ttilde \subseteq \Ftilde^{(5)}_i$ such that $\adm(\Ttilde) \leq 6L_i$, then $\Ttilde$ is connected via $\msttilde_{i}$ edge to a node in some subgraph $\mx \in \mathbb{X}_1 \cup \mathbb{X}_2\cup \mathbb{X}_4$, unless there is no subgraph formed in Steps 1-4, i.e, $ \mathbb{X}_1 \cup \mathbb{X}_2\cup \mathbb{X}_4 = \emptyset$. 
\end{observation}

We call the case where $\mathbb{X}_1 \cup \mathbb{X}_2\cup \mathbb{X}_4 = \emptyset$ the \emph{degenerate case}. When the degenerate case happens,  $\mg_i$ has a very special structure, which will be described later (in \Cref{lm:exception}); for now, we focus on the construction of the last step. 

\paragraph{Step 5.~} Let $\Ttilde$ be  a path in  $\Ftilde^{(5)}_i$ obtained by Item (5) of \Cref{lm:Clustering-Step4}. We construct two sets of subgraphs, denoted by $\mathbb{X}^{\internal}_5$ and $\mathbb{X}^{\prefix}_5$, of $\mg_i$. The construction is broken into two steps. Step 5A is only applicable when the degenerate case does not happen; Step 5B is applicable regardless of the degenerate case.  
\begin{itemize}
	\item (Step 5A)\hypertarget{5AFast}{}  If $\Ttilde$ has augmented diameter at most $6L_i$, let $\mathbf{e}$ be an $\widetilde{\mst}_i$ edge connecting $\Ttilde$  and a node in some subgraph $\mx \in \mathbb{X}_1\cup \mathbb{X}_2 \cup \mathbb{X}_4$; $\mathbf{e}$ exists by \Cref{obs:Clustering-F5}. We add both $\mathbf{e}$ and $\Ttilde$ to $\mx$.
	\item (Step 5B)\hypertarget{5BFast}{} 	Otherwise,  the augmented diameter of $\Ttilde$ is at least $6L_i$ and hence, it must be a path by Item (4) in \Cref{lm:Clustering-Step2}.  In this case, we greedily break $\Ttilde$ into subpaths of augmented diameter at least $L_i$ and at most $2L_i$. Let $\Ptilde$ be a subpath broken from $\Ttilde$. If $\Ptilde$ is connected to a node in a subgraph $\mx$ via an  edge $\mathbf{e}\in \msttilde_{i}$, we add $\Ptilde$ and $\mathbf{e}$ to $\mx$.	If $\Ptilde$ contains an endpoint of $\Ttilde$, we add $\Ptilde$ to $\mathbb{X}^{\prefix}_5$; otherwise, we add $\Ptilde$ to $\mathbb{X}^{\internal}_5$. 
\end{itemize}

\begin{lemma}\label{lm:Clustering-Step5} We can implement the construction of $ \mathbb{X}_5^{\internal}$ and $\mathbb{X}_5^{\prefix}$ in $O(|\mv_i|)$ time.	Furthermore, every subgraph $\mx \in \mathbb{X}_5^{\internal} \cup \mathbb{X}_5^{\prefix}$ satisfies:
	\begin{enumerate}[noitemsep]
		\item[(1)] $\mx$ is a subpath of $\msttilde_{i}$.
		\item[(2)] $L_i \leq \adm(\mx)\leq 2 L_i$ when $\eps \leq 1/g$.
		\item[(3)] $|\mv(\mx)| = \Theta(\frac{1}{\epsilon})$.
	\end{enumerate}
\end{lemma}
\begin{proof} 
	Items (1) and (2) follow directly from the construction. For Item (3), we observe the following facts:   $\adm(\mx)\geq L_i$,   each edge has a weight of at most $L_{i-1}$, and each node has a weight of at most $g L_{i-1}$. Thus,  $|\mv(\mx)|\geq \frac{L_i}{(1+g)L_{i-1}} = \Omega(\frac{1}{\epsilon})$. By the same argument, since each node has a weight at least $L_{i-1}$ by \hyperlink{P3'}{property (P3')},  $|\mv(\mx)|\leq \frac{2L_i}{L_{i-1}} = O(1/\eps)$. The construction time follows by implementing the algorithm greedily. 
\end{proof}

Finally, we construct the collection $\mathbb{X}$ of subgraphs of $\mg_i$ as follows:
\begin{equation}\label{eq:MathbbX}
	\mathbb{X} = \mathbb{X}_1 \cup \mathbb{X}_2\cup \mathbb{X}_4 \cup \mathbb{X}_5^{\internal} \cup \mathbb{X}_5^{\prefix}.
\end{equation}

To complete the proof of \Cref{lm:ClusteringFast}, we need to:
\begin{enumerate}
	\item show that subgraphs in  $\mathbb{X}$ satisfies three properties: \hyperlink{P1'}{(P1')}, \hyperlink{P2'}{(P2')}, and \hyperlink{P3'}{(P3')}, and that $|\me_i\cap \me(\mx)| = O(|\mv(\mx)|)$. This implies Item (5) of \Cref{lm:ClusteringFast}. We present the proof in \Cref{subsec:PropX}.
	\item  construct a partition $\{\mv_i^{\high},\mv_i^{\lowp}, \mv_i^{\lowm}\}$  of $\mv_i$, show Items (1)-(4) and the running time bound as claimed by \Cref{lm:ClusteringFast}. We present the proof in \Cref{subsec:PartitionMvi}
\end{enumerate}

\subsection{Properties of $\mathbb{X}$}\label{subsec:PropX}

In this section, we prove the following lemma.

\begin{lemma}\label{lm:XProp} Let $\mathbb{X}$ be the set of subgraph as defined in \Cref{eq:MathbbX}. For every subgraph $\mx \in \mathbb{X}$, $\mx$ satisfies the three properties (\hyperlink{P1'}{P1'})-(\hyperlink{P3'}{P3'}) with $g = 31$ and $\eps \leq \frac{1}{8(g+1)}$, and $|\me(\mx)\cap \me_i| = O(|\mv(\mx)|)$. Furthermore, $\mathbb{X}$ can be constructed in $O((|\mv_i| + |\me_i|)\eps^{-1})$ time.
\end{lemma}
\begin{proof} We observe that \hyperlink{P1'}{property (P1')} follows directly from the construction. Additionally,  \hyperlink{P2'}{property (P2')} follows from Item (4) of \Cref{lm:Clustering-Step1}, Items (3) of \Cref{lm:Clustering-Step2}, \Cref{lm:Clustering-Step4}, and \Cref{lm:Clustering-Step5}. The lower bound $L_i$ on the augmented diameter  of a subgraph $\mx \in \mathbb{X}$ follows from Item (3) of \Cref{lm:Clustering-Step1}, Items (2) of \Cref{lm:Clustering-Step2}, \Cref{lm:Clustering-Step4}, and \Cref{lm:Clustering-Step5}. Thus, to complete the proof of \hyperlink{P3'}{property (P3')}, it remains to show that $\adm(\mx) \leq gL_i$ with $g = 31$ and $\eps \leq \frac{1}{8(g+1)}$. Observe that  the condition that $\eps \leq \frac{1}{8(g+1)}$ follows by considering all constraints on $\eps$ in \Cref{lm:Clustering-Step1,lm:Clustering-Step2,lm:Clustering-Step3,lm:Clustering-Step4,lm:Clustering-Step5}.

	If $\mx$ is formed in Step 5B, that is $\mx \in \mathbb{X}^{\internal}_5\cup \mathbb{X}^{\prefix}_5$,  then $\adm(\mx)~\leq~ 2L_i$ by \Cref{lm:Clustering-Step5}. Otherwise, excluding any augmentation to $\mx$ due to Step 5, \Cref{lm:Clustering-Step1},  \Cref{lm:Clustering-Step2} and \Cref{lm:Clustering-Step3} yield $\adm(\mx) \leq 13L_i + 4L_i \leq 17 L_i$ where $+4L_i$ is due to the augmentation in Step 3 (see  \Cref{lm:Clustering-Step3}). By \Cref{lm:Clustering-Step4}, $\adm(\mx) \leq \max(17L_i,5L_i) = 17L_i$.

	 We then may augment $\mx$ with trees of diameter at most $6L_i$ (Step 5A) and/or with subpaths of diameter at most $2L_i$ (Step 5B). As the augmentation is star-like and via  $\widetilde{\mst}_i$ edges, if we denote the resulting subgraph by $\mx^+$, then 
	\begin{equation*}
		\adm(\mx^+) \leq \adm(\mx) + 2\bar{w} + 12 L_i \leq \adm(\mx) + 14L_i  \leq 31L_i. \end{equation*}
	 \hyperlink{P3'}{Property (P3')} now follows.
	
	The fact that $|\me(\mx)\cap \me_i| = O(|\mv(\mx)|)$ and the running time bound follow directly from  \Cref{lm:Clustering-Step1}, \Cref{lm:Clustering-Step2}, \Cref{lm:Clustering-Step3}, \Cref{lm:Clustering-Step4} and \Cref{lm:Clustering-Step5}. Recall that the augmentation in Step 3 is in a star-like way and hence, no cycle is formed in subgraphs of $\mathbb{X}_1\cup \mathbb{X}_2$ after the augmentation.  
	\qed
\end{proof}

\subsection{Constructing  a Partition of $\mv_i$}\label{subsec:PartitionMvi}

We first consider the degenerate case where $\mathbb{X}_1\cup \mathbb{X}_2\cup \mathbb{X}_4 = \emptyset$. 

\begin{lemma}[Structure of Degenerate Case]\label{lm:exception}
	If  $\mathbb{X}_1\cup \mathbb{X}_2\cup \mathbb{X}_4 = \emptyset$, then
	$\Ftilde^{(5)}_i =  \msttilde_{i}$, and $\msttilde_{i}$  is a single (long) path.   	Moreover, every edge $\mbe \in \me_i$ must be incident to a  node in $\Ptilde_1\cup \Ptilde_2$,
	where $\Ptilde_1$ and $\Ptilde_2$ are the prefix and suffix subpaths of $\msttilde_{i}$ of augmented diameter at most $L_i$. Consequently, we have that $|\me_i| = O(1/\epsilon^2)$.
\end{lemma}
\begin{proof}	By the assumption of the lemma, no subgraph is formed in Steps 1-4.
	
	\begin{wrapfigure}{r}{0.5\textwidth}
		\vspace{-15pt}
		\begin{center}
			\includegraphics[width=0.4\textwidth]{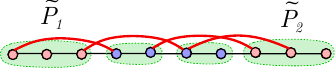}
		\end{center}
		\caption{Red edges are edges in $\me_i$; every edge is incident to at least one red node.}
		\vspace{-15pt}
		\label{fig:exception}
	\end{wrapfigure}
	
	Since no subgraph is formed in Step 1, $\Ftilde^{(2)}_i = \msttilde_i$.  Since no subgraph is formed in Step 2, there is no branching node in $\Ftilde^{(2)}_i$; thus $\Ftilde^{(3)}_i = \Ftilde^{(2)}_i$ and it is a  single (long) path. 	Since $\mathbb{X}_1 \cup \mathbb{X}_2  = \emptyset$, there  is no augmentation in Step 3.  Since no subgraph is formed in Step 4, $\Ftilde^{(5)}_i = \Ftilde^{(4)}_i$ and both are equal to $\msttilde_i$, which is a long path (see Figure~\ref{fig:exception}).

	By Item (5) in \Cref{lm:Clustering-Step4}, any edge $\mbe \in \me_i$ must be incident to a red node. The augmented distance from any red node to at least one endpoint of $\msttilde_{i}$ is at most $L_i$ by the  definition of \hyperlink{RBColoring}{Red/Blue Coloring}, and hence every red node belongs to $\Ptilde_1\cup \Ptilde_2$. Since each node has a weight of at least $L_{i-1}$ by \hyperlink{P3'}{property (P3')}, we have:
	\begin{equation*}
		|\mv(\Ptilde_1\cup \Ptilde_2)|\leq  \frac{2L_i}{L_{i-1}} = \frac{2}{\epsilon}
	\end{equation*}
	Since each node of $\Ptilde_1\cup \Ptilde_2$ is incident to at most $\frac{2g}{\epsilon}$ edges in $\me_i$ (as there is no subgraph formed in Step 1; $\mv^{\high}_i = \emptyset$), it holds that $|\me_i| = O(1/\epsilon^2)$, as desired.	\qed
\end{proof}

We are now ready to describe the construction of the partition  $\{\mv^{\high}_i, \mv^{\lowp}_i, \mv^{\lowm}_i\}$ of $\mv_i$

\begin{tcolorbox}
	\hypertarget{PartitionV}{}
	\textbf{Construct Partition $\{\mv^{\high}_i, \mv^{\lowp}_i, \mv^{\lowm}_i\}$:} If the degenerate case happens, we define $\mv^{\lowm}_i = \mv_i$ and $\mv^{\high}_i = \mv^{\lowp}_i = \emptyset$. Otherwise, we define $\mv^{\high}_i$ to be the set of all nodes that are incident to at least $2g/\eps$ edges in $\me_i$,  $\mv^{\lowm}_i = \cup_{\mx \in \mathbb{X}^{\internal}_5}\mv(\mx)$ and $\mv^{\lowp}_i = \mv_i\setminus (\mv^{\high}_i \cup \mv^{\lowm}_i )$. 
\end{tcolorbox}

We show the following property of $\{\mv^{\high}_i, \mv^{\lowp}_i, \mv^{\lowm}_i\}$, which is equivalent to Item (4) in \Cref{lm:ClusteringFast}.

\begin{lemma}\label{lm:Xlowm} 
	\begin{enumerate}
		\item[(1)] If $\mx$ contains a node in $\mv^{\lowm}$, then $\mv(\mx)\subseteq \mv^{\lowm}$.
		\item[(2)]	There is no edge in $\me_i$ between a node in $\mv^{\high}_i$ and a node in $\mv^{\lowm}_i$.
		\item[(3)]  If there exists an edge   $(\varphi_{C_u},\varphi_{C_v}) \in \me_i$ such that both $\varphi_{C_u}$ and $\varphi_{C_v}$ are in 
		$\mv_i^{\lowm}$, then the degenerate case happens.
	\end{enumerate}
\end{lemma}
\begin{proof}
	Item (1) follows directly from the construction. By the construction of Step 1 (\Cref{lm:Clustering-Step1}), any neighbor, say $\varphi$, of a node in $\mv^{\high}_i$ is in $\mv^{\high+}_i$. Thus, $\varphi$ will not be considered after Step 1. It follows that there is no edge between a node in $\mv^{\high}_i$ and a node in $\mv^{\lowm}_i$ since nodes in $\mv^{\lowm}_i$ are in Step 5; Item (2) follows. To show Item (3), we observe that every node, say $\varphi_{C_u}$, in $\mv^{\lowm}$ is a blue node of some long path $\Ptilde$ in $\Ftilde^{(5)}_i$. If the degenerate case does not happen, then by Item (5) of \Cref{lm:Clustering-Step4}, every edge  $(\varphi_{C_u},\varphi_{C_v})$ must have the node $\varphi_{C_v}$ being a red node of $\Ptilde$. But then by the construction of Step \hyperlink{5BFast}{5B}, $\varphi_{C_v}$ belongs to some subgraph of $\mathbb{X}_5^{\prefix}$ and hence is not in $\mv^{\lowm}$.\qed
\end{proof}

Next, we focus on bounding the corrected potential change $\Delta^+_i(\mx)$ of every cluster $\mx\in \mathbb{X}$. Specifically, we show that: 
	\begin{itemize}[noitemsep]
		\item if $\mx \in \mathbb{X}_1$, then $\Delta^+_{i+1}(\mx) = \Omega(|\mv(\mx)|L_i\epsilon)$; the proof is in \Cref{lm:Step1Potential}.
		\item if $\mx \in \mathbb{X}_2$, then $\Delta^+_{i+1}(\mx) = \Omega(|\mv(\mx)|L_i\epsilon^2)$; the proof is in \Cref{lm:Step2Potential}. 
		\item  if $\mx \in \mathbb{X}_4$, then $\Delta^+_{i+1}(\mx) = \Omega(|\mv(\mx)|L_i\epsilon^2)$; the proof is in \Cref{lm:Step4Potential}. 
		\item the corrected potential change is non-negative and we provide a lower bound  of the average corrected potential change for subgraphs in $\mathbb{X}\setminus \mathbb{X}^{\lowm}$ in \Cref{lm:Step5PotentialChange}. 
	\end{itemize}

\begin{lemma}\label{lm:Step1Potential} For every subgraph $\mx\in \mathbb{X}_1$, it holds that $ \Delta^+_{i+1}(\mx)\geq   \frac{|\mv(\mx)|L_i\epsilon}{2}$. 
\end{lemma}
\begin{proof}
	 Let $\mx \in \mathbb{X}_1$ be a subgraph  formed in Step 1. By Item (4) of \Cref{lm:Clustering-Step1}, $|\mv(\mx)| \geq \frac{2g}{\epsilon}$.   By definition of $\Delta^i_L(\mx)$ (\Cref{lm:ClusteringFast}), we have:
	\begin{equation}\label{eq:Step1-potential}
		\begin{split}
			\Delta^+_{i+1}(\mx) &\geq \sum_{\varphi \in \mv(\mx)}\omega(\varphi)  - \adm(\mx) \stackrel{\mbox{\hyperlink{P3'}{\tiny{(P3')}}}}{\geq} \sum_{\varphi \in \mx}L_{i-1} - gL_i  = \frac{|\mv(\mx)| L_{i-1}}{2} + \underbrace{(\frac{|\mv(\mx)| L_{i-1}}{2} - gL_i)}_{\geq 0 \mbox{ since }|\mv(\mx)|\geq (2g)/\epsilon}\\
			&\geq  \frac{|\mv(\mx)| L_{i-1}}{2}  =  \frac{|\mv(\mx)|\epsilon L_{i}}{2},
		\end{split}
	\end{equation} 
as claimed. \qed
\end{proof}

\begin{lemma}\label{lm:Step2Potential}For every subgraph $\mx\in \mathbb{X}_2$,  it holds that 
		$\Delta^+_{i+1}(\mx)\ = \Omega\left( |\mv(\mx)|L_i\epsilon^2\right)$.
\end{lemma}
\begin{proof}
	Let $\mx$ be a subgraph that is initially formed in Step 2 and could   possibly be augmented in Steps  3 and 5.   Recall that in the augmentation done in Step 3, we add to $\mx$ nodes of $\mv_i$ via  $\widetilde{\mst}_i$ edges, and in the augmentation done in Step 5, we add to $\mx$ subtrees of $\widetilde{\mst}_i$ via $\widetilde{\mst}_i$ edges. Thus, the resulting subgraph after the augmentation remains, as prior to the augmentation, a subtree of $\widetilde{\mst}_i$. That is, $\me(\mx)\subseteq \msttilde_{i}$. Letting $\md$ be an augmented diameter path of $\mx$, we have by definition of augmented diameter that
	\begin{equation*}
		\adm(\mx)  = \sum_{\varphi \in \md}\omega(\varphi) + \sum_{\mbe \in \me(\md)} \omega(e)
	\end{equation*}
	Let $\my = \mv(\mx)\setminus \mv(\md)$. Then $|\my| >  0$  since $\mx$ has a $\mx$-branching node by Item (1) of \Cref{lm:Clustering-Step2} and that 
	\begin{equation}\label{eq:Step2-Potential}
		\begin{split}
			\Delta^+_{i+1}(\mx) =  \left(\sum_{\varphi \in \mx}\omega(\varphi) + \sum_{ e\in \me(\mx)} \omega(e)\right) - \adm(\mx) \geq \sum_{\varphi \in \my}\omega(\varphi) \stackrel{\mbox{\hyperlink{P3'}{\tiny{(P3')}}}}{\geq} |\my|L_{i-1}
		\end{split}
	\end{equation} 
	As  $\adm(\md) \leq gL_i$, it holds that $	|\mv(\md)|  =O(1/\eps) = O(\frac{|\my|}{\epsilon})$. Thus,
	\begin{equation*}
		\Delta^+_{i+1}(\mx) \geq \frac{|\my| L_{i-1}}{2} +  \Omega(\epsilon|\mv(\md)|L_{i-1}) = \Omega((|\my| + \mv(\md))\epsilon L_{i-1}) = \Omega(|\mv(\mx)|\epsilon^2 L_i),
	\end{equation*}
as claimed.	\qed
\end{proof}

\begin{lemma}\label{lm:Step4Potential} For every subgraph $\mx\in \mathbb{X}_4$,  it holds that 
	$\Delta^+_{i+1}(\mx) = \Omega\left( |\mv(\mx)|L_i\epsilon^2\right)$.
\end{lemma}
\begin{proof}  Let $\mx\in \mathbb{X}_4$ be a subgraph initially formed in Step 4; $\mx$ is possibly augmented in Step 5. Let $\mx^+$ be $\mx$ after the augmentation (if any).  Let $\md^+$ be the augmented diameter path of $\mx^+$ and $\md = \md^+\cap \mx$.  By the same argument in \Cref{lm:Step2Potential},
 \begin{equation}\label{eq:Step4PotentialXplus}
 	\Delta_{i+1}^{+}(\mx)  = \Omega(|\mv(\mx)|\eps^2 L_i) =  \Omega(|\mv(\mx) \cup \mv(\md^+)|\eps^2 L_i).
 \end{equation}
Furthermore,
 \begin{equation*}
 	\begin{split}
 		\Delta_{i+1}^{+}(\mx^+) & = \sum_{\varphi \in \mx^+}\omega(\varphi) + \sum_{\mathbf{e}\in \me(\mx^+)\cap \msttilde_i} \omega(\mathbf{e}) - \omega(\md^+)\\
 		&\geq  \sum_{\varphi \in \my}\omega(\varphi) + \sum_{\varphi \in \mx}\omega(\varphi) + \sum_{\mathbf{e}\in \me(\mx)\cap \msttilde_i} \omega(\mathbf{e}) - \omega(\md) \\
 		& \geq \Omega(L_{i}\eps |\my|) +  \Delta_{i+1}^{+}(\mx) \stackrel{\mbox{\tiny{~\cref{eq:Step4PotentialXplus}}}}{=} \Omega(|\my|\eps L_i)  + \Omega(|\mv(\mx) \cup \mv(\md^+)|\eps^2 L_i)\\
 		&= \Omega(|\mv(\mx) \cup \mv(\md^+) \cup \my|\eps^2 L_i) = \Omega(|\mv(\mx^+)|\eps^2 L_i),\\
 	\end{split}
 \end{equation*} 
 as claimed. \qed
\end{proof}

Next, we show Item (3)  of \Cref{lm:ClusteringFast} regarding the corrected potential changes of subgraphs in $\mathbb{X}$.

\begin{lemma}\label{lm:Step5PotentialChange}  $\Delta_{i+1}^+(\mx) \geq 0$ for every $\mx \in \mathbb{X}$, and
	\begin{equation*}
		\sum_{\mx \in \mathbb{X}\setminus \mathbb{X}^{\lowm}} \Delta_{i+1}^+(\mx) = \sum_{\mx \in \mathbb{X}\setminus \mathbb{X}^{\lowm}} \Omega(|\mv(\mx)|\eps^2 L_i). 
	\end{equation*}
\end{lemma}
\begin{proof}
	If $\mx \in \mathbb{X}_1\cup \mathbb{X}_2\cup \mathbb{X}_4$, then $\Delta^+_{i+1}(\mx)\geq 0$ by \Cref{lm:Step1Potential,lm:Step2Potential,lm:Step4Potential}. Otherwise, $\mx \in \mathbb{X}^{\prefix}_5\cup \mathbb{X}^{\internal}_5$, and hence is a subpath of $\msttilde_{i}$. Thus, by definition, $\Delta^+_{i+1}(\mx) = \sum_{\varphi\in \mx}\omega(\varphi) + \sum_{\mathbf{e}\in \me(\mx)\cap \msttilde_{i}}\omega(\mathbf{e}) - \adm(\mx) = 0$. That is, $\Delta^+_{i+1}(\mx)\geq 0$ in every case.
	
	We now show a lower bound on the average potential change of subgraphs in $\mathbb{X}\setminus \mathbb{X}^{\lowm}$. We assume that the degenerate case does not happen; otherwise, $\mathbb{X}\setminus \mathbb{X}^{\lowm} = \emptyset$ and there is nothing to prove. 	By Item (1) of \Cref{lm:Xlowm}, $\mathbb{X}^{\lowm} =\mathbb{X}^{\internal}_5$ and only subgraphs in $\mathbb{X}^{\prefix}_5$ may not have positive potential change. By \Cref{lm:Step1Potential,lm:Step2Potential,lm:Step4Potential}, on average, each node $\varphi$ in any subgraph $\mx \in \mathbb{X}_1\cup \mathbb{X}_2\cup \mathbb{X}_4$ has $\Omega(\eps^2 L_i)$ corrected potential change, denoted by $\overline{\Delta}(\varphi)$.
	
	By construction, a subgraph in $\mathbb{X}^{\prefix}_{5}$ is a prefix (or suffix), say $\Ptilde_1$, of a long path $\Ptilde$. The other suffix, say $\Ptilde_2$, of $\Ptilde$ is augmented to a subgraph, say $\mx \in \mathbb{X}_1\cup \mathbb{X}_2\cup \mathbb{X}_4$ by the construction of Step \hyperlink{5BFast}{5BFast} and Item (2) \Cref{lm:Clustering-Step3}. Since $|\mv(\Ptilde_2)| = \Omega(1/\eps)$ by Item (3)  of \Cref{lm:Clustering-Step5}, $\sum_{\varphi \in \Ptilde_2}\overline{\Delta}(\varphi) = \Omega(1/\eps)(\eps^2 L_i) = \Omega(\eps L_i)$. We distribute half this corrected potential change to all the nodes in $\Ptilde_1$, by Item (3)  of \Cref{lm:Clustering-Step5}, each gets $\Omega(\frac{\eps L_i}{1/\eps}) = \Omega(\eps^2 L_i)$. This implies:
		\begin{equation*}
			\sum_{\mx \in \mathbb{X}\setminus \mathbb{X}^{\lowm}} \Delta_{i+1}^+(\mx) = \sum_{\varphi \in \mv_i\setminus \mv^{\lowm}_i} \Omega(\eps^2 L_i) =  \sum_{\mx \in \mathbb{X}\setminus \mathbb{X}^{\lowm}} \Omega(|\mv(\mx)|\eps^2 L_i),
		\end{equation*}
	as desired.
	 \qed	
\end{proof}

We are now ready to prove \Cref{lm:ClusteringFast}.

\begin{proof}[Proof of \Cref{lm:ClusteringFast}]
	We observe that Items (1), (2) and (4) follow directly \Cref{lm:exception} and  \Cref{lm:Xlowm}. Item (5) follows from \Cref{lm:XProp}. Item (3) follows from \Cref{lm:Step5PotentialChange}. The construction time is asymptotically the same as the construction time of $\mathbb{X}$, which is $O((|\mv_i| + |\me_i|)\eps^{-1})$ by \Cref{lm:XProp}. 
	
	Finally, we compute the augmented diameter of each subgraph $\mx \in \mathbb{X}$. We observe that the augmentations in Step 3 and Step 5 do not create any cycle. Thus, if $\mx$ is initially formed in Steps 1, 2 or 5B, then finally $\mx$ is a tree. It follows that the augmented diameter of $\mx$ can be computed in $O(|\mv(\mx)|)$ time by a simple tree traversal. If $\mx$ is formed in Step 4, then it has exactly one edge $\mbe$ not in $\msttilde_{i}$ by Item (1) in \Cref{lm:Clustering-Step4}  and that  $\mx$ contains at most one cycle. Let $\mz$ be such a cycle (if any); $\mz$ has $O(1/\eps)$ edges by Item (3) in \Cref{lm:Clustering-Step4}. Thus, we can reduce computing the diameter of $\mx$ to computing the diameter of trees by guessing an edge of $\mz$ that does not belong to the diameter path of $\mx$ and remove this edge from $\mx$; the resulting graph is a tree. There are $O(\frac{1}{\eps})$ guesses and each for each guess, computing the diameter takes $O(|\mv(\mx)|)$ time, which implies $O(|\mv(\mx)|\eps^{-1})$ time to compute $\adm(\mx)$\footnote{It is possible to compute the augmented diameter of $\mx$ in $O(|\mv(\mx)|)$ time using a more involved approach.}. Thus, the total running time is $\sum_{\mx\in \mathbb{X}} O(|\mv(\mx)|\eps^{-1}) = O(|\mv_i|\eps^{-1})$.  
	\qed
\end{proof}

\section{Light Spanners for Minor-free Graphs in Linear Time}\label{sec:minor-linear}

In \Cref{sec:applications-fast}, we show a construction of a light spanner for $K_r$-minor-free graphs with running time $O(nr\sqrt{r}\alpha(nr\sqrt{r},n))$. The extra factor $\alpha(nr\sqrt{r},n)$ is due to  \textsc{Union-Find} data structure in the proof of \Cref{lm:framework}. To remove this factor, we do not use  \textsc{Union-Find}. Instead, we follow the idea of Mare{\v{s}}~\cite{Mare04} that was  applied to construct a minimum spanning tree for $K_r$-minor-free graphs. Specifically, after the construction of level-$(i+1)$ clusters, we prune the set of edges that are involved in the construction of levels at least $i+1$, which is $\cup_{j\geq i+1}E^{\sigma}_{j}$, as follows.

Let $E^{\sigma}_{\geq i} = \cup_{j\geq i}E^{\sigma}_{j}$. We inductively maintain a set of edges $\me_{\geq i}$, where each edge in $\me_{\geq i}$ corresponds to an edge in $E^{\sigma}_{\geq i}$. (Note that only those in $\me_{i}$ are involved in the construction of spanner at level $i$.) Furthermore, we inductively guarantee that $|\me_{\geq i}| = O(r\sqrt{\log r})|\mv_i|$; we call this the \emph{size invariant}.  Upon completing the construction of level-$(i+1)$ clusters, we construct the set of nodes $\mv_{i+1}$. We now consider the set of edges $\me'_{\geq i+1} = \me_{\geq i}\setminus \me$. Let $\tilde{\me}_{\geq i+1}$ be  obtained from  $\me'_{\geq i+1}$ by removing parallel edges: two edges $(\varphi_{1},\varphi_{2})$ and  $(\varphi'_{1},\varphi'_{2})$ are \emph{parallel} if there exist two subgraphs $\mx,\my \in \mathbb{X}$ such that,~w.l.o.g, $\varphi_1, \varphi_1' \in \mv(\mx)$ and  $\varphi_2, \varphi_2' \in \mv(\my)$. (Among all parallel edges, we keep the edge with minimum weight in $\tilde{\me}_{i+1}$.)  We construct the edge set  $\me_{\geq i+1}$ (between vertices in $\mv_{i+1}$) at level $(i+1)$ from $\tilde{\me}_{\geq i+1}$ by creating one edge $(\mx,\my)\in \me_{\geq i+1}$  for each corresponding edge $(\varphi_x,\varphi_y) \in \tilde{\me}_{\geq i+1}$ where $\varphi_x \in \mv(\mx)$ and $\varphi_y \in \mv(\my)$; $\omega(\mx,\my) = \omega(\varphi_x,\varphi_y)$.

Observe that $\me_{i+1}$ corresponds to a subset of edges of  $E^{\sigma}_{\geq i+1}$ since $\me'_{\geq i+1}$, by definition, corresponds to a subset of edges of  $E^{\sigma}_{\geq i+1}$.  The stretch is in check (at most $(1+O(\eps))$), since we only remove parallel edges and that level-$(i+1)$ clusters have diameter  $O(\eps)$ times the weight of level-$(i+1)$ edges by \hyperlink{P3}{property (P3)}.  Furthermore,  since $\me_{\geq i} = O(r\sqrt{\log r} |\mv_i|)$  by the size invariant, the construction of $\me_{i+1}$ can be done in $O(|\mv_i|)$ time. Since the graph $(\mv_{i+1}, \me_{\geq i+1})$ is a minor of $G$ and hence, is $K_r$-minor-free, we conclude that $| \me_{\geq i+1}| = O(r\sqrt{\log r})|\mv_{i+1}|$ by the sparsity of minor-free graphs, which implies the size invariant for level $i+1$. 

By the size invariant, we do not need \textsc{Union-Find} data structure, as $\me_{\geq i}$ now has  $O(r\sqrt{\log r}|\mv_i|) = O(r\sqrt{\log r}|\mc_i|)$ edges. Thus, the running time to construct $\mg_i$ in \Cref{lm:G_i-construction} becomes $O_{\eps}(|\mc_i| + |\me_{i}|) = O_{\eps}(r\sqrt{\log r}|\mc_i|)$, and the running time to  construct $\msttilde_{i+1}$  in \Cref{lm:MSTiPlus1} also becomes $O((r\sqrt{\log r}|\mc_i|)$. The rest of the proof is the same as the proof in \Cref{subsec:minor-free}.

\section{Fine-Grained Optimal Lightness: Proof of \Cref{lm:framework}(2)}\label{sec:LightProof}

Our goal is to construct a cluster graph $\mg_i$ and a collection  $\mathbb{X}$ of subgraphs of $\mg_i$ satisfying properties  \hyperlink{P1'}{(P1')}-\hyperlink{P3'}{(P3')}. We set  $\psi = 1/250$ where $\psi$ is the parameter in \Cref{eq:Esigmaixdef}.

By \Cref{lm:PropEquiv}, the set of level-$(i+1)$ obtained from subgraphs in $\mathbb{X}$ obtained by applying the transformation in \Cref{eq:XtoCluster} will satisfy properties \hyperlink{P1}{(P1)}-\hyperlink{P3}{(P3)}. To be able to bound the set of edges  in $H_i$ (constructed in \Cref{sec:stretch2,sec:stretch1E}), we need to guarantee that subgraphs in $\mathbb{X}$ have sufficiently large potential changes. This indeed is the crux of our construction. We assume that  $\eps > 0$ is a sufficiently small constant, i.e., $\eps \ll 1, \eps = \Omega(1)$.

\paragraph{Constructing $\mathcal{G}_i$.~}  We shall assume inductively on $i, i \ge 1$ that:
\begin{itemize}[noitemsep]
	\item The set of edges $\widetilde{\mst}_i$ is given by the construction of the previous level $i$ in the hierarchy; for the base case $i = 1$ (see \Cref{subsec:DesignPotential}), $\widetilde{\mst}_1$ is simply a set of edges of $\widetilde{\mst}$ that are not in any level-$1$ cluster. 
	\item The weight $\omega(\varphi_C )$ on each node $\varphi_C \in \mv_i$ is the potential value of cluster $C \in \mathcal{C}_i$; for the base case $i = 1$, the  potential values of level-$1$ clusters were set in \Cref{eq:Level1Poten}.
\end{itemize}

After completing the construction of $\mathbb{X}$, we can compute the weight of each node of $\mg_{i+1}$ by computing the augmented diameter of each subgraph in $\mathcal{X}$; the running time is clearly polynomial. By the \hyperlink{MSTiPlus1}{end of this section}, we show to compute the spanning tree  $\msttilde_{i+1}$ for $\mg_{i+1}$ for the construction of the next level.

\paragraph{Realization of a path.~}  Let $\mp = ( \varphi_0, (\varphi_0,\varphi_1), \varphi_1, (\varphi_1,\varphi_2), \ldots, \varphi_{p})$ be a path of $\mg_i$, written as an alternating sequence of vertices and edges. Let $C_i$ be the cluster corresponding to $\varphi_i$, $0\leq i \leq p$. Let $u$ and $v$ be two vertices such that $u$ is in the cluster corresponding to $\varphi_0$ and $v$ is in the cluster corresponding to $\varphi_p$. See \Cref{fig:path} for an illustration.

Let $\{y_i\}_{i=0}^p$ and $\{z_i\}_{i=0}^p$ be sequences of vertices of $G$ such that (a) $z_0 = u$ and $y_p = v$ and (b) $(y_{i-1}, z_i)$ is the edge on $G$ corresponds to edge $(\varphi_{i-1},\varphi_i)$ in $\mathcal{P}$ for $1\leq i\leq p$. Let $Q_i$, $0\leq i \leq p$, be a shortest path in $H_{< L_{i-1}}[C_i]$ between $z_i$ and $y_i$ where $C_i$ is the cluster corresponding to $\varphi_i$.  Let $P = Q_0\circ (y_0,z_1)\circ \ldots\circ Q_p$ be a (possibly non-simple) path from $u$ to $v$. We call $P$ a \emph{realization of $\mathcal{P}$ with respect to $u$ and $v$}. The following observation follows directly from the definition of the weight function of $\mg_i$.

\begin{observation}\label{obs:realization} Let $P$ be a realization of $\mp$ w.r.t two vertices $u$ and $v$. Then $w(P) \leq \omega(\mp)$.
\end{observation}

Next, we show that to construct $H_i$, it suffices to focus on the edges of $E^{\sigma}_i$ that correspond to edges in $\me_i$ of $\mg_i$. 

\begin{lemma}\label{lm:G_i-construction-Light}Let $\psi = 1/250$. We can construct a cluster graph 
	$\mathcal{G}_i = (\mathcal{V}_i,\mathcal{E}_i\cup \widetilde{\mst}_i,\omega)$ in  polynomial time 
	such that $\mathcal{G}_i$ satisfies all properties in \Cref{def:GiProp}. Furthermore, let $F^{\sigma}_i$ be the set of edges in $E^{\sigma}_i$ that correspond to $\mathcal{E}_i$. If every edge in $F^{\sigma}_i$ has a stretch $t(1+s\cdot \eps)$ in $H_{<L_i}$ for some constant $s\geq 1$, then every edge in $E^{\sigma}_i$ has stretch $t(1+ (2s+16g+1)\eps)$ when $\eps < \frac{1}{2(12g+1)}$.
\end{lemma}
\begin{proof} Since $\msttilde_{i}$ is given  at the  outset of the construction of $\mg_i$, we only focus on constructing $\me_i$. For each edge $e = (u,v) \in E^{\sigma}_i$, we add an edge $(\varphi_{C_u}, \varphi_{C_v})$ to $\mg_i$. Next, we remove edges from $\mg_i$. (Step 1) we remove self-loops and parallel edges from $\mg_i$; we only keep the edge of minimum weight in $\mg_i$ among parallel edges. (Step 2) If $t\geq 2$, we remove every edge  $(\varphi_{C_u},\varphi_{C_v})$ from $\mg_i$ such that $\omega(\msttilde_i[\varphi_{C_u},\varphi_{C_v}]) \leq t(1 + 6g\eps)\omega(\varphi_{C_u},\varphi_{C_v})$; the remaining edges of  $\mg_i$ not in $\msttilde_{i+1}$ are $\me_i$. If $t = 1+\eps$, we  apply the  $\pathg$ algorithm  to $\mg_i$ with stretch $t(1 + 6g\eps)$ to obtain $\ms_i$. (Note that we use augmented distances rather than normal distances when apply the greedy algorithm.) It was shown~\cite{ADDJS93} that the $\pathg$ algorithm contains the minimum spanning tree of the input graph. Thus, $\ms_i$ contains $\msttilde_{i}$ as a subgraph. We then set $\me_i = \me(\ms_i)\setminus \msttilde_{i}$; this completes the construction of $\mg_i$.
	
	We now show  the second claim: the stretch of $E^{\sigma}_i$ in $H_{<L_i}$ is  $t(1+ \max\{s+4g,10g\}\eps)$. Let $(u',v')$ be any edge in $E^{\sigma}_i\setminus F^{\sigma}_i$.  Recall that  $(u',v')$ is not in $ F^{\sigma}_i$ because (a) both $u'$ and $v'$ are in the same level-$i$ cluster in the construction of the cluster graph in \Cref{lm:G_i-construction-Light}, or (b) $(u',v')$ is parallel with another edge $(u,v)$, or (c) the edge $(\varphi_{C_{u'}},\varphi_{C_{v'}})$ corresponding to $(u',v')$ is removed from $\mg_i$ in Step 2. 
	
	Case (a) does not happen since otherwise, there is a path in$H_{<L_i}$ of length at most $gL_{i-1} ~=~ g\eps L_i ~\leq \frac{L_i}{1+\psi}~\leq~w(u',v')$ when $\eps < \frac{1}{(1+\psi)g}$, contradicting that every edge is a shortest path between its endpoints. 
	
	For case (c), observe that if $t\geq 2$, then by  construction, 	$d_{H_{< L_{i-1}}}(u',v') \leq t(1+ 6g\eps)w(u',v')$. Otherwise ($t = 1+\eps$), let $\mp'$ be the shortest path between $\varphi_{C_{u'}}$ and $\varphi_{C_{v'}}$ in $\ms_i$. Since $\ms_i$ is a $t(1+6g\eps)$-spanner of $\mg_i$, we have: 
	\begin{equation} \label{eq:P-vs-uv}
		\begin{split}
			\omega(\mp') &\leq (1+\eps)(1+6g\eps)\omega(\varphi_{C_{u'}}, \varphi_{C_{v'}})\leq (1 + (12g+1)\eps)w(u',v')
		\end{split}
	\end{equation}
	Observe that $\mp'$ contains  at most one edge in $\me_i$. Let $P'$ be a realization of $\mp'$ w.r.t $u'$ and $v'$.  If $\mp'$ contains no edge in $\me_i$, then $P'$ is a path in $H_{< L_{i-1}}$. This implies that $d_{H\leq i}(u',v') \leq (1 + (12g+1)\eps)w(u',v') \leq t(1 + (12g+1)\eps)w(u',v')$ since $t\geq 1$. Otherwise,  $P'$ contains exactly one edge $(x,y) \in F^{\sigma}_i$. Let $Q'$ be obtained from $P'$ by replacing edge $(x,y)$ by a shortest path from $x$ to $y$ in $H_{<L_i}$. Since $d_{H_{<L_i}}(x,y) \leq t(1+s\cdot \eps)w(x,y)$. Then  by \Cref{eq:P-vs-uv}, we have:
	\begin{equation*}
			w(Q') \leq t(1+s\cdot \eps) w(P') \leq  t(1+(2s+12g+1)\cdot \eps)w(u',v') \qquad \mbox{(since $(12g+1)\eps \leq 1$)}
	\end{equation*}
	Thus, in all cases, $d_{H_i}(u',v')\leq t(1+(2s+12g+1)\cdot \eps)w(u',v')$.
	
	We now consider case (b); that is, $(u',v')$  is not in $F^{\sigma}_i $ because  it is parallel with another edge $(u,v)$. 	Let $C_u$ and $C_v$ be two level-$i$ clusters containing $u$ and $v$, respectively. W.l.o.g, we assume that $u' \in C_u$  and $v' \in C_v$. Since we only keep the edge of minimum weight among all parallel edges, $w(u,v) \leq w(u',v')$.By property (\hyperlink{P3}{P3}), $\dm(H_{<L_i}[C_u]),\dm(H_{<L_i}[C_v]) \le g\epsi L_i$. 
	\begin{equation*}
		\begin{split}
			d_{H_{<L_i}}(u',v') &\leq 	d_{H_{<L_i}}(u,v) + \dm(H_{<L_i}[C_u]) + \dm(H_{<L_i}[C_v])\\ &\leq t(1+(2s+12g+1)\eps)w(u,v) +   2g\epsi L_i\\
		&\leq (1+(2s+16g+1)\eps) w(u',v') \qquad \mbox{(since $t\geq 1$)}~,
		\end{split}
	\end{equation*}
Since $w(u',v') \geq L_i/(1+\psi) \geq L_i/2$.
	\qed
\end{proof}

To construct the set of subgraphs $\mathbb{X}$ of $\mg_i$, we distinguish between two cases: (a) $t = 1+\eps$ and (b) $t\geq 2$. Subgraphs in $\mathbb{X}$ constructed for the case $t=1+\eps$ have properties similar to those of subgraphs constructed in \Cref{sec:FastProof}; the key difference is that subgraphs constructed in our work have a larger \emph{average potential change}, which ultimately leads to an optimal dependency on $\eps$ of the lightness. When the stretch $t\geq 2$, we show that one can construct a set of subgraphs $\mathbb{X}$ of $\mg_i$ with much larger potential change, which reduces the dependency of the lightness on $\eps$  by a factor $1/\eps$ compared to the case $t = 1+\eps$. Our construction uses \hyperlink{SPSSO}{$\sso$} as a black box. The following lemma summarizes our construction.

\begin{restatable}{lemma}{HiConstructionLight}
	\label{lm:ConstructClusterHi} Given \hyperlink{SPSSO}{$\sso$}, we can construct in polynomial time a set of subgraphs $\mathbb{X}$ such that every subgraph $\mx \in \mathbb{X}$ satisfies the three properties (\hyperlink{P1'}{P1'})-(\hyperlink{P3'}{P3'}) with constant $g=223$, and graph $H_i$ such that:
	\begin{equation*}
		d_{H_{<L_i}}(u,v) \leq t(1+ \max\{s_{\sso}(2g),6g\}\eps)w(u,v) \quad \forall (u,v)\in F^{\sigma}_{i}
	\end{equation*}
	where $ F^{\sigma}_{i}$ is the set of edges defined in \Cref{lm:G_i-construction-Light}. Furthermore,  $w(H_i) \leq  \lambda \Delta_{i+1} + a_i$ such that
	\begin{enumerate}[noitemsep]
		\item \textbf{when $t \geq 2$:}  $\lambda = O(\chi \eps^{-1} )$, and $A = O(\chi \eps^{-1} )$.
		\item \textbf{when $t = 1+\eps$:} $\lambda = O(\chi \eps^{-1} + \epsilon^{-2})$, and $A = O(\chi \eps^{-1} + \epsilon^{-2})$. 
	\end{enumerate}
	Here  $A \in \mathbb{R}^+$ such that $\sum_{i\in \mathbb{N}^+}a_i \leq A \cdot w(\mst)$.
\end{restatable}

The proof of \Cref{lm:ConstructClusterHi} is deferred to \Cref{sec:stretch2} for the case $t\geq 2$ and  \Cref{sec:stretch1E} for the case $t = 1+\eps$.   

\paragraph{Constructing $\msttilde_{i+1}$.~}\hypertarget{MSTiPlus1}{}  Let $\msttilde^{out}_{i} = \msttilde_{i}\setminus (\cup_{\mx \in \mathbb{X}}(\me(\mx)\cap \msttilde_{i}))$ be the set of $\msttilde_{i}$ edges that are not contained in any subgraph $\mx \in \mathbb{X}$. Let $\msttilde_{i+1}'$ be the graph with vertex set $\mv_{i+1}$ and there is an edge between two nodes $(\mx,\my)$ in $\mv_{i+1}$ of there is at least one edge in $\msttilde^{out}_{i}$ between two nodes in the two corresponding subgraphs $\mx$ and $\my$. Note that $\msttilde_{i+1}'$ could have parallel edges (but no self-loop).  Since $\msttilde_{i}$ is  a spanning tree of $\mg_i$, $\msttilde_{i+1}'$ must be connected. $\msttilde_{i+1}$ is then a spanning tree of $\msttilde_{i+1}'$.

We are now ready to prove Item (2) of \Cref{lm:framework}. 
\begin{proof}[Proof of Item (2)  of \Cref{lm:framework}] We apply \Cref{lm:framework-technical} to construct a light spanner $H$ for $G$ where each graph $H_i$, $i \in \mathbb{N}^+$, is constructed using \Cref{lm:ConstructClusterHi}.   
	
	When $t\geq 2$, by Item (1) of \Cref{lm:ConstructClusterHi} and \Cref{lm:framework-technical}, the lightness of $H$ is $O((\frac{O(\chi \eps^{-1}) +  O(\chi \eps^{-1}) + 1}{1+\psi})\log(\frac{1}{\eps}) + \frac{1}{\eps}) = O_{\eps}(\chi \eps^{-1})$. When $t = 1+\eps$, by Item (2) of \Cref{lm:ConstructClusterHi} and \Cref{lm:framework-technical}, the lightness of $H$ is $O((\frac{O(\chi \eps^{-1}) +  O(\chi \eps^{-1}) + 1}{1+\psi})\log(\frac{1}{\eps}) + \frac{1}{\eps^2}) = O_{\eps}(\chi \eps^{-1} + \eps^{-2})$. 
	
	We now bound the stretch of $H$. By \Cref{lm:ConstructClusterHi} and \Cref{lm:G_i-construction-Light}, the stretch of edges in $E^{\sigma}_i$ in the graph $H_{<L_i}$  is $t(1 + (2s_{\sso}(2g) +  16g+1)\eps)$ with $g  = 223$. Thus, by \Cref{lm:framework-technical}, the stretch of $H$ is $t(1 +(2s_{\sso}(2g) +  16g+1)\eps) = t(1 + (2s_{\sso}(O(1)) +  O(1))\eps)$ as claimed.
	\qed
\end{proof}

\begin{comment}
\subsection{Summary of Notation}
\renewcommand{\arraystretch}{1.3}
\begin{longtable}{| l | l|} 
\hline
\textbf{Notation} & \textbf{Meaning} \\ \hline
$E^{light}$ &$ \{e \in E(G) : w(e)\le w/\varepsilon\}$\\ \hline 
$E^{heavy}$ & $E \setminus E^{light}$ \\\hline
$E^{\sigma} $ & $\bigcup_{i \in \mathbb{N}^{+}}E_{i}^{\sigma}$\\\hline
$E_{i}^{\sigma} $ & $\{e \in E(G) : \frac{L_i}{1+\psi} \leq w(e) < L_i\}$\\\hline
$g$ & constant in \hyperlink{P3}{property (P3)}, $g = 223$ \\\hline
$\mathcal{G}_i = (V_i, \msttilde_{i} \cup \mathcal{E}_i, \omega)$ & cluster graph; see \Cref{def:ClusterGraphNew}. \\\hline
$\me_i$ & corresponds to a subset of edges of $E^{\sigma}_i$\\\hline
$\mathbb{X}$ & a collection of subgraphs of $\mathcal{G}_i$\\\hline
$\mx, \mv(\mx), \me(\mx)$ & a subgraph in $\mathbb{X}$, its vertex set, and its edge set\\\hline
$\Phi_i$ & $\sum_{c \in C_i}\Phi(c)$ \\\hline
$\Delta_{i+1} $&$ \Phi_i - \Phi_{i+1}$\\\hline
$\Delta_{i+1}(\mx)$ & $(\sum_{\phi_C\in \mx }\Phi(C) ) - \Phi(C_{\mx})$\\\hline
$C_\mx$ & $\bigcup_{\phi_C \in \mx}C$ \\\hline
$s_{\sso}$ & the stretch constant of \hyperlink{SPSSO}{$\sso$}\\\hline
\caption{Notation introduced in \Cref{sec:framework}.}
\label{table:notationLight}
\end{longtable}
\renewcommand{\arraystretch}{1}
\end{comment}

\section{Clustering for Stretch $t\geq 2$: Proof of \Cref{lm:ConstructClusterHi}(1)}\label{sec:stretch2}

In this section, we prove Item (1) of \Cref{lm:ConstructClusterHi} (when the stretch $t$ is at least 2). The general idea is to construct a set $\mathbb{X}$ of subgraphs  of $\mg_i$ such that each subgraph in $\mathbb{X}$ has a sufficiently large local potential change, and carefully choose a subset of edges of $\mg_i$, with the help from \hyperlink{SPSSO}{$\sso$}, such that the total weight could be bounded by the potential change of subgraphs in $\mathbb{X}$ and distances between endpoints of edges in $\me_i$ are preserved. (By \Cref{lm:G_i-construction}, it is sufficient to preserve distances between the endpoints of edges in $\me_i$.)  In \Cref{lm:Clustering2} below, we state desirable properties of subgraphs in $\mathbb{X}$. Recall that $H_{< L_{i-1}}$ is the spanner constructed for edges of $G$ of weight less than $L_{i-1}$.

\begin{restatable}{lemma}{Clustering2}
	\label{lm:Clustering2} Let $\mg_i = (\mv_i,\me_i)$ be the cluster graph. We can construct in polynomial time  (i) a collection $\mathbb{X}$ of subgraphs of $\mg_i$ and its partition into two sets $\{\mathbb{X}^{+}, \mathbb{X}^{-}\}$ and (ii) a partition of $\me_i$ into three sets $\{\me_i^{\take}, \me_i^{\reduce}, \me_i^{\redunt}\}$ such that:
	\begin{enumerate}
		\item[(1)] For every subgraph $\mx \in \mathbb{X}$,  $\deg_{\mg^{\take}_i}(\mv(\mx)) = O(|\mv(\mx)|)$ where $\mg^{\take}_i = (\mv_i,\me_i^{\take})$, and $\me(\mx)\cap \me_i \subseteq \me^{\take}$. Furthermore, if $\mx \in \mathbb{X}^{-}$, there is no edge in $\me_i^{\reduce}$ incident to a node in $\mx$.
		
		\item[(2)] Let $H_{< L_i}^{-}$ be a subgraph obtained by adding corresponding edges of $\me_i^{\take}$ to $H_{< L_{i-1}}$.  Then for every edge $(u,v)$ that corresponds to an edge in $\me^{\redunt}$, $d_{H_{< L_i}^{-}}(u,v)\leq 2d_G(u,v)$. 
		
		\item[(3)] Let $\Delta_{i+1}^+(\mx) = \Delta(\mx) + \sum_{\mbe \in \msttilde_i\cap \me(\mx)}w(\mbe)$ be the \emph{corrected potential change} of $\mx$. Then, $\Delta_{i+1}^+(\mx) \geq 0$ for every $\mx \in \mathbb{X}$ and 
		\begin{equation}\label{eq:averagePotential-t2}
			\sum_{\mx \in \mathbb{X}^{+}} \Delta_{i+1}^+(\mx) = \sum_{\mx \in \mathbb{X}^{+}} \Omega(|\mv(\mx)|\eps L_i). 
		\end{equation}
		\item[(4)] For every edge $(\varphi_1,\varphi_2)\in \me_i$ such that $\varphi_1 \in \mx, \varphi_2 \in \my$ for some subgraphs $\mx,\my \in \mathbb{X}^{-}$, then $(\varphi_1,\varphi_2)\in \me^{\redunt}_i$, unless a \emph{degenerate case} happens, in which  $\me^{\reduce}_i = \emptyset$ and  $\me_i^{\take} = O(\frac{1}{\eps})$.
		
		\item[(5)] For every subgraph $\mx \in \mathbb{X}$, $\mx$ satisfies the three properties (\hyperlink{P1'}{P1'})-(\hyperlink{P3'}{P3'}) with constant $g=223$. Furthermore, if $\mx \in \mathbb{X}^{-}$, then $|\me(\mx)\cap \me_i| = 0$.
	\end{enumerate}	
\end{restatable}

\Cref{lm:Clustering2} is analogous to \Cref{lm:ClusteringFast}. Here we point out two major differences, which ultimately lead to the optimal dependency on $\eps$ of the lightness. In  \Cref{lm:ClusteringFast}, roughly $O(1/\eps)$ edges are added to $H_i$ per node of $\mv_i$. Furthermore,  each node has $\Omega(L_i \eps^2)$ average potential change. These two facts together incur a factor of $\Omega(1/\eps^3)$ in the lightness. Another factor of $1/\eps$ is due to $\psi = \eps$ for the purpose of obtaining a fast construction. The overall lightness has a factor of $1/\eps^4$ dependency on $\eps$. Our goal is to reduce this dependency all the way down to $1/\eps$. By choosing $\psi = 1/250$, we already eliminate one factor of $1/\eps$. By carefully partitioning $\me_i$ into three set of edges  $\{\me_i^{\take}, \me_i^{\reduce}, \me_i^{\redunt}\}$, and only taking  edges of $\me_i^{\take}$ to $H_i$, we essentially reduce the number of edges we take per node in every subgraph $\mx$ from $O(1/\eps)$ to $O(1)$ (by Item (1) in \Cref{lm:Clustering2}), thereby saving another factor of $1/\eps$. Finally, we show that (by Item (3) in \cref{lm:Clustering2}), each node in $\mathbb{X}^+$ has $\Omega(L_i \eps)$ average potential change, which is larger than the average potential change of nodes in \Cref{lm:ClusteringFast}  by a factor of $1/\eps$. We crucially use the fact that $t\geq 2$ in bounding   the average potential change here. All of these ideas together reduce the dependency on $\eps$ from $1/\eps^{4}$ to $1/\eps$ as desired.

Next we show to construct $H_i$ given that we can construct a set of subgraphs $\mathbb{X}$ as claimed in \Cref{lm:Clustering2}. The proof of \Cref{lm:Clustering2} is deferred to \Cref{subsec:clusteringT2}.

\subsection{Constructing $H_i$: Proof of \Cref{lm:ConstructClusterHi} for $t\geq 2$.} \label{subsec:ConstructHiT2}

In this section, we construct graph $H_i$ as described in \Cref{lm:ConstructClusterHi} in two steps. In Step 1, we take every edge  in $\me^{\take}_i$ to $H_i$. In Step 2, we use \hyperlink{SPSSO}{$\sso$} to construct a subset of edges $F$ to provide a good stretch for edges in $\me^{\reduce}_i$. Note that edges in $F$ may not correspond to edges in $\me^{\reduce}_i$.  As the implementation of  \hyperlink{SPSSO}{$\sso$} depends on the input graph, this is the only place in our framework where the structure of the input graph plays an important role in the construction of the light spanner.

\begin{tcolorbox}
	\hypertarget{HiConstT2}{}
	\textbf{Constructing $H_i$:} We construct $H_i$ in two steps; initially $H_i$ contains no edges.
	\begin{itemize}[noitemsep]
		\item \textbf{(Step 1).~} We  add to $H_i$ every edge of $E^{\sigma}_{i}$ corresponding to an edge in $\me^{\take}_i$. 

		\item \textbf{(Step 2).~} Let 
		$\mathcal{J}_i$ be a subgraph of $\mg_i$ induced by $\me_i^{\reduce}$. Observe that  $\mathcal{J}_i$ is a $(L_i/(1+\psi),\eps,\beta,\Upsilon = 2)$-cluster graph w.r.t  $H_{< L_{i-1}}$. We  run \hyperlink{SPSSO}{$\sso$} on $\mathcal{J}_i$ to obtain a set of edges $F$. We then add every edge in $F$ to $H_i$.
	\end{itemize}
\end{tcolorbox}

\paragraph{Analysis.~} Recall that $F^\sigma_{i}$ is the set of edges in $E^{\sigma}_i$ that correspond to $\mathcal{E}_i$.

\begin{lemma}\label{lm:Hi-StretchT2} For every edge $(u,v) \in F^\sigma_{i}$, $d_{H_{< L_i}}(u,v) \leq t(1+ s_{\sso}(2g)\eps)w(u,v)$.
\end{lemma}
\begin{proof}
	By construction, edges in $F^{\sigma}_i$ that correspond to $\me_i^{\take}$ are added to $H_i$ and hence have stretch $1$. By Item (2) of \Cref{lm:Clustering2}, edges in $F^{\sigma}_i$ that correspond to $\me_i^{\redunt}$ have stretch $2 \leq t$ in $H_{< L_i}$. Thus, it remains to focus on edges corresponding to $\me_i^{\reduce}$. Let $(\varphi_{C_u},\varphi_{C_v}) \in \me^{\reduce}_i$ be the edge corresponding to an edge $(u,v \in F^{\sigma}_i$. 	Since we add all edges of $F$ to $H_i$, by property (2) of \hyperlink{SPSSO}{$\sso$}, the stretch of edge $(u,v)$ in $H_{< L_i}$ is at most $t(1+s_{\sso}(\beta)\eps) = t(1+s_{\sso}(2g)\eps)$ since $\beta  = 2g$.\qed
	\end{proof}

 Let $\msttilde^{in}_i(\mx) = \me(\mx)\cap \msttilde_i$ for each $\mx \in \mathbb{X}$. Let $\msttilde^{in}_i = \cup_{\mx \in \mathbb{X}}(\me(\mx)\cap \msttilde_i)$ be the set of $\msttilde_i$ edges that are contained in subgraphs in $\mathbb{X}$.  We have the following observations.

\begin{observation}\label{obs:supportingPropHiT2}
	\begin{enumerate}[noitemsep]
	\item[(1)]  $\sum_{\mx \in\mathbb{X}} \Delta^+_{i+1}(\mx) = (\Delta_{i+1} + w(\msttilde^{in}_i))$. Furthermore, $(\Delta_{i+1} + w(\msttilde^{in}_i))\geq 0$.
	\item[(2)] $\sum_{i\in \mathbb{N}^+} \msttilde^{in}_i\leq w(\mst)$.
	\end{enumerate}
\end{observation}

\begin{lemma}\label{lm:Hi-WeightT2}  $w(H_i) \leq \lambda \Delta_{i+1} + a_i$ for $\lambda = O(\chi \eps^{-1} )$  and $a_i =   O(\chi \eps^{-1} )w(\msttilde^{in}_i) + O(L_i/\eps)$.  
\end{lemma}
\begin{proof}  First, we consider the non-degenerate case. Note by the \hyperlink{HiConstT2}{construction of $H_i$} that we do not add any edge  corresponding to an edge in $\me^{\redunt}_i$ to $H_i$. Thus, we only need to consider edges in $\me^{\take}_i \cup \me^{\reduce}_i$. Let $\mv_i^{+} = \cup_{\mx \in \mathbb{X}^+}\mv(\mx)$ and  $\mv_i^{-} = \cup_{\mx \in \mathbb{X}^-}\mv(\mx)$. By \Cref{obs:supportingPropHiT2}, any edge in $\me^{\take}_i$ incident to a node in $\mv_i^{-}$ is also incident to a node in $\mv_i^{+}$.   Let $F^{(a)}_i$ be the set of edges added to $H_i$ in the construction in Step $a$, $a\in \{1,2\}$.
	
		By Item (3) of \Cref{obs:supportingPropHiT2}, $\me(\mx)\cap \me_i  = \emptyset$ if $\mx \in \mathbb{X}^{-}$. By the construction in Step 1, $F^{(1)}_i$ includes edges in $E^{\sigma}_{i}$ corresponding to $\me_i^{\take}$. By Item (1) in \Cref{lm:Clustering2}, the total weight of the edges added to $H_i$ in Step 1 is:
	\begin{equation}\label{eq:Fi1T2}
		\begin{split}
			w(F^{(1)}_i)  &=  \sum_{\mx \in \mathbb{X}^{+}} O(|\mv(\mx)|) L_i \stackrel{\mbox{\tiny{\cref{eq:averagePotential-t2}}}}{=}  O(\frac{1}{\eps})\sum_{\mx \in \mathbb{X}^{+}} \Delta^+_{i+1}(\mx) = O(\frac{1}{\eps})(\Delta_{i+1} + w(\msttilde^{in}_i)).
		\end{split}
	\end{equation}

	Next, we bound $w(F^{(2)}_i)$.  By Item (1) of \Cref{lm:Clustering2}, there is no edge in $\me^{\reduce}_i$ incident to a node in $\mv_i^{-}$. Thus, $\mv(\mathcal{J}_i) \subseteq \mv_i^{+}$.	 By property (1) of \hyperlink{SPSSO}{$\sso$}, it follows that
	\begin{equation}\label{eq:Fi3T2}
		\begin{split}
			w(F^{(2)}_i)  &~\leq~  \chi |\mv(\mathcal{J}_i)|  L_i \leq \chi|\mv_i^{+}| L_i =  \chi\sum_{\mx \in \mathbb{X}^{+}}|\mv(\mx)| L_i = O(\chi/\eps)(\Delta_{i+1} + w(\msttilde^{in}_i))~.
		\end{split}
	\end{equation} 
	
	By \Cref{eq:Fi1T2,eq:Fi3T2}, we conclude that:
	\begin{equation}\label{eq:Hi-nondegenT2}
		\begin{split}
			w(H_i) &=  O(\chi/\eps) (\Delta_{i+1} + w(\msttilde^{in}_i)) \leq \lambda(\Delta_{i+1} + w(\msttilde^{in}_i))
		\end{split}
	\end{equation} 
	for some $\lambda =  O(\chi/\eps) $.

	It remains to consider the degenerate case. By Item (4) of \Cref{lm:Clustering2}, we only add to $H_i$ edges corresponding to $\me^{\take}_i$, and there are $O(1/\eps)$ such edges. Thus, we have:
	\begin{equation}\label{eq:Hi-degenT2}
		w(H_i) = O(\frac{L_i}{\eps}) \leq  \lambda\cdot (\Delta_{i+1} + w(\msttilde^{in}_i)) + O(\frac{L_i}{\eps}), 
	\end{equation} 
since $\Delta_{i+1} + w(\msttilde^{in}_i) \geq 0$ by Item (1) in \Cref{obs:supportingPropHiT2}. Thus, the lemma follows from \Cref{eq:Hi-degenT2,eq:Hi-nondegenT2}. \qed
\end{proof}

We are now ready to prove \Cref{lm:ConstructClusterHi} for the case $t\geq 2$, which we restate below.

\HiConstructionLight*
\begin{proof}[Proof of Item 1.] The fact that subgraphs in $\mathbb{X}$ satisfy the three properties (\hyperlink{P1'}{P1'})-(\hyperlink{P3'}{P3'}) with constant $g=223$ follows from Item (5) of \Cref{lm:Clustering2}.  The stretch in $H_{< L_i}$ of edges in $F^{\sigma}_{i}$ follows from \Cref{lm:Hi-StretchT2}.
	
	By  \Cref{lm:Hi-WeightT2}, $w(H_i) \leq \lambda \Delta_{i+1} + a_i$ where  $\lambda = O(\chi \eps^{-1} )$  and $a_i =   O(\chi \eps^{-1} )w(\msttilde^{in}_i) + O(L_i/\eps)$. It remains to show that $A = \sum_{i\in \mathbb{N}^+}a_i = O(\chi \eps^{-1} )$.  Observe that
	\begin{equation*}
		\sum_{i\in \mathbb{N}^+}O(\frac{L_i}{\eps}) ~=~  O(\frac{1}{\epsilon}) \sum_{i=1}^{i_{\max}} \frac{L_{i_{\max}}}{\epsilon^{i_{\max}-i}} ~=~ O(\frac{L_{i_{\max}}}{\epsilon(1-\epsilon)}) ~=~ O(\frac{1}{\epsilon}) w(\mst)~;
	\end{equation*}
	here $i_{\max}$ is the maximum level. The last equation is due to that $\eps \leq 1/2$  and every edge has weight at most $w(\mst)$ since the weight of every is the shortest distance between its endpoints. By Item (2) of \Cref{obs:supportingPropHiT2},  $\sum_{i\in \mathbb{N}^+} \msttilde^{in}_i\leq w(\mst)$.  Thus, $A = O(\chi/\eps) + O(1/\eps) = O(\chi /\eps)$ as desired.   \qed
\end{proof}

\subsection{Clustering} \label{subsec:clusteringT2}

In this section, we give a construction of the set of subgraphs $\mathbb{X}$ of the cluster graph $\mg_i$ as claimed in \Cref{lm:Clustering2}. Our construction builds on the construction in \Cref{sec:ClusteringDetails}. However, there are two specific goals we would like to achieve: the total degree of nodes in each subgraph $\mx$ in $\mg_i^{\take}$  is $O(|\mv(\mx)|)$, and the average potential change of each node (up to some edge cases) is $\Omega(\eps L_i)$ (instead of $\Omega(\eps^2 L_i)$ as achieved in \Cref{sec:ClusteringDetails}),

Our construction  has 6 main steps (Steps 1-6). The first five steps are similar to the first five steps in the construction in \Cref{sec:ClusteringDetails}. The major differences are in Step 2 and Step 4. In particular, in Step 2, we need to apply a clustering procedure of~\cite{LS19} to guarantee that the formed clusters have large average potential change. In Step 4, by using the fact that the stretch is at least 2, we form subgraphs in such a way that the potential change of the formed subgraphs is large. Step 6 is new in this paper. The idea is to post-process clusters formed in Steps 1-5 to form larger subgraphs that are trees, and hence, the average degree of nodes is $O(1)$. For those that are not grouped in the larger subgraphs,  the total degree of the nodes in each subgraph  is $O(1/\eps)$, which is at most the number of nodes.   In this step, we also rely on the fact that the stretch $t\geq 2$.

Now we give the details of the construction. Recall that $g$ is a constant defined in \hyperlink{P3}{property (P3)} ($g = 223$ in \Cref{lm:Clustering2}), and that $\msttilde_{i}$ is a spanning tree of $\mg_i$ by Item (2) in \Cref{def:GiProp}. We reuse the construction in  \Cref{lm:Clustering-Step1} for Step 1 which applies to the subgraph $\mk_i$ of $\mg_i$ with edges in $\me_i$, as described by the following lemma.

\begin{lemma}\label{lm:Clustering2-Step1T2} Let $\mv^{\high}_i = \{\varphi_{C} \in \mv: \varphi_{C} \mbox{ is incident to at least $\frac{2g}{\zeta\eps}$ edges in } \me_i\}$. Let $\mv_i^{\high+}$ be obtained from $\mv^{\high}_i$  by adding all neighbors that are connected to nodes in $\mv^{\high}_i$ via edges in $\me_i$. We can construct in polynomial time a collection of node-disjoint subgraphs $\mathbb{X}_1$ of $\mk_i =(\mv_i, \me_i)$ such that:
	\begin{enumerate}[noitemsep]
		\item[(1)] Each subgraph $\mx \in \mathbb{X}_1$ is a tree.
		\item[(2)] $\cup_{\mx \in \mathbb{X}_1}\mv(\mx) = \mv^{\high+}_i$.
		\item[(3)] $L_i \leq \adm(\mx) \leq (6+7\eta)L_i$, assuming that every node of $\mv_i$ has weight at most $\eta L_i$.
		\item[(4)] $\mx$ contains a node in $\mv^{\high}_i$ and all of its neighbors in $\mk_i$. In particular,  this implies $|\mv(\mx)|\geq \frac{2g}{\zeta\eps}$.
	\end{enumerate}
\end{lemma}

We note \Cref{lm:Clustering2-Step1T2} is slightly more general than \Cref{lm:Clustering-Step1} in that we parameterize the weights of nodes in $\mv_i$ by $\eta L_i$. Clearly, we can choose $\eta = g\eps \leq 1$ when $\eps \leq 1/g$ since every node in $\mv_i$ has a weight at most $g\eps L_i$ by property \hyperlink{P3'}{(P3')} for level $i-1$. By parameterizing the weights, it will be more convenient for us to use the same construction again in Step 6 below.

Given a tree $T$, we say that a node $x\in T$ is \emph{$T$-branching} if it has degree at least 3 in $T$.  For brevity, we shall omit the prefix $T$ in ``$T$-branching'' whenever this does not lead to  confusion.  Given a forest $F$, we say that $x$ is \emph{$F$-branching} if it is $T$-branching for some tree $T\subseteq F$. Our construction of Step 2 uses the following lemma by \cite{LS19}.

\begin{lemma}[Lemma 6.12, full version~\cite{LS19}]\label{lm:tree-clustering} Let $\mt$ be a tree with vertex weights and edge weights. Let $L, \eta, \gamma,\beta$  be parameters where $\eta \ll \gamma \ll 1$ and $\beta\geq 1$. Suppose that for any vertex $v\in \mt$ and any edge $e\in \mt$, $w(e) \leq w(v) \leq \eta L$ and $w(v)\geq \eta L/\beta$. There is a polynomial-time algorithm that finds a collection of vertex-disjoint subtrees $\mathbb{U} = \{\mt_1,\ldots, \mt_k\}$ of $\mt$ such that:
	\begin{enumerate}[noitemsep]
		\item[(1)] $\adm(\mt_i) \leq 190\gamma L$ for any $1\leq i \leq k$.
		\item[(2)] Every branching node is contained in some tree in $\mathbb{U}$. 
		\item[(3)] Each tree $\mt_i$ contains a $\mt_i$-branching node $b_i$ and three internally node-disjoint paths $\mp_1,\mp_2, \mp_3$ sharing $b_i$ as the same endpoint, such that $\adm(\mp_1\cup \mp_2) = \adm(\mt_i)$ and $\adm(\mp_3 \setminus \{b_i\})= \Omega(\adm(\mt_i)/\beta)$. We call $b_i$ the \emph{center} of $\mt_i$.
		\item[(4)] Let $\overline{\mt}$ be obtained by contracting each subtree of $\mathbb{U}$ into a single node. Then each $\overline{\mt}$-branching node corresponds to a sub-tree  of augmented diameter at least $\gamma L$.
	\end{enumerate}
\end{lemma}

\begin{figure}[!h]
	\begin{center}
		\includegraphics[width=0.8\textwidth]{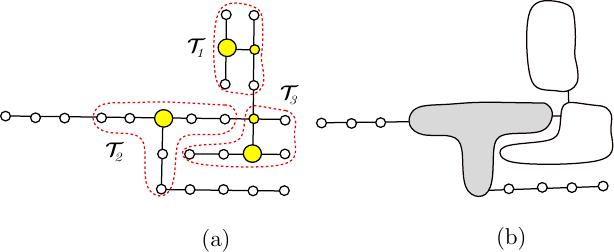}
	\end{center}
	\caption{(a) A collection $\mathbb{U} = \{\mt_1,\mt_2,\mt_3\}$ of a tree $\mt$ as in Lemma~\ref{lm:tree-clustering}. Yellow nodes are $\mt$-branching nodes. Big yellow nodes are the centers of their corresponding subtrees in $\mathbb{U}$. (b)  The shaded node in $\overline{\mt}$ is a $\overline{\mt}$-branching node and has an augmented diameter of at least $\gamma L$.}
	\label{fig:T-clustering}
\end{figure}

Let $\treeClustering(\mt,L,\eta,\gamma,\beta)$ be the output of \Cref{lm:tree-clustering} for input $\mt$ and parameters $L,\eta,\gamma,\beta$. 

\noindent See an illustration of Lemma~\ref{lm:tree-clustering}  in Figure~\ref{fig:T-clustering}. We are now ready to describe Step 2. Recall that $\zeta = 1/250$ is the constant in property \hyperlink{P3'}{(P3')}

\begin{lemma}[Step 2]\label{lm:Clustering2-Step2T2} Let $\Ftilde^{(2)}_i$ be the forest obtained from $\msttilde_{i}$ by removing every node in $\mv^{\high+}_i$ (defined in \Cref{lm:Clustering2-Step1T2}). Let $\mathcal{U} = \cup_{\tilde{T}\in \Ftilde^{(2)}_i} \treeClustering(\tilde{T},L_i,g\eps,\zeta,g/\zeta)$ and  $\mathbb{X}_2 = \{\tilde{T} \in \mathcal{U}: \adm(\tilde{T})\geq \zeta L_i\}$. Then, for every $\mx \in \mathbb{X}_2$, 
	\begin{enumerate}[noitemsep]
		\item[(1)] $\mx$ is a subtree of $\msttilde_{i}$.
		\item[(2)] $\zeta L_i \leq \adm(\mx)\leq L_i$.
		\item[(3)] $|\mv(\mx)| = \Omega(\frac{1}{\epsilon})$  when $\epsilon \leq 2/g$. 
		\item[(4)] $\Delta^+_{i+1}(\mx)  = \Omega(L_i)$.
 	\end{enumerate}

Furthermore, let $\Fbar^{(3)}_i$ be obtained from $\Ftilde^{(2)}_i$ by removing every tree in $\mathbb{U}$ that is added to $\mathbb{X}_2$, and contracting each remaining tree in $\mathbb{U}$ into a single node.  Then every tree $\Tbar \subseteq \Fbar^{(3)}_i$ is a path.
\end{lemma}
\begin{proof} We observe that Items (1), (2), and (3) follows directly from the construction. We focus on showing Item (4). Let $\varphi_b$ be the center node of $\mx$. By Item (3) in \Cref{lm:tree-clustering}, there are three internally node-disjoint paths $\mp_1,\mp_2, \mp_3$ sharing $\varphi_b$ as the same endpoint. There must be an least one path, say $\mp_1$, such that $\mp_1\cap \md \subseteq \{\varphi_b\}$. That is, $\mp_1$ is internally disjoint from the diameter path $\md$. Also by Item (3) in \Cref{lm:tree-clustering}, $\adm(\mp_1\setminus \{\varphi_{b}\}) = \Omega(\adm(\mx)/\beta) = \Omega(\zeta L_i/(g/\zeta)) = \Omega(L_i)$.  Thus, $\Delta^+_{i+1}(\mx) \geq \adm(\mp_1\setminus \{\varphi_{b}\})  = \Omega(L_i),$ as claimed.	\qed
\end{proof}

By Item (4) of \Cref{lm:Clustering2-Step2T2}, the amount of potential change of subgraphs in $\mathbb{X}_2$ is $\Omega(L_i)$, while in subgraphs in $\mathbb{X}_2$ in the construction in \Cref{sec:ClusteringDetails} only have $\Omega(\eps L_i)$ potential change. 

We note that there might be isolated nodes in $\Fbar^{(3)}_i$, which we still consider as paths.  We refer to nodes in $\Fbar^{(3)}_i$ that are contracted from $\mathcal{U}$ as \emph{contracted nodes},  and nodes that correspond to original nodes of $\Ftilde^{(2)}_i$ as \emph{uncontracted nodes}. For each node $\bar{\varphi} \in \Fbar^{(3)}_i$, we abuse notation by denoting $\bar{\varphi}$ the subtree of $\Ftilde^{(2)}_i$ corresponding to the node $\bar{\varphi}$; $\bar{\varphi}$  could be a single node in $\Ftilde^{(2)}_i$ for the uncontracted case. We then define the weight function of $\bar{\varphi}$ as follows:
\begin{equation}\label{eq:weightContractedNode}
	\omega(\bar{\varphi}) = \adm(\bar{\varphi})
\end{equation}

In the RHS of \Cref{eq:weightContractedNode}, we interpret $\bar{\varphi}$ as a subtree of $\Ftilde^{(2)}_i$ with weights on nodes an edges. 
\begin{observation}\label{obs::weightContractedNode} $\omega(\bar{\varphi}) \leq  \zeta L_i $ for every node $\bar{\varphi} \in \Fbar^{(3)}$.
\end{observation}

For each subpath $\Pbar \subseteq \Fbar^{(3)}_i$, let $\tilde{P}^{\uncontract}$ be the subtree of $\msttilde_{i}$ obtained by uncontracting the contracted nodes in $\Pbar$. We say that a node $\bar{\varphi} \in \Fbar^{(3)}_i$ is \emph{incident to an edge} $\mbe \in \msttilde_{i}\cup \me_i$ if one endpoint of $\mbe$ belongs to $\bar{\varphi}$.

\paragraph{Step 3: Augmenting $\mathbb{X}_1\cup \mathbb{X}_2$.~}\hypertarget{S3T2}{}   Let $\Fbar^{(3)}_i$ be the forest obtained in Item (4b) in \Cref{lm:Clustering2-Step2T2}. Let $\bar{A}$ be the set of all nodes $\bar{\varphi}$ in $\Fbar^{(3)}_i$ such that there is (at least one)  $\msttilde_i$ edge $\mbe = (\varphi_1,\varphi_2)$ between a node $\varphi_1 \in \bar{\varphi}$, and a node $\varphi_2 \in \mx$ for some subgraph $\mx \in\mathbb{X}_1\cup \mathbb{X}_2$. Then, for each node  $\bar{\varphi}\in \bar{A}$, we augment $\mx$ by adding $\bar{\varphi}$ and $\mbe$ to $\mx$.

The following lemma follows directly from the construction.

\begin{lemma}\label{lm:Clustering2-Step3} The augmentation in Step 3 increases the augmented diameter of each subgraph in  $\mathbb{X}_1\cup \mathbb{X}_2$ by at most $4L_i$ when $\eps \leq 1/g$. \\
	Furthermore, let $\Fbar^{(4)}_i$ be the forest obtained from $\Fbar^{(3)}_i$ by removing every node in $\bar{A}$. Then, for every path $\Pbar \subseteq \Fbar^{(4)}_i$,  at least one endpoint $\bar{\varphi} \in \Pbar$ has an $\msttilde_{i}$ edge to a subgraph of $\mathbb{X}_1\cup \mathbb{X}_2$, unless $\mathbb{X}_1\cup \mathbb{X}_2 = \emptyset$. 
\end{lemma}

\paragraph{Required definitions/preparations for Step 4.~} Let $\Fbar^{(4)}_i$ be the forest obtained from $\Fbar^{(3)}_i$ as described in \Cref{lm:Clustering2-Step3}. We call every path of augmented diameter at least $6L_i$ of $\Fbar^{(4)}_i$ a \emph{long path}. We use red/blue coloring, which is analogous to \hyperlink{RBColoring}{red/blue coloring} in \Cref{sec:ClusteringDetails}.

\begin{quote}
	\textbf{Red/Blue Coloring.~}\hypertarget{RBColoringT2}{}  Given a path $\Pbar\subseteq \Fbar^{(4)}_i$, we color their nodes red or blue. If a node has augmented distance at most $L_i$ from at least one of the path's endpoints, we color it red; otherwise, we color it blue. Observe that each red node belongs to the suffix or prefix of $\Pbar$; the other nodes are colored blue. 
\end{quote}

For each blue node $\bnu$ in a long path $\Pbar$, we denote by $\Ibar(\bnu)$ the subpath of $\Pbar$ containing every node  within an augmented distance (in $\Pbar$) at most $(1-\psi)L_i$ from $\bnu$. We call $\Ibar(\bnu)$ the \emph{interval} of $\bnu$. Recall that $\psi = 1/250$ is the constant defined in \Cref{eq:Esigmaixdef}.

We define the following set of edges between nodes of $\Fbar^{(4)}_i$.
	\begin{equation}\label{eq:Ebar-i}
		\bar{\me}_i = \{(\bmu,\bnu) | \exists \mu \in \bar{\mu}, \nu \in \bar{\nu} \mbox{ and }(\mu,\nu)\in \me_i\}.
	\end{equation}
We note that there is no edge in $\me_i$ whose nodes belong to the same tree, say $\bar{\mu}$, that corresponds to a node in $\Fbar^{(4)}_i$, because such an edge, say $\mbe$, will have weight at most  $\omega(\bar{\mu}) \leq \zeta L_i < L_i/2 < \omega(\mbe)$, a contradiction. 

Next, we define the weight: 
\begin{equation}\label{eq:Ebar-weight}
	\omega(\bmu,\bnu) = \min_{\substack{\mu\in\bar{\mu}, \nu\in\bar{\nu}\\(\mu,\nu)\in \me_i}}\omega(\mu,\nu)
\end{equation}

That is, the weight of edges $(\bmu,\bnu)$ is the minimum weight over all edges between two trees $\bar{\mu}$ and $\bar{\nu}$. We then denote $(\mu,\nu)$ the edge in $\me_i$ corresponding to an edge $(\bmu,\bnu) \in \bar{\me}_i$. Next, we define:

\begin{equation}\label{eq:Ebar-farclose}
	\begin{split}	
		\bar{\me}^{far}_i(\Fbar^{(4)}) &= \{(\bnu,\bmu) \in \bar{\me}_i | color(\bnu) = color({\bmu}) = blue  \mbox{ and }\Ibar(\bnu)\cap \Ibar(\bmu) = \emptyset\}\\
		\bar{\me}^{close}_i(\Fbar^{(4)})  &= \{(\bnu,\bmu) \in \bar{\me}_i | color(\bnu) = color({\bmu}) = blue  \mbox{ and }\Ibar(\bnu)\cap \Ibar(\bmu)\not= \emptyset\}
	\end{split}
\end{equation}
We note that the definition of $\bar{\me}^{far}_i(\Fbar^{(4)})$ and $\bar{\me}^{close}_i(\Fbar^{(4)})$ depends on the underlying forest $\Fbar^{(4)}$.

\begin{lemma}[Step 4]\label{lm:Clustering2-Step4} Let $\Fbar^{(4)}_i$ be the forest obtained from $\Fbar^{(3)}_i$ as described in \Cref{lm:Clustering2-Step3}. We can construct a collection $\mathbb{X}_4$ of subgraphs of $\mg_i$ such that every $\mx\in \mathbb{X}_4$:
	\begin{enumerate}[noitemsep]
		\item[(1)] $\mx$ is a tree and  contains a single edge in $\me_i$.
		\item[(2)] $L_i \leq \adm(\mx)\leq 5L_i$.
		\item[(3)]  $|\mv(\mx)| = \Omega(1/\eps)$ when $\epsilon \leq 1/8$. 
		\item[(4)] $\Delta_{i+1}^{+}(\mx) = \Omega(L_i)$.
	\end{enumerate}
 Let $\Fbar^{(5)}_i$ be obtained from $\Fbar^{(4)}_i$ by removing every node whose corresponding tree is contained in subgraphs of $\mathbb{X}_4$. If we apply \hyperlink{RBColoringT2}{Red/Blue Coloring} to each path of augmented diameter at least $6L_i$ in $\Fbar^{(5)}_i$, then $\bar{\me}^{far}_i(\Fbar^{(5)}) = \emptyset$. Furthermore,  for every path $\Pbar \subseteq \Fbar^{(5)}_i$,  at least one endpoint of $\Pbar$ has an $\msttilde_{i}$ edge to a subgraph of $\mathbb{X}_1\cup \mathbb{X}_2\cup \mathbb{X}_4$, unless $\mathbb{X}_1\cup \mathbb{X}_2 \cup \mathbb{X}_4 = \emptyset$. 
\end{lemma}
\begin{proof} We only apply the construction to long paths of $\Fbar^{(4)}_i$; those that have  augmented diameter at least $6L_i$. We use the following claim which is analogous to  \Cref{clm:Interval-node}.
	
	\begin{claim}\label{clm:Interval-nodeT2}
		For any blue node $\nu$, it holds that
		\begin{itemize}[noitemsep]
			\item[(a)] $ (2 - 3\zeta - 2 \epsilon-2\psi)L_i \leq  \adm(\overline{\mathcal{I}}(\bar{\nu}))\leq 2(1-\psi)L_i $.
			\item[(b)]   	Denote by  $\overline{\mi}_1$ and $\overline{\mi}_2$  the two subpaths obtained by removing $\bnu$ from the path $\overline{\mathcal{I}}(\bar{\nu})$. 
			Each of these subpaths has augmented diameter at least $(1-2\zeta - \epsilon -\psi)L_i$.
		\end{itemize}
	\end{claim}
	
	We now construct $\mathbb{X}_4$, which initially is empty. 
	
	\begin{itemize}
		\item  Pick an edge $(\bnu,\bmu)$ with both blue endpoints and  form a subgraph $\bmx = \{(\bnu,\bmu)\cup \overline{\mi}(\bnu) \cup \overline{\mi}(\bmu)\}$. We remove  all nodes in  $\overline{\mi}(\bnu) \cup \overline{\mi}(\bmu) $ from the path or two paths containing $\bnu$ and $\bmu$, update the color of nodes in the new paths to satisfy \hyperlink{RBColoringT2}{Red/Blue Coloring}. We then uncontract nodes in $\bmx$ to obtain a subgraph $\mx$ of $\mg_i$, add $\mx$ to $\mathbb{X}_4$, and  repeat this step until it no longer applies.
	\end{itemize}
	
	Items (1), (2) and (3) follows from the same argument in \Cref{lm:Clustering-Step4}.  We only focus on Item (4). Let $\overline{\mi}_1, \overline{\mi}_2, \overline{\mi}_3, \overline{\mi}_4$ be four paths obtained from $\overline{\mi}(\bmu)$ and $\overline{\mi}(\bnu)$ by removing $\bmu$ and $\bnu$. Let $\overline{\md}$ be the diameter path of $\bar{\mx}$. Then $\overline{\md}$ contains at most 2 paths among the four paths, and possibly contains edge $(\bnu,\bmu)$ as well. Since each path has augmented diameter at most $2L_i$ and $\omega(\bnu,\bmu) \leq L_i$, we have that:
	
	\begin{equation*}
		\begin{split}
			 Delta^+_{i+1}(\mx)  \geq  \left(\sum_{\bar{\varphi} \in \bar{\mx}}\omega(\bar{\varphi}) + \sum_{ \mbe\in \me(\bar{\mx})\cap \msttilde_{i}} \omega(\mbe)\right) - \adm(\bar{\md}) \geq (1-8\zeta -4\eps - 4\psi)L_i  = \Omega(L_i)~,
		\end{split}
	\end{equation*}
when $\eps \leq 1/8$. \qed		
\end{proof}

As each node has weight at most $g\eps L_{i-1}$, we have:
\begin{observation}\label{lm:size-MSTsubree} Let $\overline{P} \subseteq \Fbar^{(3)}_i$ be a path of augmented diameter $\Omega(L_i)$. Then $|\mv(\tilde{P}^{\uncontract})| =  \Omega(1/\eps)$.
\end{observation}

\paragraph{Step 5.~} Let $\Pbar$ be  a path in  $\Fbar^{(5)}_i$ obtained by Item (5) of \Cref{lm:Clustering2-Step4}. We construct two sets of subgraphs, denoted by $\mathbb{X}^{\internal}_5$ and $\mathbb{X}^{\prefix}_5$, of $\mg_i$. The construction is broken into two steps. Step 5A is only applicable when $\mathbb{X}_1 \cup \mathbb{X}_2\cup \mathbb{X}_4 \not= \emptyset$.

\begin{itemize}
	\item (Step 5A)\hypertarget{5AT2}{}  If $\Pbar$ has augmented diameter at most $6L_i$, let $\mbe$ be an $\widetilde{\mst}_i$ edge connecting $\Ptilde^{\uncontract}$  and a node in some subgraph $\mx \in \mathbb{X}_1\cup \mathbb{X}_2 \cup \mathbb{X}_4$; $\mbe$ exists by \Cref{lm:Clustering2-Step4}. We add both $\mbe$ and $\Ptilde^{\uncontract}$ to $\mx$.
	\item (Step 5B)\hypertarget{5BT2}{} 	Otherwise,  the augmented diameter of $\Pbar$ is at least $6L_i$. In this case, we greedily break $\Pbar$ into subpaths $\{\Qbar_1,\ldots, \Qbar_k\}$ such that for each $j\in [1,k]$, $\tilde{Q}^{\uncontract}_j$ has augmented diameter at least $L_i$ and at most $2L_i$.  If $\Qbar_j$ is connected to a node in a subgraph $\mx \in \mathbb{X}_1 \cup \mathbb{X}_2\cup \mathbb{X}_4$ via an  edge $e\in \msttilde_{i}$, we add $\tilde{Q}^{\uncontract}_j$ and $e$ to $\mx$.	If $\Qbar_j$ contains an endpoint of $\Pbar$, we add $\Qtilde_j^{\uncontract}$ to $\mathbb{X}^{\prefix}_5$; otherwise, we add  $\Qtilde_j^{\uncontract}$ to $\mathbb{X}^{\internal}_5$. 
\end{itemize}

In Step 5B, we want $\tilde{Q}_j^{\uncontract}$ to have augmented diameter at least $L_i$ (to satisfy property\hyperlink{P3'}{(P3')})  instead of  requiring $\adm(\Qbar_j) \geq L_i$ because a lower bound on the augmented diameter of $\Qbar_j$ does not translate to a lower bound on the augmented diameter of $\tilde{Q}^{\uncontract}_j$.

\begin{lemma}\label{lm:Clustering2-Step5}  Every subgraph $\mx \in \mathbb{X}_5^{\internal} \cup \mathbb{X}_5^{\prefix}$ satisfies:
	\begin{enumerate}[noitemsep]
		\item[(1)] $\mx$ is a subtree of $\msttilde_{i}$.
		\item[(2)] $L_i \leq \adm(\mx)\leq 2 L_i$.
		\item[(3)] $|\mv(\mx)| = \Omega(1/\eps)$.
	\end{enumerate}
Furthermore, if $\mx \in \mathbb{X}_5^{\prefix}$, then $\mx$ the uncontraction of a prefix subpath $\Qbar$ of a long path $\Pbar$, and additionally, the (uncontraction of) other suffix $\Qbar'$ of   $\Pbar$ is augmented to a subgraph in $\mathbb{X}_1 \cup \mathbb{X}_2\cup \mathbb{X}_4$, unless $\mathbb{X}_1 \cup \mathbb{X}_2\cup \mathbb{X}_4 = \emptyset$.
\end{lemma}
\begin{proof} 
	Items (1) and (2) follow directly from the construction. Item (3) follows  from \Cref{lm:size-MSTsubree}.  The last claim about subgraphs in  $\mathbb{X}_5^{\prefix}$ follows from \Cref{lm:Clustering2-Step4}.
	 \qed
\end{proof}

\begin{lemma}\label{lm:Adm-Xprime}Let $\mathbb{X}' = \mathbb{X}_1 \cup \mathbb{X}_2\cup \mathbb{X}_4 \cup \mathbb{X}_5^{\internal} \cup \mathbb{X}_5^{\prefix}$. Every node of $\mv_i$ is grouped to some subgraph in $\mathbb{X}'$. Furthermore, for every $\mx \in \mathbb{X}'$,
		\begin{enumerate}[noitemsep]
		\item[(1)] $\mx$ is a tree. Furthermore, if $\mx \not\in \mathbb{X}_4$, it is a subtree of $\msttilde_{i}$. 
		\item[(2)]  $\zeta L_i \leq \adm(\mx) \leq 31 L_i$ when $\eps \leq 1/g$.
		\item[(3)] $|\mv(\mx)| = \Omega(1/\eps)$.
	\end{enumerate}
\end{lemma}
\begin{proof}
 The fact that  every node of $\mv_i$ is grouped to some subgraph in $\mathbb{X}'$ follows directly from the construction. Observe that only subgraphs  in $\mathbb{X}'$ formed in Step 4 contain edges in $\me_i$, and such subgraphs are trees by Item (1)  of \Cref{lm:Clustering2-Step4};  this implies Item (1). Item 3 follows directly from \Cref{lm:Clustering2-Step1T2,lm:Clustering2-Step2T2,lm:Clustering2-Step4,lm:Clustering2-Step5}. 
 
 We now focus on bounding $\adm(\mx)$. The lower bound on $\adm(\mx)$ follows directly from Item (3) of \Cref{lm:Clustering2-Step1T2}, Items (2) of \Cref{lm:Clustering2-Step2T2,lm:Clustering2-Step4,lm:Clustering2-Step5}. For the upper bound, we observe that if $\mx$ is formed in Step 1, it could be augmented further in Step 3, and hence, by Item (3) of \Cref{lm:Clustering2-Step1T2} (here $\eta = g\eps$), and \Cref{lm:Clustering2-Step3}, $\adm(\mx) \leq (6 + 7g\eps)L_i + 4L_i \leq 17L_i$ since $\eps \leq 1/g$. By Items (2) of \Cref{lm:Clustering2-Step2T2,lm:Clustering2-Step4,lm:Clustering2-Step5}, $\adm(\mx) \leq 5L_i$ if $\mx$ is not initially formed in Step 1. Furthermore,  the augmentation in Step 5A and 5B increases $\adm(\mx)$ by at most $2(\bar{w} + 6L_i)\leq 14L_i$. This implies that, in any case, $\adm(\mx)\leq \max\{17L_i, 5L_i\} + 14L_i = 31L_i$. \qed
\end{proof}

Except for subgraphs in $\mathbb{X}_5^{\internal} \cup \mathbb{X}_5^{\prefix}$, we can show every subgraph $\mx \in \mathbb{X}_1 \cup \mathbb{X}_2\cup \mathbb{X}_4$ has large potential change: $\Delta_{i+1}(\mx) = \Omega(L_i)$. The last property that we need to complete the proof of \Cref{lm:Clustering2} is to guarantee that the total degree of vertices in  $\mx \in  \mathbb{X}_2\cup \mathbb{X}_4\cup \mathbb{X}_5^{\internal} \cup \mathbb{X}_5^{\prefix}$ in $\mg^{\reduce}$ is $O(1/\eps)$ (we have not defined $\mg^{\reduce}$ yet). To this end, we need Step 6. The basic idea is that if any subgraph has many out-going edges in $\bar{\me}_i$ (defined in \Cref{eq:Ebar-i}), then we apply the clustering procedure in Step 1 to group it to a larger subgraph.

\paragraph{Required definitions/preparations for Step 6.~} We construct a graph $\doverline{\mk}_i(\doverline{\mv}_i, \doverline{\me}_i, \doverline{\omega})$ as follows. Each node $\doverline{\varphi}_{\mx} \in \doverline{\mv_i}$ corresponds to a subgraph $\mx \in \mathbb{X}'$.  We then set $\doverline{\omega}(\doverline{\varphi}_{\mx}) = \adm(\mx)$.  There is an edge $(\doverline{\varphi}_{\mx},\doverline{\varphi}_{\my}) \in \doverline{\me}_i$ between two \emph{different nodes} $\doverline{\varphi}_{\mx},\doverline{\varphi}_{\my}$  if there exists an edge $(\varphi_1,\varphi_2) \in \me_i$ between a node $\varphi_1 \in \mx$ and a node $\varphi_2 \in \my$. We set the weight $\doverline{\omega}(\doverline{\varphi}_{\mx},\doverline{\varphi}_{\my})$ to be the minimum weight over all edges in $\me_i$ between $\mx$ and $\my$. We call nodes of $\doverline{\mk}_i$ \emph{supernodes}.

We call $\doverline{\varphi}_{\mx}$ a \emph{heavy} supernode if $|\mv(\mx)|\geq \frac{2g}{\zeta\eps}$ or $\doverline{\varphi}_{\mx}$ is incident to at least $\frac{2g}{\zeta\eps}$ edges in $\doverline{\mk}_i$. Otherwise, we call  $\doverline{\varphi}_{\mx}$ a \emph{light} supernode. By definition of a heavy supernode and by Item (4) in \Cref{lm:Clustering2-Step1T2}, if $\mx$ is formed in Step 1, then  $\doverline{\varphi}_{\mx}$  is a heavy supernode. We then do the following.

\begin{quote}
	We apply the construction in \Cref{lm:Clustering2-Step1T2} to graph $\doverline{\mk}_i(\doverline{\mv}_i, \doverline{\me}_i, \doverline{\omega})$, where $\doverline{\mv}_i^{\high}$ is the set of heavy supernodes in $\doverline{\mk}$ and  $\doverline{\mv}_i^{\highp}$ is obtained from  $\doverline{\mv}_i^{\high}$ by adding neighbors in $\doverline{\mk_i}$. Let $\doverline{\mathbb{X}}_6$ be the set of subgraphs of $\doverline{\mk}_i(\doverline{\mv}_i, \doverline{\me}_i, \doverline{\omega})$ obtained by the construction. Every subgraph $\doverline{\mx} \in \doverline{\mathbb{X}}_6$ satisfies all properties in \Cref{lm:Clustering2-Step1T2} with $\eta = 31$.
\end{quote}

 Let $\mathbb{X}_6$ be obtained from $\doverline{\mathbb{X}}_6$ by uncontracting supernodes. This completes our Step 6.
 
 By the construction and a simple calculation, we have:
 
 \begin{lemma}\label{lm:Step6-T2-Prop} Every subgraph $\mx \in \mathbb{X}_6$ has $\zeta L_i \leq \adm(\mx) \leq 223L_i$.
 \end{lemma}

In \Cref{subsec:X-T2} we construct the set of subgraphs $\mathbb{X}$, and show several properties of subgraphs in $\mathbb{X}$. In \Cref{subsec:E-T2}, we construct a partition of $\me_i$ into three sets, and prove \Cref{lm:Clustering2}.

\subsubsection{Constructing $\mathbb{X}$}\label{subsec:X-T2}
 
 For each $i\in \{2,4,5\}$ let $\mathbb{X}_i^{-}$ be obtained from $\mathbb{X}_i$ by removing subgraphs corresponding to nodes in  $\doverline{\mv}_i^{\highp}$ (which then form subgraphs in $\mathbb{X}_6$).  We now define $\mathbb{X}$ and a partition of $\mathbb{X}$ into two sets $\mathbb{X}^{+}$ and 	$\mathbb{X}^{-}$ $\mathbb{X}^{\lowm}$  as claimed in \Cref{lm:Clustering2}. We distinguish two cases:
 
\paragraph{Degenerate Case.~} The degenerate case is the case where   $\mathbb{X}^{-}_1\cup \mathbb{X}^{-}_2\cup \mathbb{X}^{-}_4 = \mathbb{X}^{\internal}_5 =  \emptyset$. In this case, we set $\mathbb{X} = \mathbb{X}^{-} =  \mathbb{X}_5^{\internal} \cup \mathbb{X}_5^{\prefix}$, and $	\mathbb{X}^{+} = 	 \emptyset$. 

\paragraph{Non-degenerate case.~} If $\mathbb{X}^{-}_1\cup \mathbb{X}^{-}_2\cup \mathbb{X}^{-}_4 = \mathbb{X}_6 \not=  \emptyset$, we call this the non-degenerate case. In this case, we define.
\begin{equation}\label{eq:MathbbXT2}
	\begin{split}
		\mathbb{X}^{+} &=    \mathbb{X}^{-}_2\cup \mathbb{X}^{-}_4 \cup \mathbb{X}_5^{\prefix-} \cup \mathbb{X}_6, \quad
		\mathbb{X}^{-} = \mathbb{X}_5^{\internal -} \\
		\mathbb{X} &= \mathbb{X}^{+}\cup \mathbb{X}^{-}
	\end{split}
\end{equation}

We note that every subgraph in $\mathbb{X}_1$ corresponds to a heavy supernode in $\doverline{\mk_i}$ and hence, it will be grouped in some subgraph in $\mathbb{X}_6$.

In the analysis below, we only explicitly  distinguish the degenerate case from the non-degenerate case when it is necessary, i.e, in the proof Item (4) of \Cref{lm:Clustering2}. Otherwise, which case we are in is either implicit from the context, or does not matter.

\begin{lemma}\label{lm:XPropT2} Let $\mathbb{X}$ be the subgraph as defined in \Cref{eq:MathbbXT2}. For every subgraph $\mx \in \mathbb{X}$, $\mx$ is a tree and satisfies the three properties (\hyperlink{P1'}{P1'})-(\hyperlink{P3'}{P3'}) with $g = 223$. Consequently, Item (5) of \Cref{lm:Clustering2} holds.
\end{lemma}
\begin{proof} We observe that property \hyperlink{P1'}{(P1')} follows directly from the construction.  Property \hyperlink{P2'}{(P2')} follows from Item (3) of \Cref{lm:Adm-Xprime}.  Property \hyperlink{P3'}{(P3')} follows from \Cref{lm:Step6-T2-Prop}. 
	
	By Item (1) of \Cref{lm:Adm-Xprime}, every subgraph $\mx \in \mathbb{X}'$ is a tree. Since subgraphs in $\doverline{\mx}_6$ in the construction of Step 6 are trees, subgraphs in $\mathbb{X}$ are also trees. Thus, $|\me(\mx) \cap \me_i|  = O(|\mv(\mx)|)$. Furthermore, if $\mx \in \mathbb{X}^{-}$, by the definition $\mathbb{X}^{-}$, $\mx \not\in \mathbb{X}_4$. Thus, $\mx$ is a subtree of $\msttilde_{i}$ by Item (1) of \Cref{lm:Adm-Xprime}. That implies $\me(\mx)\cap \me_i =  \emptyset$,  which implies Item (5) of \Cref{lm:Clustering2}. \qed
\end{proof}

Our next goal is to show Item (3) of \Cref{lm:Clustering2}. \Cref{lm:manynodes} below  implies that if $\mx \in \mathbb{X}$ is formed in Steps 2,4, and 6, then $\Delta^+_{i+1}(\mx) = \Omega(\eps L_i |\mv(\mx)|)$.

\begin{lemma}\label{lm:manynodes} For any subgraph $\mx \in \mathbb{X}$ such that $|\mv(\mx)|\geq \frac{2g}{\zeta\eps}$ or $\Delta^+_{i+1}(\mx) = \Omega(L_i)$, then $\Delta^+_{i+1}(\mx) = \Omega(\eps L_i |\mv(\mx)|)$.
\end{lemma}
\begin{proof} We fist consider the case where $|\mv(\mx)|\geq \frac{2g}{\zeta\eps}$.	By definition of corrected potential change in Item (3) of \Cref{lm:Clustering2}, we have:
	\begin{equation*}
		\begin{split}
			\Delta^+_{i+1}(\mx) &\geq 	\sum_{\varphi \in \mv(\mx)}\omega(\varphi) -  \adm(\mx) \geq (\zeta \eps L_i |\mv(\mx)|)  - \adm(\mx)  \\
			&\geq  (\zeta \eps L_i |\mv(\mx)|)/2   - gL_i + (\zeta \eps L_i |\mv(\mx)|)/2  = \Omega(\eps L_i |\mv(\mx)|)~.
		  	\end{split}
	\end{equation*}
Next, we consider the case where $\Delta^+_{i+1}(\mx) = \Omega(L_i)$. If $|\mv(\mx)|\geq \frac{2g}{\zeta \eps}$, then  $\Delta^+_{i+1}(\mx) = \Omega(\eps L_i |\mv(\mx)|)$ as we have just shown. Otherwise, we have:
\begin{equation*}
	\Delta^+_{i+1}(\mx) = \Omega(L_i) = \Omega(\eps L_i \frac{2g}{\zeta \eps}) = \Omega(\epsilon L_i |\mv(\mx)|),
\end{equation*}
as claimed.	\qed
\end{proof}

\begin{lemma}\label{lm:Item3Clustering}   $\Delta_{i+1}^+(\mx) \geq 0$ for every $\mx \in \mathbb{X}$ and 
	\begin{equation*}
		\sum_{\mx \in \mathbb{X}^{+}} \Delta_{i+1}^+(\mx) = \sum_{\mx \in\mathbb{X}^{+}} \Omega(|\mv(\mx)|\eps L_i). 
	\end{equation*}
Consequently, Item (3) of \Cref{lm:Clustering2} holds.
\end{lemma}
\begin{proof} The fact that $	\Delta^+_{i+1}(\mx) \geq 0$ follows directly from the definition. By \Cref{lm:manynodes} for every $\mx \in \mathbb{X}^{-}_2\cup \mathbb{X}^{-}_4 \cup \mathbb{X}_6$, it holds that 	
	\begin{equation} \label{eq:Delta-246}
		\Delta^+_{i+1}(\mx) = \Omega(\epsilon L_i |\mv(\mx)|)~,
	\end{equation}
By the definition of $\mathbb{X}^{+}$ in \Cref{eq:MathbbXT2}, the only case where $\Delta^+_{i+1}(\mx)$ could be $0$ is $\mx \in  \mathbb{X}_5^{\prefix-}$.  Next, we  use an averaging argument to assign potential change to $\mx$. Observe that $\mx$ is an uncontraction of some prefix $\overline{Q}$ of some path $\Pbar \in \Fbar^{(5)}$. By \Cref{lm:Clustering2-Step5}, the uncontraction of the other suffix  $\Qbar'$ of $\Pbar$, say $\Qtilde'$, is augmented to a subgraph in $\mathbb{X}_1\cup  \mathbb{X}_2\cup  \mathbb{X}_4$. It follows that $\Qtilde'$ is a subgraph of some graph $\my \in \mathbb{X}^{-}_2\cup \mathbb{X}^{-}_4 \cup \mathbb{X}_6$.  If we distribute the corrected potential change $\Delta^+_{i+1}(\my)$ to nodes in $\my$, each node gets $\Omega(\eps L_i)$ potential change. Thus, the total potential change of nodes in $\Qtilde'$  is $\Omega(\eps L_i|\mv(\Qtilde')|)$. By Item (3) of \Cref{lm:Clustering2-Step5}, $|\mv(\Qtilde')| = \Omega(1/\eps)$. Thus the potential change of nodes in $\Qtilde'$  is $\Omega( L_i|)$. We distribute \emph{half} of the potential change to $\mx$. Thus,  $\mx$ has $\Omega(L_i)$ potential change, and by \Cref{lm:manynodes}, the potential change of $\mx$ is $\Omega(\eps L_i |\mv(\mx)|)$. This, with \Cref{eq:Delta-246}, implies that:
	\begin{equation*}
	\sum_{\mx \in \mathbb{X}^{+}} \Delta_{i+1}^+(\mx) = \sum_{\mx \in\mathbb{X}^{+}} \Omega(|\mv(\mx)|\eps L_i), 
\end{equation*}
as desired.\qed
\end{proof}

\subsubsection{Constructing the partition of of $\me_i$: Proof of \Cref{lm:Clustering2}}\label{subsec:E-T2}

In this section, we construct a partition of $\me$ and prove \Cref{lm:Clustering2}. Items (3) and (5) of \Cref{lm:Clustering2} were proved in \Cref{lm:Item3Clustering} and \Cref{lm:XPropT2}, respectively. In the following, we prove Items (1), (2) and (4). Indeed, Item (2) follows directly from the construction (\Cref{obs:Item2Clustering}). Item (1) is proved in \Cref{lm:Item1Clustering} and Item (4) is proved in \Cref{lm:Item4-Nonde} and \Cref{lm:degenerate}. 

Recall that we define $\mathbb{X}' = \mathbb{X}_1\cup \mathbb{X}_2\cup \mathbb{X}_4\cup \mathbb{X}^{\prefix}_5 \cup \mathbb{X}^{\internal}_5$ in \Cref{lm:Adm-Xprime}. We say that a subgraph $\mx\in \mathbb{X}'$ is \emph{light} if it corresponds to a light supernode in $\doverline{\mk_i}$ (defined in Step 6); otherwise, we say that $\mx$ is \emph{heavy}. We construct $\me_i^{\take}$ and $\me^{\redunt}_i$ in two steps below;  $\me_i^{\reduce} = \me_i \setminus (\me_i^{\take}\cup \me_i^{\redunt})$. Initially, both sets are empty.

\begin{tcolorbox}
	\hypertarget{EiPartition}{}
	\textbf{Constructing $\me_i^{\take}$ and $\me^{\redunt}_i$:} Let $\mathbb{X}^{\light}$ be the set of light subgraphs in $\mathbb{X}'$.
	\begin{itemize}
		\item \textbf{Step 1:} For each subgraph $\mx\in \mathbb{X}$,  we add all edges of $\me_i$ in $\mx$ to $\me_i^{\take}$. That is, $$\me_i^{\take} \leftarrow \me_i^{\take} \cup (\me_i\cap \me(\mx)).$$
		\item \textbf{Step 2:} We construct a graph $\mh_i = (\mv_i, \msttilde_{i}\cup \me_i^{\take}, \omega)$. We then consider every edge $\mbe = (\nu\cup \mu) \in \me_i$, where both endpoints are in subgraphs in  $\mathbb{X}^{\light}$, in the non-decreasing order of the weight. If:
		\begin{equation}\label{eq:greedy-Hi}
			d_{\mh_i }(\nu,\mu) > 2 \omega(\mbe)~,
		\end{equation}
		then we add $\mbe$ to $\me_i^{\take}$ (and hence, also to $\mh_i$). Otherwise, we add $\mbe$ to $\me_i^{\redunt}$. Note that the distance in $\mh_i$ in \Cref{eq:greedy-Hi} is the augmented distance. 
	\end{itemize}
\end{tcolorbox}

The construction in Step 2 is the $\pathg$ algorithm. We observe that:

\begin{observation}\label{obs:Ereduce} For every edge $\mbe \in \me^{\reduce}_i$, at least one endpoint of $\mbe$ is in a heavy subgraph.
\end{observation}

\begin{observation}\label{obs:Item2Clustering}  Let $H_{< L_i}^{-}$ be a subgraph obtained by adding corresponding edges of $\me_i^{\take}$ to $H_{< L_{i-1}}$.  Then for every edge $(u,v)$ that corresponds to an edge in $\me^{\redunt}$, $d_{H_{< L_i}^{-}}(u,v)\leq 2d_G(u,v)$. 
\end{observation}

We now focus on proving Item (1) of \Cref{lm:Clustering2}. The key idea is the following lemma.

\begin{lemma}\label{lm:partitionX} Any subgraph $\mx \in \mathbb{X}'\setminus \mathbb{X}_1$ can be partitioned into $ k = O(1/\zeta)$ subgraphs $\{\my_1,\ldots, \my_k\}$ such that $\adm(\my_j)\leq 9 \zeta L_i$ for any $1\leq j\leq k$ when $\eps \leq \frac{\zeta}{g}$.
\end{lemma}
\begin{proof}
	Let $\varphi$ be a branching node in $\Ftilde^{(2)}$, the tree in \Cref{lm:Clustering2-Step2T2}. We say that $\varphi$ is \emph{special} if there exists three internally node disjoint paths $\Ptilde_1,\Ptilde_2,\Ptilde_3$ of  $\Ftilde^{(2)}$ sharing the same node $\varphi$ such that $\adm(\Ptilde_j \setminus \{\varphi\})\geq \zeta L_i$. Observe by the construction in \Cref{lm:Clustering2-Step2T2} that
	\begin{observation}\label{clm:special}
		Any special node $\varphi$ of $\Ftilde^{(2)}$ is contained in a subgraph  in $\mathbb{X}_2$.
	\end{observation}
	By \Cref{lm:Adm-Xprime}, $\mx$ is a tree. Let $\mx'$ be a maximal subtree of $\mx$ such that $\mx'$ is a subtree of $\msttilde_{i}$. If $\mx$ is in $\mathbb{X}_2 \cup \mathbb{X}_5^{\prefix}\cup \mathbb{X}_5^{\internal}$ then $\mx' = \mx$. Otherwise, $\mx \in \mathbb{X}_4$, and thus it has a single edge in $\me_i$ by Item (1) of \Cref{lm:Clustering2-Step4}. That is, $\mx$ has exactly two such maximal subtrees $\mx'$. Thus, to complete the lemma, we show that $\mx'$ can be partitioned into $O(1/\zeta)$ subtrees as claimed in the lemma. 
	
	Let $\md$ be the path in $\mx'$ of maximum augmented diameter. Let $\mathcal{J}$ be the forest obtained from  $\mx'$ by removing nodes of $\md$.  Then,
	
	\begin{observation}\label{clm:diameter-J} $\adm(\mt) \leq 2\zeta L_i \quad \forall \mbox{ tree } \mt \in \mathcal{J} $
	\end{observation}

	Now we greedily partition $\md$ into $k = O(1/\zeta)$ subpaths $\{\mp_1,\ldots, \mp_k\}$, each of augmented diameter at least $\zeta L_i$ and at most $3\zeta L_i$. This is possible because each node/edge has a weight at most $\max\{g\eps L_i,\bar{w}\} \leq \max\{g\eps L_i,\eps L_i\} \leq \zeta L_i$ when $\eps \leq \zeta/g$. Next, for every tree $\mt \in \mathcal{J}$, if $\mt$ is connected to a node $\varphi \in \mp_j$ via some $\msttilde_{i}$ edge $\mbe$ for some $j \in [1,k]$, we augment $\mbe$ and $\mt$ to $\mp_j$.   By  \Cref{clm:diameter-J}, the augmentation increases the diameter of $\mp$ by at most $2(\bar{w} + 2\zeta L_i)\leq 6\zeta L_i$ additively.\qed
\end{proof}

\begin{lemma}\label{lm:Const-Edge}Let $\mx, \my$ be two (not necessarily distinct) subgraphs in $ \mathbb{X}^{\light}$. Then there are $O(1)$ edges  in $\me_i^{\take}$ between nodes in $\mx$ and nodes in $\my$.
\end{lemma}
\begin{proof} Let  $\{\ma_1,\ldots, \ma_x\}$ ($\{\mb_1,\ldots, \mb_{y}\}$) be a partition of $\mx$ ($\my$) into $x = O(1/\zeta)$ ($y = O(1/\zeta)$) subgraphs of augmented diameter at most $9\zeta L_i$ as guarantee by \Cref{lm:partitionX}. Observe that by \Cref{eq:greedy-Hi}, there is at most one edge in $\me^{\take}$ between $\ma_j$ and $\mb_k$ for any $1\leq j\leq x, 1\leq k \leq y$. Thus, the number of edges in $\me^{\take}$ between $\mx$ and $\my$ is at most $x\cdot y = O(1/\zeta^2) = O(1)$.\qed 
\end{proof}

We obtain the following corollary of \Cref{lm:Const-Edge}.

\begin{corollary}\label{cor:bounded-DegXprime}For any $\mx \in  \mathbb{X}^{\light}$,  $\deg_{\mg^{\take}_i}(\mv(\mx)) = O(1/\eps) = O(|\mv(\mx)|)$ where $\mg^{\take}_i = (\mv_i,\me_i^{\take})$. 
\end{corollary}

We now prove Item (1) of \Cref{lm:Clustering2}.

\begin{lemma}\label{lm:Item1Clustering}For every subgraph $\mx \in \mathbb{X}$,  $\deg_{\mg^{\take}_i}(\mx) = O(|\mv(\mx)|)$ where $\mg^{\take}_i = (\mv_i,\me_i^{\take})$, and $\me(\mx)\cap \me_i \subseteq \me^{\take}$.  Furthermore, if $\mx \in \mathbb{X}^{-}$, there is no edge in $\me_i^{\reduce}$ incident to a node in $\mx$.
\end{lemma}
\begin{proof} Let $\mx$ be a subgraph in $\mathbb{X}$. Observe by the construction of $\me^{\take}_i$ in Step 1, $\me\cap \me(\mx)\subseteq \me^{\take}_i$.  Clearly, the number of edges incident to nodes in $\mx$ added in Step 1 is $O(|\mv(\mx)|)$ since  every subgraph in $\mathbb{X}$ is a tree by \Cref{lm:Adm-Xprime}.  Thus, it remains to bound the number of edges added in Step 2.	
	
	If $\mx \in \mathbb{X}^{-}_2\cup \mathbb{X}^{-}_4 \cup \mathbb{X}^{\prefix-}_5\cup \mathbb{X}^{\internal-}_5$, then $\mx$ corresponds to a light supernode in $\mk_i$. Thus, $\deg_{\mg^{\take}_i}(\mv(\mx)) = O(|\mv(\mx)|)$  by \Cref{cor:bounded-DegXprime}. Otherwise, $\mx \in \mathbb{X}_6$. By construction in Step 6, $\mx$ is the union  heavy subgraphs and light subgraphs  (and some edges in $\me_i$). By construction of $\me_i^{\take}$, only light subgraphs have nodes incident to edges in $\me_i^{\take}$. Let $\{\my_1,\ldots, \my_p\}$ be the set of light subgraphs constituting $\mx$. Then, by \Cref{cor:bounded-DegXprime}, we have that:
	\begin{equation*}
		\deg_{\mg^{\take}_i}(\mx)  \leq \sum_{k=1}^{p}\deg_{\mg^{\take}_i}(\my) =  \sum_{k=1}^{p} (|\mv(\my_k)|) = O(|\mv(\mx)|)~.
	\end{equation*}

We now show that there is no edge in $\me_i^{\reduce}$ incident to a node in $\mx \in \mathbb{X}^-$. Suppose otherwise, let $\mbe$ be such an edge.  
By \Cref{obs:Ereduce}, $\mbe$ is incident to a node in a heavy subgraph, say $\my$. That is, $\doverline{\varphi}_{\my}\in \doverline{\mv}^{\high}_i$.  By the construction in Step 6, $\doverline{\varphi}_{\mx} \in \doverline{\mv}^{\highp}_i$ and hence $\mx$ is grouped to a larger subgraph in $\mathbb{X}_6$, contradicting that  $\mx \in \mathbb{X}^-$. 
 \qed
\end{proof}

We now focus on proving Item (4) of \Cref{lm:Clustering2}. In \Cref{lm:Item4-Nonde}, we consider the non-degenerate case, and in \Cref{lm:degenerate} we consider the degenerate case. 

\begin{lemma}\label{lm:Item4-Nonde}Let $(\varphi_1,\varphi_2)$ be any edge in $\me_i$ between nodes of two light subgraphs $\mx ,\my$ in $\mathbb{X}_5^{\internal}$. Then,  $(\varphi_1,\varphi_2) \in \me_i^{\redunt}$.
\end{lemma}
\begin{proof} By the construction of Step 5, $\mx$ and $\my$ correspond to two subpaths $\bar{{\mx}}$ and $\bar{\my}$ of two paths $\Pbar$ and $\Qbar$ in $\Fbar^{(5)}$. Note that all nodes in $\bar{{\mx}}$ and $\bar{\my}$  have a blue color since the suffix/prefix of  $\Pbar$ and $\Qbar$ are either in  $\mathbb{X}_5^{\prefix}$ or are augmented to existing subgraphs in Step 5B.

	Since there is an edge in $\me_i$ between $\mx$ and $\my$, there must be an edge in $\bar{\me}_i$, say $(\bar{\mu},\bar{\nu})$ between a node of $\bar{\mu} \in \bar{{\mx}}$ and a node of $\bar{\nu} \in \bar{\my}$ by the definition of $\bar{\me}_i$ (in \Cref{eq:Ebar-i}) such that $\varphi_1 \in \bmu, \varphi_2 \in \bnu$.
	
	As $\bar{\mu}$ and $\bar{\nu}$ both have a blue color, either $(\bar{\mu},\bar{\nu}) \in \me_i^{far}(\Fbar^{(5)})$ or $(\bar{\mu},\bar{\nu}) \in \me_i^{close}(\Fbar^{(5)})$ by the definition in \Cref{eq:Ebar-farclose}. By \Cref{lm:Clustering2-Step4}, $\me_i^{far}(\Fbar^{(5)}) = \emptyset$. Thus,  $(\bar{\mu},\bar{\nu}) \in \me_i^{close}(\Fbar^{(5)})$. This implies $\Ibar(\bnu)\cap \Ibar(\bmu)\not= \emptyset$, and hence, $\bar{{\mx}}$ and $\bar{\my}$ are broken from the same path, say $\bar{P} \in \Fbar^{(5)}$, in Step 5B.  
	
	Furthermore,  by the definition of $\Ibar(\bnu)$, every node $\bar{\varphi} \in \Ibar(\bnu)$ is within an augmented distance (along $\Pbar$) of at most $(1-\psi)L_i$ from $\bnu$. This means, $\adm(\bar{P}[\bnu,\bmu]) \leq 2(1-\psi)L_i$. Note that the uncontraction of $\bar{P}[\bnu,\bmu]$ is a subtree of $\msttilde_{i}$. Thus, $d_{\msttilde_{i}}(\varphi_1,\varphi_2) \leq \adm(\bar{P}[\bnu,\bmu]) \leq 2(1-\psi)L_i \leq \frac{2L_i}{1+\psi} \leq 2 \omega(\varphi_1,\varphi_2)$. As $\msttilde_{i}$ is a subgraph of $\mh_i$, $(\varphi_1,\varphi_2)$ will be added to $\me_i^{\redunt}$ in Step 2, \Cref{eq:greedy-Hi}.	\qed
\end{proof}

\begin{lemma}[Structure of Degenerate Case]\label{lm:degenerate}
	If the degenerate case happens, then
	$\Fbar^{(5)}_i = \Fbar^{(4)}_i = \Fbar^{(3)}_i$, and $\Fbar^{(5)}_i$  is a single (long) path. Moreover, $|\me^{\take}_i| = O(1/\epsilon)$.
\end{lemma}
\begin{proof} Recall that the degenerate case happens when $\mathbb{X}^{-}_1\cup \mathbb{X}^{-}_2\cup \mathbb{X}^{-}_4 = \mathbb{X}_6 =  \emptyset$. This implies $\mathbb{X}_1\cup \mathbb{X}_2\cup \mathbb{X}_4 = \emptyset$. Thus, $\Fbar^{(5)}_i = \Fbar^{(4)}_i = \Fbar^{(3)}_i$. Furthermore, $\Fbar^{(5)}_i$  is a single (long) path since $\Fbar^{(3)}_i$ is a path by \Cref{lm:Clustering2-Step2T2}. This gives  $|\mathbb{X}_5^{\prefix}| = 2$. By \Cref{lm:Item4-Nonde}, there is no edge in $\me_i^{\take}$ between two subgraphs in $\mathbb{X}_5^{\internal}$. Thus, any edge in $\me_i^{\take}$ must be incident to a node in a subgraph of $\mx \in\mathbb{X}_5^{\prefix}$. By \Cref{cor:bounded-DegXprime}, there are $O(1/\eps)$ such edges.	\qed
\end{proof}

\section{Clustering for Stretch $t = 1+ \eps$}\label{sec:stretch1E}

In this section, we prove \Cref{lm:ConstructClusterHi} when the stretch $t = 1+\eps$. The key technical idea is the following clustering lemma, which is analogous to \Cref{lm:Clustering2} in \Cref{sec:stretch2}; the highlighted texts below are the major differences.  Recall that $H_{< L_{i-1}}$ is the spanner constructed for edges of $G$ of weight less than $L_{i-1}$.

\begin{restatable}{lemma}{ClusteringE}
	\label{lm:ClusteringE} Let $\mg_i = (\mv_i,\me_i)$ be the cluster graph. We can construct in polynomial time  (i) a collection $\mathbb{X}$ of subgraphs of $\mg_i$ and its partition into  two sets $\{\mathbb{X}^{+}, \mathbb{X}^{-}\}$ and (ii) a partition of $\me_i$ into three sets $\{\me_i^{\take}, \me_i^{\reduce}, \me_i^{\redunt}\}$ such that:
	\begin{enumerate}
		\item[(1)] For every subgraph $\mx \in \mathbb{X}$,  \hl{$\deg_{\mg^{\take}_i}(\mv(\mx)) = O(|\mv(\mx)|/\eps)$}  where $\mg^{\take}_i = (\mv_i,\me_i^{\take})$, and $\me(\mx)\cap \me_i \subseteq \me^{\take}$. Furthermore, if $\mx \in \mathbb{X}^{-}$, there is no edge in $\me_i^{\reduce}$ incident to a node in $\mx$.
		
		\item[(2)] Let $H_{< L_i}^{-}$ be a subgraph obtained by adding corresponding edges of $\me_i^{\take}$ to $H_{< L_{i-1}}$.  Then for every edge $(u,v)$ that corresponds to an edge in $\me^{\redunt}$, $d_{H_{< L_i}^{-}}(u,v)\leq(1+6g\eps)2d_G(u,v)$. 
		
		\item[(3)] Let $\Delta_{i+1}^+(\mx) = \Delta(\mx) + \sum_{\mbe \in \msttilde_i\cap \me(\mx)}w(\mbe)$ be the \emph{corrected potential change} of $\mx$. Then, $\Delta_{i+1}^+(\mx) \geq 0$ for every $\mx \in \mathbb{X}$ and 
		\begin{equation}\label{eq:averagePotential-t1E}
			\sum_{\mx \in \mathbb{X}^{+}} \Delta_{i+1}^+(\mx) = \sum_{\mx \in \mathbb{X}^{+}} \Omega(|\mv(\mx)|\eps L_i). 
		\end{equation}
		\item[(4)] \hl{There exists an orientation of edges in $\me_i^{\take}$ such that for every subgraph $\mx \in \mathbb{X}^{-}$, if $\mx$ has $t$ out-going edges for some $t\geq 0$, then $\Delta^+_{i+1}(\mx) =\Omega(|\mv(\mx)|t\eps^2 L_i)$}, unless a \emph{degenerate case} happens, in which  $\me^{\reduce}_i = \emptyset$ and  
		\begin{center}
			\hl{$\omega(\me_i^{\take}) = O(\frac{1}{\eps^2})(\sum_{\mx \in \mathbb{X}} \Delta_{i+1}^+(\mx) + L_i).$}
		\end{center}
		\item[(5)] For every subgraph $\mx \in \mathbb{X}$, $\mx$ satisfies the three properties (\hyperlink{P1'}{P1'})-(\hyperlink{P3'}{P3'}) with constant $g=31$. 
	\end{enumerate}	
\end{restatable}

The total node degree of $\mx$ in $\mg_i^{\take}$ in \Cref{lm:ClusteringE} is worst than the total node degree of $\mx$  in \Cref{lm:Clustering2} by a factor of $1/\eps$.  Furthermore, Item (4) of \Cref{lm:ClusteringE} is qualitatively different from  Item (4) of \Cref{lm:Clustering2} and we no longer can bound the size of $\me_i^{\take}$ in the degenerate case. All of these are due to the fact that the stretch $t = 1+\eps < 2$ when $\eps < 1$.

Next we show to construct $H_i$ given that we can construct a set of subgraphs $\mathbb{X}$ as claimed in \Cref{lm:ClusteringE}. The proof of \Cref{lm:ClusteringE} is deferred to \Cref{subsec:clusteringTE}.

\subsection{Constructing $H_i$: Proof of \Cref{lm:ConstructClusterHi} for $t = 1+\eps$.} \label{subsec:ConstructHiTE}

Let $\msttilde^{in}_i(\mx) = \me(\mx)\cap \msttilde_i$ for each $\mx \in \mathbb{X}$. Let $\msttilde^{in}_i = \cup_{\mx \in \mathbb{X}}(\me(\mx)\cap \msttilde_i)$ be the set of $\msttilde_i$ edges that are contained in subgraphs in $\mathbb{X}$.   The construction of $H_i$ is exactly the same as the construction of $H_i$ in \Cref{subsec:ConstructHiT2}: first, add every edge of $\me_i^{\take}$ to $H_i$, and then apply \hyperlink{SPSSO}{$\sso$} on the subgraph of $\mg_i$ induced by $\me_i^{\reduce}$. Furthermore, \Cref{clm:Ki-clustergraph} and \Cref{obs:supportingPropHiT2} hold here.  

 Recall that $F^\sigma_{i}$ is the set of edges in $E^{\sigma}_i$ that correspond to $\mathcal{E}_i$. By the same proof in \Cref{lm:Hi-StretchT2} we have:

\begin{lemma}\label{lm:Hi-StretchT1E} For every edge $(u,v) \in F^\sigma_{i}$, $d_{H_{< L_i}}(u,v) \leq t(1+ \max\{s_{\sso}(2g),6g\}\eps)w(u,v)$.
\end{lemma}
Next, we bound the total weight of $H_i$.  

\begin{lemma}\label{lm:Hi-WeightTE}  $w(H_i) \leq \lambda \Delta_{i+1} + a_i$ for $\lambda = O(\chi \eps^{-1}  + \eps^{-2})$  and $a_i =   O(\chi \eps^{-1} +\eps^{-2})w(\msttilde^{in}_i) + O(L_i/\eps^2)$.  
\end{lemma}
\begin{proof}  First, we consider the non-degenerate case. Note that edges in $\me^{\redunt}_i$ are not added to $H_i$. Let $\mv_i^{+} = \cup_{\mx \in \mathbb{X}^{+}}\mv(\mx)$ and  $\mv_i^{-} = \cup_{\mx \in \mathbb{X}^{-}}\mv(\mx)$. 
	Let $F^{(a)}_i$ be the set of edges added to $H_i$ in the construction in Step $a$, $a\in \{1,2\}$. 
	
By the construction in Step 1, $F^{(1)}_i$ includes edges in $\me_i^{\take}$. 	Let $A^{(1)}\subseteq F^{(1)}_i$  be the set of edges incident to at least one node in $\mv_i^{+}$ and $A^{(2)}=F^{(1)}_i\setminus A^{(1)}$.  By Item (1) in \Cref{lm:ClusteringE}, the total weight of the edges added to $H_i$ in Step 1 is:
	\begin{equation}\label{eq:A1TE}
		\begin{split}
			w(A^{(1)}_i)  &=  \sum_{\mx \in \mathbb{X}^{+}} O(|\mv(\mx)|/\eps) L_i \stackrel{\mbox{\tiny{\cref{eq:averagePotential-t1E}}}}{=}  O(\frac{1}{\eps^2})\sum_{\mx \in \mathbb{X}^{+}} \Delta^+_{i+1}(\mx) = O(\frac{1}{\eps^2})(\Delta_{i+1} + w(\msttilde^{in}_i))~.
		\end{split}
	\end{equation}   
By definition 	$A^{(2)}$ is the set of edges with both endpoints in subgraphs of $\mv_i^{-}$. Consider the orientation of $\me^{\take}_i$ as in \Cref{lm:ClusteringE}. Then, every edge of $A^{(2)}$ is an out-going edge from some node in a  graph in $\mathbb{X}^{-}$. For each graph $\mx \in \mathbb{X}^{-}$, by Item (4) of \Cref{lm:ClusteringE}, the total weight of incoming edges of $\mx$ is $O(t L_i) = O(1/\eps^2)\Delta^{+}_{i+1}(\mx)$. Thus,  we have:
\begin{equation}\label{eq:A2TE}
	\begin{split}
		w(A^{(2)}_i)  &= O(\frac{1}{\eps^2})\sum_{\mx \in\mathbb{X}} \Delta^+_{i+1}(\mx)  = O(\frac{1}{\eps^2})(\Delta_{i+1} + w(\msttilde^{in}_i))~.
	\end{split}
\end{equation}  
 Thus, by \Cref{eq:A1TE,eq:A2TE}, we have $w(F^{(1)}) = O(\frac{1}{\eps^2})(\Delta_{i+1} + w(\msttilde^{in}_i))$. By the exactly the same argument in \Cref{lm:Hi-WeightTE}, we have that $w(F^{(2)}) =O(\chi/\eps)(\Delta_{i+1} + w(\msttilde^{in}_i)) $. This gives: 
\begin{equation}\label{eq:Hi-nondegenT1E}
		\begin{split}
			w(H_i) &=  O(\chi/\eps + 1/\eps^2) (\Delta_{i+1} + w(\msttilde^{in}_i)) \leq \lambda(\Delta_{i+1} + w(\msttilde^{in}_i))
		\end{split}
	\end{equation} 
	for some $\lambda =  O(\chi/\eps + 1/\eps^2) $.

	It remains to consider the degenerate case, and in which case, we only add to $H_i$ edges corresponding to $\me^{\take}_i$. Thus, by Item (4) of \Cref{lm:ClusteringE}, we have:
	\begin{equation}\label{eq:Hi-degenT1E}
		w(H_i) = O(\frac{L_i}{\eps^2}) \leq  \lambda\cdot (\Delta_{i+1} + w(\msttilde^{in}_i)) + O(\frac{L_i}{\eps^2}), 
	\end{equation} 
	since $\Delta_{i+1} + w(\msttilde^{in}_i) = \sum_{\mx\in \mathbb{X}}\Delta^+_{i+1}(\mx)$ by Item (1) in \Cref{obs:supportingPropHiT2}. Thus, the lemma follows from \Cref{eq:Hi-degenT1E,eq:Hi-nondegenT1E}. \qed
\end{proof}

We are now ready to prove \Cref{lm:ConstructClusterHi} for the case $t = 1+\eps$.

\begin{proof}[Proof of Item 2 of \Cref{lm:ConstructClusterHi}] The fact that subgraphs in $\mathbb{X}$ satisfy the three properties (\hyperlink{P1'}{P1'})-(\hyperlink{P3'}{P3'}) with constant $g=31$ follows from Item (5) of \Cref{lm:ClusteringE}.  The stretch in $H_{< L_i}$ of edges in $F^{\sigma}_{i}$ follows from \Cref{lm:Hi-StretchT1E}.

	By  \Cref{lm:Hi-WeightTE}, $w(H_i) \leq \lambda \Delta_{i+1} + a_i$ where  $\lambda = O(\chi \eps^{-1} +\eps^{-2})$  and $a_i =   O(\chi \eps^{-1} + \eps^{-2})w(\msttilde^{in}_i) + O(L_i/\eps^2)$. It remains to show that $A = \sum_{i\in \mathbb{N}^+}a_i = O(\chi \eps^{-1} + \eps^{-2})$.  Observe that
	\begin{equation*}
		\sum_{i\in \mathbb{N}^+}O(\frac{L_i}{\eps^2}) ~=~  O(\frac{1}{\epsilon^2}) \sum_{i=1}^{i_{\max}} \frac{L_{i_{\max}}}{\epsilon^{i_{\max}-i}} ~=~ O(\frac{L_{i_{\max}}}{\epsilon^2(1-\epsilon)}) ~=~ O(\frac{1}{\epsilon^2}) w(\mst)~;
	\end{equation*}
	here $i_{\max}$ is the maximum level. The last equation is due to that $\eps \leq 1/2$  and every edge has weight at most $w(\mst)$ since the weight of every is the shortest distance between its endpoints. By Item (2) of \Cref{obs:supportingPropHiT2},  $\sum_{i\in \mathbb{N}^+} \msttilde^{in}_i\leq w(\mst)$.  Thus, $A = O(\chi/\eps^2) + O(1/\eps^2)$ as desired.   \qed
\end{proof}

\subsection{Clustering} \label{subsec:clusteringTE}

In this section, we prove \Cref{lm:ClusteringE}. The construction of $\mathbb{X}$ has 5 steps. The first four steps are exactly the same as the first four steps in the construction in \Cref{sec:stretch2}. In Step 5, we construct $\mathbb{X}^{\internal}_5$ differently, taking into account of edges in $\bar{\me}_i^{close}$ in \Cref{eq:Ebar-farclose}. Recall that when the stretch parameter $t\geq 2$, we show that edges in $\me_i$ corresponding to $\bar{\me}_i^{close}$ are added to $\me_i^{\redunt}$ (implicitly in \Cref{lm:Item4-Nonde}). However, when $t = 1+\eps$, we could not afford to do so, and the construction in Step 5 will take care of these edges.

\paragraph{Steps 1-4.~} The construction of Steps 1 to 4 are exactly the same as Steps 1-4 in \Cref{subsec:clusteringT2} to obtain three sets of clusters $\mathbb{X}_1,\mathbb{X}_2$ and $\mathbb{X}_4$ whose properties are described in \Cref{lm:Clustering2-Step1T2,lm:Clustering2-Step2T2,lm:Clustering2-Step3,lm:Clustering2-Step4}. After the four steps, we obtain the forest $\Fbar^{(5)}_i$, where every tree is a path. In particular, for every edge $(\bmu,\bnu)\in \bar{\me}_i$ with both endpoints in $\Fbar^{(5)}_i$, either (i) the edge is in $ \bar{\me}^{close}_i(\Fbar^{(5)}_i)$, or (ii) at least one of the endpoints must belong to a low-diameter tree of $\Fbar^{(5)}_i$ or (iii) in a (red) suffix of a long path in $\Fbar^{(5)}_i$  of augmented diameter at most $L_i$.

Before moving on to Step 5, we need a preprocessing step in which we find all edges in $\me^{\redunt}_i$. The construction of Step 5 relies on edges that are not in  $\me^{\redunt}_i$. 

\paragraph{Constructing $\me_i^{\redunt}$ and $\me_i^{\take-}$.~} Let $\Ftilde^{(5)}_i$ be obtained from $\Fbar^{(5)}_i$ by uncontracting the contracted nodes.  We apply the greedy algorithm. Initially, both $\me_i^{\redunt}$ and $\me_i^{\take-}$ are empty sets.  We construct a graph $\mh_i = (\mv_i, \msttilde_{i}\cup \me_i^{\take-}, \omega)$, which initially only include edges in $\msttilde_{i}$. We then consider every edge $\mbe = (\nu\cup \mu) \in \me_i$, where both endpoints are in $\mv(\Ftilde^{(5)}_i)$, in the non-decreasing order of the weight. If:
\begin{equation}\label{eq:greedy-HiE}
	d_{\mh_i }(\nu,\mu) \leq (1+6g\eps) \omega(\mbe)~,
\end{equation}
then we add $\mbe$ to $\me_i^{\redunt}$. Otherwise, we add $\mbe$ to $\me_i^{\take-}$ (and hence to $\mh_i$).  Note that the distance in $\mh_i$ in \Cref{eq:greedy-HiE} is the augmented distance.  We have the following observation which follows directly from the greedy algorithm.

\begin{observation}\label{obs:greedy-HiE} For every edge $\mbe = (\nu,\mu)  \in \me_i^{\take-}$, $d_{\mh_i }(\nu,\mu) \geq (1+6g\eps) \omega(\mbe)$.
\end{observation}

\paragraph{Step 5.~}  
Let $\Pbar$ be  a path in  $\Fbar^{(5)}_i$ obtained by Item (5) of \Cref{lm:Clustering2-Step4}. We construct two sets of subgraphs, denoted by $\mathbb{X}^{\internal}_5$ and $\mathbb{X}^{\prefix}_5$, of $\mg_i$. The construction is broken into two steps. Step 5A is only applicable when $\mathbb{X}_1 \cup \mathbb{X}_2\cup \mathbb{X}_4 \not= \emptyset$. In Step 5B, we need a more involved construction by ~\cite{LS19}, as described in \Cref{lm:Clustering2-Step5B1E}. 

\begin{itemize}
	\item (Step 5A)\hypertarget{5AT1E}{}  If $\Pbar$ has augmented diameter at most $6L_i$, let $\mbe$ be an $\widetilde{\mst}_i$ edge connecting $\Ptilde^{\uncontract}$  and a node in some subgraph $\mx \in \mathbb{X}_1\cup \mathbb{X}_2 \cup \mathbb{X}_4$; $\mbe$ exists by \Cref{lm:Clustering2-Step4}. We add both $\mbe$ and $\Ptilde^{\uncontract}$ to $\mx$.
	\item (Step 5B)\hypertarget{5BT1E}{} 	Otherwise,  the augmented diameter of $\Pbar$ is at least $6L_i$.  Let $\{\Qbar_1, \Qbar_2\}$ be the suffix and prefix of $\Pbar$ such that $\Qbar^{\uncontract}_1$ and $\Qbar^{\uncontract}_2$ have augmented diameter at least $L_i$ and at most $2L_i$. If $\Qbar_j$, $j\in \{1,2\}$ is connected to a node in a subgraph $\mx \in \mathbb{X}_1 \cup \mathbb{X}_2\cup \mathbb{X}_4$ via an  edge $e\in \msttilde_{i}$, we add $\tilde{Q}^{\uncontract}_j$ and $e$ to $\mx$. 	If $\Qbar_j$ contains an endpoint of $\Pbar$, we add $\Qtilde_j^{\uncontract}$ to $\mathbb{X}^{\prefix}_5$. 
	
	Next, denote by $\Pbar^{'}$ the path obtained by removing $\Qbar_1, \Qbar_2$  from $\Pbar$. We then apply the construction in \Cref{lm:Clustering2-Step5B1E} to $\Pbar^{'}$ to obtain a set of subgraphs $\mathbb{C}_5(\Pbar^{'})$ and an orientation of edges in  $\me^{\take-}_i(\Pbar')$, the set edges of $\me^{\take-}_i$	with both endpoints in  the uncontraction of $\Pbar^{'}$. We add all edges of  $\me^{\take-}_i(\Pbar')$ to a set $\me^{(5B)}_{i}$ (which is initially empty).  We then add all subgraphs in $\mathbb{C}_5(\Pbar^{'})$ to  $\mathbb{X}^{\internal}_5$.
\end{itemize}

\begin{figure}[!htb]
	\center{\includegraphics[width=0.9\textwidth]{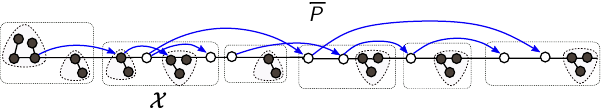}}
	\caption{A path $\Pbar$, a cluster $\mx$, and a set of (blue) edges in $\me^{\take-}_i(\Pbar)$.
		 White nodes are uncontracted nodes and black nodes are those in contracted nodes (triangular shapes). $\Delta^+_{i+1}(\mx) $ is proportional to the number of out-going edges from nodes in $\mx$, which is 3 in this case; there could be edges with both endpoints in $\mx$.}
	\label{fig:5B}
\end{figure}

The construction of Step 5B is described in the following lemma, which is a slight adaption of Lemma 6.17 in~\cite{LS19}. See \Cref{fig:5B} for an illustration. The construction crucially exploit the fact that  	$d_{\mh_i }(\nu,\mu) \leq (1+6g\eps) \omega(\mbe)$.  

\begin{lemma}[Step 5B, Lemma 6.17 in \cite{LS19}]\label{lm:Clustering2-Step5B1E} Let $\Pbar$ be a path in $\Fbar^{(5)}_i$.  
	Let  $\me^{\take-}_i(\Pbar)$ be the edges of $\me^{\take-}_i$
	with both endpoints in $\Ptilde^{\uncontract}$. We can construct a set of subgraphs $\mathbb{C}_5(\Pbar)$ such that: 	
	\begin{enumerate}
		\item[(1)]  Subgraphs in $\mathbb{C}_5(\Pbar)$ contain every node in $\Ptilde^{\uncontract}$.
		\item[(2)] For every subgraph $\mx\in \mathbb{C}_4(\Pbar)$, $\zeta L_i \leq \adm(\mx)\leq 5L_i$. Furthermore, $\mx$ is a subtree of $\Ptilde^{\uncontract}$ and some edges in  $\me^{\take-}_i(\Pbar)$  whose both endpoints are in $\mx$.
		\item[(3)]  There is an orientation of edges in $\me^{\take-}_i(\Pbar)$  
		such that, for any subgraph $\mx \in  \mathbb{C}_5(\Pbar)$,  if the total number of out-going edges incident to nodes in $\mx$ is $t$ for any $t\geq 0$, then:
		\begin{equation}\label{eq:Step5Bpotential}
			\Delta^+_{i+1}(\mx) =  \Omega(t\epsilon^2) L_i
		\end{equation}
	\end{enumerate}
\end{lemma}

We observe the following from the construction.
\begin{observation}\label{obs:MEiTakeMinus} For every edge $\mbe \in \mathcal{E}_i^{\take-}$, either at least one endpoint of $\mbe$ is in a subgraph in $\mathbb{X}_5^{\prefix}$, or both endpoints of $\mbe$ are in $\me_i^{(5B)}$.
\end{observation}

The following lemma is analogous to \Cref{lm:Clustering2-Step5}.

\begin{lemma}\label{lm:Clustering2-Step51E}  Every subgraph $\mx \in \mathbb{X}_5^{\internal} \cup \mathbb{X}_5^{\prefix}$ satisfies:
	\begin{enumerate}[noitemsep]
		\item[(1)] $\mx$ is a subtree of $\msttilde_{i}$ if $\mx\in \mathbb{X}_5^{\prefix}$.
		\item[(2)] $\zeta L_i \leq \adm(\mx)\leq 20 L_i$.
		\item[(3)] $|\mv(\mx)| = \Omega(1/\eps)$.
	\end{enumerate}
	Furthermore, if $\mx \in \mathbb{X}_5^{\prefix}$, then $\mx$ the uncontraction of a prefix/suffix subpath $\Qbar$ of a long path $\Pbar$, and additionally, the (uncontraction of) other suffix $\Qbar'$ of   $\Pbar$ is augmented to a subgraph in $\mathbb{X}_1 \cup \mathbb{X}_2\cup \mathbb{X}_4$, unless $\mathbb{X}_1 \cup \mathbb{X}_2\cup \mathbb{X}_4 = \emptyset$.
\end{lemma}

In the next section, we prove \Cref{lm:ClusteringE}.

\subsubsection{Constructing $\mathbb{X}$ and the partition of $\me_i$:  Proof of \Cref{lm:ClusteringE}}\label{subsec:X-T1E}
  We distinguish two cases:
 
 \paragraph{Degenerate Case.~} The degenerate case is the case where   $\mathbb{X}_1\cup \mathbb{X}_2\cup \mathbb{X}_4 =  \emptyset$. In this case, we set $\mathbb{X} = \mathbb{X}^{-} =  \mathbb{X}_5^{\internal} \cup \mathbb{X}_5^{\prefix}$, and $	\mathbb{X}^{+} = 	 \emptyset$. 
 
 \paragraph{Non-degenerate case.~} We define:
 \begin{equation}\label{eq:MathbbX1E}
 	\begin{split}
 		\mathbb{X}^{+} &=    \mathbb{X}_1\cup \mathbb{X}_2\cup \mathbb{X}_4 \cup \mathbb{X}_5^{\prefix}, \quad
 		\mathbb{X}^{-} = \mathbb{X}_5^{\internal}\\
 		\mathbb{X} &= 	\mathbb{X}^{+}\cup \mathbb{X}^{-} 
 	\end{split}
 \end{equation}

 Next, we construct the partition of $\{\me^{\take}_i, \me^{\redunt}_i, \me_i^{\reduce}\}$ of $\me_i$. Recall that we constructed two edge sets $\me_i^{\redunt}$  and $\me^{\take-}_i$ above (\Cref{eq:greedy-HiE}). We then construct $\me_i^{\take}$ as described below. It follows that $\me^{\reduce}_i = \me_i\setminus (\me_i^{\take}\cup \me_i^{\redunt})$.
 
 \begin{tcolorbox}
 	\hypertarget{EiPartition1E}{}
 	\textbf{Constructing $\me_i^{\take}$:} Let $\mv_i^{+} = \cup_{\mx\in \mathbb{X}^{+}}\mv(\mx)$ and $\mv_i^{-} = \cup_{\mx\in \mathbb{X}^{-}}\mv(\mx)$. 	First, we add all edges in  $\me^{\take-}_i$ to $\me^{\take}_i$. Next, we add $ (\cup_{\mx\in \mathbb{X}}\me(\mx)\cap \me_i)$ to $\me_i^{\take}$. Finally, for every edge $\mbe \in \me_i\setminus \me_i^{\redunt}$ such that $\mbe$ is incident to at least one node in $\mv_i^{-}$, we add $\mbe$ to   $\me_i^{\take}$.
 \end{tcolorbox}

In the analysis below, we only explicitly  distinguish the degenerate case from the non-degenerate case when it is necessary, i.e, in the proof Item (4) of \Cref{lm:ClusteringE}. Otherwise, which case we are in is either implicit from the context, or does not matter.

We observe that Item (2) in \Cref{lm:ClusteringE} follows directly from the construction of $\me_i^{\redunt}$. Henceforth, we focus on proving other items of \Cref{lm:ClusteringE}.  We first show Item (5).

\begin{lemma}\label{lm:XProp1E} Let $\mathbb{X}$ be the subgraph as defined in \Cref{eq:MathbbX1E}. For every subgraph $\mx \in \mathbb{X}$, $\mx$ satisfies the three properties (\hyperlink{P1'}{P1'})-(\hyperlink{P3'}{P3'}) with $g = 31$. Consequently, Item (5) of \Cref{lm:ClusteringE} holds.
\end{lemma}
\begin{proof} 	We observe that property \hyperlink{P1'}{(P1')} follows directly from the construction.  Property \hyperlink{P2'}{(P2')} follows directly from \Cref{lm:Clustering2-Step1T2,lm:Clustering2-Step2T2,lm:Clustering2-Step4,lm:Clustering2-Step51E}. We now bound  $\adm(\mx)$. The lower bound on $\adm(\mx)$ follows directly from Item (3) of \Cref{lm:Clustering2-Step1T2}, Items (2) of \Cref{lm:Clustering2-Step2T2,lm:Clustering2-Step4,lm:Clustering2-Step51E}. For the upper bound, by the same argument in \Cref{lm:XProp}, if $\mx$ is initially formed in Steps 1-4, then $\adm(\mx)\leq 31L_i$. Otherwise, by \Cref{lm:Clustering2-Step51E}, $\adm(\mx) \leq 5L_i$, which implies  property \hyperlink{P3'}{(P3')} with $g= 31$. \qed 
\end{proof}

 We observe that \Cref{lm:manynodes} and  \Cref{lm:Item3Clustering} holds for $\mathbb{X}^{+}$, which we restate below in \Cref{lm:manynodes1E} and \Cref{lm:Item3Clustering1EHigh}, respectively. In particular,  \Cref{lm:Item3Clustering1EHigh} implies Item (3) of \Cref{lm:ClusteringE}.

\begin{lemma}\label{lm:manynodes1E} For any subgraph $\mx \in \mathbb{X}$ such that $|\mv(\mx)|\geq \frac{2g}{\zeta\eps}$ or $\Delta^+_{i+1}(\mx) = \Omega(L_i)$, then $\Delta^+_{i+1}(\mx) = \Omega(\eps L_i |\mv(\mx)|)$.
\end{lemma}

\begin{lemma}\label{lm:Item3Clustering1EHigh}   $\Delta_{i+1}^+(\mx) \geq 0$ for every $\mx \in \mathbb{X}$ and 
	\begin{equation*}
		\sum_{\mx \in \mathbb{X}^{+}} \Delta_{i+1}^+(\mx) = \sum_{\mx \in\mathbb{X}^{+}} \Omega(|\mv(\mx)|\eps L_i). 
	\end{equation*}
\end{lemma}

We now prove Item (1) of \Cref{lm:ClusteringE}, which we restate here for convenience.

\begin{lemma}\label{lm:Item1Clustering1E} For every subgraph $\mx \in \mathbb{X}$,  $\deg_{\mg^{\take}_i}(\mv(\mx)) = O(|\mv(\mx)|/\eps)$  where $\mg^{\take}_i = (\mv_i,\me_i^{\take})$, and $\me(\mx)\cap \me_i \subseteq \me^{\take}$. Furthermore, if $\mx \in \mathbb{X}^{-}$, there is no edge in $\me_i^{\reduce}$ incident to a node in $\mx$.
\end{lemma}
\begin{proof} 	Let $\mv^{\highp}_i = \cup_{\mx\in \mathbb{X}_1}\mx$. Note by the construction in Step 1 (\Cref{lm:Clustering2-Step1T2}), nodes in $\mv_i\setminus \mv^{\highp}_i$ have degree $O(\frac{1}{\eps})$. Let $\me_i^{(1)}$  be the set of edges in $\me_i^{\take}$ with both endpoints in $\mv^{\highp}_i$ and $\me_i^{2} =\me_i^{\take}\setminus \me_i^{(1)}$. Also by the construction in Step 1 (\Cref{lm:Clustering2-Step1T2}), both endpoints of every edge in $\me_i^{2}$ have degree $O(1/\eps)$. Thus, for any $\mx\in \mathbb{X}$, the number of edges in $\me_i^{(2)}$ incident to nodes in $\mx$ is $O(|\mv(\mx)|/\eps)$. 
	
	Next, we consider  $\me_i^{(1)}$. Observe by the construction of $\me_i^{\take}$ that there is no edge in  $\me_i^{(1)}$  with two endpoints in two \emph{different graphs} of $\mathbb{X}_1$. Furthermore, since $\mx$ is a tree for every subgraph $\mx\in \mathbb{X}_1$, the number of edges in  $\me_i^{(1)}$   incident to nodes in $\mx$ is $O(|\mv(\mx)|)$. This bound also holds for every subgraph $\mx$ not in $\mathbb{X}_1$ since the number of incident edges in $\me_i^{(1)}$ is 0; this implies the claimed bound on $\deg_{\mg^{\take}_i}(\mv(\mx))$.
	
	For the last claim, we observe that nodes in subgraphs of  $\mathbb{X}^{-}$ are in $\mv_i^{-}$. Thus, by the construction of $\me_i^{\take}$, every edge incident to a node of  $\mx \in \mathbb{X}^{-}$ is either in $\me_i^{\take}$ or $\me_i^{\redunt}$.\qed
\end{proof}

We now focus on proving Item (4) of \Cref{lm:ClusteringE} which we restate below.

\begin{lemma}\label{lm:Item4Clustering1E}  There exists an orientation of edges in $\me_i^{\take}$ such that for every subgraph $\mx \in \mathbb{X}^{-}$, if $\mx$ has $t$ out-going edges for some $t\geq 0$, then $\Delta^+_{i+1}(\mx) =\Omega(|\mv(\mx)|t\eps^2 L_i)$, unless a \emph{degenerate case} happens, in which  $\me^{\reduce}_i = \emptyset$ and  $$\omega(\me_i^{\take}) = O(\frac{1}{\eps^2})(\sum_{\mx \in \mathbb{X}} \Delta_{i+1}^+(\mx) + L_i).$$
\end{lemma}
\begin{proof} First, we consider the non-degenerate case. Recall that $\{\mv^+_i,\mv^-_i\}$ is a partition of $\mv_i$ in the \hyperlink{EiPartition1E}{construction of $\me_i^{\take}$}. We orient edges of $\me_i^{\take}$ as follows.
	
	First, for any $\mbe = (\mu,\nu)\in \me_i^{\take}$ such that at least one endpoint, say $\mu \in \mv_i^+$, we orient $\mbe$ as out-going from $\mu$. (If both $\mu,\nu$ are in $\mv_i^+$, we orient $\mbe$ arbitrarily).  Remaining  edges are subsets  of $\me_i^{(5B)}$ by \Cref{obs:MEiTakeMinus}. We orient edges in  $\me_i^{(5B)}$ as in the construction of Step 5B.	For every subgraph  $\mx \in \mathbb{X}^{-}$, by construction, out-going edges incident to nodes in $\mx$ are in $\me_i^{(5B)}$. By Item (3) of \Cref{lm:Clustering2-Step5B1E}, $\Delta^+_{i+1}(\mx) =\Omega(|\mv(\mx)|t\eps^2 L_i)$.

	It remains to consider the degenerate case. In this case,  by the same argument in \Cref{lm:degenerate}, 	$\Fbar^{(5)}_i = \Fbar^{(4)}_i = \Fbar^{(3)}_i$, and $\Fbar^{(5)}_i$  is a single (long) path. Furthermore, $\me_i^{\take} = \me_i^{\take-}$, and $|\mathbb{X}_5^{\prefix}| = 2$.  We orient edges in  $\me_i^{(5B)}$ as in the construction of Step 5B, and other edges of $\me_i^{\take}$, which must be incident to nodes in subgraphs of $\mathbb{X}_5^{\prefix}$, are oriented as out-going from subgraphs in  $\mathbb{X}_5^{\prefix}$. 	By Item (3) of \Cref{lm:ClusteringE}, for any subgraph $\mx \in \mathbb{X}^{\internal}_5$ that has $t$ out-going edges, the total weight of the out-going edges is at  most $tL_i = O(1/\eps) \Delta^+_{i+1}(\mx)$. Thus, $\omega(\me_i^{(5B)}) =  O(1/\eps) \sum_{\mx \in \mathbb{X}^{\internal}_5}\Delta^+_{i+1}(\mx) = O(1/\eps^2)\sum_{\mx \in \mathbb{X}}\Delta^+_{i+1}(\mx) $.
	
	It remains to consider edges incident to at leas one node in a subgraph in $\mathbb{X}_5^{\prefix}$ by \Cref{obs:MEiTakeMinus}. Let $\mx \in \mathbb{X}_5^{\prefix}$. Observe that if $|\mv(\mx)|\geq \frac{2g}{\zeta\eps}$, then
	\begin{equation*}
		\begin{split}
			\Delta_{i+1}^+(\mx) &= O(|\mv(\mx)|\eps L_i)  \qquad \mbox{(by \Cref{lm:manynodes1E})}\\
			& =  O(|\deg_{\mg_i^{\take}}(\mx)|\eps^2 L_i)   \qquad \mbox{(by Item (1) of \Cref{lm:ClusteringE})}
		\end{split}
	\end{equation*}
	Otherwise, $|\mv(\mx)| \leq \frac{2g}{\zeta\eps}$, and hence $|\deg_{\mg_i^{\take}}(\mx)| = O(1/\eps^2)$ by Item (1) of \Cref{lm:ClusteringE}. This implies that the total weight of edges incident to $\mx$ is at most $|\deg_{\mg_i^{\take}}(\mx)| L_i = O(1/\eps^2) (\Delta_{i+1}^+(\mx) + L_i)$. Since $|\mathbb{X}_5^{\prefix}| = 2$ and $\Delta^+_{i+1}(\mx)\geq 0$ for every $\mx \in \mathbb{X}$ by Item (3) of \Cref{lm:ClusteringE}, we have that the total weight of edges ncident to at leas one node in a subgraph in $\mathbb{X}_5^{\prefix}$ is $O(\frac{1}{\eps^2})(\sum_{\mx \in \mathbb{X}} \Delta_{i+1}^+(\mx) + L_i)$. The lemma now follows.  \qed
\end{proof}

\paragraph{Acknowledgement.~}  Hung Le is supported by a start up funding of University of Massachusetts at Amherst  and by the National Science Foundation under Grants No. CCF-2121952 and No. CCF-2237288. Shay Solomon is partially supported by the Israel Science Foundation grant No.1991/19. We thank Oded Goldreich for his suggestions concerning the presentation of this work and we thank Lazar Milenković for his support.
\bibliographystyle{plain}
\bibliography{spanner}
\appendix
\end{document}